%% file: manuscript.tex
\documentclass[12pt]{article}

\usepackage{etoolbox}
\usepackage[utf8]{inputenc}
\usepackage{geometry}
\usepackage{setspace}
\usepackage{ragged2e} 
\usepackage{amsmath}
\usepackage{amsthm}
\usepackage{amssymb}
\usepackage{amsfonts}
\usepackage{listings}
\usepackage{tcolorbox} 
\usepackage{soul} 
\usepackage{placeins} 
\usepackage{booktabs}
\usepackage[bf]{caption}
\usepackage{subcaption}
\usepackage{multirow} 
\usepackage{makecell}
\usepackage{datetime}
\usepackage{rotating}   
\usepackage{siunitx}    
\usepackage{graphicx}
\usepackage{float}
\usepackage{adjustbox}
\usepackage{tabularx}
\usepackage{pdflscape}
\usepackage{enumitem}
\usepackage[bottom]{footmisc}

\usepackage{epstopdf}
\usepackage{svg}

\usepackage{xparse}    
\usepackage{array}     
\usepackage{xstring}   

\usepackage{filecontents,catchfile}
\usepackage{dsfont}

\usepackage{titlesec}
\titlespacing*{\paragraph}
{0pt}          
{2.5ex plus 1ex minus .2ex} 
{1em}          

\usepackage{tikz}
\usetikzlibrary{calc,shapes.geometric,intersections}

\definecolor{DemConsColor}{HTML}{43AA8B} 
\definecolor{SupConsColor}{HTML}{F8C471}
\definecolor{UnconsColor}{HTML}{3A86FF} 

\definecolor{CoolRed}{HTML}{CC3333}
\definecolor{CoolMix}{HTML}{865C74}
\definecolor{CoolBlue}{HTML}{4682B4}

\definecolor{ColorPortugal}{HTML}{000080}
\definecolor{ColorNorway}{HTML}{008000}
\definecolor{ColorColombia}{HTML}{808080}

\input{autofill_lms}

\input{autofill_var_constraints}
\input{autofill_welfare}

\usepackage[round,authoryear,comma]{natbib}  
\usepackage{xcolor}  
\usepackage[backref=page]{hyperref}

\usepackage{bibunits}

\makeatletter
\patchcmd{\@startbibunit}
  {\@input{\@bibunitname.aux}}
  {\makeatletter\@input{\@bibunitname.aux}}
  {}
  {\PackageError{bibunits-fix}{Cannot patch startbibunit}{}}

\patchcmd{\@finishstartedbibunit}
  {\@input{\@bibunitname.aux}}
  {\makeatletter\@input{\@bibunitname.aux}}
  {}
  {\PackageError{bibunits-fix}{Cannot patch finishstartedbibunit}{}}
\makeatother

\renewcommand*{\backref}[1]{}
\renewcommand*{\backrefalt}[4]{%
  \begingroup
  \let\backrefsep\empty
  \let\backreftwosep\empty
  \let\backreflastsep\empty
  \def\MainBackrefSep{\footnotesize(Cited on p.~}%
  \let\MainBackrefEnd\empty
  \let\OriginalBackref\backrefxxx
  \def\backrefxxx##1##2##3{%
    \IfBeginWith{##1}{[Appendix-}{}{%
      \MainBackrefSep
      \OriginalBackref{##1}{##2}{##3}%
      \def\MainBackrefSep{, }%
      \def\MainBackrefEnd{)}%
    }%
  }%
  #2\MainBackrefEnd
  \endgroup
}

\setcitestyle{authoryear,open={(},close={)},aysep={}}

\usepackage{stackengine,scalerel} 

\newdateformat{monthyeardate}{\monthname[\THEMONTH] \THEYEAR}
\hypersetup{
    colorlinks=true,
    linkcolor={red},    
    citecolor={blue},   
    urlcolor={blue}     
}

\makeatletter
\patchcmd{\@maketitle}{\LARGE}{\Large}{}{}
\makeatother

\newcommand{\W}{\underline{W}}
\newcommand{\upperPhi}{\overline{\Phi}}
\newcommand{\lowerPhi}{\underline{\Phi}}
\newcommand{\J}{\mathcal{J}}

\newcommand{\aggW}{\mathbb{W}}

\newcommand{\VA}{V\!A}

\newcommand{\CPset}{\mathcal{P}}  

\DeclareMathOperator*{\E}{\mathbb{E}}

\DeclareMathOperator*{\pre}{\text{pre}}
\DeclareMathOperator*{\post}{\text{post}}

\newtheorem{definition}{Definition}

\newtheorem{lemma}{Lemma}
\newtheorem{corollary}{Corollary}

\newtheorem{assumption}{Assumption}

\newlist{assumpinline}{enumerate*}{1}
\setlist[assumpinline]{
    label=\textbf{(\alph*)},   
    ref=(\alph*)               
}

\newcommand{\assumpfullref}[2]{\ref{#1}\ref{#1:#2}}

\newenvironment{assumptionprime}[2][]{
    \renewcommand{\theassumption}{$\neg$\ref*{#2}}
    \addtocounter{assumption}{-1}
    \begin{assumption}[#1]
}{
    \end{assumption}
}

\newtheorem{proposition}{Proposition}

\newlist{proplist}{enumerate*}{1}
\setlist[proplist]{
    label=\textbf{(\textit{\roman*})},   
    ref=(\textit{\roman*})               
}

\newcommand{\propfullref}[2]{\ref{#1}\ref{#1:#2}}

\newtheorem{result}{Result}

\newlist{reslist}{enumerate*}{1}
\setlist[reslist]{
    label=\textbf{(\alph*)},   
    ref=(\alph*)               
}

\title{\Large
    \textbf{Wage-Setting Constraints and Firm Responses \\
            to Demand Shocks}%
    \thanks{
        \textit{Acknowledgments}:
        This project has received funding from the European Union’s Horizon Europe Research and Innovation Programme under Grant Agreement No. 101043127. 
        Santiago Hermo has also received funding from the Norwegian Research Council under Grant No. 335380.
        We thank David Card, Mons Chan, Simon Jäger, Attila Lindner, Nina Roussille, Raffaele Saggio, Bradley Setzler and Micole De Vera, as well as seminar and workshop participants at the RFBerlin-CEPR Annual Symposium in Labour Economics 2026, the 2025 European Winter Meeting of the Econometric Society, the Princeton University, the University of Glasgow, the University of Warwick, the Stockholm School of Economics, the University of Stavanger, the Norwegian Institute for Social Research, the 2025 Nordic International Trade Seminars, the ifo Institute, the Academia Sinica, the Universidad Los Andes Chile, the BI Norwegian Business School, the 2026 International Symposium on Applied Economics at National Chengchi University, the Workshop on Industrial Relations at Collegio Carlo Alberto, the 2026 Labor Chair Workshop at Paris School of Economics, the 2026 Workshop on Public Policy Design at the University of Girona, the 2026 Nordic Symposium in Macroeconomics, the 1st Women in International Trade European Conference, the 4th Annual Oslo Labor Workshop, the Monash Brownbag Seminar, the University of Oslo Trade Coffee, the Oslo Applied Lunch Seminar, and the CUFE Forum in Beijing for helpful comments. 
        This paper won the Best Paper Award at the CUFE Forum in Beijing 2025.
    }

    \vspace{2mm}
}

\author{
  \begin{tabular}{cccc}
    Manudeep Bhuller\thanks{University of Oslo, CEPR, CESifo, IZA, RFBerlin. E-mail: \href{mailto:manudeep.bhuller@econ.uio.no}{manudeep.bhuller@econ.uio.no}.} & &
    &
    Lukas Delgado-Prieto\thanks{University of Oslo, RFBerlin. E-mail: \href{mailto:laprieto@econ.uio.no}{laprieto@econ.uio.no}.} \\
    \\
    Santiago Hermo\thanks{Monash University, RFBerlin. E-mail: \href{mailto:santiago.hermo@monash.edu}{santiago.hermo@monash.edu}.} & &
    &
    Linnea Lorentzen\thanks{University of Oslo, CESifo, RFBerlin. E-mail: \href{mailto:linnea.lorentzen@econ.uio.no}{linnea.lorentzen@econ.uio.no}.}
  \end{tabular}
}
\date{}

\begin{document}

\maketitle

\begin{center}
    \vspace{-1cm}
    First Version: December 2025 \\
    This Version: November 2026
\end{center}

\begin{abstract} 
    This paper investigates how institutional wage-setting constraints, such as a national minimum wage or collectively bargained wages, affect firm responses to demand shocks. 
    We develop a framework to interpret heterogeneous shock responses that depend on the constraints firms face, and provide empirical evidence on the relevance of these constraints in shaping firm behavior in Portugal and Norway, two countries with different institutional settings.
    We discuss the implications of our findings for conventional measures of employer wage-setting power and rent-sharing.
\end{abstract}

\vfill

\vspace{-0.1cm}


\thispagestyle{empty}
\clearpage

\onehalfspacing

\setcounter{page}{1}

\section{Introduction}

\input{m_intro}

\section{Framework}
\label{sec:m_framework}

\input{m_framework}

\section{Institutional Settings and Data}
\label{sec:m_setting_data}

\input{m_data}

\section{Empirical Strategy}
\label{sec:m_emp_strategy}

\input{m_emp_strategy}

\section{Results}
\label{sec:m_main_evidence}

\input{m_evidence}


\section{Policy Implications}
\label{sec:m_policy}

\input{m_policy}

\section{Discussion and Conclusion}
\label{sec:m_conclusion}

\input{m_conclusions}


\setlength{\bibsep}{3.4pt}

\bibliographystyle{ecta}
\bibliography{literature}

\begin{bibunit}[ecta]

\clearpage
\appendix

\section*{\centering Appendix for ``Wage-Setting Constraints and Firm Responses to Demand Shocks''}

\setcounter{page}{1}
\global\long\def\thepage{[Appendix-\arabic{page}]}

\setcounter{footnote}{0}

\renewcommand\thetable{\thesection.\arabic{table}}    
\renewcommand\thefigure{\thesection.\arabic{figure}}  
\renewcommand{\tablename}{Appendix Table}
\renewcommand{\figurename}{Appendix Figure}
\setcounter{table}{0}
\setcounter{figure}{0}

\setcounter{equation}{0}
\renewcommand{\theequation}{\Alph{section}.\arabic{equation}}


\input{m_appendix_theory}

\input{m_appendix_empirical}

\input{m_appendix_policy}


\setcounter{table}{0}
\setcounter{figure}{0}

\section{Supplementary Figures and Tables}
\label{asec:bace}

\begin{singlespace}
    \input{tables/cohort_size}
\end{singlespace}
\clearpage
\input{tables/exporter_descriptives}

\clearpage
\input{figures/model_variation_constraints_design}
\clearpage
\input{figures/CBA_wagefloors_portugal}
\clearpage
\input{figures/mw_evolution_portugal}
\clearpage
\input{figures/institutions_world}
\clearpage
\input{figures/firm_wages_portugal}
\clearpage
\input{figures/export_countries}

\clearpage
\input{figures/lms_stayers_incumbents}
\clearpage
\input{figures/global_lselast_cba}
\clearpage
\input{figures/lms_by_constraints_va}
\input{figures/elast_vary_threshold}
\clearpage
\input{figures/lms_by_constraints_portugal_combined}
\clearpage
\input{figures/wage_floor_effects}

\clearpage
\input{figures/export_shock_by_constraints}
\clearpage
\input{figures/elast_const_external}
\clearpage
\input{figures/tripleDiff_portugal_emp_VA}
\clearpage
\input{figures/tripleDiff_portugal_var_thresh}
\clearpage
\input{figures/struc_model_res_sensitivity.tex}

\begingroup
\renewcommand{\refname}{Appendix References}

\renewcommand*{\backrefalt}[4]{%
  \begingroup
  \let\backrefsep\empty
  \let\backreftwosep\empty
  \let\backreflastsep\empty
  \def\AppendixBackrefSep{\footnotesize(Cited on p.~}%
  \let\AppendixBackrefEnd\empty
  \let\OriginalBackref\backrefxxx
  \def\backrefxxx##1##2##3{%
    \IfBeginWith{##1}{[Appendix-}{%
      \AppendixBackrefSep
      \OriginalBackref{##1}{##2}{##3}%
      \def\AppendixBackrefSep{, }%
      \def\AppendixBackrefEnd{)}%
    }{}%
  }%
  #2\AppendixBackrefEnd
  \endgroup
}

\putbib[literature]
\endgroup
\end{bibunit}

\end{document}

%% file: autofill_lms.tex
\newcommand{\PtWageConsBeta}{0.018}

\newcommand{\PtWageUnconsBeta}{0.030}

\newcommand{\PtLSelastAllBeta}{5.90}
\newcommand{\PtLSelastAllSE}{0.394}
\newcommand{\PtLSelastConsBeta}{8.06}

\newcommand{\PtLSelastUnconsBeta}{5.41}

\newcommand{\PtRentShAllBeta}{0.089}
\newcommand{\PtRentShAllSE}{0.005}
\newcommand{\PtRentShConsBeta}{0.064}

\newcommand{\PtRentShUnconsBeta}{0.098}

\newcommand{\NoWageConsBeta}{0.032}

\newcommand{\NoWageUnconsBeta}{0.055}

\newcommand{\NoLSelastAllBeta}{3.29}
\newcommand{\NoLSelastAllSE}{0.234}
\newcommand{\NoLSelastConsBeta}{5.31}

\newcommand{\NoLSelastUnconsBeta}{2.79}

\newcommand{\NoRentShAllBeta}{0.185}
\newcommand{\NoRentShAllSE}{0.007}
\newcommand{\NoRentShConsBeta}{0.128}

\newcommand{\NoRentShUnconsBeta}{0.208}

\newcommand{\CoLSelastAllBeta}{4.47}
\newcommand{\CoLSelastAllSE}{0.724}

\newcommand{\CoRentShAllBeta}{0.072}
\newcommand{\CoRentShAllSE}{0.009}

\newcommand{\PtLSelastConsVsUncons}{49}
\newcommand{\PtRentShConsVsUnconsAbs}{34}
\newcommand{\NoLSelastConsVsUncons}{90}
\newcommand{\NoRentShConsVsUnconsAbs}{38}

\newcommand{\PtLSelastAllVsUncons}{9}
\newcommand{\NoLSelastAllVsUncons}{18}

\newcommand{\MaxPctDiffUnconsVsAllLSelast}{15}

\newcommand{\MaxPctDiffUnconsVsAllRentSh}{12}

\newcommand{\ImpliedDemConsSharePtPct}{3.9}
\newcommand{\ImpliedDemConsShareNoPct}{5.9}

\newcommand{\ImpliedDemConsSharePt}{0.039}
\newcommand{\ImpliedDemConsShareNo}{0.059}
\newcommand{\ImpliedDemConsShareNoLocBargLambdaLow}{0.071}
\newcommand{\ImpliedDemConsShareNoLocBargLambdaHigh}{0.094}


%% file: autofill_var_constraints.tex
\newcommand{\VcShareSC}{7.7}
\newcommand{\VcShareUCC}{9.6}
\newcommand{\VcShareUUC}{6.2}
\newcommand{\VcShareUUU}{17.0}
\newcommand{\VcShareSU}{59.5}

\newcommand{\VcNfirmsSC}{3,939}
\newcommand{\VcNfirmsUCC}{4,922}
\newcommand{\VcNfirmsUUC}{3,150}
\newcommand{\VcNfirmsUUU}{8,714}
\newcommand{\VcNfirmsSU}{30,446}
\newcommand{\VcNfirmsAll}{51,171}

\newcommand{\VcRatioMWSC}{1.00}

\newcommand{\VcRatioMWUUU}{1.17}

\newcommand{\VcWageSCPct}{0.7}

\newcommand{\VcWageUUUPct}{2.8}

\newcommand{\VcMWWageSCSixteenPct}{2.4}
\newcommand{\VcMWWageSCSeventeenPct}{4.0}

\newcommand{\VcLSelastSC}{15.32}

\newcommand{\VcRentShSCBeta}{0.025}

\newcommand{\VcLSelastUUU}{5.34}

\newcommand{\VcRentShUUUBeta}{0.099}

\newcommand{\VcSlopeWageBeta}{0.012}

\newcommand{\VcSlopeInvLSelastBeta}{0.074}

\newcommand{\VcSlopeRentShBeta}{0.042}

\newcommand{\VcSlopeLSelastBeta}{5.90}

\newcommand{\VcPvalWageSCvsUUU}{0.008}
\newcommand{\VcPvalInvLSelastSCvsUUU}{0.080}
\newcommand{\VcPvalRentShSCvsUUU}{0.003}

\newcommand{\VcRFDemandWageUUUPct}{2.8}
\newcommand{\VcRFDemandEmpUUUPct}{15.1}
\newcommand{\VcRFDemandVAUUUPct}{28.0}
\newcommand{\VcRFJointWageUUUPct}{2.8}
\newcommand{\VcRFJointWageSCTotalPct}{0.7}
\newcommand{\VcRFJointWageUCCTotalPct}{1.0}

%% file: autofill_welfare.tex
\newcommand{\WelfDemandConstrainedData}{3.9}
\newcommand{\WelfLaborShare}{0.592}
\newcommand{\WelfAlpha}{0.711}
\newcommand{\WelfEmploymentRate}{0.62}
\newcommand{\WelfParticipationElasticity}{0.3}

\newcommand{\WelfOptimalDeviationPctEtaFive}{1.5}
\newcommand{\WelfDemandConstrainedObservedEtaFive}{8.5}

\newcommand{\WelfDemandConstrainedOptimalEtaFive}{11.5}

\newcommand{\WelfBiasPct}{15}
\newcommand{\WelfGainNotRealizedEtaThree}{0.9}

\newcommand{\WelfGainNotRealizedEtaFive}{7.6}

\newcommand{\WelfChosenDeviationPctEtaFive}{1.6}
\newcommand{\WelfDemandConstrainedChosenEtaFive}{15.6}

\newcommand{\WelfGainNotRealizedEtaSeven}{18.2}


%% file: m_intro.tex
A growing body of literature documents substantial heterogeneity in wages among observably similar workers, a pattern often linked to employer wage-setting power \citep{CCHK2018, Manning2021, LMS2022}.
To quantify this phenomenon, recent research has focused on estimating two key parameters: firm-level labor supply and rent-sharing elasticities \citep{CaldwellDubeNaidu2026}. 
A common approach exploits firm-level demand shocks and estimates these elasticities from employment and wage adjustments. 
The labor supply elasticity is then used to quantify measures of employer wage-setting power, such as ``wage markdowns'', i.e., the ratio of wages to the marginal revenue product of labor \citep{SokolovaSorensen2021, AzarMarinescu2024, Kline2025}. 
The rent-sharing elasticity, in turn, informs how wages respond to changes in firm profitability \citep{AbowdLemieux1993, CCHK2018, JagerEtAl2020}.
These parameters are central to policy evaluation, informing analyses of mergers and acquisitions \citep[e.g.,][]{Arnold2019, PragerSchmitt2021} and the effects of minimum wage policies \citep[e.g.,][]{BergerEtAl2025}, among others.

In practice, however, many labor markets operate under wage-setting constraints---such as national minimum wage laws and sectoral wage bargaining agreements---that limit how firms adjust in response to demand shocks.
These constraints are prevalent across institutional contexts \citep{OECD2019collective, BhullerEtAl2022, DubeLindner2024, JagerEtAl2024} and can shape the observed pass-through of firm-level shocks to wages and employment.%
    \footnote{Besides institutional constraints, firms may also face informal constraints, e.g., due to fairness norms \citep{DubeEtAl2019, HazellEtAl2025} or pressures against nominal wage cuts \citep{GrigsbyEtAl2021}.}
When binding, they may lead constrained firms to adjust primarily along the employment margin, leaving wages less affected. 
This has direct implications for the estimation of labor supply and rent-sharing elasticities and, if unaccounted for, can distort the estimates of the \textit{potential} wage-setting power that firms hold.%
    \footnote{The distinction between potential and exercised wage-setting power dates to \citet{Bronfenbrenner1956} \citep[see also][]{Manning2021}.
    Illustrating this distinction, \citet[p.~4]{CaldwellDubeNaidu2026} note that ``policy constrains the exercise of monopsony power, but monopsony power exists'', describing a firm that pays workers their marginal revenue product under a minimum wage while facing a finite labor supply elasticity.} 
Although the literature acknowledges that a minimum wage may affect the relationship between firm productivity, wages and employment \citep[e.g.,][]{Manning2003Ch12, Kline2024}, we lack a clear understanding of what conventional measures of labor supply and rent-sharing elasticities capture under binding constraints, as well as systematic evidence on their empirical relevance across institutional contexts.

This paper fills that gap by developing a theoretical and empirical framework to study how wage-setting constraints affect the responses of firms to demand shocks. 
We lay out a model in which firms set wages and employment facing an upward-sloping labor supply curve as well as an institutional wage-setting constraint, covering distinct institutions such as a national minimum wage, sectoral bargaining agreements with negotiated wage floors, or wage demands from a local union. 
As in \cite{Manning2003Ch12}, a firm operates in one of three regimes: it may operate on the labor demand curve where the wage-setting constraint is binding (``demand-constrained''), it may operate on the labor supply curve but still face a binding wage-setting constraint (``supply-constrained''), or it may operate on the labor supply curve without being bound by the wage-setting constraint (``unconstrained''). 
Our model shows that, when constraints bind, conventional estimators mismeasure the underlying structural elasticities as some constrained firms adjust wages less and employment more in response to demand shocks.
Conventional measures will thus tend to overstate (understate) the structural labor supply (rent-sharing) elasticities that firms face. 
Crucially, the share of demand-constrained firms serves as a sufficient statistic for the extent of bias in conventional labor supply estimates. 
We take this framework to data from Portugal and Norway---two countries with distinct wage-setting institutions---to quantify the share of demand-constrained firms and the magnitude of bias in conventional elasticity estimates.

To empirically test the predictions of our framework, we must circumvent several methodological challenges. 
First, a firm's constraint regime is not directly observable in standard employer-employee datasets. 
We address this by classifying firms using predetermined indicators of exposure to minimum wages, sectoral bargaining, or local unions.
Second, identification requires isolating exogenous variation in labor demand.
We employ two approaches: an ``internal'' shock design following \citet{LMS2022}, which uses short-run fluctuations in value added and covers a broad sample of firms across the productivity distribution, and an ``external'' shock design based on export-demand shifts \citep{hjmx2014, garin2024responsive}.%
    \footnote{The internal shock approach in \citet{LMS2022} builds on \citet{Guiso2005}, who used changes in value added to capture idiosyncratic demand shocks at the firm level. 
    By comparison, the external shock approach relies on plausibly exogenous changes in observed factors that correlate with firm value added. Prominent examples of the latter approach include \citet{VanReenen1996} using technological innovations, \citet{KlineEtAl2019} using patent allowances, and \citet{KroftEtAl2025} using public procurement winners.}
Third, direct cross-country comparisons may confound wage-setting institutions with broader macroeconomic or structural differences.
Thus, we compare shock responses across differentially exposed firms within each country.
Finally, to address the concern that differences in adjustments across firms may reflect factors other than their exposure to wage-setting constraints, we develop a novel empirical strategy that exploits both exposure to demand shocks and within-firm changes in exposure to constraints over time.

Our study focuses on Portugal and Norway, two countries that are notable examples of institutional settings prevalent across the globe \citep{OECDVisser2023}.
In Portugal, we focus on firms' exposure to the national minimum wage, which is the most prevalent wage constraint, though we also consider collectively bargained wage floors. 
In Norway, which has no statutory minimum wage, we examine coverage by sectoral collective bargaining agreements as well as the presence of local unions.
As in several European countries, local unions engage in firm-level bargaining in Norway, while their role is fairly limited in Portugal.

We begin by documenting firm-level responses to demand shocks in each country.
Using the internal shock design, our estimates of the implied labor supply elasticities (the ratio of employment to wage responses to demand shocks) are $\PtLSelastAllBeta$ in Portugal and $\NoLSelastAllBeta$ in Norway, whereas
rent-sharing elasticities (the ratio of wage to value-added responses) are $\PtRentShAllBeta$ and $\NoRentShAllBeta$, respectively.
The external shock design reveals similar, though less precise, estimates.
These estimates reveal meaningful cross-country differences in how firms adjust to demand shocks, which could reflect differences in the bite of wage-setting constraints or underlying differences in employers' structural wage-setting power.%

Moving beyond cross-country differences, our results point to a consistent within-country pattern: wage responses are significantly lower among constrained firms. 
As a result, the implied labor supply elasticities for wage-constrained firms are $\PtLSelastConsVsUncons\%$ larger in Portugal and $\NoLSelastConsVsUncons\%$ larger in Norway than those for unconstrained firms, while the implied rent-sharing elasticities are $\PtRentShConsVsUnconsAbs\%$ smaller in Portugal and $\NoRentShConsVsUnconsAbs\%$ smaller in Norway.

This evidence suggests that wage-setting constraints can make constrained firms appear to face a flatter labor supply curve, seemingly closer to the competitive benchmark.
We find that conventional estimands overstate the firm-level labor supply elasticity by \PtLSelastAllVsUncons\% in Portugal and \NoLSelastAllVsUncons\% in Norway, understating firms’ potential wage-setting power.
Interpreted through the lens of our model, these estimates suggest that $\ImpliedDemConsSharePtPct\%$ of all firms in Portugal and $\ImpliedDemConsShareNoPct\%$ in Norway are demand-constrained (around 13\% and 20\% of wage-constrained firms).

These patterns also hold when we focus on firms that become constrained following minimum wage increases in Portugal.
We show that wage responsiveness to demand shocks diminishes significantly when firms become wage-constrained.
This changes the implied elasticity estimates, suggesting that the heterogeneity we observe across constrained and unconstrained firms partly reflects causal effects of the wage constraints. 
Consistent with our theoretical predictions, we also find lower rent-sharing and higher labor supply elasticities for Norwegian firms with high local union densities, likely reflecting that firms facing strong local unions are more likely to be demand-constrained.

We quantify the policy implications in a general equilibrium model calibrated to Portugal. 
We extend our model to allow for multiple local labor markets and extensive-margin employment responses to the minimum wage, and study the consequences of a planner choosing an employment-maximizing minimum wage with an incorrect firm-level labor supply elasticity.
In the baseline, overestimating the elasticity by \WelfBiasPct\% leads the planner to set the minimum wage too high, forgoing \WelfGainNotRealizedEtaFive\% of the attainable employment gain of the optimal policy.

This paper contributes to several strands of literature.
First, we show both theoretically and empirically that firm-level responses to demand shocks are shaped by wage-setting constraints, thereby contributing to our understanding of the role of labor market institutions \citep{DubeLindner2024, JagerEtAl2024}. 
Related literature finds that wage responses to demand shocks are stronger for firms bound by firm-level wage bargaining relative to sectoral bargaining \citep[e.g.,][]{Gurtzgen2009, RusinekRycx2013, CardDevicientiMaida2013}.%
    \footnote{Other articles study rent-sharing by focusing on wages negotiated in union contracts directly \citep[e.g.,][]{HolmlundZetterberg1991, AbowdLemieux1993, CardCardoso2022, Hermo2025}.} 
More recently, \cite{Olsson2024} shows that Swedish firms whose workers were covered by predetermined union wage contracts adjusted primarily through employment during the Great Recession.%
    \footnote{Similarly, \citet{balsvik2015} find that Chinese import competition reduced employment without wage adjustments across Norwegian manufacturing regions, which they attribute to wage-setting constraints.}
Our study complements this line of research in three ways: we provide comparable evidence of heterogeneity in firm responsiveness across multiple settings; we use direct, firm-level measures of exposure to wage-setting constraints--the ``bite'' of the minimum wage, collective bargaining coverage, and the presence of strong local unions--rather than comparing institutional regimes; and we exploit within-firm variation in constraints.%
    \footnote{\citet{AmodioEtAl2025} show cross-country estimates of (inverse) labor supply elasticities as well; however, they do not compare constrained and unconstrained firms within countries. 
    An earlier example showing cross-country evidence is \citet{HolmlundZetterberg1991}, who estimate industry-level rent-sharing models in Sweden, Norway, Finland, Germany, and the US, and discuss the role of wage-setting practices.}

More precisely, our work relates to recent studies that quantify labor market power in the presence of wage-setting constraints.
\citet{TortaroloZarate2018} estimate higher labor supply elasticities for Colombian manufacturing firms employing more minimum-wage workers.
\citet{Hermo2025} finds that wage and employment responses to wage-floor shocks in Argentina vary with firms' exposure to constraints.
\citet{BassierBudlender2025} estimate smaller rent-sharing elasticities among constrained firms in South Africa, and \cite{FaiaAl2026} test the implications of imposing a minimum wage in a monopsony framework.%
    \footnote{Related work by \cite{DiegmannAl2026} suggests that departures from a standard monopsony model may better explain the size-wage premium in large German firms.}
Our paper makes several distinct contributions. 
First, we explicitly discuss the implications of wage-setting constraints for the identification of firm-level elasticities, informing a discussion of potential remedies. 
Second, we provide comparable quasi-experimental evidence from multiple settings and exploit variation in wage-floor constraints.
Third, we develop a framework that combines insights from models of monopsony and firm-level bargaining, while allowing for wage-floor constraints.%
    \footnote{Our framework connects to recent work explicitly incorporating unions and firm-level bargaining in monopsony settings, including \citet{AzkarateAskasuaZerecero2025} and \citet{Wong2025}.} 
This framework rationalizes the presence of rent-sharing among constrained firms, unlike stylized models of firm-level bargaining.%
    \footnote{In stylized models of firm-union Nash bargaining, wages depend on the average revenue product of labor and an outside option. However, as first pointed out by \citet{McDonaldSolow1981}, the equilibrium level of the average revenue product of labor does not depend on demand shifters when the production function is homogeneous, implying the absence of rent sharing. 
    By contrast, our framework generates non-zero rent-sharing elasticities also among constrained firms, even when the production function is homogeneous.}
The model can also rationalize that wages in some firms lie strictly above the marginal revenue product of labor, i.e., the presence of ``wage markups'', which is ruled out in static monopsony models.%
    \footnote{Recent evidence by \citet{YehMacalusoHershbein2022} and \citet{ChanMattanaSalgadoXu2023} shows that a significant fraction of firms have markdowns above one, a feature that can be rationalized in models of firm-level bargaining.} 

Our paper also relates to a broader literature quantifying monopsony power which later informs policy-relevant discussions \citep[e.g.,][]{BHM2022, LMS2022, DebEtAl2024}.
Our work suggests that estimating structural elasticities by fitting structural labor supply or rent-sharing relationships when the economy operates under wage-setting constraints may result in biased estimates of the structural parameters. 
Furthermore, we show that relying on instruments will not solve the problem.
This insight complements recent work on how labor adjustment frictions \citep{DhyneEtAl2024, ChanMattanaSalgadoXu2023} and long-term employment relationships \citep{BalkeLamadon2022, AgostinelliEtAl2025} shape measured employer wage-setting power.

Finally, our paper relates to recent work on optimal minimum wage policy.
\citet{BergerEtAl2025} study the efficiency implications of the minimum wage in the US, and
\citet{AhlfeldtEtAl2023} study the spatial implications in Germany.
Our contribution is to quantify the forgone gains from using an incorrect elasticity estimate to choose optimal policy.
This exercise also connects to the broader literature on using estimated parameters for policy evaluation and decision-making \citep[e.g.,][]{Chetty2009, Manski2021, AndrewsBarnhardCarlson2026}.

The paper is structured as follows.
Section \ref{sec:m_framework} introduces the framework we use to interpret firm responses to demand shocks under wage constraints. 
Section \ref{sec:m_setting_data} describes the context and data sources for Portugal and Norway.
Section \ref{sec:m_emp_strategy} describes our empirical strategy. 
Section \ref{sec:m_main_evidence} presents our findings.
Section \ref{sec:m_policy} discusses policy implications, and Section \ref{sec:m_conclusion} concludes.

%% file: m_framework.tex
This section proceeds in three steps.
First, we lay out an economic model in which firms maximize profits facing a labor supply constraint and a wage-setting constraint and discuss how the wage constraint affects the responses to demand shocks.
Second, we discuss what is identified by usual approaches to estimate labor supply and rent-sharing elasticities, relying on the potential outcomes from the model.
Finally, we introduce a novel strategy for causally testing the role of constraints in shaping firms' responses to demand shocks.

\subsection{Theoretical Model} 
\label{sec:m_framework_theory}


Firms $j \in \J$ produce a homogeneous good but differ in their total factor productivity $\Phi_j$.
Workers' preferences result in upward-sloping labor supply curves to the firm.
Additionally, firms face an institutional wage-setting constraint.
Labor market clearing determines the overall level of wages and, as a result, the constrained status of different firms.

\subsubsection{Labor Supply} 

We let the measure of workers willing to work for firm $j$ at wage $W_{j}$ be
\begin{equation}\label{eq:labor_supply}
    H(\psi_{j}W_{j},\aggW;A_j) 
      = A_j \left(\frac{\psi_{j}W_{j}}{\aggW}\right)^{\eta},
\end{equation}
where 
$A_j$ is the amenity value of working for the firm,
$\aggW$ is an aggregate wage index that will be determined in equilibrium,
$\psi_{j}$ is the firm's hiring probability (an endogenous quantity that allows us to generate labor rationing), and
$\eta>0$ is the elasticity of labor supply.
Appendix \ref{asec:micr_labor_supply} provides a micro-foundation of this labor supply equation assuming that workers have idiosyncratic tastes for different firms \citep{CCHK2018}. 
The quantity $\psi_{j}W_{j}$ can be interpreted as the expected wage workers receive when choosing to work for firm $j$. 

The structure in Equation \eqref{eq:labor_supply} imposes several assumptions for notational convenience that we later relax.
First, the assumption that the labor supply elasticity $\eta$ is homogeneous across firms simplifies the interpretation of our empirical analysis of demand shocks. We explicitly relax this assumption in Section \ref{sec:m_framework_constraints} when we discuss our design exploiting within-firm variation in wage-setting constraints over time. Second, for notational simplicity, we also abstract from market-level adjustments. However, our empirical analysis exploits within-market variation in firm-level shocks by flexibly controlling for market-by-year effects.

\subsubsection{Firm Optimization Problem}

We assume that firm $j$'s revenue is $P \Phi_{j} f_j(L_{j})$, where 
$P$ is the output price,
$\Phi_{j}$ is total factor productivity,
$L_{j}$ is the number of workers employed by the firm, and
$f_j(\cdot)$ is a standard production function with $f_j'(\cdot)>0$ and $f_j''(\cdot)<0$.%
    \footnote{While we exclude inputs other than labor from the production process, flexible inputs such as capital or materials can be added. 
    For instance, let $M_j$ be a flexible input with exogenous price $P_M$ and let $\tilde f_j(M_j, L_j)$ be the relevant production function. 
    The optimal level of the input for any $L_j$ is $M^*_j = M_j(P_M,L_j)$. 
    Then, we can study the optimal choice of labor using the production function $f_j(L_j) = \tilde f_j(M_j(P_M,L_j),L_j)$.}
Additionally, we assume that the firm takes $P$ as exogenous, and set $P=1$ as the numeraire of the economy.

Firm $j$ chooses the wage and employment level to maximize profits $\Pi_{j} = \Phi_{j} f_j(L_{j}) - W_{j} L_{j}$.
Following the recent monopsony literature, we let the firm face the labor supply constraint
\begin{equation}\label{eq:labor_supply_constraint}
    L_{j} \leq H\left(\psi_{j}W_{j},\aggW;A_j\right) .
\end{equation}
We assume that the firm is atomistic, so it ignores its influence on the aggregate wage index~$\aggW$.%
    \footnote{\citet[Section II.A]{BHM2022} discuss how strategic interactions between firms pose challenges to the identification of firm-level labor supply elasticities using the conventional approach that relies on firm-level demand shocks. 
    Notably, the identification problem that we discuss in Section \ref{sec:m_framework_ident} is distinct and carries over in an oligopsonistic setting, but arises even in settings without such strategic interactions.}
Additionally, given a wage floor $\W>0$, the firm faces a wage-setting constraint
\begin{equation}\label{eq:wage_setting_constraint}
    W_{j} \geq g_j\left(\Upsilon_j,\W\right),
\end{equation}
where $\Upsilon_j(L_j)=\Phi_jf_j(L_j)/L_j$ is the average revenue product of labor, and the wage-setting constraint $g_j\left(\Upsilon_j,\W\right) \in [\W,\Upsilon_j]$ is weakly increasing in both arguments.%
    \footnote{We note that $g_j\left(\Upsilon_j,\W\right) \in [\W,\Upsilon_j]$ is a natural restriction in this setting. 
    A negotiated wage above the average product $\Upsilon_j$ would imply that the firm operates at a loss, whereas a wage below $\W$ would violate the institutional constraint.
    In this spirit, \citet{MoeneWallerstein1997} present a model of local bargaining where they impose that the negotiated wage is at least as large as the market wage faced by the firm.}

\paragraph{The unconstrained solution.} 

The profit-maximizing wage is given by $W_{j}^* = \mu \Phi_{j} f_j'(L_{j}^*)$, where $\Phi_{j} f_j'(L_{j}^*)$ is the equilibrium marginal revenue product of labor (MRPL) and $\mu = \eta / (\eta+1)$ is the markdown factor (or what \citealt{robinson1933economics} defined as the ``exploitation index'').
For any wage, it is optimal to hire all workers supplying labor, i.e., $\psi^*=1$ and $L_{j}^* = H_{j}\left(W_{j}^*,\aggW;A_j\right)$.
Panel (a) of Figure \ref{fig:model_solutions} shows the equilibrium, relying on the inverse labor supply curve $W(H)$.
The firm sets employment to equate the MRPL and the marginal cost of labor (MCL), and the wage is determined by the labor supply curve.

\paragraph{Production technology.} 

It will be informative to discuss special cases for $f_j(\cdot)$ that will allow us to draw sharper comparative statics from the framework and compare with existing work.
To do so, we rely on the output elasticity of labor $\alpha_j(L) \equiv L f'(L)/f(L)>0$.

\begin{assumption}[Production Technologies]
    \label{assu:production_function}
    \begin{assumpinline}
        \item\label{assu:production_function:a}
        The output elasticity of labor is constant, $\alpha_j(L)=\alpha$ for $\alpha\in(0,1)$. 
        Equivalently, the production function takes the Cobb-Douglas form $f_j(L_j)=L_j^{\alpha}$.
        \item\label{assu:production_function:b} The output elasticity of labor is decreasing, $\alpha_j'(L_j) < 0$ for all $L_j$.
    \end{assumpinline}
\end{assumption}

Assumption \assumpfullref{assu:production_function}{a} corresponds to the familiar Cobb-Douglas case.
Assumption \assumpfullref{assu:production_function}{b} requires that the output elasticity of labor declines with employment.
This property naturally arises in the presence of fixed factors or organizational constraints.
We use these assumptions when studying comparative statics with respect to $\Phi$.

Next, we analyze cases where the wage-setting constraint binds.
We discuss pure wage floor constraints, where $g_j(\Upsilon_j,\W) = \W$, and firm-level bargaining between a local union and the firm, assuming $g_j(\Upsilon_j,\W) = \kappa_j \Upsilon_j + \left(1-\kappa_j\right)\W$, where $\kappa_j\in(0,1)$ is a parameter reflecting union bargaining power. 
Our key insights also hold for general forms of $g_j(\Upsilon_j,\W)$.

\subsubsection{Wage Floor Constraint} 
\label{subsec:m_framework_theory_floor}

Motivated by the global prevalence of minimum wage policies and sectoral wage floors between employer associations and sectoral unions, we consider the behavior of constrained firms when the constraint takes the form $W_j\geq \W$.
Appendix \ref{asec:derivations_wage_floor_const} provides derivations.

\paragraph{Solution to the firm's problem.}

The solution $(W^*_{j}, L^*_{j})$ can be characterized by three distinct regimes \citep{Manning2003Ch12}. 
A high-productivity firm will be \textit{unconstrained} and choose $W^*_j>\W$, as shown in Panel (a) of Figure \ref{fig:model_solutions}.
Low-productivity firms will be wage constrained, in which case $W_{j}^* = \W$, and may operate under two regimes.

\input{figures/model_solutions}

First, \textit{supply-constrained} firms hire all workers that are willing to work at the wage floor, i.e., $L_{j}^* = H_{j}\left(\W,\aggW;A_j\right)$, so employment is determined by labor \textit{supply}.
As shown in Panel (b) of Figure \ref{fig:model_solutions}, these firms would prefer to pay a wage below $\W$ and hire fewer workers, but because of the constraint, they choose a higher wage-employment pair.
The constraint creates a wedge between the MRPL and the wage floor $\W$ (the relevant MCL). 

Second, \textit{demand-constrained} firms choose employment $L^*_{j}$ to equate the MRPL to the wage floor, i.e., $\Phi_{j} f_j'(L^*_{j}) = \W$, so employment is determined by labor \textit{demand}.
Panels (c) and (d) of Figure \ref{fig:model_solutions} illustrate the equilibrium for firms that differ in the level at which the MRPL intersects with the wage floor $\W$, which determines employment in the latent unconstrained equilibrium.
In Panel (c), the intersection occurs above $\W$, and employment is larger than the latent unconstrained level, whereas Panel (d) shows the alternative case.
While this distinction matters for the response of the labor market to the introduction of a wage floor \citep[e.g.,][]{AhlfeldtEtAl2023,FaiaAl2026}, we note that it does not affect firm responses to demand shocks (shifts in $\Phi$), which is the focus of our analysis.

Demand-constrained firms ration employment, i.e., $L_{j}^* < H_{j}\left(\psi_{j}\W,\aggW;A_j\right)$.
This is illustrated by the difference between $L^*$ and $H^*$ in Panels (c) and (d) of Figure \ref{fig:model_solutions}.
To determine the value of $\psi_{j}$, we assume that workers correctly anticipate the firm's choice, and so the equilibrium hiring probability $\psi_{j}^*$ will be implicitly determined by $\psi_{j}^*=L_{j}^* / H_{j}\left(\psi_{j}^*\W,\aggW;A_j\right)$.

\paragraph{Threshold productivity levels.}

The solution is characterized by two threshold productivity levels, denoted by $\upperPhi_{j}$ and $\lowerPhi_{j}$, where firms with $\Phi_j>\upperPhi_{j}$ are unconstrained, $\Phi_j\in(\lowerPhi_j,\upperPhi_{j}]$ are supply-constrained, and $\Phi_j\leq\lowerPhi_j$ are demand-constrained.
The upper threshold can be derived by equating the unconstrained wage to the wage floor, resulting in $\upperPhi_{j} = \W/\left[\mu f_j'\left(H(\W,\aggW;A_j)\right)\right]$.
The lower threshold is given by the productivity that makes the demand-constrained employment equal to the supply-constrained employment $H(\W,\aggW;A_j)$, so that $\lowerPhi_{j} = \W/\left[f_j'\left(H(\W,\aggW;A_j)\right)\right]$.

\paragraph{Potential outcomes.}

The framework delivers potential outcomes for firm $j$, depending on its productivity, its amenity value, the wage floor, and the aggregate wage index, which we denote by 
$L^*_j\left(\Phi_j, A_j, \W, \aggW\right)$ and $W^*_j\left(\Phi_j, A_j, \W, \aggW\right)$.
Panel (e) of Figure~\ref{fig:model_solutions} plots $W^*$ and $L^*$ as productivity $\Phi$ changes, holding constant the remaining parameters. 
When $\Phi_{j} > \upperPhi_{j}$ the firm is unconstrained, and both employment and wages increase with productivity.
When $\Phi_{j} \in \left(\lowerPhi_{j}, \upperPhi_{j}\right]$ the firm is supply-constrained, and employment and wages are invariant to productivity.
Finally, for $\Phi_{j} \leq \lowerPhi_{j}$, the firm is demand-constrained, so employment increases with productivity along the labor demand curve.

The wage-setting constraints also affect the observed markdown factor and labor share.
Panel (f) of Figure~\ref{fig:model_solutions} shows the potential outcomes for these objects, assuming a homogeneous production technology (Assumption \assumpfullref{assu:production_function}{a} holds) so that the labor share of unconstrained firms is constant.
We highlight two results.
First, we observe a markdown factor below 1 for supply-constrained firms.
Second, exposure to the wage floor increases the labor share. 

\subsubsection{Beyond Pure Wage Floor Constraints}
\label{subsec:m_framework_theory_general}

Motivated by settings with firm-level bargaining, where $\W$ can be interpreted as an outside wage, or two-tier bargaining, where $\W$ is a floor determined in sectoral negotiations, we consider the constraint $g_j(\Upsilon_j,\W) = \kappa_j \Upsilon_j + \left(1-\kappa_j\right)\W$.%
    \footnote{In many settings, wage floors are determined in sectoral negotiations between employer associations and sectoral unions.
    Thus, we assume the wage floor is unrelated to $j$'s productivity and only depends on sectoral factors. 
    We can allow wage floors to vary across firms, i.e., $\W_j$, without losing any of the insights developed here. 
    In the empirical analysis, we find no effects of firm-level demand shocks on average wage floors $\W_j$.}
This constraint can be derived from a Nash bargaining problem in which the firm first commits to an employment level, and the local union then negotiates over the wage, treating employment as given.%
    \footnote{More precisely, the Nash bargaining problem is $\max_{\{W\}} \left[(W - \W) L \right]^{\kappa_j} \left[\Phi f(L) - WL\right]^{1-\kappa_j}$.
    See \citet{Holden1988} and \citet{Moene1988} for models with similar structure.
    These papers, however, neither incorporate a firm-level labor supply curve nor allow for unconstrained firms for whom the wage constraint does not bind.}
We interpret $\kappa_j\in(0,1)$ as the bargaining power of the local union.
Appendix \ref{asec:derivations_local_bargaining} provides derivations.

This setting yields unconstrained, supply-constrained, and demand-constrained firms as well. 
Panels (a) through (d) of Appendix Figure \ref{fig:model_solutions_local_bargaining} illustrate the regime under a homogeneous production function (Assumption \assumpfullref{assu:production_function}{a}).
Unconstrained firms only exist when the unconstrained markdown factor $\mu$ is large relative to $\kappa_j$.
Strong union power ($\kappa_j\to1$) or employer power ($\mu\to0$) results in all firms being constrained.
Thus, unlike stylized firm-union bargaining models, we allow for unconstrained firms---high-productivity firms facing a relatively weak union---who choose wages above union demands to expand.

Our formulation allows for equilibria with firm wages above the MRPL, a fact that has been documented recently \citep[e.g.,][]{ChanMattanaSalgadoXu2023}.
Appendix Figure \ref{fig:model_solution_markdown_lshare} shows this result, which holds for demand-constrained firms and lower-productivity supply-constrained firms.
Appendix \ref{asec:derivations_general_wconstraint} generalizes to the general constraint $W_j\geq g_j(\Upsilon_j,\W)$, showing a \textit{positive wedge} in the FOC of demand-constrained firms, as firms expand in excess of the MRPL to dilute the average product of labor and reduce the wage.

Appendix \ref{asec:equilibrium_def} closes our model, showing that an equilibrium wage index $\aggW^*$ that equates labor demand and supply exists and is unique under general $g_j(\cdot)$.
The wage $\aggW^*$ depends on the wage floor $\W$ and the productivities $\Phi_j$, underscoring the importance of accounting for general equilibrium effects in the empirical analyses later when we shift these quantities.

\subsubsection{Comparative Statics and Testable Predictions}
\label{sec:framework_predictions}

We now study the wage, employment, and value-added responses to a marginal productivity shock $d\ln\Phi_j$ under the constrained and unconstrained regimes, for each type of wage-setting constraint. In our framework, the implications of a marginal productivity shock are identical to those of a marginal product demand shock.%
    \footnote{Alternatively, one could allow for price heterogeneity and analogously consider changes in output prices.}
We derive the ratio of employment to wage responses, typically used to measure labor supply elasticities \citep[e.g.,][]{SokolovaSorensen2021}, and the ratio of wage to value-added responses, used as a measure of rent-sharing elasticities \citep[e.g.,][]{CCHK2018}.
Note that, as we abstract from inputs, value added equals total revenue, i.e., $\VA_j = \Phi_j f(L_j)$.
See Appendix \ref{asec:comparative_statics} for formal derivations.

We start with unconstrained firms.
Our results depend on the curvature of the production function, which we summarize by the output elasticity of labor $\alpha_j(L)$ and the absolute value of the employment elasticity of the MRPL $\gamma_j(L)\equiv -L f''(L)/f'(L)$.

\begin{result}[Response to Shocks among Unconstrained Firms]
    \label{imp:cstats_uncons_firm}
    \begin{reslist}
        \item\label{imp:cstats_uncons_firm:a} The ratio of the employment response to the wage response to a shock is $\frac{d\ln L_j/d\ln\Phi_j}{d\ln W_j/d\ln\Phi_j} = \eta$.
        \item\label{imp:cstats_uncons_firm:b} The ratio of the wage response to the value-added response to a shock is $\frac{d\ln W_j/d\ln\Phi_j}{d\ln \VA_j/d\ln\Phi_j}= \frac{1}{1 + \eta(\alpha_j(L_j^*) + \gamma_j(L_j^*))} \equiv \theta_j $.
    \end{reslist}
\end{result}    

Result \ref{imp:cstats_uncons_firm} shows that unconstrained firms' responses reveal the structural labor supply elasticity $\eta$ and the structural rent-sharing elasticity $\theta_j$.%
    \footnote{Note that we defined the rent-sharing elasticity as the elasticity of wages to value added.
    Alternatively, one can use the elasticity with respect to productivity $\Phi$ or value added per worker $\VA/L$ \citep[e.g.,][]{Kline2025}.}
Rent-sharing depends on $\eta$ and the sum of elasticities $\alpha_j(L^*)+\gamma_j(L^*)$, and is thus heterogeneous across firms.
Interestingly, $\theta_j$ is inversely related to $\eta$.
As $\eta\to\infty$, the model converges to a perfectly competitive benchmark where workers are paid their MRPL and the rent-sharing elasticity is zero.

Further assumptions on the production technology yield sharper predictions for the rent-sharing elasticity.
Under Assumption \assumpfullref{assu:production_function}{a} we have $\alpha_j(L)+\gamma_j(L)=1$, implying $\theta_j=1/(1+\eta)$ for all $j$.
Under Assumption \assumpfullref{assu:production_function}{b} instead $\alpha_j(L)+\gamma_j(L)>1$, implying $\theta_j<1/(1+\eta)$.%
    \footnote{
        Let us show more formally how Assumption \ref{assu:production_function} relates to the elasticity sum $\alpha_j(L)+\gamma_j(L)$.
        First, for $f_j(L)=L^\alpha$ we have $\alpha_j(L)=\alpha$ and $\gamma_j(L)=1-\alpha$, thus $\alpha_j(L)+\gamma_j(L)=1$.
        Second, we show that $\alpha_j(L)+\gamma_j(L)>1$ for $\alpha_j'(L)<0$. Differentiating $\alpha_j(L) \equiv L f_j'(L)/f_j(L)$ with respect to $L$ yields
        $\alpha_j'(L) = \frac{\alpha_j(L)}{L} \left[ 1 - (\alpha_j(L) + \gamma_j(L)) \right]$.
        Since $\alpha_j(L)>0$ and $L>0$, the condition $\alpha_j'(L) < 0$ implies $\alpha_j(L) + \gamma_j(L) > 1$.}
Next, we discuss the behavior of constrained firms. 

\begin{result}[Employment to Wages]
    \label{imp:lsupply}
    Let the ratio of employment to wage responses be $\tilde \eta_j = \left(d\ln L_j/d\ln\Phi_j\right) / \left(d \ln W_j/d\ln\Phi_j\right)$.
    \begin{reslist}
        \item\label{imp:lsupply:a} Under a pure wage floor constraint, $\tilde \eta_j$ is undefined for supply-constrained firms and $\tilde \eta_j \to \infty$ for demand-constrained firms.
        \item\label{imp:lsupply:b} Under local bargaining, $\tilde \eta_j = \eta$ for supply-constrained firms and $\tilde \eta_j = \left[\lambda_j(L)(\alpha_j(L)+\gamma_j(L)-1)\right]^{-1}$ for demand-constrained firms, where $\lambda_j(L) = \kappa_j\Upsilon_j/(\kappa_j\Upsilon_j+(1-\kappa_j)\W)\in(0,1)$.
    \end{reslist}
\end{result}

The ratio $\tilde \eta_j$ may differ from the structural elasticity $\eta$.
In particular, for demand-constrained firms, we expect $\tilde \eta_j$ to be large.
To see this for the local bargaining constraint note that, under Assumption \assumpfullref{assu:production_function}{a}, we have $\alpha_j(L)+\gamma_j(L)=1$ and thus $\tilde\eta_j$ diverges as well.
More generally, with ``weak curvature'' ($\alpha_j(L)+\gamma_j(L)$ close to 1), we expect a large $\tilde\eta_j$ satisfying $\tilde\eta_j>\eta$.
Finally, we discuss the ratio of wage to value-added responses to shocks.

\begin{result}[Wages to Value Added]
    \label{imp:rentsharing}
    Let the ratio of wage to value-added responses be $\tilde \theta_j = \left(d \ln W_j/d\ln\Phi_j\right) / \left(d \ln \VA_j/d\ln\Phi_j\right)$.
    \begin{reslist}
        \item\label{imp:rentsharing:a} Under a wage floor constraint, $\tilde \theta_j=0$ for constrained firms.
        \item\label{imp:rentsharing:b} Under local bargaining, $\tilde \theta_j = \lambda_j(L)/(1+\eta\lambda_j(L))$ for supply-constrained firms and $\tilde \theta_j = \lambda_j(L)\left[1 - (\alpha_j(L)+\gamma_j(L))^{-1}\right]$ for demand-constrained firms.
    \end{reslist}
\end{result}

The rent-sharing elasticity is zero under a pure wage floor constraint for constrained firms.
Under local bargaining, we also expect low rent-sharing elasticities (as $\lambda_j(L)\in(0,1)$ and, under Cobb-Douglas, $\alpha_j(L)+\gamma_j(L) = 1$).
Our model generates positive rent-sharing for supply-constrained firms even under homogeneous production functions, a result that has eluded previous literature.%
    \footnote{\label{fn:rent_sharing_bargaining}
    Prior work has noticed that, under a homogeneous production function and no labor supply constraint, the equilibrium level of the average revenue product of labor is constant, which results in a constant wage (see, e.g., \citealt[][p.\ 899]{McDonaldSolow1981}, \citealt{Manning1987}, and \citealt[][p.\ 987]{AbowdLemieux1993}).
    This holds in our model for demand-constrained firms only, i.e., when the labor supply constraint does not bind.}
Note that, while a wage constraint reduces the observed rent-sharing \textit{elasticity}, it increases the labor share or ``rent-sharing \textit{level}'' (see Panel (f) of Figure \ref{fig:model_solutions} for wage floor constraints and Appendix Figure \ref{fig:model_solution_markdown_lshare} for local bargaining).

\paragraph{Testable implications.}

Consider studying the average responses across firms that are plausibly constrained and those that are plausibly unconstrained.
We expect a smaller ratio of wage to employment responses for firms that are plausibly constrained by a wage floor relative to unconstrained firms, resulting in larger estimates of labor supply elasticities and smaller estimates of rent-sharing elasticities. 
Similarly, we expect smaller wage responses among firms that are plausibly constrained by stronger unions, as reflected in $\kappa_j$, resulting in larger estimates of labor supply and smaller estimates of rent-sharing elasticities as well.
\paragraph{Heterogeneous labor supply elasticities.}

Our assumption of homogeneous labor supply elasticity $\eta$ is not necessary for these results.
We can instead allow for a firm-specific elasticity $\eta_j$, and Results \ref{imp:cstats_uncons_firm} through \ref{imp:rentsharing} would hold with $\eta$ reinterpreted as a firm-specific parameter.
Our framework can also accommodate multiple worker types with isoelastic labor supply curves for each type.%
    \footnote{For instance, we can interpret $j$ as the combination of firm and occupation.
    Then, a firm-occupation cell may be unconstrained, supply-constrained, or demand-constrained.
    Such a model extension would need to consider that there may be spillovers across occupations within the firm \citep{AdamopoulouEtAlForthcoming}.}
We assume, however, that such firm- or type-specific elasticities are invariant to shocks.
This rules out, for example, non-iso-elastic preference structures in which the elasticity changes with the wage \citep[e.g.,][]{Kline2025} or the firm's workforce composition \citep[e.g.,][]{Volpe2026}.
While our derivations do not apply directly to such cases, we expect similar effects of wage-setting constraints on the responsiveness to demand shocks.

\subsection{Identification} 
\label{sec:m_framework_ident}

We consider the identification of labor supply and rent-sharing elasticities using wage, employment, and value added data before ($t=0$) and after ($t=1$) a demand shock $Z_j$.
We denote our data using first differences, $\Delta \ln X = \ln X_1 - \ln X_0$ for some $X$.
Our data are $\{\Delta\ln L_{j}, \Delta\ln W_{j}, \Delta\ln \VA_{j}, Z_j\}_{j\in\J}$, where for simplicity we assume $Z_j$ is binary.
Later, we use $Z_j$ as an instrument in our identification discussion.
We focus on wage floor constraints, though similar results would hold for other types of constraints allowed by our framework.

\paragraph{Target parameters.}

We define target parameters in terms of counterfactual outcomes without constraints, $\{L_{jt}^*(\Phi,\W=0),W_{jt}^*(\Phi,\W=0)\}_{j\in\J}$ for $t\in\{0,1\}$.
Then, we denote our counterfactual data as $\{\Delta\ln W^*_{j}, \Delta\ln L^*_{j}, \Delta\ln \VA^*_{j}, Z_j\}_{j\in\J}$.

\begin{definition}[Target Parameters]
    \label{def:target_param}
    The target parameters are
    $$
    \bar{\eta} = \frac{\E\left[\Delta\ln L^*_{j}|Z_j=1\right] - \E\left[\Delta\ln L^*_{j}|Z_j=0\right] }
                  {\E\left[\Delta\ln W^*_{j}|Z_j=1\right] - \E\left[\Delta\ln W^*_{j}|Z_j=0\right]  } ,
    $$
    $$
    \bar{\theta} = \frac{\E\left[\Delta\ln W^*_{j}|Z_j=1\right] - \E\left[\Delta\ln W^*_{j}|Z_j=0\right]}
                    {\E\left[\Delta\ln \VA^*_{j}|Z_j=1\right] - \E\left[\Delta\ln \VA^*_{j}|Z_j=0\right]  } .
    $$
\end{definition}

The parameters indicate the average firm-level responsiveness of employment to wage changes ($\bar{\eta}$) and of wages to value-added changes ($\bar{\theta}$) in a hypothetical economy without wage constraints. 
The parameter $\bar{\eta}$ reveals the average potential monopsony power that firms hold and is informative about, for example, the extent to which minimum wage policies may result in employment losses.
The parameter $\bar{\theta}$ is informative about, for example, how government policies that influence a firm's demand are transmitted to workers' wages.

\paragraph{Connection of target parameters with theoretical model.}

We assume that the shock $Z_j$ affects the firm productivity $\Phi_j$, and denote the (unobserved) vector of productivity changes as $\{\Delta\ln\Phi_{j}\}_{j\in\J}$. 
The following lemma establishes the relationship between the parameters in Definition \ref{def:target_param} and the model's structural parameters.

\begin{lemma}[Identification of Structural Parameters]
    \label{lem:ident_target_params}
    The target parameter $\bar{\eta}$ identifies the structural labor supply elasticity, i.e., $\bar{\eta} = \eta$.
    The target parameter $\bar{\theta}$ identifies a weighted average of the firm-specific rent-sharing elasticities $\theta(\alpha_j, \gamma_j)$, i.e.,
    $$
    \bar{\theta} = \E\left[\omega(\alpha_j, \gamma_j)\, \theta(\alpha_j, \gamma_j)\right],
    $$
    where $(\alpha_j, \gamma_j)$ summarize firm $j$'s technology, $\theta(\alpha_j, \gamma_j) = (1 + \eta[\alpha_j + \gamma_j])^{-1}$ is the structural rent-sharing elasticity, and the weight
    $\omega(\alpha_j, \gamma_j) = \delta(\alpha_j, \gamma_j)\, \pi(\alpha_j, \gamma_j)/\E\left[\delta(\alpha_j, \gamma_j)\, \pi(\alpha_j, \gamma_j)\right]$
    is proportional to $\delta(\alpha_j, \gamma_j) = (1 + \eta[\alpha_j + \gamma_j])/(1 + \eta\gamma_j)$, which is the coefficient on $\Delta\ln\Phi_j$ in the unconstrained value-added response, and $\pi(\alpha_j, \gamma_j) = \E[\Delta\ln\Phi_j|Z_j=1, \alpha_j, \gamma_j] - \E[\Delta\ln\Phi_j|Z_j=0, \alpha_j, \gamma_j]$, which is the technology-specific first stage.
    Under homogeneous production functions (Assumption \assumpfullref{assu:production_function}{a}), $\theta(\alpha_j, \gamma_j)$ is constant across firms and $\bar{\theta} = \frac{1}{1+\eta}$.
\end{lemma}

Proofs are available in Appendix \ref{asec:proofs}.

The result that $\bar{\eta}=\eta$ is general and holds for general production technology.
However, to obtain a constant rent-sharing elasticity, we need a stronger assumption on the production technology. 
For simplicity, in what follows we use $\eta$ and $\theta$ to denote the target parameters, with the caveat that the interpretation of the rent-sharing elasticity parameter depends on the assumptions on the production technologies.

\paragraph{Conventional estimands.}

The conventional approach uses the estimands
$$
\eta^{CE} 
  = \frac{\E\left[\Delta \ln L_{j} |Z_j=1\right] - \E\left[\Delta \ln L_{j} |Z_j=0\right] }
         {\E\left[\Delta \ln W_{j} |Z_j=1\right] - \E\left[\Delta \ln W_{j} |Z_j=0\right]} ,
$$
$$
\theta^{CE} 
  = \frac{\E\left[\Delta \ln W_{j} | Z_j=1\right] - \E\left[\Delta \ln W_{j} | Z_j=0\right] }
         {\E\left[\Delta \ln \VA_{j} | Z_j=1\right] - \E\left[\Delta \ln \VA_{j} | Z_j=0\right]} .
$$
These estimands can be motivated by 2SLS systems.
For $\eta^{CE}$, we treat $Z_j$ as an instrument for wages in a regression of employment changes on wage changes.
Similarly, for $\theta^{CE}$, we use $Z_j$ as an instrument for value added (VA) in a regression of wage changes on VA changes.
We seek to clarify the behavior of these estimands when wage-setting constraints bind.

\paragraph{Transitions across constrained states.}

As the demand shock is discrete, we may observe firms changing their constrained status.
We introduce notation to reflect those cases: $\rho_{xy}\in[0,1]$ denotes the share of firms that had status $x$ at $t=0$ and $y$ at $t=1$.
For instance, starting from being unconstrained ($u$) at $t=0$, 
$\rho_{uu}$ denotes the share of firms that continue being unconstrained, 
$\rho_{us}$ the share that move to supply-constrained ($s$) in $t=1$, and
$\rho_{ud}$ the share that change to demand-constrained ($d$).
By definition, $\sum_{x\in\{u,s,d\}}\sum_{y\in\{u,s,d\}}\rho_{xy}=1$.

\subsubsection{Absence of Wage-Setting Constraints}

We study firm responses allowing for shocks to both productivity, $\Delta\ln\Phi_j$, and labor supply, $\Delta\ln A_j$.
We formalize the absence of binding wage-setting constraints as follows.

\begin{assumption}[Absence of Wage-Setting Constraints]
    \label{assu:absence_constraints}
    \begin{assumpinline}
        \item\label{assu:absence_constraints:a} All firms in the economy are unconstrained, before and after the shock. Formally, $\rho_{uu} = 1$.
        \item\label{assu:absence_constraints:b} There are no demand-constrained firms in the economy. Formally, $\rho_{du} + \rho_{ds} + \rho_{dd} + \rho_{ud} + \rho_{sd} = 0$. 
    \end{assumpinline}
\end{assumption}

Assumption \assumpfullref{assu:absence_constraints}{a} is stronger in the sense that it imposes that all firms in the economy are unconstrained.
Assumption \assumpfullref{assu:absence_constraints}{b} is weaker as it also allows for the presence of supply-constrained firms, which is consistent with moderate wage floor levels.%
    \footnote{More precisely, Assumption \assumpfullref{assu:absence_constraints}{b} allows for a wage floor level $\W$ such that $\Phi_j<\upperPhi_j$ for some $j\in\J$ but $\Phi_j>\lowerPhi_j$ for all $j\in\J$, both before and after the shock. Notably, Assumption \assumpfullref{assu:absence_constraints}{a} implies Assumption \assumpfullref{assu:absence_constraints}{b}.}
This distinction is useful since we can allow for supply-constrained firms in some of our identification results.

The presence of labor supply shifters creates a well-known simultaneity problem: because equilibrium wages and employment are determined by these shifters, naive regressions of employment changes on wage changes, or of wage changes on value-added changes, will yield biased estimates of labor supply and rent-sharing elasticities.
To address this, we introduce the key assumptions required for identification using the demand shock $Z_j$.

\begin{assumption}[Relevance]\label{assu:relevant_shock}
    The shock $Z_j$ is correlated with changes in productivity $\Phi_j$, i.e.,  
    $\E\left[\Delta \ln \Phi_{j} | Z_j=1\right] \neq \E\left[\Delta \ln \Phi_{j} | Z_j=0\right]$.
\end{assumption}

\begin{assumption}[Exogeneity]\label{assu:labor_exogeneity}
    The shock $Z_j$ is independent of changes in the labor supply shifters $A_j$, i.e.,
    $Z_j\perp\Delta\ln A_j$.
\end{assumption}

We require the shock $Z_j$ to predict changes in productivity while remaining unrelated to changes in workers' preferences. 
We can then establish the following result.

\begin{proposition}[Identification without Wage-Setting Constraints]
    \label{prop:id_no_constraints}
    Let $Z_j$ satisfy Assumptions \ref{assu:relevant_shock} and \ref{assu:labor_exogeneity}. 
    \begin{proplist}
        \item\label{prop:id_no_constraints:eta} Under Assumption \assumpfullref{assu:absence_constraints}{b} (no demand-constrained firms), $\eta^{CE}$ identifies $\eta$.
        \item\label{prop:id_no_constraints:theta} Under Assumption \assumpfullref{assu:absence_constraints}{a} (no constrained firms), $\eta^{CE}$ identifies $\eta$ and $\theta^{CE}$ identifies $\theta$.
    \end{proplist}
\end{proposition}

Proposition \ref{prop:id_no_constraints} establishes a crucial benchmark for our analysis,
confirming that the conventional approach recovers the structural elasticities in the absence of wage-setting constraints.
We next discuss the implications of binding wage constraints.

\subsubsection{Binding Wage-Setting Constraints}
\label{sec:m_framework_ident_failure}

We now turn to the case where wage-setting constraints are present.
We formalize two assumptions on the presence of constrained firms. 

\begin{assumptionprime}[Presence of Wage-Setting Constraints]{assu:absence_constraints}\label{assu:presence_constraints}
    \begin{assumpinline}
        \item\label{assu:presence_constraints:a} (Weak Constraints) There are constrained firms in the economy. Formally, $\rho_{uu} < 1$.
        \item\label{assu:presence_constraints:b} (Strong Constraints) There are demand-constrained firms in the economy. Formally, $\rho_{du} + \rho_{ds} + \rho_{dd} + \rho_{ud} + \rho_{sd} > 0$. 
    \end{assumpinline}
\end{assumptionprime}

Assumption \ref{assu:presence_constraints} defines the regimes where wage-setting constraints are binding, indicating the negation of Assumption
\ref{assu:absence_constraints}.
We can now establish the following identification result.

\begin{proposition}[Identification Failure with Wage-Setting Constraints]
    \label{prop:id_failure}
    Let Assumptions \ref{assu:relevant_shock} and \ref{assu:labor_exogeneity} hold.
    \begin{proplist}
        \item\label{prop:id_failure:i} Under Assumption \assumpfullref{assu:presence_constraints}{b} (strong constraints), $\eta^{CE}$ does not identify $\eta$.
        \item\label{prop:id_failure:ii} Under Assumption \assumpfullref{assu:presence_constraints}{a} (weak constraints), $\theta^{\text{CE}}$ does not identify $\theta$.
    \end{proplist}
\end{proposition}

This result shows that, even in the presence of the usual IV assumptions, identification fails when constraints are present.
Proposition \propfullref{prop:id_failure}{i} follows from demand-constrained firms responding in employment but not wages, leading to the numerator in $\eta^{CE}$ being ``too large''.
We can see this as an \textit{exclusion restriction failure}. 
The employment response of demand-constrained firms is not driven by changes in the wage, but rather the demand shock directly affects the employment level.%
    \footnote{Some studies focus instead on estimating the \textit{inverse} labor supply elasticity by regressing wages on employment and instrumenting employment with a demand shifter \citep[see, e.g., ][]{AmodioEtAl2025}. 
    This approach faces a similar identification challenge: as the wage is determined by the wage constraint, the estimate of the inverse labor supply elasticity will be ``too small''. }
Proposition \propfullref{prop:id_failure}{ii} follows from the muted wage response from wage-constrained firms, which attenuates the rent-sharing coefficient to zero.

Next, we derive an explicit expression for the bias of the conventional estimand of the labor supply elasticity. We focus on the labor supply elasticity as it is directly related to the wage markdown, which is the most commonly used measure of labor market power.

\begin{corollary}[Bias Decomposition Formula]
    \label{coro:bias_formula}
    Let Assumptions \assumpfullref{assu:production_function}{a}, \ref{assu:relevant_shock}, and \ref{assu:labor_exogeneity} hold for all firm types.
    Furthermore, assume no transition across groups as a result of the shock, so $\rho_{uu}+\rho_{ss}+\rho_{dd}=1$.
    Denote the $uu$-, $ss$-, and $dd$-specific average changes in productivity $\ln\Phi$ between treated ($Z_j=1$) and control ($Z_j=0$) firms by $\zeta_{uu}$, $\zeta_{ss}$, and $\zeta_{dd}$, respectively, and further assume that  $\operatorname{sgn}(\zeta_{uu})=\operatorname{sgn}(\zeta_{ss})=\operatorname{sgn}(\zeta_{dd})$.
    Then,
    \begin{equation}\label{eq:bias_lsupply}
        \eta^{CE} - \eta 
                         = \frac{\rho_{dd}}{\rho_{uu}}
                           \left(\frac{1+\eta(1-\alpha)}{1-\alpha}\right)
                           \frac{\zeta_{dd}}{\zeta_{uu}} 
                         \ge 0.
    \end{equation}
\end{corollary}

This formula shows more explicitly that the conventional estimand $\eta^{CE}$ may be upward-biased, where the bias vanishes in the absence of demand-constrained firms ($\rho_{dd}=0$).
More complicated formulas can be derived for cases with transitions, though with a common theme: demand-constrained firms before or after the shock will respond proportionally more via employment relative to wages, resulting in ``excess'' employment responses.

Corollary \ref{coro:bias_formula} provides insights into the drivers of the bias.
First, the bias increases in $\rho_{dd}$, which is expected as demand-constrained firms respond to shocks with employment and not wages.
Second, the bias increases in $\eta$. 
A large $\eta$ makes the wage of unconstrained firms less responsive relative to employment.
Hence, the average employment response across all firms is divided by a smaller wage response, leading to a larger estimate.
Finally, the bias increases in the ratio $\zeta_{dd}/\zeta_{uu}$.
A stronger shock among demand-constrained firms would lead to more employment responses but still no wage responses, leading again to a larger estimate.

We can also use Corollary \ref{coro:bias_formula} to derive an expression for the share of demand-constrained firms as a function of $\eta^{CE}$ and $\eta$.
Assuming $\zeta_{dd}/\zeta_{uu}=1$ and solving for $\rho_{dd}$ in Equation \eqref{eq:bias_lsupply},
\begin{equation}\label{eq:bias_ratio_rhoD_rhoS}
    \rho_{dd} = \rho_{uu}(\eta^{\mathrm{CE}} - \eta) \left(\frac{1-\alpha}{1+\eta(1-\alpha)}\right) .
\end{equation}
Using this equation, we can plug in values for $\eta^{CE}$, $\eta$, and $\alpha$ to quantify the presence of demand-constrained firms, given the share of unconstrained firms $\rho_{uu}$.
Appendix \ref{asec:additional_results_local_bargaining} derives an analogous corollary and formula for the case of local bargaining constraints.

\subsubsection{Heterogeneity Analysis}

Our discussion implies that observed responses to demand shocks will systematically differ based on whether firms are affected by wage-setting constraints.
In our empirical analysis, we will split the sample into plausibly constrained and plausibly unconstrained firms and estimate labor supply and rent-sharing elasticities separately, exploring these predictions.

\subsection{Exploiting Variation in Exposure to Wage-Setting Constraints} 
\label{sec:m_framework_constraints}

Differences beyond the bite of wage constraints may also lead to heterogeneous shock responses.
For instance, high-productivity firms may face less elastic labor supply curves $\eta_j$ \citep[e.g.,][]{Kline2025}.
In turn, a negative correlation between $\eta_j$ and productivity $\Phi_j$ will result in larger observed employment-wage elasticities among constrained firms.%
    \footnote{Another possibility is that differences in product market power correlate with productivity \citep[e.g.,][]{Vera2022} and, thus, with the bite of wage-setting constraints.
    If so, firms with greater product-market power would exhibit lower rent-sharing elasticities, as they restrict output expansion to maintain high prices, thereby dampening increases in employment and wages, resulting in an \textit{underestimate} of the true role of wage-setting constraints.
    Our conclusions regarding labor supply elasticity estimates would not be affected, as high-productivity firms still operate on the labor supply curve.
    Yet another possibility is that firms face different labor adjustment costs \citep[e.g.,][]{ChanMattanaSalgadoXu2023}, limiting their labor adjustment to shocks.
    Notably, if firms facing wage-setting constraints also face higher labor adjustment costs, then the latter is expected to counteract the role of the former, which may again yield an underestimate of the true role of constraints.}
We now propose a strategy to causally test for the role of constraints in affecting firm responses to shocks, relaxing the assumption of homogeneous $\eta$ implicit in a heterogeneity analysis.

Our identification strategy relies on within-firm variation in exposure to constraints.
We rely on two sources of variation at time~$t^*$: a tightening in the wage floor affecting the constrained status of some firms and an orthogonal demand shock $Z_j$.
Denote by $\W^{\pre}$ and $\W^{\post}$ the wage floors before and after $t^*$, where $\W^{\post}>\W^{\pre}$.
Relying on pre-event firm wages $W_{j,t<t^*}$, we classify firms in three groups:
``CC'' firms are constrained both at the old and new wage floors ($W_{j,t<t^*} < \W^{\pre} < \W^{\post}$),
``UC'' firms are not constrained but would be under the new floor ($\W^{\pre} < W_{j,t<t^*} < \W^{\post}$), 
and ``UU'' are firms ``just'' unconstrained at both wage floor levels ($\W^{\pre} < \W^{\post} < W_{j,t<t^*} < (1+c)\W^{\post}$ for a small $c$).

Before presenting this new identification strategy, let us clarify the predictions of the model 
in the case of homogeneous~$\eta$.
Start with wages.
Panel (a) of Figure \ref{fig:model_outcomes_mw_change} illustrates the potential outcomes, which now depend on both productivity $\Phi$ and the wage floor $\W$.%
    \footnote{For simplicity, we hold $\aggW$ constant. In practice, we would expect a wage floor hike to affect $\aggW$. However, as this effect is common to all firms, it will be netted out when differencing $Z_j=1$ and $Z_j=0$ firms.}
Productivity $\Phi$ will not change for control firms ($Z_j=0$), and thus wages will change only for constrained firms due to the wage floor hike.
Productivity $\Phi$ will increase for firms that receive a shock ($Z_j=1$); thus, observed wage changes will reflect the demand shock and, for constrained firms, the wage floor hike.
By comparing firms with $Z_j=0$ and $Z_j=1$ within each group (CC, UC, UU), we can evaluate the group-specific responses to the demand shock controlling for the wage-floor hike.
Appendix Figure \ref{fig:model_variation_constraints_design} illustrates this procedure in an event study design.
The left column in Panels (a) and (b) shows the evolution of wages for firms with $Z_j=0$ and $Z_j=1$, respectively, whereas Panel (c) shows their difference.
Similarly, we can construct an estimate of the employment response to the shock for each group, as illustrated in Panel (b) of Figure \ref{fig:model_outcomes_mw_change} and the right column of Appendix Figure \ref{fig:model_variation_constraints_design}.

\input{figures/model_outcomes_mw_change}

Let us now describe our new strategy comparing firms for which constraints tightened (UC) to firms that remained ``just'' unconstrained (UU), again starting with wage responses.
We only assume homogeneous $\eta$ between these groups.
Firms that are always unconstrained respond the most, as the full demand shock is passed through to wages.
Firms in the UC group respond less, as the wage floor increases the wage of control firms ($Z_j=0$) in the post-period.
The assumption on $\eta$ implies that UU firms reveal the counterfactual response of UC firms had they not received the constraint, and thus the causal effect of the constraint on the wage responsiveness to shocks.
Then, if the demand shock is similarly strong across groups, we expect to find a smaller rent-sharing elasticity among UC firms.%
    \footnote{One could also compare the wage paths of UC and CC firms, maintaining the assumption of homogeneous $\eta$ between them. In this case, we expect larger wage responses in the UC group.}

Employment responses depend on the supply- and demand-constrained status of different firms.
Note that, among UC firms, there will be a mix of supply- and demand-constrained firms after the wage floor hike ($\lowerPhi^{\post}>\upperPhi^{\pre}$ is possible, as illustrated in Panel (b) of Figure \ref{fig:model_outcomes_mw_change}), implying that strong employment responsiveness is not ruled out.%
    \footnote{Appendix Figure \ref{fig:model_variation_constraints_design} illustrates an event study design with an example where we assume that the employment response for UU firms is in between the responses of CC and UC firms.}
Given this ambiguity, we let the data inform us about the strength of employment responses across groups.
To the extent that employment responses are not too different, we would expect the labor supply elasticity obtained from UC responses to be larger than the one obtained from UU responses.

This experiment reveals the causal effect of wage-setting constraints under two assumptions.
First, and most importantly, we only require that the firms being compared face the same labor supply elasticity.
To discuss the second assumption, note that this design is tantamount to a triple difference: the difference-in-differences compares shocked firms within each group over time, while the triple difference compares the groups.
As is well known in triple-differences designs, we can allow for a weaker exogeneity assumption requiring only that the relative expected $\Delta\ln A_j$ between $Z_j=1$ and $Z_j=0$ firms is equal between UU and UC firms \citep{OldenMoen2022}.
That is, we can allow the treated and control firms to be trending differently, though such ``differential trends'' should be similar across UU and UC firms.
We note, however, that we find no evidence of differential pre-trends in any group, suggesting that the shocks we use are indeed orthogonal to changes in supply shifters. 

To summarize, we use variation in wage constraints to test the mechanisms behind our model.
The idea is to compare the responsiveness to shocks of firms that experience an increase in the wage floor (UC) to firms that do not (UU).
Comparing treated ($Z_j=1$) and control ($Z_j=0$) firms \textit{within} each group differences out the effect of the wage floor increase, and comparing responsiveness to shocks \textit{across} groups reveals the effects of imposing the constraint.
This approach relies on weaker assumptions than a heterogeneity analysis across constrained and unconstrained firms, and thus provides a stronger test of our model.

%% file: figures/model_solutions.tex
\begin{figure}[p]
    \centering
    \caption{Solution to Firm Optimization Problem with Wage Floor Constraint.}
    \label{fig:model_solutions}

    \pgfmathsetmacro{\A}{1}
    \pgfmathsetmacro{\lsupelast}{2.6}
    \pgfmathsetmacro{\alphaParam}{0.48}
    \pgfmathsetmacro{\minW}{0.9}
    \pgfmathsetmacro{\phiMax}{4.4}
    \pgfmathsetmacro{\phiZero}{3.4}
    \pgfmathsetmacro{\phiOne}{1.8}
    \pgfmathsetmacro{\phiTwo}{1.3}
    \pgfmathsetmacro{\phiThree}{0.65}
    \pgfmathsetmacro{\phiMin}{0.65}
    \pgfmathsetmacro{\mkdwnfactor}{\lsupelast / (\lsupelast+1)}

    \pgfmathsetmacro{\lstarZero}{((\phiZero*\alphaParam/\A) * (\mkdwnfactor))^(\lsupelast/(1+\lsupelast*(1-\alphaParam)))}
    \pgfmathsetmacro{\wstarZero}{\A * (\lstarZero)^(1/\lsupelast)}
    \pgfmathsetmacro{\mclstarZero}{\A * ((\lsupelast+1)/\lsupelast) * (\lstarZero)^(1/\lsupelast)}
    
    \pgfmathsetmacro{\lstarOne}{(\minW/\A)^(\lsupelast)}
    \pgfmathsetmacro{\lstarOneLatent}{((\phiOne*\alphaParam/\A) * (\mkdwnfactor))^(\lsupelast/(1+\lsupelast*(1-\alphaParam)))}
    \pgfmathsetmacro{\wstarOne}{\minW}
    \pgfmathsetmacro{\wstarOneLatent}{\A * (\lstarOneLatent)^(1/\lsupelast)}
    \pgfmathsetmacro{\mclstarOne}{\A * ((\lsupelast+1)/\lsupelast) * (\lstarOne)^(1/\lsupelast)}
    \pgfmathsetmacro{\mclstarOneLatent}{\A * ((\lsupelast+1)/\lsupelast) * (\lstarOneLatent)^(1/\lsupelast)}

    \pgfmathsetmacro{\lstarTwo}{(\phiTwo*\alphaParam/\minW)^(1/(1-\alphaParam))}
    \pgfmathsetmacro{\lstarTwoLatent}{((\phiTwo*\alphaParam/\A) * (\lsupelast/(\lsupelast+1)))^(\lsupelast/(1+\lsupelast*(1-\alphaParam)))}
    \pgfmathsetmacro{\wstarTwo}{\minW}
    \pgfmathsetmacro{\wstarTwoLatent}{\A * (\lstarTwoLatent)^(1/\lsupelast)}
    \pgfmathsetmacro{\mclstarTwoLatent}{\A * ((\lsupelast+1)/\lsupelast) * (\lstarTwoLatent)^(1/\lsupelast)}
    
    \pgfmathsetmacro{\lstarThree}{(\phiThree*\alphaParam/\minW)^(1/(1-\alphaParam))}
    \pgfmathsetmacro{\lstarThreeLatent}{((\phiThree*\alphaParam/\A) * (\lsupelast/(\lsupelast+1)))^(\lsupelast/(1+\lsupelast*(1-\alphaParam)))}
    \pgfmathsetmacro{\wstarThree}{\minW}
    \pgfmathsetmacro{\wstarThreeLatent}{\A * (\lstarThreeLatent)^(1/\lsupelast)}
    \pgfmathsetmacro{\mclstarThreeLatent}{\A * ((\lsupelast+1)/\lsupelast) * (\lstarThreeLatent)^(1/\lsupelast)}

    \pgfmathsetmacro{\phiLowerThresh}{(\minW/\A)^(1+\lsupelast*(1-\alphaParam)) * \A / (\alphaParam)}
    \pgfmathsetmacro{\phiUpperThresh}{\phiLowerThresh / \mkdwnfactor}
    
    \tikzset{
        solutionplot/.style={
            scale=2.3, 
            font=\small, 
            declare function={
                invSupply(\l) = \A * (\l)^(1/\lsupelast);
                Supply(\x) = (\x/\A)^(\lsupelast);
                mrpl0(\l) = \phiZero * \alphaParam * (\l)^(\alphaParam - 1);
                mrpl1(\l) = \phiOne * \alphaParam * (\l)^(\alphaParam - 1);
                mrpl2(\l) = \phiTwo * \alphaParam * (\l)^(\alphaParam - 1);
                mrpl3(\l) = \phiThree * \alphaParam * (\l)^(\alphaParam - 1);
                mcl(\l) = \A * ((\lsupelast+1)/\lsupelast) * (\l)^(1/\lsupelast);
                optimalLstar(\phi) = ((\phi*\alphaParam/\A) * \mkdwnfactor)^(\lsupelast/(1+\lsupelast*(1-\alphaParam)));
                optimalWage(\phi) = \A * (optimalLstar(\phi))^(1/\lsupelast);
            }
        }
    }
    
    \tikzset{
        wageplot/.style={
            scale=2.35,
            font=\small,
            declare function={
                lstarPhi(\p)=((\p*\alphaParam/\A) * (\mkdwnfactor))^(\lsupelast/(1+\lsupelast*(1-\alphaParam)));
                wstarPhi(\p)=\A*(lstarPhi(\p))^(1/\lsupelast);
                lstarConstrainedPhi(\p)=(\minW/\A)^(\lsupelast);
                lstarDemandPhi(\p)=(\p*\alphaParam/\minW)^(1/(1-\alphaParam));
                lbar = (\minW/\A)^(\lsupelast);
                mrplPhi(\p) = \p * \alphaParam * (lbar)^(\alphaParam-1);
                markdownPhi(\p) = \minW / mrplPhi(\p);
                laborSharePhi(\p) = \alphaParam * markdownPhi(\p);
            }
        }
    }

    \def\xmax{2.8}
    \def\ymax{1.8}
    \def\plotdomstart{0.12}

    \begin{subfigure}[t]{0.5\textwidth}
        \centering
        \caption{Unconstrained, $\textcolor{UnconsColor}{\Phi_0}$}
        \label{fig:unconstrained}
        \begin{tikzpicture}[solutionplot]
            \draw[->, thick] (0,0) -- (\xmax, 0) node[below] {$L,H$};
            \draw[->, thick] (0,0) -- (0, \ymax) node[left] {$W$};

            \draw[dashed, color=CoolRed] (0, \minW) -- (\xmax-0.4, \minW) node[font=\small, anchor=west] {$\W$};
            \draw[domain=\plotdomstart:\xmax-0.4, smooth, variable=\l, thick] 
                plot ({\l}, {invSupply(\l)}) node[font=\small, anchor=west] {$W(H)$};
            \draw[domain=\plotdomstart:\xmax-0.7, smooth, variable=\l, thick, color=orange] 
                plot ({\l}, {mcl(\l)}) node[font=\small, anchor=west] {$MCL$};
            \draw[domain=0.85:\xmax-0.7, smooth, variable=\l, thick, color=UnconsColor]
                plot ({\l}, {mrpl0(\l)}) node[font=\small, anchor=west] {$MRPL(\Phi_0)$};
            
            \fill[black] (\lstarZero, \mclstarZero) circle (1.1pt);
            \fill[black] (\lstarZero, \wstarZero) circle (1.1pt);
            \draw[dashed] (\lstarZero, 0) node[below] {$L^*_0$} -- (\lstarZero, \mclstarZero);
            \draw[dashed] (0, \mclstarZero) -- (\lstarZero, \mclstarZero);
            \draw[dashed] (0, \wstarZero) node[left] {$W^*_0$} -- (\lstarZero, \wstarZero);
        \end{tikzpicture}
    \end{subfigure}%
    \begin{subfigure}[t]{0.5\textwidth}
        \centering
        \caption{Supply-Constrained, $\textcolor{SupConsColor}{\Phi_1}<\textcolor{UnconsColor}{\Phi_0}$}
        \label{fig:supply_constrained}
        \begin{tikzpicture}[solutionplot]
            \draw[->, thick] (0,0) -- (\xmax, 0) node[below] {$L,H$};
            \draw[->, thick] (0,0) -- (0, \ymax) node[left] {$W$};
            
            \draw[dashed, color=CoolRed] (0, \minW) -- (\xmax-0.4, \minW) node[font=\small, anchor=west] {$\W$};
            \draw[domain=\plotdomstart:\xmax-0.4, smooth, variable=\l, thick] 
                plot ({\l}, {invSupply(\l)}) node[font=\small, anchor=west] {$W(H)$};
            \draw[domain=\plotdomstart:\xmax-0.7, smooth, variable=\l, thick, color=orange] 
                plot ({\l}, {mcl(\l)}) node[font=\small, anchor=west] {$MCL$};
            \draw[domain=0.3:\xmax-0.8, smooth, variable=\l, thick, color=SupConsColor]
                plot ({\l}, {mrpl1(\l)}) node[font=\small, anchor=west]  {$MRPL(\Phi_1)$};

            \node[fill=gray, shape=diamond, minimum size=6pt, inner sep=0pt] at (\lstarOneLatent, \mclstarOneLatent) {};
            \node[fill=gray, shape=diamond, minimum size=6pt, inner sep=0pt] at (\lstarOneLatent, \wstarOneLatent) {};
            \draw[dashed, gray] (\lstarOneLatent, 0) -- (\lstarOneLatent, \mclstarOneLatent);
            \draw[dashed, gray] (0, \mclstarOneLatent) -- (\lstarOneLatent, \mclstarOneLatent);
            \draw[dashed, gray] (0, \wstarOneLatent) -- (\lstarOneLatent, \wstarOneLatent);

            \fill[black] (\lstarOne, \wstarOne) circle (1.1pt);
            \draw[dashed] (\lstarOne, 0) node[below] {$L^*_1$} -- (\lstarOne, \wstarOne);
            \draw[dashed] (0, \wstarOne) node[left] {$W^*_1$} -- (\lstarOne, \wstarOne);
        \end{tikzpicture}
    \end{subfigure} \\
    \begin{subfigure}[t]{0.5\textwidth}
        \centering
        \caption{Demand-Constrained, $\textcolor{DemConsColor}{\Phi_2}<\textcolor{SupConsColor}{\Phi_1}<\textcolor{UnconsColor}{\Phi_0}$}
        \label{fig:demand_constrained_A}
        \begin{tikzpicture}[solutionplot]
            \draw[->, thick] (0,0) -- (\xmax, 0) node[below] {$L,H$};
            \draw[->, thick] (0,0) -- (0, \ymax) node[left] {$W$};

            \draw[dashed, color=CoolRed] (0, \minW) -- (\xmax-0.4, \minW) node[font=\small, anchor=west] {$\W$};
            \draw[domain=\plotdomstart:\xmax-0.4, smooth, variable=\l, thick] 
                plot ({\l}, {invSupply(\l)}) node[font=\small, anchor=west] {$W(H)$};
            \draw[domain=\plotdomstart:\xmax-0.7, smooth, variable=\l, thick, color=orange] 
                plot ({\l}, {mcl(\l)}) node[font=\small, anchor=west] {$MCL$};
            \draw[domain=\plotdomstart+.05:\xmax-0.8, smooth, variable=\l, thick, color=DemConsColor, samples=200]
                plot ({\l}, {mrpl2(\l)}) node[font=\small, anchor=west]  {$MRPL(\Phi_2)$};

            \node[fill=gray, shape=diamond, minimum size=6pt, inner sep=0pt] at (\lstarTwoLatent, \mclstarTwoLatent) {};
            \node[fill=gray, shape=diamond, minimum size=6pt, inner sep=0pt] at (\lstarTwoLatent, \wstarTwoLatent) {};
            \draw[dashed, gray] (\lstarTwoLatent, 0) -- (\lstarTwoLatent, \mclstarTwoLatent);
            \draw[dashed, gray] (0, \mclstarTwoLatent) -- (\lstarTwoLatent, \mclstarTwoLatent);
            \draw[dashed, gray] (0, \wstarTwoLatent) -- (\lstarTwoLatent, \wstarTwoLatent);

            \fill[gray] (\lstarOne, \wstarOne) circle (1.1pt);
            \draw[dashed] (\lstarOne, 0) node[below] {$H^*_2$} -- (\lstarOne, \wstarOne);
            
            \fill[black] (\lstarTwo, \wstarTwo) circle (1.1pt);
            \draw[dashed] (\lstarTwo, 0) node[below] {$L^*_2$} -- (\lstarTwo, \wstarTwo);
            \draw[dashed] (0, \wstarTwo) node[left] {$W^*_2$} -- (\lstarTwo, \wstarTwo);
        \end{tikzpicture}
    \end{subfigure}%
    \begin{subfigure}[t]{0.5\textwidth}
        \centering
        \caption{Demand-Constrained, $\textcolor{DemConsColor}{\Phi_3}<\textcolor{DemConsColor}{\Phi_2}$}
        \label{fig:demand_constrained_B}
        \begin{tikzpicture}[solutionplot]
            \draw[->, thick] (0,0) -- (\xmax, 0) node[below] {$L,H$};
            \draw[->, thick] (0,0) -- (0, \ymax) node[left] {$W$};

            \draw[dashed, color=CoolRed] (0, \minW) -- (\xmax-0.4, \minW) node[font=\small, anchor=west] {$\W$};
            \draw[domain=\plotdomstart:\xmax-0.4, smooth, variable=\l, thick] 
                plot ({\l}, {invSupply(\l)}) node[font=\small, anchor=west] {$W(H)$};
            \draw[domain=\plotdomstart:\xmax-0.7, smooth, variable=\l, thick, color=orange] 
                plot ({\l}, {mcl(\l)}) node[font=\small, anchor=west] {$MCL$};
            \draw[domain=\plotdomstart-0.05:\xmax-1.1, smooth, variable=\l, thick, color=DemConsColor, samples=200]
                plot ({\l}, {mrpl3(\l)}) node[font=\small, anchor=west]  {$MRPL(\Phi_3)$};

            \node[fill=gray, shape=diamond, minimum size=6pt, inner sep=0pt] at (\lstarThreeLatent, \mclstarThreeLatent) {};
            \node[fill=gray, shape=diamond, minimum size=6pt, inner sep=0pt] at (\lstarThreeLatent, \wstarThreeLatent) {};
            \draw[dashed, gray] (\lstarThreeLatent, 0) -- (\lstarThreeLatent, \mclstarThreeLatent);
            \draw[dashed, gray] (0, \mclstarThreeLatent) -- (\lstarThreeLatent, \mclstarThreeLatent);
            \draw[dashed, gray] (0, \wstarThreeLatent) -- (\lstarThreeLatent, \wstarThreeLatent);
    
            \fill[gray] (\lstarOne, \wstarOne) circle (1.1pt);
            \draw[dashed] (\lstarOne, 0) node[below] {$H^*_3$} -- (\lstarOne, \wstarOne);
            
            \fill[black] (\lstarThree, \wstarThree) circle (1.1pt);
            \draw[dashed] (\lstarThree, 0) node[below] {$L^*_3$} -- (\lstarThree, \wstarThree);
            \draw[dashed] (0, \wstarThree) node[left] {$W^*_2$} -- (\lstarThree, \wstarThree);
        \end{tikzpicture}
    \end{subfigure} \\
    \begin{subfigure}[t]{0.5\textwidth}
        \centering
        \caption{Optimal Wage and Labor Input}
        \label{fig:wage_productivity}
        \begin{tikzpicture}[wageplot]
            \draw[->, thick] (0,0) -- (\xmax,0) node[below] {$\Phi$};
            \draw[->, thick] (0,0) -- (0,\ymax-0.2) node[left] {$W,L$};
    
            \foreach \p/\lab in {
                \phiMin/$\Phi_{\min}$,
                \phiLowerThresh/$\lowerPhi$,
                \phiUpperThresh/$\upperPhi$
            }{
                \draw (\p/2,0) -- ++(0,-2pt) node[below] {\lab};
                \draw[dashed, gray] (\p/2, 0) -- (\p/2, \ymax-.3);
            }
            \node[color=CoolRed, anchor=east] at (0, \minW) {$\W$};
    
            \draw[domain=\phiMin:\phiUpperThresh, smooth, variable=\p, thick, gray] 
                plot ({\p/2},{\minW*.95});
    
            \draw[domain=\phiUpperThresh:\phiMax+.05, smooth, variable=\p, thick, gray] 
                plot ({\p/2},{wstarPhi(\p)*.95});
            \node[right, font=\small, gray] at ({\phiMax/2-.1},{wstarPhi(\phiMax)*.81}) {$W^*(\Phi,\cdot)$};

            \draw[domain=\phiMin:\phiLowerThresh, smooth, variable=\p, thick, dashed] 
                plot ({\p/2},{lstarDemandPhi(\p)*1.0});
            
            \draw[domain=\phiLowerThresh:\phiUpperThresh, smooth, variable=\p, thick, dashed] 
                plot ({\p/2},{lstarConstrainedPhi(\p)*1.0});
            
            \draw[domain=\phiUpperThresh:\phiMax, smooth, variable=\p, thick, dashed] 
                plot ({\p/2},{lstarPhi(\p)*1.0});
            \node[right, font=\small] at ({\phiMax/2},{lstarPhi(\phiMax-.06)*1.0}) {$L^*(\Phi,\cdot)$};

        \end{tikzpicture}
    \end{subfigure}%
\begin{subfigure}[t]{0.5\textwidth}
        \centering
        \caption{Markdown Factor and Labor Share}
        \label{fig:markdown_productivity}
        \begin{tikzpicture}[wageplot]
            
            \def\markdownShift{1.3}
            \def\lshareShift{1.6}
            
            \draw[->, thick] (0,0) -- (\xmax,0) node[below] {$\Phi$};
            \draw[->, thick] (0,0) -- (0,\ymax-0.2) node[left] {$\tilde{\mu}$,$\tilde{s}_L$};

            \foreach \p/\lab in {
                \phiMin/$\Phi_{\min}$,
                \phiLowerThresh/$\lowerPhi$,
                \phiUpperThresh/$\upperPhi$
            }{
                \draw (\p/2,0) -- ++(0,-2pt) node[below] {\lab};
                \draw[dashed, gray] (\p/2, 0) -- (\p/2, \ymax-.3);
            }

            
            \draw[dotted, gray] (0, 1*\markdownShift) -- (\xmax, 1*\markdownShift);
            \node[anchor=east, font=\small] at (0, 1*\markdownShift) {$1$};
            
            \draw[dotted, gray] (0, \mkdwnfactor*\markdownShift) -- (\xmax, \mkdwnfactor*\markdownShift);
            \node[anchor=east, font=\small] at (0, \mkdwnfactor*\markdownShift) {$\mu$};

            \draw[thick, color=black] (\phiMin/2, 1*\markdownShift) -- (\phiLowerThresh/2, 1*\markdownShift);

            \draw[domain=\phiLowerThresh:\phiUpperThresh, smooth, variable=\p, thick, color=black, samples=200] 
                plot ({\p/2}, {markdownPhi(\p)*\markdownShift});

            \draw[thick, color=black] (\phiUpperThresh/2, \mkdwnfactor*\markdownShift) -- (\phiMax/2+.2, \mkdwnfactor*\markdownShift);
            
            \node[above, font=\scriptsize] at (\phiMax/2+.13, \mkdwnfactor*\markdownShift) {Markdown Factor};
            
            
            \draw[dotted, gray] (0, \alphaParam*\lshareShift) -- (\xmax, \alphaParam*\lshareShift);
            \node[anchor=east, font=\small, color=blue] at (0, \alphaParam*\lshareShift) {$\alpha$};
            
            \draw[dotted, gray] (0, \alphaParam*\mkdwnfactor*\lshareShift) -- (\xmax, \alphaParam*\mkdwnfactor*\lshareShift);
            \node[anchor=east, font=\small, color=blue] at (0, \alphaParam*\mkdwnfactor*\lshareShift) {$\alpha\mu$};

            \draw[thick, color=blue] (\phiMin/2, \alphaParam*\lshareShift) -- (\phiLowerThresh/2, \alphaParam*\lshareShift);

            \draw[domain=\phiLowerThresh:\phiUpperThresh, smooth, variable=\p, thick, color=blue, samples=200] 
                plot ({\p/2}, {laborSharePhi(\p)*\lshareShift});

            \draw[thick, color=blue] (\phiUpperThresh/2, \alphaParam*\mkdwnfactor*\lshareShift) -- (\phiMax/2+.2, \alphaParam*\mkdwnfactor*\lshareShift);
            
            \node[below, font=\scriptsize, color=blue] at (\phiMax/2, \alphaParam*\mkdwnfactor*\lshareShift) {Labor Share};

        \end{tikzpicture}
    \end{subfigure}

    \vspace{2mm}
    \begin{singlespace}
    \begin{minipage}{1\textwidth} \footnotesize\singlespacing
        Notes: 
        The figure illustrates the equilibrium of the model of firm maximization subject to a labor supply and a wage floor constraint introduced in Section \ref{sec:m_framework_theory}.
        Panel (a) shows an unconstrained firm for which the wage floor is not binding.
        Panel (b) shows a supply-constrained firm for which both constraints bind.
        Panels (c) and (d) show demand-constrained firms for which only the wage floor is binding, and which differ in their latent monopsony equilibrium, indicated by gray-filled diamonds.
        $W(H)$ stands for the inverse labor supply curve, $MCL$ for the marginal cost of labor, and $MRPL$ for the marginal revenue product of labor.
        Panel (e) shows the optimal wage (solid line) and labor input (dashed line) as functions of productivity $\Phi$, whereas Panel (f) shows the observed markdown factor $\tilde{\mu}$ and labor share $\tilde{s}_L$.   
        $\Phi_{\min}$ indicates the minimum productivity, $\mu=\eta/(\eta+1)$ denotes the unconstrained markdown factor, and $\alpha$ is the output elasticity of labor with a homogeneous production technology.
    \end{minipage}
    \end{singlespace}
\end{figure}

%% file: figures/model_outcomes_mw_change.tex
\begin{figure}[t!]
    \centering
    \caption{Potential Outcomes Following an Increase in the Wage Floor.}
    \label{fig:model_outcomes_mw_change}

    \pgfmathsetmacro{\A}{1}
    \pgfmathsetmacro{\eta}{1.5}
    \pgfmathsetmacro{\alpha}{0.72}
    \pgfmathsetmacro{\mu}{\eta / (\eta+1)}
    
    \pgfmathsetmacro{\phiMax}{4.3}
    \pgfmathsetmacro{\phiMin}{0.4}
    
    \pgfmathsetmacro{\minWpre}{0.75} 
    \pgfmathsetmacro{\minWpost}{1.2} 

    \pgfmathsetmacro{\phiLowerThreshPre}{(\minWpre/\A)^(1+\eta*(1-\alpha)) * \A / (\alpha)}
    \pgfmathsetmacro{\phiUpperThreshPre}{\phiLowerThreshPre / \mu}
    
    \pgfmathsetmacro{\phiLowerThreshPost}{(\minWpost/\A)^(1+\eta*(1-\alpha)) * \A / (\alpha)}
    \pgfmathsetmacro{\phiUpperThreshPost}{\phiLowerThreshPost / \mu}
    
    \pgfkeys{/pgf/declare function={
        lstarPhi(\p) = ((\p*\alpha/\A) * \mu)^(\eta/(1+\eta*(1-\alpha)));
        wstarPhi(\p) = \A*(lstarPhi(\p))^(1/\eta);
        lstarConstrainedPhiPre(\p)=(\minWpre/\A)^(\eta);
        lstarDemandPhiPre(\p)=(\p*\alpha/\minWpre)^(1/(1-\alpha));
        lstarConstrainedPhiPost(\p)=(\minWpost/\A)^(\eta);
        lstarDemandPhiPost(\p)=(\p*\alpha/\minWpost)^(1/(1-\alpha));
    }}

    \pgfmathsetmacro{\lstarSupConsPre}{lstarPhi(\phiUpperThreshPre)}
    \pgfmathsetmacro{\lstarSupConsPost}{lstarPhi(\phiUpperThreshPost)}
    
    \tikzset{
        unifiedplot/.style={
            scale=2.5,
            font=\small,
        }
    }
    
    \def\xmax{2.8}
    \def\ymax{1.95}
    \def\labelY{1.8} 

    \begin{subfigure}[t]{0.5\textwidth}
        \centering
        \caption{Wage}
        \begin{tikzpicture}[unifiedplot]
            \path[use as bounding box] (-.35,-.2) rectangle (\xmax, 1.05*\ymax);
            \fill[CoolRed, opacity=0.12] (\phiMin*.6, 0) rectangle (\phiUpperThreshPre*.6, .98*\ymax);
            \node[CoolRed] at ({\phiMin*.6 + (\phiUpperThreshPre - \phiMin)*.5*.6}, \labelY) {$CC$};

            \fill[CoolMix, opacity=0.12] (\phiUpperThreshPre*.6, 0) rectangle (\phiUpperThreshPost*.6, .98*\ymax);
            \node[CoolMix] at ({\phiUpperThreshPre*.6 + (\phiUpperThreshPost - \phiUpperThreshPre)*.5*.6}, \labelY) {$UC$};

            \fill[CoolBlue, opacity=0.12] (\phiUpperThreshPost*.6, 0) rectangle (.98*\xmax, .98*\ymax);
            \node[CoolBlue] at ({\phiUpperThreshPost*.6 + (\xmax/.6 - \phiUpperThreshPost)*.5*.6 - .02}, \labelY) {$UU$};
            
            \draw[->, thick] (0,0) -- (\xmax,0) node[below] {$\Phi$};
            \draw[->, thick] (0,0) -- (0,\ymax) node[left] {$W^*(\Phi)$};

            \draw (\phiMin*.6, -2pt) node[below] {$\Phi_{\min}$} -- (\phiMin*.6, 2pt);
            \draw[densely dotted, gray] (\phiUpperThreshPre*.6, 0) node[below] {$\upperPhi^{\pre}$} -- (\phiUpperThreshPre*.6, \minWpre);
            \draw[densely dotted] (\phiUpperThreshPost*.6, 0) node[below] {$\upperPhi^{\post}$} -- (\phiUpperThreshPost*.6, \minWpost);

            \draw[domain=\phiMin:\phiUpperThreshPre, smooth, variable=\p, thick, gray]
                plot ({\p*.6}, {\minWpre});
            \draw[domain=\phiUpperThreshPre:\phiMax, smooth, variable=\p, thick, gray]
                plot ({\p*.6}, {wstarPhi(\p)});
            \node[below, font=\small, gray] at (.875*\phiMax*.6, 1.02*\minWpost) {$W^*(\Phi; \W^{\pre})$};
            \node[anchor=east] at (0, \minWpre) {$\W^{\pre}$};

            \draw[domain=\phiMin:\phiUpperThreshPost, smooth, variable=\p, thick, densely dashed, black]
                plot ({\p*.6}, {\minWpost});
            \draw[domain=\phiUpperThreshPost:\phiMax, smooth, variable=\p, thick, densely dashed, black]
                plot ({\p*.6}, {1.008*wstarPhi(\p)});
            \node[above, font=\small, black] at (.67*\phiMax*.6, 1.13*\minWpost) {$W^*(\Phi; \W^{\post})$};
            \node[anchor=east] at (0, \minWpost) {$\W^{\post}$};

        \end{tikzpicture}
    \end{subfigure}%
    \begin{subfigure}[t]{0.5\textwidth}
        \centering
        \caption{Employment}
        \begin{tikzpicture}[unifiedplot]
            \path[use as bounding box] (-.35,-.2) rectangle (\xmax, 1.05*\ymax);
            \fill[CoolRed, opacity=0.12] (\phiMin*.6, 0) rectangle (\phiUpperThreshPre*.6, .98*\ymax);

            \fill[CoolMix, opacity=0.12] (\phiUpperThreshPre*.6, 0) rectangle (\phiUpperThreshPost*.6, .98*\ymax);

            \fill[CoolBlue, opacity=0.12] (\phiUpperThreshPost*.6, 0) rectangle (.98*\xmax, .98*\ymax);
            
            \draw[->, thick] (0,0) -- (\xmax,0) node[below] {$\Phi$};
            \draw[->, thick] (0,0) -- (0,\ymax) node[left] {$L^*(\Phi)$};
    
            \draw (\phiMin*.6, -2pt) node[below] {$\Phi_{\min}$} -- (\phiMin*.6, 2pt);
            
            \draw[dotted, gray] (\phiLowerThreshPre*.6, 0) -- (\phiLowerThreshPre*.6, .9*\ymax) node[above, gray, xshift=6pt] {$\lowerPhi^{\pre}$};
            \draw[densely dotted, gray] (\phiUpperThreshPre*.6, 0) node[below, gray] {$\upperPhi^{\pre}$} -- (\phiUpperThreshPre*.6, .9*\lstarSupConsPre);
            
            \draw[dotted, black] (\phiLowerThreshPost*.6, 0) -- (\phiLowerThreshPost*.6, .9*\ymax) node[above, black, xshift=6pt] {$\lowerPhi^{\post}$};
            \draw[densely dotted, black] (\phiUpperThreshPost*.6, 0) node[below, black] {$\upperPhi^{\post}$} -- (\phiUpperThreshPost*.6, .9*\lstarSupConsPost);
            
            \draw[domain=\phiMin:\phiLowerThreshPre, smooth, variable=\p, thick, gray]
                plot ({\p*.6},{lstarDemandPhiPre(\p)*.9});
            
            \draw[domain=\phiLowerThreshPre:\phiUpperThreshPre, smooth, variable=\p, thick, gray]
                plot ({\p*.6},{lstarConstrainedPhiPre(\p)*.9});
            
            \draw[domain=\phiUpperThreshPre:\phiMax, smooth, variable=\p, thick, gray]
                plot ({\p*.6},{lstarPhi(\p)*.9});
            \node[below, font=\small, gray] at (.87*\phiMax*.6, .91*\lstarSupConsPost) {$L^*(\Phi; \W^{\pre})$};

            \draw[domain=\phiMin:\phiLowerThreshPost, smooth, variable=\p, thick, densely dashed, black]
                plot ({\p*.6},{lstarDemandPhiPost(\p)*.9});
            
            \draw[domain=\phiLowerThreshPost:\phiUpperThreshPost, smooth, variable=\p, thick, densely dashed, black]
                plot ({\p*.6},{lstarConstrainedPhiPost(\p)*.9});
            
            \draw[domain=\phiUpperThreshPost:\phiMax, smooth, variable=\p, thick, densely dashed, black]
                plot ({\p*.6},{1.008*lstarPhi(\p)*.9}) ;
            \node[anchor=east] at (1.0*\phiMax*.6, 1.36*\lstarSupConsPost) {$L^*(\Phi; \W^{\post})$};

        \end{tikzpicture}
    \end{subfigure}
    
    \vspace{2mm}
    \begin{singlespace}
    \begin{minipage}{1\textwidth} \footnotesize\singlespacing
        Notes: 
        The figure illustrates the potential outcomes following an increase in the wage floor from 
        $\W^{\pre}$ to $\W^{\post}$ at time $t^*$.
        Panels (a) and (b) show the potential outcomes for the wage and employment, respectively.
        The shaded colors highlight firms constrained before and after $t^*$ (CC), those unconstrained before but constrained after the increase (UC), and those unconstrained at either wage floor (UU).
    \end{minipage}
    \end{singlespace}
\end{figure}

%% file: m_data.tex
\subsection{Institutional Settings}
\label{subsec:m_setting_data_setting}

Portugal has a long-standing system of sectoral bargaining and a national minimum wage (MW). 
Although union membership is only around 10\%, CBA coverage reached roughly 90\% of workers between 2010 and 2013 \citep{addison2023union}.
The reason is that sectoral wage floors, negotiated between employer associations and sectoral unions, are automatically extended to non-covered firms and non-affiliated workers. 
The national MW sets a floor for negotiated floors, though in practice negotiated wage floors are usually close to the MW (Appendix Figure~\ref{fig:cba_rel_floor_cushion_hist}). 
The MW has become increasingly binding over time: the share of MW workers rose from 8\% in 2006 to 24\% in 2017 (Appendix Figure~\ref{fig:mw_workers_time}).

In Norway, approximately half of private-sector workers are covered by CBAs, and, unlike Portugal, there is no national MW. 
The system is two-tiered, featuring centralized sectoral negotiations that determine CBA wage floors and general wage increments, followed by firm-level bargaining between employers and local unions \citep{BhullerEtAl2022}. 
The sectoral agreements are not automatically extended to all firms; thus, both covered and non-covered firms are present within narrowly defined industries.
However, since 2004, the national labor board has allowed for limited extensions of collective agreements to non-covered firms, reaching around 10\% of private-sector workers in 2023.

Another common institutional arrangement corresponds to a high national MW and limited collective bargaining.
In Appendix~\ref{asec:colombia}, we conduct our analyses for one such country: Colombia.
We obtain similar results for Colombia, though the data only cover a sample of manufacturing firms and are thus less representative, and our estimates are less precise.

The wage-setting practices prevalent in these countries represent institutional settings that are widespread across the globe. 
As shown in Appendix Figure~\ref{fig:institutions_world}, most OECD and non-OECD countries in the ICTWSS database \citep{OECDVisser2023} can be categorized into one of the wage-setting systems represented by the countries we consider.
Although we emphasize that the examples of wage-setting constraints that we study are prevalent around the globe, we do not argue that our evidence is directly applicable to other contexts.

\subsection{Data Sources}

For Portugal, our main data source is \textit{Quadros de Pessoal} (QP): a matched employer-employee dataset that captures information on all private-sector employers each October. 
We link QP to the \textit{Sistema de Contas Integradas das Empresas} (SCIE) dataset, thereby obtaining information on value added.
We add annual imports and exports data at the HS6 product–destination-country level from customs data. 
The resulting dataset covers the years 2004 through 2017. 

For Norway, our main data source is population-wide matched employer-employee registers (Amelding/ATMLTO), which cover all employers, both in the private and public sectors.
We link the employer-employee registers to firms' balance sheet data, and add information at the firm level on CBA coverage and labor union density.%
    \footnote{We define a firm as being covered by a CBA if the firm or one of its establishments has adopted a sectoral CBA, which was negotiated between a union confederation and an employer association, or if any of the workers in the firm were covered by a policy-based extension of wage floors in sectoral CBAs. We define labor union density as the share of a firm's workers that are formally members of a labor union.} 
Customs data deliver information on international trade, containing exported values at the 8-digit HS product code and destination country level.
We study private sector firms from 1997 through 2019.

We define our main variables as follows.
The firm wage equals the mean hourly wage of ``stayers'', i.e., workers that remain in the firm before and after the shock, though we also provide evidence for all incumbent workers, including ``movers''. 
Employment is simply the count of wage workers in the year.
Value added equals revenues minus costs of materials and services purchased from other firms, and it includes profits, labor costs, and taxes.

We use publicly available bilateral trade flow data from the BACI-CEPII dataset to measure import demand worldwide \citep{gaulier2010baci}. 
This provides annual traded values between country pairs using the HS6 classification of products.

Appendix \ref{asec:data} provides further details and sample restrictions.

\subsection{Measuring Demand Shocks}

We construct two firm-level demand-shock measures: an internal instrument based on changes in firm value added and an external instrument based on changes in international demand for firms' exports.

\subsubsection{Internal Shock}

We construct an internal demand-shock measure following the logic of \citet{LMS2022}. 
Observed changes in firm-level value added reflect both persistent changes in productivity or demand and transitory fluctuations, including measurement error. 
As the productivity changes are more persistent than the measurement error, short-run value-added growth contains information whose effects remain after the transitory fluctuations have dissipated. 
Changes in outcomes over long windows around a short-run value-added increase can thus reveal responses to the persistent component of firm performance.

The identifying variation comes from relating short-run value-added growth to changes in outcomes over a longer window.
Interpreting this variation as a labor-demand shift additionally requires excluding correlated changes in firm amenities or other determinants of labor supply. 
We assess this assumption using changes in outcomes before the event.

In practice, we define a demand shock indicator $Z^{\text{Int}}_{j}$ equal to one for firms with above-median growth in log value added between years $-1$ and $0$, within the same country, 2-digit industry, and event-year cohort.
Then, we compare long differences in outcomes for firms with $Z^{\text{Int}}_{j}=1$ and $Z^{\text{Int}}_{j}=0$ within each event-year.
Unlike the export-demand instrument, this measure can be constructed for all firms, including lower productivity firms that are more likely to face wage-setting constraints.

\subsubsection{External Shock}

We complement the internal instrument with an export-demand shock based on changes in international trade around the Great Recession. 
Our instrument is
$Z^{\text{Ext}}_j=\sum_{p\in\CPset}s_{jp}\Delta_p, $
where $s_{jp}$ is the share of firm $j$'s exports destined for country-product pair $p$ during 2005--2007, indicating $j$'s exposure to the country-product. 
Using BACI trade data \citep{gaulier2010baci}, we measure the change in international demand $\Delta_p$ as the symmetric growth rate of imports into that destination-product market between 2006--2007 and 2009--2010, excluding trade originating in the country under study \citep{garin2024responsive}.%
    \footnote{Trade values are deflated using the US CPI-U.
    We winsorize $Z^{\text{Ext}}_j$ at the 5th and 95th percentiles.} 

Our export shock has a shift-share structure, with its exogenous variation arising from multiple country-product shocks \citep{BorusyakEtAl2022}.
The key identification assumption is that the changes in international demand are uncorrelated with underlying trends in mean firm performance across firms exposed to each country-product.
Since the disruptions in international demand arose from the Great Recession, we argue that they are exogenous to each firm, making this assumption plausible \citep[see discussion in][]{garin2024responsive}.

\subsection{Classifying Firm Exposure to Wage-Setting Constraints}
\label{subsec:data_constr}

\subsubsection{Baseline Definition for Heterogeneity Analyses}
\label{subsec:data_constr_baseline}

Our framework motivates two empirical measures of exposure to wage-setting constraints: proximity to a wage floor and coverage by a particular wage-setting institution. 
These measures identify firms as plausibly constrained, helping us bring our stylized model to the data, without directly distinguishing the model's three regimes.

For Portugal, we classify firms as plausibly constrained if their mean wage was within 15 percent of the national MW prior to the shock, and unconstrained otherwise.%
    \footnote{More precisely, we use firms' average base wages of all incumbent workers.}
We assess the robustness of our findings by varying the 15\% threshold.
Additionally, we approximate CBA wage floors following \citet{cardoso2005contractual} and explore a constrained classification relative to the CBA wage floor, yielding similar results.%
    \footnote{Results are similar because CBA wage floors are tightly concentrated around the MW. 
    Panel (a) of Appendix Figure~\ref{fig:cba_rel_floor_cushion_hist} shows the wage-floor distribution, and Panel (b) shows the distribution of firm-level wage cushions \citep{CardCardoso2022}.
    Appendix \ref{asec:data} discusses our approach to approximating wage floors.}

In Norway, our baseline analysis classifies firms that prior to the shock were exposed to a sectoral CBA---either through formal adoption or policy-based extensions---as plausibly constrained and the remainder as unconstrained. 
CBA-covered firms face a wage floor and are thus more likely to be demand-constrained.
To study the role of local bargaining constraints, we split CBA-covered firms---which are more likely to face local unions---into groups with above- and below-median pre-shock union density.
High-union-density firms are more likely to be demand-constrained because they face a union with stronger bargaining power.

\subsubsection{Variation in Exposure to Wage-Setting Constraints}
\label{subsec:data_var_constr}

We present a dynamic definition of exposure to study the effects of imposing constraints on firms' responsiveness to shocks.
To do so, we exploit changes in Portugal's national MW between 2015 and 2017.
Following Section~\ref{sec:m_framework_constraints}, we split firms into five groups based on their mean base wage-to-MW ratio averaged over 2014--2015 ($R_j$), relative to the 2015, 2016, and 2017 MW levels, also expressed as a ratio with respect to the 2015 MW ($m_t$).%
    \footnote{We define $R_j=\frac{1}{2}\sum_{t=2014}^{2015}\bar{W}^{\mathrm{base}}_{jt}/MW_t$, where $\bar{W}^{\mathrm{base}}_{jt}$ is the mean monthly base wage of eligible full-time workers.
    We express the MW cutoffs relative to 2015 as $m_t = MW_t^{\mathrm{real}}/MW_{2015}^{\mathrm{real}}$, using real MW levels.}
Appendix Figure~\ref{fig:firm_wages_portugal} shows the firm wage distribution and illustrates our definition.

The first group, which we denote as SC, consists of strongly constrained firms with a baseline wage ratio no more than 1\% above the 2015 MW
($R_j \leq 1.01$).
They comprise \VcShareSC\% of firms in 2015.
The second group, denoted UCC, includes firms above the SC cutoff but no more than 2.5\% above the 2016 MW
($1.01 < R_j \leq \delta m_{2016}$) and represents \VcShareUCC\% of firms.%
    \footnote{We set $\delta = 1.025$. The 2.5\% margin is about half of the average nominal MW increase in each hike between 2014 and 2017 (equal to 4.7\%).}
The third group, denoted UUC, contains firms above the 2016 cutoff but no more than 2.5\% above the 2017 MW
($\delta m_{2016} < R_j \leq \delta m_{2017}$) and represents \VcShareUUC\% of firms.
The fourth group, denoted UUU, comprises firms above the 2017 cutoff but no more than 15\% above the 2017 MW
($\delta m_{2017} < R_j \leq 1.15 m_{2017}$).
This ``just'' unconstrained group accounts for \VcShareUUU\% of the sample.
The remaining \VcShareSU\% of firms, denoted SU, are strongly unconstrained firms with baseline wage ratios more than 15\% above the 2017 MW
($R_j > 1.15 m_{2017}$).

\subsection{Descriptive Statistics}

Table~\ref{tab:lms_descriptives} presents pre-shock descriptive statistics for firm-cohorts for which we define the internal instrument, separately for all firms and by constrained status.
Appendix Table \ref{tab:cohort_size} shows the number of firms we use in each cohort.
In Portugal, constrained firms have lower mean wages, employment, and value added, as expected for firms paying close to the MW. 
In Norway, the pattern is reversed: constrained firms are larger and have higher value added, while mean wages are very similar across groups.
This aligns with the institutional setting, where CBAs are more common among more productive firms. 

\input{tables/lms_descriptives}

We provide descriptive context for our external shock in Appendix Table~\ref{tab:exporter_descriptives}.
Exporting firms account for 8 percent and 9 percent of firms in Portugal and Norway, respectively. 
Value added, employment, and wages are higher among exporting firms. 
Appendix Figure \ref{fig:exports_growth_countries} shows the evolution of aggregate exports around the Great Recession.
Portugal experienced a decline in 2009, followed by a relatively rapid recovery, similar to the EU and US.
By comparison, Norway faced a smaller though more persistent decline.

%% file: tables/lms_descriptives.tex
\begin{table}[hbt!]
    \centering
    \caption{Descriptive Statistics in the Pre-Period by Country and Constrained Status.}
    \label{tab:lms_descriptives}
    
    \begin{tabular}{@{}lcccccc@{}}
        \toprule
        & \multicolumn{3}{c}{Portugal} & \multicolumn{3}{c}{Norway} \\
        \cmidrule(lr){2-4} \cmidrule(lr){5-7}
        &  & Con- & Uncon- &  & Con- & Uncon- \\
        & All & strained & strained & All & strained & strained \\
        \cmidrule(lr){2-7}
        & (1) & (2) & (3) & (4) & (5) & (6) \\
        \midrule
        \textit{Number of firms} & 83,054 & 34,456 & 63,674 & 68,484 & 23,270 & 54,845 \\
        \textit{Number of firm-cohorts} & 370,957 & 109,448 & 261,478 & 517,400 & 149,111 & 368,184 \\
        \hspace{3mm} (Share constrained) &   & (0.30) & (0.70) &   & (0.29) & (0.71) \\
        \addlinespace[1mm]
        \textit{Log value added} & & & & & & \\
        \hspace{3mm} Mean & 12.12 & 11.36 & 12.43 & 15.42 & 16.09 & 15.14 \\
        \hspace{3mm} SD & 1.30 & 0.93 & 1.30 & 1.23 & 1.39 & 1.04 \\
        \hspace{3mm} N & 370,952 & 109,445 & 261,476 & 517,400 & 149,111 & 368,184 \\
        \addlinespace[1mm]
        \textit{Log Employment} & & & & & & \\
        \hspace{3mm} Mean & 2.14 & 1.81 & 2.28 & 2.31 & 2.98 & 2.03 \\
        \hspace{3mm} SD & 1.05 & 0.84 & 1.10 & 1.09 & 1.25 & 0.88 \\
        \hspace{3mm} N & 370,957 & 109,448 & 261,478 & 517,400 & 149,111 & 368,184 \\
        \addlinespace[1mm]
        \textit{Log Wages} & & & & & & \\
        \hspace{3mm} Mean & 1.64 & 1.30 & 1.79 & 5.37 & 5.38 & 5.37 \\
        \hspace{3mm} SD & 0.41 & 0.21 & 0.39 & 0.35 & 0.27 & 0.37 \\
        \hspace{3mm} N & 370,956 & 109,448 & 261,477 & 517,400 & 149,111 & 368,184 \\
        \addlinespace[1mm]
        \bottomrule
    \end{tabular}

    \vspace{2mm}
    \begin{minipage}{1\textwidth} \footnotesize
        Notes: 
        The table shows descriptive statistics for the internal instrument sample.
        Constrained firms pay an average wage within 15\% of the national MW for Portugal, and are covered by a CBA in Norway.
        The statistics are computed as the average over the periods $-3$ and $-2$, weighted by the number of firms in the cohort.
        The sum of constrained and unconstrained firm-cohorts does not equal the total because constraint status is missing for a small number of observations.
        Hourly wages use stayers (workers who stayed in the firm for at least 7 years around the shock), and are
        measured in Euros and Norwegian Kroner, respectively.
    \end{minipage}
\end{table}

%% file: m_emp_strategy.tex
This section presents the empirical strategy we use to study the responsiveness to demand shocks and the role of wage-setting constraints, building on Section \ref{sec:m_framework}.

\subsection{Baseline Specification Exploiting Demand Shocks} 

Let $L_{jt}$, $W_{jt}$, and $\VA_{jt}$ denote the level of employment, wages, and value added, respectively, of firm $j$ in year $t$. 
We consider the following set of reduced-form equations:
\begin{subequations}
    \label{eq:reduced_forms}
    \begin{align}
        \ln L_{jt} & = \beta_L \ Z_j \ \text{Post}_t + \delta_{k(j)t}^{L} + \omega_{j}^{L} + v_{jt}^{L}, \label{eq:rf_l} \\
        \ln W_{jt} & = \beta_W \ Z_j \ \text{Post}_t + \delta_{k(j)t}^{W} + \omega_{j}^{W} + v_{jt}^{W}, \label{eq:rf_w} \\
        \ln \VA_{jt} & = \beta_{\VA} \ Z_j \ \text{Post}_t + \delta_{k(j)t}^{\VA} + \omega_{j}^{\VA} + v_{jt}^{\VA}, \label{eq:rf_va}
    \end{align}
\end{subequations}
where $Z_j$ is the demand shock measure (either the internal or external instrument), $\text{Post}_t$ is an indicator for the post-shock period, $k(j)$ is the local labor market of $j$, $\omega_j^Y$ are firm fixed effects, and $v_{jt}^{Y}$ are residuals, for $Y\in\{W, L, \VA\}$. 
The coefficients $\beta_L$, $\beta_W$, and $\beta_{\VA}$ capture the effects of the shock on employment, wages, and value added. 
The fixed effects $\delta_{k(j)t}^Y$ control for common changes for all firms in a local labor market, such as those driven by the wage index $\aggW$ in the theoretical model.
The firm fixed effects $\omega_j^Y$ control for time-invariant differences across firms, such as those driven by different baseline amenity values $A_j$ in the labor supply case, or productivity levels $\Phi_j$ in the wage case.

The conventional estimands of labor supply and rent-sharing elasticities are constructed as the ratios of the reduced-form coefficients.
Specifically, we estimate $\eta^{CE}=\beta_{L}/\beta_{W}$ and $\theta^{CE}=\beta_{W}/\beta_{\VA}$.
This is equivalent to an instrumental variables (IV) approach, where the shock $Z_{j}$ instruments for the wage change $\Delta \ln W_{j}$ in the labor supply case, or the change in value added $\Delta \ln \VA_{j}$ in the rent-sharing case.
For this approach to be valid, the standard IV assumptions must hold.
Relevance requires a nonzero first-stage effect (either $\beta_W\neq 0$ or $\beta_{\VA}\neq 0$), which maps to Assumption \ref{assu:relevant_shock}. 
The shock $Z_j$ must be uncorrelated with unobserved determinants of employment, when we estimate the labor supply elasticity, or wages, when we estimate the rent-sharing elasticity.
In our framework, this relates to Assumption \ref{assu:labor_exogeneity}.
As discussed in Section \ref{sec:m_framework_ident}, however, exclusion may not hold, leading to an identification failure.

We estimate reduced-form effects using the dynamic specification
\begin{equation}
    \label{eq:dynamic_did}
    \ln Y_{jt}  = \sum_{\tau\in \mathcal{T}} \beta_{Y,\tau} \ Z_j \ \mathbf{1}\{\tau = t - t^*\} 
           + \delta^Y_{k(j)t} + \omega^Y_j + v^Y_{jt},
\end{equation}
where $Y\in\{W, L, \VA\}$ is the outcome, $\tau$ indexes relative years and $\mathcal{T}$ is the set of relative years included, $\mathbf{1}\{\tau = t - t^*\}$ is an indicator for years relative to the event year $t^*$, $v^Y_{jt}$ is the residual, and $\omega^Y_j$ and $\delta^Y_{k(j)t}$ are defined as before.
In practice, we define the local labor market $k$ as the 2-digit sector reported by the firm in year $t^*-1$.
The coefficients $\{\beta_{Y,\tau}\}_{Y\in\{W, L, \VA\},\tau\in \mathcal{T}}$ measure the dynamic response of the outcomes to each instrument.
In particular, pre-shock coefficients allow us to assess the parallel trends assumption underlying our design.

The reduced-form parameters $\{\beta_L, \beta_W, \beta_{\VA}\}$ are defined as the mean post-period coefficient from estimates of \eqref{eq:dynamic_did}, and the elasticities are computed as $\eta^{CE}=\beta_{L}/\beta_{W}$ and $\theta^{CE}=\beta_{W}/\beta_{\VA}$. 
Standard errors for the elasticities are estimated with the delta method.

\paragraph{Internal shock.}

Our dataset is indexed by firm $j$, event-year cohort $c$, and calendar year $t$.
Within sector $k$ and cohort $c$, $Z_{jc}^{\text{Int}}$ equals one for firms with above-median value-added growth from relative year $-1$ to $0$.
We estimate a version of Equation \eqref{eq:dynamic_did} where we control for firm-by-cohort and cohort-by-sector-by-year fixed effects.
For each cohort $c$, we keep years from $c-4$ to $c+3$ so that $\mathcal{T}=\{-4,...,3\}$.
We omit the category $-2$ following \citet{LMS2022}.
Reduced-form effects are defined as the mean of coefficients for relative years 1, 2, and 3, and standard errors are clustered at the two-digit industry level.

\paragraph{External shock.}

The export shock $Z_j^{\text{Ext}}$ is defined as the mean growth in world import demand for each firm $j$, so we fit Equation \eqref{eq:dynamic_did} in the sample of exporting firms for which the shock can be defined.
The event year $t^*$ is set to 2007. 
As our data for Portugal start in 2004, we set $\mathcal{T}=\{-3,...,4\}$ excluding $-1$.
Reduced-form effects are defined as the mean of coefficients for years 2009, 2010, and 2011, and standard errors are clustered at the firm level.
As the sample of firms is significantly smaller, we expect the estimates to be less precise.

\paragraph{Heterogeneity.}

We split the sample between ``plausibly constrained'' and ``plausibly unconstrained'' firms, following Section \ref{subsec:data_constr}, and estimate Equation \eqref{eq:dynamic_did} for each sample.
Briefly, in Portugal, we classify firms that pay close to the MW as constrained, whereas in Norway, we classify firms covered by a sectoral CBA as constrained. Within the group of CBA firms in Norway, we further split the sample into high- and low-union-density firms.

\subsection{Specification Exploiting Variation in Constraints} 
\label{subsec:emp_var_constr}

This section presents a design to test the causal effects of wage-setting constraints on the responsiveness to demand shocks, building on Section \ref{sec:m_framework_constraints}.
The setting is the increase in the MW between 2015 and 2017 in Portugal.
We classify firms into five groups based on the changing ``bite'' of constraints using 2014--2015 data: SC (constrained at any MW level), UCC (unconstrained at the 2015 MW but constrained at the 2016 MW), UUC (unconstrained at the 2016 MW but constrained at the 2017 MW), UUU (``just'' unconstrained at the 2017 MW), and SU (remaining firms).
We define an internal instrument using value-added growth between 2015 and 2016, and study differential shock responsiveness across these groups.%
    \footnote{A concern with these definitions is that the wage bill mechanically increased for groups that became constrained in 2016 (SC and UCC), increasing both their workers' wages and their value added, making it more likely they are classified as ``treated'' by the internal instrument.
    However, we find no evidence of differential effects of the internal instrument on value added within each group.}

The following specification estimates the effect of the shock~$Z_j$ \textit{within} different groups:
\begin{equation}
    \label{eq:wage_effect_group_G}
    \ln Y_{jt} = \beta^G_{Y} \ Z_j \ \text{Post}_t 
           + \delta_{k(j)t}^{Y,G} + \omega_j^{Y,G} + v_{jt}^{Y,G} 
\end{equation}
where $G\in\{ {SC} , {UCC} , {UUC} , {UUU} , {SU} \}$ denotes the groups and $Y\in\{W, L, \VA\}$ denotes the outcomes, and the other objects are defined as before.
The coefficients $\beta^G_{W}$, $\beta^G_{L}$, and $\beta^G_{\VA}$ give the group-specific response of wages, employment, and value added to the shock.

We can then compare the shock responsiveness across groups.
While such comparisons are straightforward given group-specific reduced-form estimates, we implement them formally in a triple-differences specification comparing, say, UCC and UUU firms:
\begin{equation}
    \label{eq:wage_effect_triple_diff}
    \begin{split}        
        \ln Y_{jt} & = \beta^{UCC,UUU}_{Y} \ Z_j \ \text{Post}_t \ \mathbf{1} \{G_j=\text{UCC}\} 
                 + \beta^{UUU}_{Y} \ Z_j \ \text{Post}_t \\
               & + \gamma_{Y} \ \text{Post}_t \ \mathbf{1} \{G_j=\text{UCC}\} + \lambda_{Y} \ Z_j \ \mathbf{1} \{G_j=\text{UCC}\}  \\
               & + \delta_{k(j)t}^{Y} + \omega_j^{Y} + v_{jt}^{Y} .
    \end{split}
\end{equation}
Let us unpack \eqref{eq:wage_effect_triple_diff} step by step.
The baseline category, with $Z_j = 0$ and $G_j\neq\text{UCC}$, corresponds to UUU firms that have not been shocked. 
Thus, $\beta^{UUU}_{Y}$ gives the effect of the shock $Z_j$ among UUU firms.
Setting $G_j=\text{UCC}$ with $Z_j=0$ instead, we note that $\gamma_{Y}$ gives the differential evolution of the outcome for UCC firms \textit{without} a shock. 
This coefficient effectively captures the MW hike.
Finally, setting $Z_j=1$ reveals that the effect of the shock among UCC firms is $\beta^{UUU}_{Y} + \beta^{UCC,UUU}_{Y}$.
Thus, the triple-differences coefficient $\beta^{UCC,UUU}_{Y}$ gives the \textit{differential} response to the shock among UCC firms relative to UUU firms.%
    \footnote{We expect the following signs for the reduced-form triple-differences coefficients.
    First, we expect $\beta^{UCC,UUU}_{W} < 0$ as the theoretical model predicts muted wage pass-through for constrained firms.
    Second, the model's prediction for employment is ambiguous; thus, we let the data inform us about $\beta^{UCC,UUU}_{L}$.
    Finally, by construction of the internal shock, we expect $\beta^{UCC,UUU}_{\VA}$ to be close to zero.}

We construct elasticities from these reduced-form effects.
The rent-sharing elasticity, for example, can be constructed as $(\beta^{UUU}_{W} + \beta^{UCC,UUU}_{W}) / (\beta^{UUU}_{\VA} + \beta^{UCC,UUU}_{\VA})$ for UCC firms and $\beta^{UUU}_{W} / \beta^{UUU}_{\VA}$ for UUU firms.
Following our theoretical framework, we expect larger labor supply and smaller rent-sharing elasticities for UCC than UUU firms.
We construct elasticities for all five groups to test whether a more intense ``treatment'' (i.e., being constrained by the MW for a longer period) leads to a stronger attenuation of the wage response.

The interpretation that this design reveals the causal effect of wage constraints on firms' responsiveness relies on two assumptions about the two groups being compared.
First, they should have a similar structural labor supply elasticity.
Second, other determinants of shock responsiveness should evolve in parallel across the two groups.
Under these assumptions, the group that did not experience a change in constraints is a valid counterfactual for the group that did.
As discussed in Section \ref{sec:m_framework_constraints}, these assumptions are weaker than the ones underlying the heterogeneity analysis between constrained and unconstrained firms.

%% file: m_evidence.tex
This section presents our empirical results and interprets them using our framework. 
First, we provide descriptive evidence on the wage-productivity relationship for each country. 
Next, we show responses to demand shocks and obtain implied elasticities across all firms and by a firm's constrained status.
Finally, we study within-firm variation in the exposure to constraints in Portugal and variation in union density across firms in Norway.

\subsection{Descriptive Evidence}
\label{subsec:desc}

Our theoretical framework predicts that the relationships of both wages and employment with productivity should be kinked in the presence of wage-setting constraints.%
    \footnote{This can be seen in Panel (e) of Figure~\ref{fig:model_solutions} for the wage floor constraint and Panel (d) of Appendix Figure~\ref{fig:model_solutions_local_bargaining} for the local bargaining constraint.}
As shown in Figure~\ref{fig:hockey_stick_country}, the firm-level relationships of wages and employment with value added are indeed kinked in both Portugal and Norway.%
    \footnote{Similarly, in Appendix Figure \ref{fig:colombia_wages_empl_VA}, we find evidence of a kink for Colombian manufacturing firms.}
Similar evidence on the ``hockey-stick-like'' relationship between firm productivity and wages is found in several countries \citep[see, e.g.,][p.\ 121]{Kline2024}.
Interpreting this evidence as direct evidence in favor of the model has an important caveat:
the model predictions require holding parameters other than productivity fixed.
The cross-sectional relationships may confound differences in firm productivity and unobserved correlates (e.g., amenities).%
    \footnote{Recently, \cite{BassierBudlender2025} estimate the theoretically implied thresholds in the cross-sectional wages-productivity relationship within bargaining councils in South Africa, using estimated productivity from a production-function approach.
    Then, they use these thresholds to classify firms as minimum-wage-constrained or unconstrained.
    Since in our framework such thresholds are firm-specific, varying with unobserved amenities and production technologies, the mapping to firms' constraint status is less clear.
    Our classifications use observable exposure measures and do not require estimating productivity thresholds.}

\input{figures/hockey_sticks}

\subsection{Firm Responses to Demand Shocks}
\label{subsec:resu_base}

We now present our main evidence showing how firms respond to demand shocks.

\input{figures/baseline_estimates}

The reduced-form wage and employment effects reveal significant differences in responsiveness across countries.
Figure~\ref{fig:baseline_estimates} reports the results for both the internal (left column) and external (right column) instruments discussed in Section \ref{sec:m_emp_strategy}.
Panels (a) and (b) show that both shocks predict changes in value added, showing that the shocks are relevant. 
Panels (c) and (d) show that wage responses appear stronger in Norway.%
    \footnote{To mitigate concerns of compositional changes, we measure mean wages only for stayers. 
    We also provide evidence on mean wage impacts for incumbents in Appendix Figure \ref{fig:lms_stayers_incumbents}, where we allow incumbents to potentially work in a different firm after the shock. Wage impacts are similar for incumbents and stayers.}
Panels (e) and (f) show that employment responses are similar in Norway and Portugal with the internal shock, and with the external shock, they appear larger for Norway.
We find pre-treatment coefficients that are mostly insignificant or close to zero, supporting the parallel trends assumption.%
    \footnote{For the internal shock design, we find evidence of mean reversion, which is expected given how the demand shock is defined \citep[see discussions in][]{LMS2022}. However, trends remain stable at the start (for $\tau<-1$) and end (for $\tau>1$) of the time window, consistent with the parallel trends assumption.} 


The reduced-form responses result in different implied elasticities across countries.
To compute these elasticities we obtain $\{\beta_L,\beta_W,\beta_{\VA}\}$ by averaging the post-period coefficients and compute $\eta^{CE}=\beta_L/\beta_W$ and $\theta^{CE}=\beta_W/\beta_{\VA}$. 
Columns (1) and (4) of Table \ref{tab:lms_evidence} show the estimates for the internal shock design.
The implied labor supply elasticities are $\PtLSelastAllBeta$ ($\text{SE}=\PtLSelastAllSE$) in Portugal and $\NoLSelastAllBeta$ ($\text{SE}=\NoLSelastAllSE$) in Norway.
The implied rent-sharing elasticities are $\PtRentShAllBeta$ ($\text{SE}=\PtRentShAllSE$) in Portugal and $\NoRentShAllBeta$ ($\text{SE}=\NoRentShAllSE$) in Norway. 
The differences in implied elasticities are consistent with the theoretical implications of key differences in institutional settings. 
As discussed in Section \ref{sec:m_setting_data}, the national MW is high and applies to all workers in Portugal, whereas in Norway around 50\% of private sector workers are not covered by CBAs and thus not affected by wage floors. 
Although other explanations can account for differences in estimated elasticities across Portugal and Norway, we view this as evidence motivating the within-country heterogeneity analysis in the next sections.%
    \footnote{We also examine the cross-country relationship between labor supply elasticities, from \citet{AmodioEtAl2024}, and the share of workers covered by collective bargaining. 
    Appendix Figure~\ref{fig:global_ls_elast_cba_cov} shows a positive relationship between CBA coverage and estimated labor supply elasticities, consistent with our framework.}

\input{tables/lms_evidence}

\subsection{The Role of Wage-Setting Constraints}
\label{subsec:resu_byconst}

We now examine differences in responsiveness to demand shocks depending on firms' exposure to wage-setting constraints.
Following Section \ref{sec:m_framework}, we expect the observed labor supply elasticities to be larger for constrained firms, and the rent-sharing elasticities to be smaller.

\paragraph{Reduced-form evidence.}

We find significant differences in shock responses across plausibly constrained (C) and unconstrained (U) firms using the internal shock design.
Figure~\ref{fig:lms_by_constraints} presents reduced-form wage and employment evidence by firms’ constrained status. 
In line with our theoretical predictions, wage responses are smaller among constrained firms. 
Table~\ref{tab:lms_evidence} shows significant differences in wage responses for both Portugal (C$=\PtWageConsBeta$ vs.\ U$=\PtWageUnconsBeta$) and Norway (C$=\NoWageConsBeta$ vs.\ U$=\NoWageUnconsBeta$).
For Norway, employment responses are somewhat larger for constrained firms. 
The heterogeneity in shock responses does not reflect the intensity of shocks: Appendix Figure \ref{fig:va_by_country} shows similar value-added responses across groups.

In our model, wages do not respond to marginal demand shocks among firms constrained by wage floors, yet we find positive wage effects among constrained firms in Portugal.
This may reflect both our proxy classification and transitions out of constrained status: some classified firms may already be unconstrained, while others may become so following a positive shock.
Consistent with this interpretation, Appendix Figure~\ref{fig:elast_vary_threshold} shows that defining constrained firms more narrowly around the wage floor yields smaller wage responses.
Moreover, Appendix Figure~\ref{fig:lms_by_constraints_portugal_combined} shows that classifying firms relative to CBA wage floors yields results similar to the MW-based definition.%
    \footnote{We find no evidence that the average wage floors $\W_j$ that CBA firms face change in response to firm-level demand shocks (see Appendix Figure \ref{fig:wage_floor_effects}), consistent with such floors being externally determined.}

\input{figures/lms_by_constraints}

\paragraph{Implied elasticities.}

The elasticities we estimate differ significantly by constrained status within each country.
Figure~\ref{fig:elast_const_internal} presents the results visually, and Table~\ref{tab:lms_evidence} shows exact estimates.
In Portugal, labor supply elasticities are higher for constrained (C$=\PtLSelastConsBeta$) than for unconstrained firms (U$=\PtLSelastUnconsBeta$). 
A similar pattern holds in Norway (C$=\NoLSelastConsBeta$ vs.\ U$=\NoLSelastUnconsBeta$).
For the implied rent-sharing elasticities, constrained firms exhibit smaller values in Portugal (C$=\PtRentShConsBeta$ vs.\ U$=\PtRentShUnconsBeta$) and Norway ($C=\NoRentShConsBeta$ vs.\ $U=\NoRentShUnconsBeta$). 
Labor supply elasticities for constrained firms are $\PtLSelastConsVsUncons\%$ larger in Portugal and $\NoLSelastConsVsUncons\%$ larger in Norway, whereas rent-sharing elasticities among constrained firms are $\PtRentShConsVsUnconsAbs\%$ smaller in Portugal and $\NoRentShConsVsUnconsAbs\%$ smaller in Norway.
Wage-setting constraints mute wage responses, making constrained firms appear to face a more elastic labor supply, seemingly close to the competitive benchmark.
Using the external shock design, our findings broadly align with the internal shock.%
    \footnote{Appendix Figures \ref{fig:export_shock_by_constraints} and \ref{fig:elast_const_external} present reduced-form effects and implied elasticities from the external instrument design.
    The elasticities are noisier because the sample is smaller. 
    For Norway, the results by constrained status align with the internal shock design. 
    For Portugal, the confidence intervals for the wage reduced-form responses include zero, and VA responses for constrained firms are not statistically significant.}

\paragraph{The ``bias'' in conventional estimands.}

Interpreting the heterogeneity in light of Proposition~\ref{prop:id_failure}, we compare the elasticity obtained using all firms, corresponding to the conventional estimand, to the elasticity using unconstrained firms, corresponding to the target parameter when elasticities are homogeneous across firms. 
Using all firms results in a labor supply elasticity up to $\MaxPctDiffUnconsVsAllLSelast\%$ higher, and a rent-sharing elasticity up to $\MaxPctDiffUnconsVsAllRentSh\%$ lower.

\paragraph{Demand-constrained share.}

Following Corollary~\ref{coro:bias_formula}, we quantify the share of demand-constrained firms.
We set $\eta$ as the unconstrained labor supply elasticity and  $\eta^{CE}$ as the value for all firms.
The unconstrained shares come from Table~\ref{tab:lms_descriptives}, and we set $\zeta_{dd}/\zeta_{uu}=\zeta_{ss}/\zeta_{uu}=1$ given similar VA effects across groups.
We estimate a demand-constrained share of $\ImpliedDemConsSharePt$ in Portugal and $\ImpliedDemConsShareNo$ in Norway. These imply that around 13\% ($0.039/0.30\approx0.13$) and 20\% ($0.059/0.29\approx0.20$) of plausibly constrained firms are demand-constrained, respectively.

\input{figures/elast_const_internal}

\subsection{Combining Demand Shocks with Variation in Constraints}  
\label{subsec:variation_constraints}

We provide evidence from two complementary analyses that combine demand shocks with variation in constraints. 
First, building on the design that exploits within-firm variation in constraints (Section \ref{sec:m_framework_constraints}), we use MW hikes in Portugal to study how responsiveness to demand shocks changes as firms become constrained, helping distinguish the causal effects of constraints from other differences across firms. 
Second, following our local bargaining model (Section \ref{sec:m_framework_theory}), we study differential shock responsiveness for Norwegian firms facing stronger local unions, relying on cross-sectional variation in baseline labor union densities.

\subsubsection{Evidence from Minimum Wage Changes in Portugal}

We exploit Portugal's MW increases to study changes in firms' exposure to wage-setting constraints.
The monthly MW rose from 485 euros to 505 in October 2014, 530 in 2016, and 557 in 2017, a cumulative nominal increase of 14.8\%.
We classify firms into five groups---SC, UCC, UUC, UUU, and SU---using mean base wage-to-MW ratios averaged over 2014--2015 (Section~\ref{subsec:data_var_constr}).
Appendix~\ref{asec:triple_did} shows that wages rose more at firms initially closer to the MW and provides a step-by-step illustration of our triple-differences design.

\input{figures/tripleDiff_portugal_wage}

\paragraph{Reduced-form evidence.}

We combine demand shocks with differential exposure to MW hikes to implement the design from Section~\ref{subsec:emp_var_constr}, using a balanced panel of firms between 2012 and 2017. 
We define an internal instrument using value-added growth between 2015 and 2016. 
We quantify the effects of imposing wage-setting constraints by comparing the responses to demand shocks across our five firm types.

Imposing MW constraints mutes the wage responses to shocks.
Panel (a) of Figure~\ref{fig:tripleDiff_portugal_wage} shows that firms that are constrained at the 2015, 2016, and 2017 MW (the SC group) show almost no response in wages.
On the other hand, firms that are unconstrained at the 2017 MW (the UUU and SU groups) show large and significant increases in wages due to the demand shock.
Panel (b) shows differences in wage responses relative to the ``just'' unconstrained group, UUU.
The estimates show a clear pattern: the longer the exposure to the MW constraint, the smaller the wage response to the demand shock. 
Appendix Figure~\ref{fig:tripleDiff_portugal_emp_VA} shows very similar effects across groups for employment and value added, with one exception: the employment responsiveness to shocks among SC firms is somewhat smaller.

Appendix Figure~\ref{fig:tripleDiff_portugal_var_thresh} provides a placebo test that shifts the definition of UCC and UUU firms to values farther away from the MW, preserving their relative distance.
The estimated wage-response gap fluctuates around zero farther above the actual MW thresholds, supporting the interpretation that wage-setting constraints drive the wage-effect attenuation.

\input{tables/tripleDiff_implied_elasticities_portugal}

\paragraph{Implied elasticities.}

Greater exposure to MW constraints raises the implied labor supply elasticities and lowers the rent-sharing elasticities, and these differences are statistically significant.
Table~\ref{tab:tripleDiff_implied_elasticities_portugal} shows the results.
Firms constrained at the 2015 MW (SC) have a labor supply elasticity of \VcLSelastSC\ and a rent-sharing elasticity of \VcRentShSCBeta\, compared with \VcLSelastUUU\ and \VcRentShUUUBeta\ for ``just'' unconstrained firms (UUU).
The differences are driven by muted wage responses, equal to \VcWageSCPct\% in SC firms and \VcWageUUUPct\% in UUU firms ($p$-value of \VcPvalWageSCvsUUU), while employment responds strongly in all groups.
The differences in inverse labor supply and rent-sharing elasticities are also significant ($p$-values of \VcPvalInvLSelastSCvsUUU\ and \VcPvalRentShSCvsUUU, respectively).%
    \footnote{Because the labor supply estimate divides by a near-zero wage response, the standard errors are unreliable.
    Thus, we perform inference over the inverse labor supply elasticity instead.}

The estimates change gradually with exposure to new constraints.
The wage-to-MW ratio in 2014--15 ranges from \VcRatioMWSC\ in SC firms to \VcRatioMWUUU\ in UUU firms.
Fitting a line through the SC, UCC, UUC, and UUU estimates against these means, we find that a 0.1 increase in the ratio (10\% of the MW) raises the wage effect by \VcSlopeWageBeta\ and the rent-sharing elasticity by \VcSlopeRentShBeta.
The labor supply elasticity declines by $\VcSlopeLSelastBeta$, while its inverse increases by $\VcSlopeInvLSelastBeta$.

Strongly unconstrained firms (SU) are statistically indistinguishable from UUU firms.
This indicates limited heterogeneity in responsiveness across unconstrained firms, supporting the view that constraints may explain the heterogeneity analyses in Section~\ref{subsec:resu_byconst}.

\subsubsection{Evidence from Variation in Local Union Strength in Norway}

We study heterogeneity in responses to demand shocks across high- and low-union-density firms within the set of constrained---CBA-covered---firms in Norway. 
In our model, firms facing strong local unions are more likely to be demand-constrained, all else equal.
We thus expect firms with higher baseline union density to exhibit a higher ratio of employment to wage responses and a lower ratio of wage to value added responses.%
    \footnote{When comparing firms with a large difference in union density, we expect the high-density group to have stronger unions and, thus, a mix of supply- and demand-constrained firms showing low wage responsiveness, and the low-density group to have a smaller union presence, and thus unconstrained firms with high wage responsiveness.
    See Appendix \ref{asec:derivations_local_bargaining} for related model derivations and, in particular, Panel (d) of Appendix Figure \ref{fig:model_solutions_local_bargaining} showing how wages and employment relate to productivity.}
To interpret the evidence on differential shock responsiveness across firms with varying local union strength as reflecting the differential bite of local bargaining constraints, we must assume that the underlying structural labor supply elasticity $\eta$ is the same within the group of CBA-covered firms, which is still weaker than the assumption maintained in Section \ref{subsec:resu_byconst}. 

\input{tables/norway_union_share}

Table \ref{tab:norway_union_share} shows that, although baseline wage levels are higher in high-union firms, the estimated labor supply elasticity is larger and the estimated rent-sharing elasticity is smaller. 
We interpret this evidence as showing that firms with differential exposure to local bargaining constraints also have heterogeneous shock responses.
These findings align with our theoretical model predictions for local bargaining constraints described above.%
    \footnote{While this evidence shows heterogeneity in responsiveness to demand shocks across firms with differential exposure to local bargaining constraints, we do not interpret it as causal evidence on the role of local union density as such. 
    In recent work, \cite{dodini2026} provide causal evidence using changes in the tax deductibility of union dues as sources of exogenous variation in union density. 
    Their findings show that, in the average private-sector firm, higher union density raises labor costs and lowers employment and profits.} 

\paragraph{Demand-constrained share.}

We can further quantify the share of demand-constrained firms due to local bargaining constraints, based on the evidence in Table~\ref{tab:norway_union_share} and using the formula from Appendix \ref{asec:additional_results_local_bargaining}.
We use the ``Constrained'' estimate as $\eta^{CE}$, the ``Low Local Union Share'' estimate as $\eta$, and values of 0.2 and 0.8 for the worker share $\lambda$. 
The implied shares of demand-constrained firms are $\ImpliedDemConsShareNoLocBargLambdaLow$ for $\lambda=0.2$ and $\ImpliedDemConsShareNoLocBargLambdaHigh$ for $\lambda=0.8$.

%% file: figures/hockey_sticks.tex
\begin{figure}[hbt!]
    \centering
    \caption{Relationship Between Firm Value Added and Average Wages and Employment.}
    \label{fig:hockey_stick_country}

    \begin{subfigure}{0.5\textwidth}
        \centering
        \caption{Portugal}
        \includegraphics[width=\textwidth]{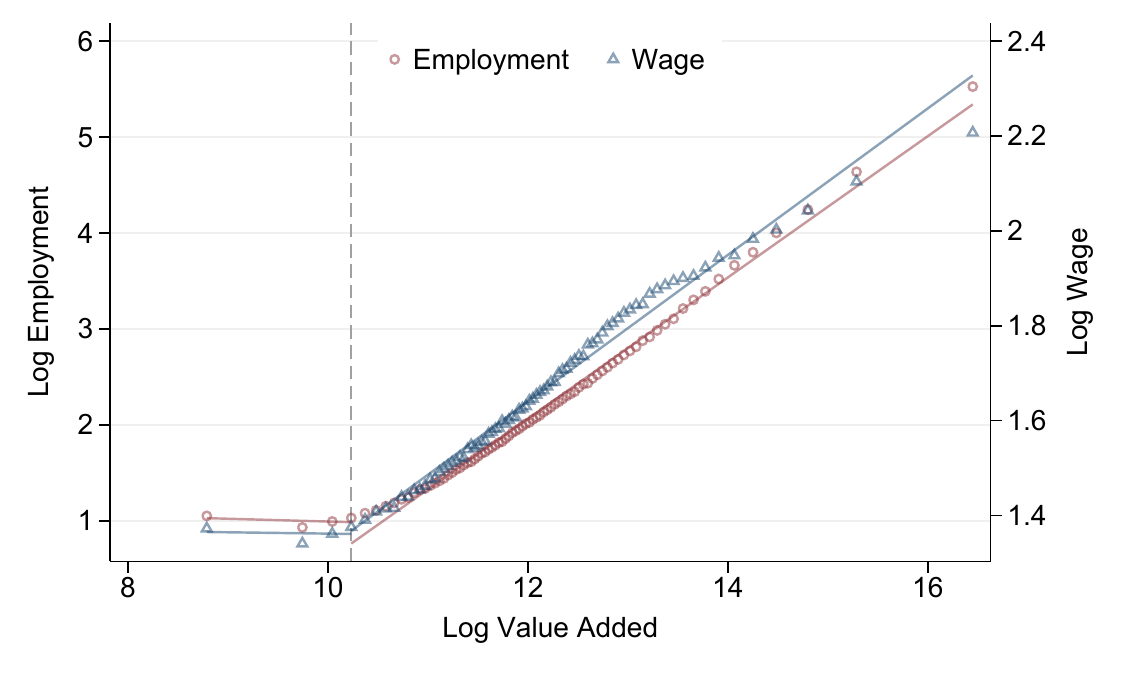}
    \end{subfigure}%
    \begin{subfigure}{0.5\textwidth}
        \centering
        \caption{Norway}
        \includegraphics[width=\textwidth]{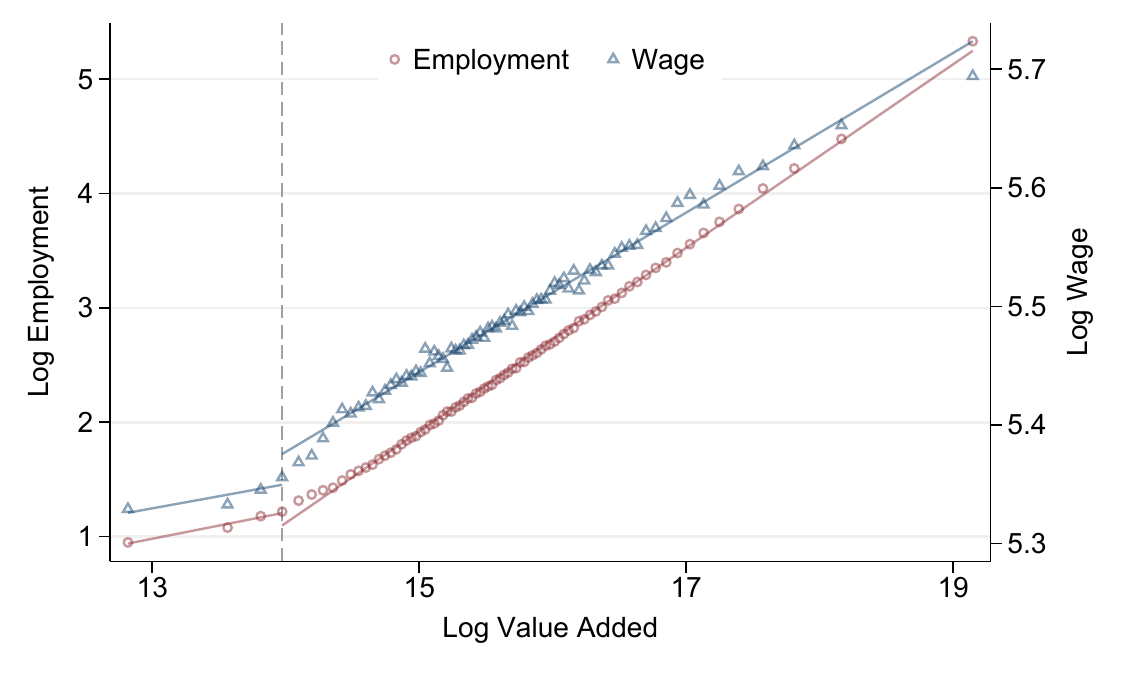}
    \end{subfigure} \\

    \vspace{2mm}
    \begin{singlespace}
    \begin{minipage}{1\textwidth} \footnotesize
        Notes: The figure shows the relationships of log wages and log employment with log value added, for Portugal (Panel a) and Norway (Panel b).
        In both countries, variables are residualized with respect to two-digit industry and year fixed effects, and in Norway, we further control for firm-level union density and collective bargaining agreement coverage dummies, interacted with dummies for exposure to sectoral wage floor extensions.
        For Portugal, we restrict the sample to the 2012 cohort, while for Norway, we use observations from 2015 onward.
        We construct 80 equal-sized bins of log value added and compute means of log employment and log wages within each bin.
        Solid lines correspond to linear fits on the binned data.
    \end{minipage}
    \end{singlespace}
\end{figure}

%% file: figures/baseline_estimates.tex
\begin{figure}
    \centering
    \caption{Firm Responses to Demand Shocks: Internal and External Shocks.}
    \label{fig:baseline_estimates}

    \begin{subfigure}{.48\textwidth}
        \centering
        \caption{Internal Shock: Value Added}
        \includegraphics[width=\textwidth]{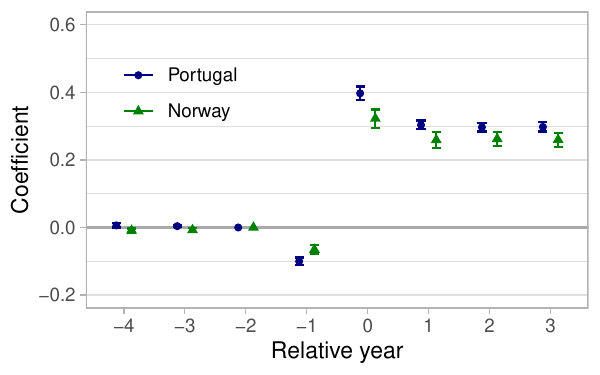}
    \end{subfigure}
    \hfill
    \begin{subfigure}{.48\textwidth}
        \centering
        \caption{External Shock: Value Added}
        \includegraphics[width=\textwidth]{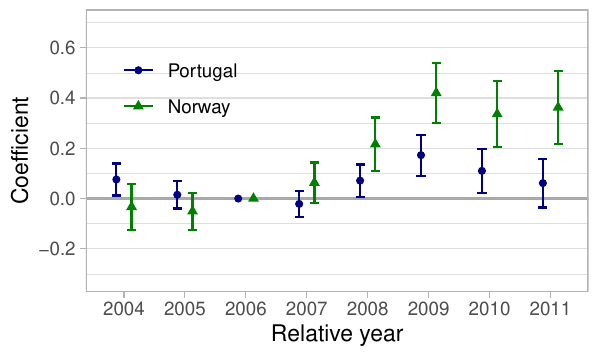}
    \end{subfigure}

    \begin{subfigure}{.48\textwidth}
        \centering
        \caption{Internal Shock: Mean Wage}
        \includegraphics[width=\textwidth]{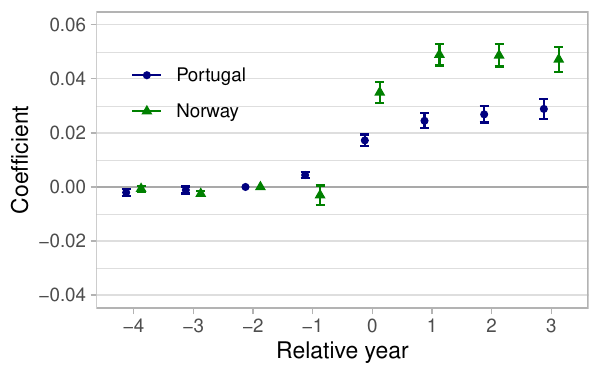}
    \end{subfigure}
    \hfill
    \begin{subfigure}{.48\textwidth}
        \centering
        \caption{External Shock: Mean Wage}
        \includegraphics[width=\textwidth]{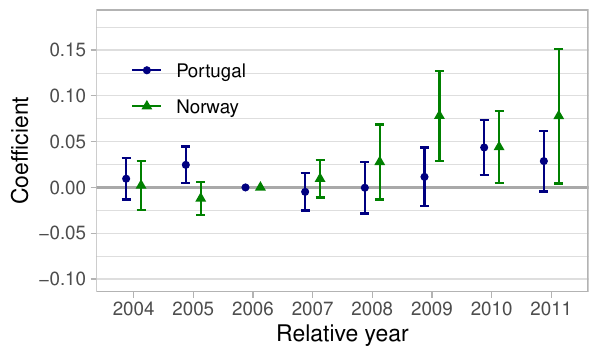}
    \end{subfigure}

    \begin{subfigure}{.48\textwidth}
        \centering
        \caption{Internal Shock: Employment}
        \includegraphics[width=\textwidth]{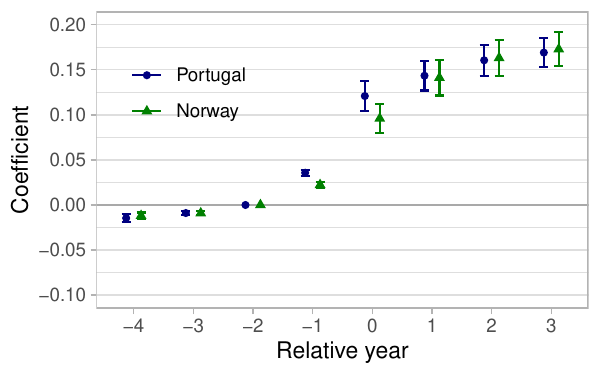}
    \end{subfigure}
    \hfill
    \begin{subfigure}{.48\textwidth}
        \centering
        \caption{External Shock: Employment}
        \includegraphics[width=\textwidth]{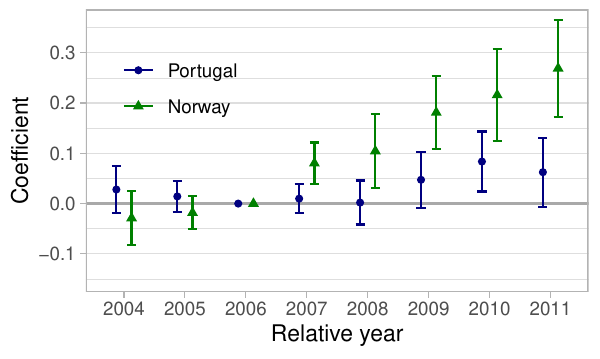}
    \end{subfigure}

    \vspace{2mm}
    \begin{singlespace}
    \begin{minipage}{1\textwidth} \footnotesize
        Notes:
        The figure shows estimates of the effects of a demand shock, using the internal and external shock designs, on firm outcomes.
        Panels (a) and (b) show log value added, Panels (c) and (d) show the log mean wage, and Panels (e) and (f) show log employment, for the internal and external shock, respectively.
        For the internal shock, we estimate the effect of a treatment indicator for having value-added growth between periods $-1$ and $0$ above the median within the firm's 2-digit industry.
        For the external shock, we estimate the effect of a change in the mean world import demand using a shift-share strategy, where the shares correspond to 2005--2007 value, and the shifts represent the symmetric growth rate in world import demand for each country-product between 2006--2007 and 2009--2010. 
        We restrict to firms exporting for at least three years up to 2007, with exports exceeding 1\% of their revenue in 2005–2007, and, for Portugal, we further exclude firms that primarily export to Angola or Spain.
    \end{minipage}
    \end{singlespace}
\end{figure}

%% file: tables/lms_evidence.tex
\begin{table}[hbt!]
    \centering
    \caption{Firm Responses to Demand Shocks and Estimated Elasticities.}
    \label{tab:lms_evidence}
    
    \begin{tabular}{@{}lcccccc@{}}
        \toprule
        & \multicolumn{3}{c}{Portugal} & \multicolumn{3}{c}{Norway} \\
        \cmidrule(lr){2-4} \cmidrule(lr){5-7}
        &  & Con- & Uncon- &  & Con- & Uncon- \\
        & All & strained & strained & All & strained & strained \\
        \cmidrule(lr){2-7}
        & (1) & (2) & (3) & (4) & (5) & (6) \\
        \midrule
        \textbf{(a)} Estimated effects & & & & & & \\[1mm]
        \hspace{3mm} Employment & 0.1579 & 0.1479 & 0.1617 & 0.1593 & 0.1716 & 0.1525 \\
         & (0.0084) & (0.0109) & (0.0083) & (0.0099) & (0.0070) & (0.0109) \\
        \hspace{3mm} Wage & 0.0268 & 0.0184 & 0.0299 & 0.0483 & 0.0323 & 0.0547 \\
         & (0.0016) & (0.0015) & (0.0016) & (0.0021) & (0.0022) & (0.0022) \\
        \hspace{3mm} Value added & 0.2995 & 0.2852 & 0.3046 & 0.2605 & 0.2521 & 0.2624 \\
         & (0.0066) & (0.0095) & (0.0071) & (0.0109) & (0.0090) & (0.0128) \\
        \textbf{(b)} Elasticities & & & & & & \\[1mm]
        \hspace{3mm} Labor Supply & 5.90 & 8.06 & 5.41 & 3.29 & 5.31 & 2.79 \\
         & (0.39) & (0.87) & (0.31) & (0.23) & (0.36) & (0.18) \\
        \hspace{3mm} Rent-Sharing & 0.089 & 0.064 & 0.098 & 0.185 & 0.128 & 0.208 \\
         & (0.005) & (0.005) & (0.005) & (0.007) & (0.012) & (0.008) \\
        \bottomrule
    \end{tabular}

    \vspace{2mm}
    \begin{minipage}{1\textwidth} \footnotesize
        Notes:
        The table shows results from the internal shock design, by constrained status of firms. 
        Constrained firms pay an average wage within 15\% of the national MW for Portugal, and are covered by a CBA in Norway.
        Each reduced-form estimate is computed as the mean among relative times 1 through 3.
        The labor supply elasticity is defined as the ratio of employment to wage effects, and the rent-sharing elasticity as the ratio of wage to value added effects.
        Standard errors clustered at the two-digit industry level are reported in parentheses.
        The standard errors of the elasticities are computed using the delta method.
    \end{minipage}
\end{table}

%% file: figures/lms_by_constraints.tex
\begin{figure}[h!]
    \centering
    \caption{Firm Responses to Demand Shocks: Exposure to Wage-Setting Constraints.}
    \label{fig:lms_by_constraints}
    
    \begin{subfigure}{.5\textwidth}
        \centering
        \caption{Portugal: Mean Wage}
        \includegraphics[width=.99\textwidth]{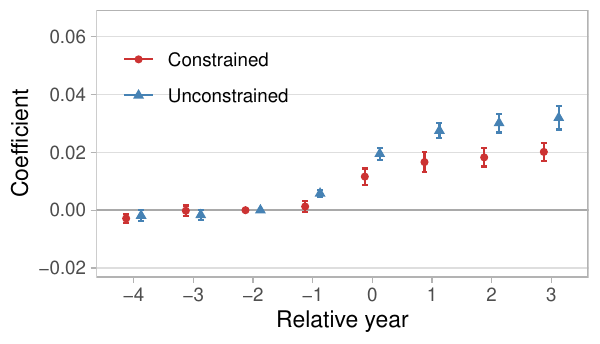}
    \end{subfigure}%
    \begin{subfigure}{.5\textwidth}
        \centering
        \caption{Portugal: Employment}
        \includegraphics[width=.99\textwidth]{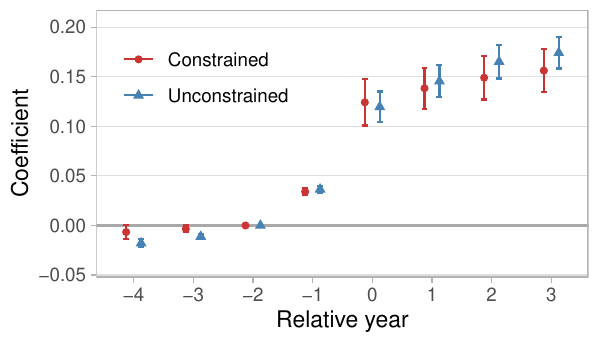}
    \end{subfigure} \\

    \begin{subfigure}{.5\textwidth}
        \centering
        \caption{Norway: Mean Wage}
        \includegraphics[width=.99\textwidth]{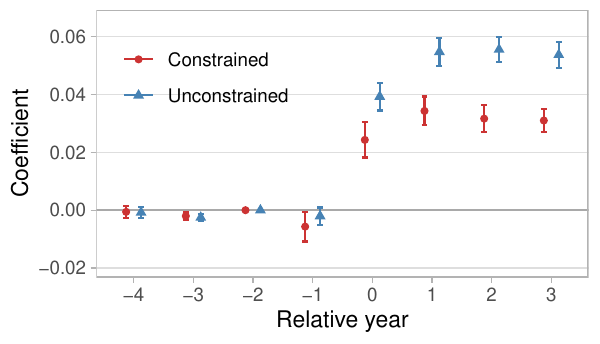}
    \end{subfigure}%
    \begin{subfigure}{.5\textwidth}
        \centering
        \caption{Norway: Employment}
        \includegraphics[width=.99\textwidth]{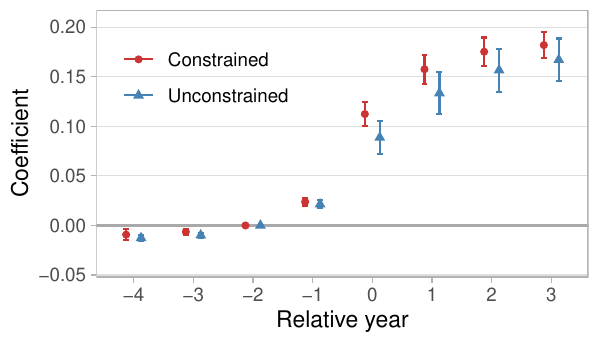}
    \end{subfigure} \\

    \vspace{2mm}
    \begin{singlespace}
    \begin{minipage}{1\textwidth} \footnotesize
        Notes:
        The figure shows reduced-form effects for the internal shock design by constraints.
        Constrained firms pay an average wage within 15\% of the national MW for Portugal, and are covered by a CBA in Norway.
        Panels (a) and (b) show estimates for Portugal, and Panels (c) and (d) for Norway, replicating the specification in Figure \ref{fig:baseline_estimates} by splitting each country's sample based on exposure to wage constraints.
    \end{minipage}
    \end{singlespace}
\end{figure}

%% file: figures/elast_const_internal.tex
\begin{figure}[h!]
    \centering
    \caption{Estimated Firm-Level Elasticities: Exposure to Wage-Setting Constraints.}
    \label{fig:elast_const_internal}

    \begin{subfigure}{.5\textwidth}
        \centering
        \caption{Labor Supply Elasticity}
        \includegraphics[width=.99\textwidth]{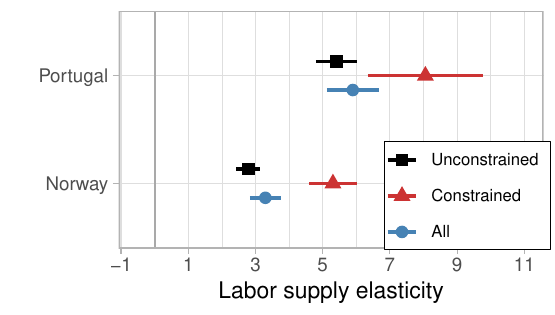}
    \end{subfigure}%
    \begin{subfigure}{.5\textwidth}
        \centering
        \caption{Rent-Sharing Elasticity}
        \includegraphics[width=.99\textwidth]{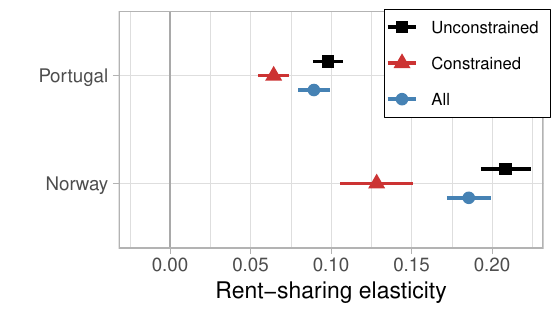}
    \end{subfigure}

    \vspace{2mm}
    \begin{singlespace}
    \begin{minipage}{1\textwidth} \footnotesize
        Notes:
        The figure shows estimated elasticities implied by reduced-form effects of a demand shock on value added, wages, and employment, using the internal shock design, by constrained status of firms, using the estimates from Table \ref{tab:lms_evidence}.
        Panel (a) shows labor supply elasticities, defined as the ratio of the employment to wage responses to the shock. 
        Panel (b) shows rent-sharing elasticities, defined as the ratio of the wage to value added responses to the shock.
    \end{minipage}
    \end{singlespace}
\end{figure}

%% file: figures/tripleDiff_portugal_wage.tex
\begin{figure}[ht!]
    \centering
    \caption{Wage Responses to Demand Shocks: Variation in Constraints in Portugal.}
    \label{fig:tripleDiff_portugal_wage}

    \begin{subfigure}{.5\textwidth}
        \centering
        \caption{Group-Specific Effects}
        \includegraphics[width=.99\textwidth]{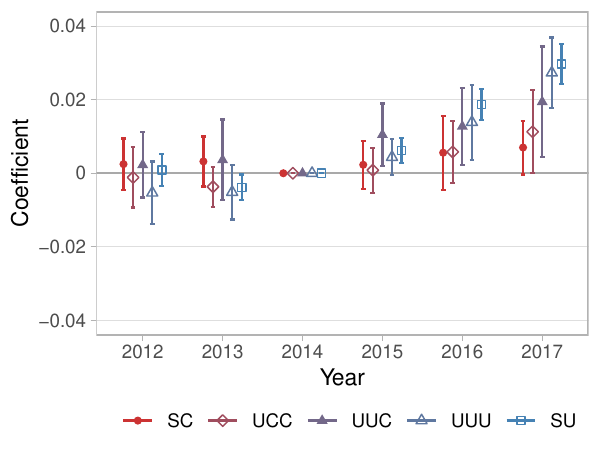}
    \end{subfigure}%
    \begin{subfigure}{.49\textwidth}
        \centering
        \caption{Relative to ``Just'' Unconstrained}
        \includegraphics[width=.99\textwidth]{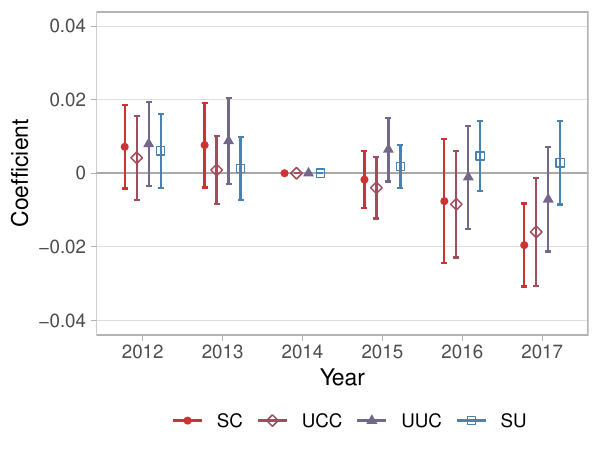}
    \end{subfigure}

    \vspace{2mm}
    \begin{singlespace}
    \begin{minipage}{1\textwidth} \footnotesize
        Notes: 
        The figure shows estimated wage responses to a demand shock for Portuguese firms, using the internal shock design, separately by exposure to minimum-wage constraints.
        We classify firms using their mean base wage-to-MW ratios averaged over 2014--2015, relative to the 2015, 2016, and 2017 MW levels expressed relative to the 2015 MW, following Section~\ref{subsec:data_var_constr}.
        Panel (a) shows event-study coefficients estimated separately for each group, and Panel (b) shows differences relative to the ``just'' unconstrained group (UUU).
    \end{minipage}
    \end{singlespace}
\end{figure}

%% file: tables/tripleDiff_implied_elasticities_portugal.tex
\begin{singlespace}
\begin{table}[!ht]
    \centering
    \caption{Estimated Firm-Level Elasticities: Variation in Constraints in Portugal.}
    \label{tab:tripleDiff_implied_elasticities_portugal}
    
    \begin{tabular}{lccccc}
        \toprule
        &  & \multicolumn{3}{c}{Elasticities} &  \\ \cmidrule(lr){3-5}
        & Wage & Labor & Inv. Labor & Rent- & Wage/MW \\
        & Effect & Supply & Supply & Sharing & 2014--15 \\
        & (1) & (2) & (3) & (4) & (5) \\
        \midrule
        (SC) Constrained at 2015 MW & 0.007 & 15.32 & 0.065 & 0.025 & 1.00 \\
         & (0.005) &  & (0.046) & (0.015) &  \\[1mm]
        (UCC) Constrained at 2016 MW & 0.011 & 12.75 & 0.078 & 0.045 & 1.04 \\
         & (0.007) &  & (0.055) & (0.027) &  \\[1mm]
        (UUC) Constrained at 2017 MW & 0.019 & 7.30 & 0.137 & 0.063 & 1.09 \\
         & (0.010) &  & (0.071) & (0.036) &  \\[1mm]
        (UUU) ``Just'' unconstrained & 0.028 & 5.34 & 0.187 & 0.099 & 1.17 \\
         & (0.006) &  & (0.053) & (0.022) &  \\[1mm]
        (SU) Strongly Unconstrained & 0.030 & 4.73 & 0.211 & 0.104 & 1.91 \\
         & (0.004) &  & (0.036) & (0.013) &  \\[1mm]
        \midrule
        $p$-value: SC $=$ UUU & 0.008 &  & 0.080 & 0.003 &  \\
        $p$-value: SU $=$ UUU & 0.829 &  & 0.678 & 0.814 &  \\
        \midrule
        Slope on wage/MW (per 0.1) & 0.012 & -5.90 & 0.074 & 0.042 &  \\
        $\quad$ excluding SU firms & (0.005) &  & (0.039) & (0.015) &  \\
        $p$-value: slope $=0$ & 0.010 &  & 0.056 & 0.007 &  \\
        \bottomrule
    \end{tabular}

    \vspace{2mm}
    \begin{minipage}{1\textwidth} \footnotesize
        Notes: 
        The table shows the estimated wage effects and elasticities for different groups of Portuguese firms, using the internal shock design, by constrained status of firms.
        Groups are based on the firm's 2014--2015 mean base wage relative to the MW: SC (\VcNfirmsSC{} firms, \VcShareSC\%), UCC (\VcNfirmsUCC{}, \VcShareUCC\%), UUC (\VcNfirmsUUC{}, \VcShareUUC\%), UUU (\VcNfirmsUUU{}, \VcShareUUU\%), and SU (\VcNfirmsSU{}, \VcShareSU\%), out of \VcNfirmsAll{} firms in 2015.
        The last column reports the group mean of this ratio.
        We estimate the 2017 wage, employment, and value added responses to the demand shock relative to 2014, $\{\beta^Y_{2017}\}$, comparing firms with and without a demand shock within each group.
        The wage effect is $\beta^{W}_{2017}$. 
        Labor supply elasticities are $\beta^{\text{L}}_{2017} / \beta^{W}_{2017}$, 
        inverses are $\beta^{W}_{2017} / \beta^{\text{L}}_{2017}$, and 
        rent-sharing elasticities are $\beta^{W}_{2017} / \beta^{\VA}_{2017}$.
        Standard errors in parentheses are clustered at the industry level and constructed using the delta method for the elasticities. 
        The bottom rows report $p$-values testing equality between SC and UUU and between SU and UUU, and the slope of a linear fit of the SC, UCC, UUC, and UUU estimates on their group mean wage/MW, per 0.1 of the ratio (10\% of the MW), with its $p$-value.
    \end{minipage}

\end{table}
\end{singlespace}

%% file: tables/norway_union_share.tex
\begin{table}[hbt!]
    \centering
    \caption{Estimated Firm-Level Elasticities: Local Union Strength in Norway.}
    \label{tab:norway_union_share}
    \begin{tabular}{@{}lccccc@{}}
        \toprule
        & \multicolumn{2}{c}{Elasticity} & \multicolumn{2}{c}{Pre-Period Mean} & \\ \cmidrule(lr){2-3} \cmidrule(lr){4-5}
        & \makecell[b]{Labor \\ Supply} & \makecell[b]{Rent- \\ Sharing} & \makecell[b]{Log \\ Wage} & \makecell[b]{Union \\ Share} & \makecell[b]{Share \\ Sample} \\ \cmidrule(lr){2-6}
        & (1) & (2) & (3) & (4) & (5) \\ \midrule
        All firms               & 3.29   & 0.185   & 5.37   & 0.196 \\
                                & (0.23) & (0.007) &  &  &  \\
        \textit{Baseline definition}    &      &      &   &     \\
    \hspace{3mm} Constrained   & 5.31   & 0.128   & 5.38   & 0.346   & 0.29 \\
                                & (0.36) & (0.012) &  &  &  \\
    \hspace{3mm} Unconstrained & 2.79   & 0.208   & 5.37   & 0.136   & 0.71 \\
                                & (0.18) & (0.008) &  &  &  \\
        \textit{Constrained firms}      &      &      &   &     \\
 \hspace{3mm} High local union share & 6.63   & 0.097   & 5.41   & 0.546   & 0.50 \\
                                & (0.57) & (0.007) &  &  &  \\
 \hspace{3mm} Low local union share  & 4.46   & 0.161   & 5.36   & 0.145   & 0.50 \\
                                & (0.27) & (0.013) &  &  &  \\
        \bottomrule
    \end{tabular}
    
    \vspace{2mm}
    \begin{minipage}{1\textwidth} \footnotesize
        Notes: 
        The table shows the elasticities estimated for different samples of Norwegian firms using the internal shock design.
        Constrained firms are those covered by a CBA. 
        For constrained firms, we split the sample into firms with above- and below-median union share.
        The labor supply elasticity is defined as the ratio of the employment to wage effects, 
        and the rent-sharing elasticity as the ratio of wage to value added effects.
        The pre-period descriptives are computed as in Table \ref{tab:lms_descriptives}.
        Standard errors clustered at the two-digit industry level are reported in parentheses.
        The standard errors for elasticities are calculated using the delta method.
    \end{minipage}
\end{table}

%% file: m_policy.tex
This section discusses the implications of using an incorrect labor supply elasticity for optimal minimum wage policy.
To do so, we extend the model in Section \ref{sec:m_framework} and calibrate it using Portuguese data.
Appendix \ref{asec:ge_model} provides details.

\subsection{General Equilibrium Model}

The economy is composed of a set of local labor markets.
We allow workers to have extensive-margin labor supply responses and differential labor supply elasticities to the firm and to each local labor market.
Workers decide whether to enter the labor market by comparing the expected value of working with the value of an outside option, subject to idiosyncratic preference shocks.
Upon entry, workers choose a labor market and, within that market, a firm, again after observing preference shocks for each market and firm.
The worker problem delivers a labor supply elasticity within markets $\eta$, a labor supply elasticity across markets $\sigma<\eta$, and the extensive-margin worker choice parameter $\zeta$.%
    \footnote{This labor supply structure imposes that log employment is proportional to $\eta$ times log firm wage.
    The cross-sectional relationship between log employment and log wages is weaker than $\eta$ for firms that pay above the minimum wage.
    Thus, we assume a parsimonious function of firm-level amenities ($A_j$ in Section \ref{sec:m_framework_theory}) that we calibrate to match the observed relationship between log employment and log wages.}

We make explicit assumptions on the distribution of firm productivity that yield smooth aggregates as a function of the MW.
Each local market is endowed with a continuum of firms with productivity drawn from a market-specific Pareto distribution, resulting in hump-shaped aggregate employment, which is maximized at a binding MW \citep{AhlfeldtEtAl2023}, and decreasing aggregate profits whenever the MW binds.
The employment gains arise from reductions in market power among supply-constrained firms, and are thus larger for smaller $\eta$.
For simplicity, we abstract from the firm exit margin.

\paragraph{Government problem.}

The planner perceives a labor supply elasticity $\tilde\eta = \eta (1+b)$, where $b$ is the ``bias'', a proportional deviation from the correct value.
We assume that the planner chooses the minimum wage to maximize aggregate employment, and study how the planner's choice of optimal minimum wage varies with $b$ given $\eta$.

\paragraph{Calibration.}

We define local labor markets using region, broad sector, and wage level of the firm.
We set $\eta = 5$ in the baseline, close to our unconstrained estimate for Portugal, and calibrate other parameters, as detailed in Appendix \ref{asec:ge_model}.%
    \footnote{In particular, the parameters of the market-specific Pareto distributions are calibrated to match the share of firms constrained by the minimum wage and the mean market wage, following \citet{Hermo2025}.}
The calibrated model matches the wage and employment distribution close to the MW well.
In particular, the model delivers a demand-constrained share of \WelfDemandConstrainedObservedEtaFive\%, whereas Section \ref{subsec:resu_byconst} implies \WelfDemandConstrainedData\%.
The employment-maximizing MW is close to, but about \WelfOptimalDeviationPctEtaFive\% above, the observed MW, resulting in a demand-constrained share of \WelfDemandConstrainedOptimalEtaFive\% under the counterfactual employment-maximizing policy.
We explore how our results change when using alternative values of the elasticity $\eta$.
When doing so, we recalibrate the parameters to match the same data moments.

\input{figures/struc_model_results}

\subsection{Results}

Misperceiving the labor supply elasticity leads the planner to forgo attainable employment gains.
Panel (a) of Figure \ref{fig:struc_model_results} expresses the gains forgone as a share of the employment increase attainable by moving from a nonbinding MW to the employment-maximizing MW.
With a \WelfBiasPct\% overestimate, the planner forgoes \WelfGainNotRealizedEtaFive\% of the attainable gain when~$\eta=5$.
Appendix Figure~\ref{fig:struc_model_res_sensitivity} shows sensitivity to the cross-market labor supply elasticity $\sigma$ and extensive-margin parameter~$\zeta$.

These unrealized gains arise because misperceiving $\eta$ distorts the chosen MW.
Panel (b) of Figure \ref{fig:struc_model_results} shows that a planner who overestimates $\eta$ sets a MW above the employment-maximizing level, whereas one who underestimates $\eta$ sets it below.
Under our benchmark \WelfBiasPct\% overestimate, the chosen MW rises by about \WelfChosenDeviationPctEtaFive\% above the optimum.
This pattern may seem counterintuitive: a higher perceived labor supply elasticity implies less monopsony power (smaller wage markdowns) and thus warrants a less aggressive policy.
Indeed, Panel (c) confirms that the planner chooses a weaker MW bite, setting the floor closer to the perceived highest nonbinding MW.
However, because other model primitives are held fixed, smaller perceived markdowns also imply higher unconstrained wages and thus a higher perceived nonbinding MW.
This baseline shift dominates the smaller desired bite, leading the planner to set a higher MW in levels.

The higher MW pushes more firms into the demand-constrained regime.
Panel (d) shows that, in the baseline, the demand-constrained share rises from \WelfDemandConstrainedOptimalEtaFive\% at the optimum ($b=0$) to \WelfDemandConstrainedChosenEtaFive\% under a \WelfBiasPct\% overestimate ($b=15\%$).
As more firms become demand-constrained, a positive deviation $b$ leads to a higher MW that reduces aggregate employment.

The share of gains forgone is larger when the true labor supply elasticity is higher.
Under a \WelfBiasPct\% overestimate, unrealized gains range from \WelfGainNotRealizedEtaThree\% when $\eta=3$ to \WelfGainNotRealizedEtaSeven\% when $\eta=7$.
A larger $\eta$ implies less monopsony power for the MW to correct, reducing the overall attainable employment gain.
The employment loss induced by the same proportional misperception of the elasticity then represents a larger share of this smaller gain.

%% file: figures/struc_model_results.tex
\begin{figure}[hbt!]
    \centering
    \caption{Minimum Wage Policy under a Mismeasured Labor Supply Elasticity.}
    \label{fig:struc_model_results}
    \begin{subfigure}{.47\textwidth}        
        \caption{Employment Gain Not Realized}
        \includegraphics[width=0.99\linewidth]{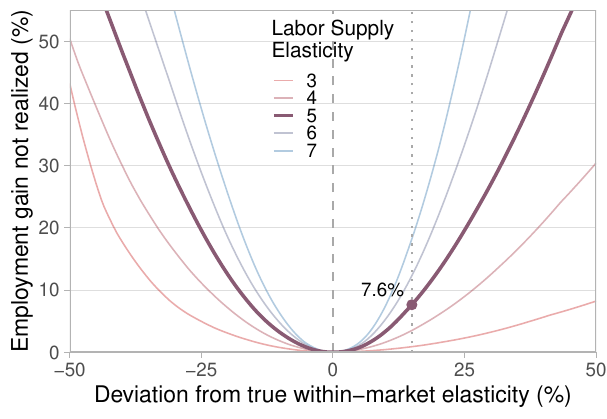}
    \end{subfigure}%
    \begin{subfigure}{.47\textwidth}
        \caption{Chosen MW Relative to Optimal MW}    
        \includegraphics[width=0.99\linewidth]{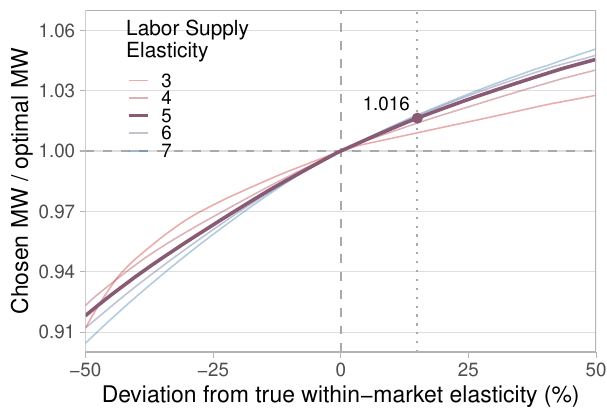}
    \end{subfigure} \\
    \begin{subfigure}{.47\textwidth}
        \caption{Perceived Bite of the Chosen MW}
        \includegraphics[width=0.99\linewidth]{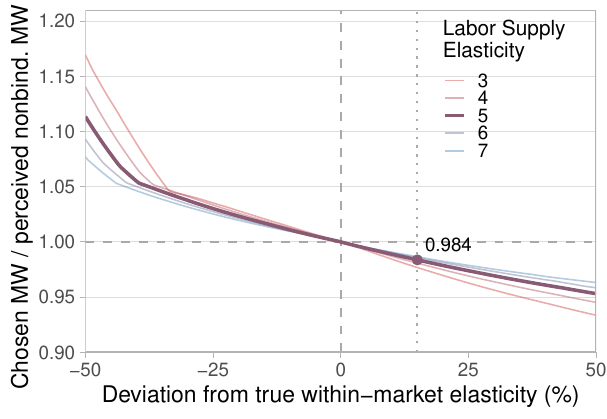}
    \end{subfigure}%
    \begin{subfigure}{.47\textwidth}
        \caption{Demand-Constrained Share}
        \includegraphics[width=0.99\linewidth]{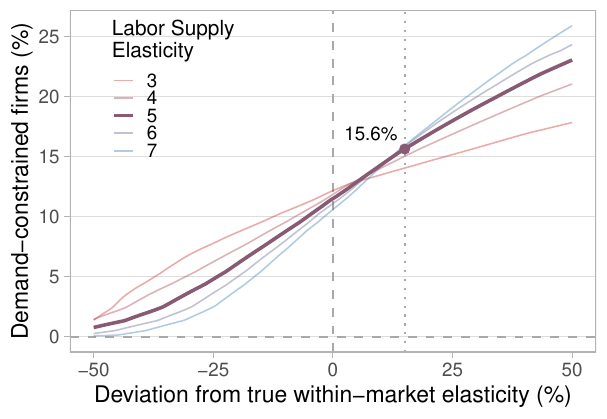}
    \end{subfigure}

    \vspace{2mm}
    \begin{minipage}{1\textwidth} \footnotesize
        Notes:
        The figure shows simulation results from the general equilibrium model calibrated to Portuguese data, for $\eta\in\{3,4,5,6,7\}$.
        The planner perceives $\tilde\eta = \eta(1+b)$ and chooses the MW that maximizes aggregate employment given $\tilde\eta$, while the chosen MW is evaluated in the economy at the true $\eta$.
        The horizontal axis shows the deviation $b$, in percent, with the dotted line marking a \WelfBiasPct\% deviation and the labeled point showing the value for $\eta=5$.
        Panel (a) shows the share of the attainable employment gain that the chosen MW does not realize, where the attainable gain is the increase in employment from the highest nonbinding MW to the employment-maximizing MW.
        Panel (b) shows the ratio of the chosen to the employment-maximizing MW.
        Panel (c) shows the ratio of the chosen MW to the highest nonbinding MW under the planner's perception $\tilde\eta$, normalized to one at $b=0$.
        Panel (d) shows the share of demand-constrained firms.
    \end{minipage}
\end{figure}

%% file: m_conclusions.tex
We show how wage-setting constraints shape firms' responses to demand shocks.
Our theoretical framework predicts that binding constraints suppress wage responsiveness to demand shocks, a pattern supported by comparable evidence from Portugal and Norway.
Exploiting minimum wage increases in Portugal, we also provide evidence that changes in exposure to constraints causally affect wage responsiveness.
These findings identify wage-setting constraints as a key source of heterogeneity in firms' adjustments to demand shocks.

Wage-setting constraints change the interpretation of conventional labor supply and rent-sharing elasticity estimates.
We quantify the resulting gap between conventional estimands and structural elasticities using differences in shock responses across constrained and unconstrained firms.
This distinction also matters for production-function approaches: measured wage markdowns describe realized wedges between wages and marginal revenue product, and need not reveal firms' potential wage-setting power when constraints bind.

Beyond highlighting the measurement challenges, we provide an approach to quantify the share of demand-constrained firms, we assess the bias in conventional elasticity estimates, and we show the consequences of using incorrect elasticities for minimum wage policy.
In our general equilibrium model calibrated to Portugal, overestimating the labor supply elasticity by \WelfBiasPct\% leads the planner to choose a minimum wage above the employment-maximizing level, forgoing \WelfGainNotRealizedEtaFive\% of the attainable employment gain.

We next discuss potential remedies.
One approach to recovering structural elasticities is to restrict estimation to plausibly unconstrained firms.
Two recent papers follow this approach to estimate the labor supply elasticity: \citet{AhlfeldtEtAl2023} do so in Germany, focusing on the period before the introduction of the minimum wage, and \citet{JaniezDelgadoPrieto2024} in Portugal, using a sample of firms whose wages exceed the minimum wage by a specified threshold.
Assuming that the parameters governing firm responses are homogeneous across firms, such restrictions help address the potential distortions highlighted here.

A second approach to this identification problem is to incorporate wage-setting constraints explicitly into structural estimation.
Our framework combines monopsony and bargaining to accommodate different forms of constraints.
Related approaches in \citet{Wong2025}, \citet{AzkarateAskasuaZerecero2025}, and \citet{Hermo2025} exploit model structure to identify elasticities and bargaining parameters that determine the responsiveness of firms to demand shocks.
Accounting for institutional constraints complements efforts to incorporate adjustment frictions and long-term employment relationships into the measurement of labor market power \citep{DhyneEtAl2024, ChanMattanaSalgadoXu2023, AgostinelliEtAl2025}.

Our analysis allows structural labor supply elasticities to differ across firms but assumes that they remain invariant to demand shocks.
In richer settings, firm-level labor supply elasticities may vary with oligopsonistic interactions \citep{BHM2022}, worker composition \citep{Volpe2026}, or non-isoelastic preferences \citep{Kline2025}.
In these settings, elasticities estimated among unconstrained firms need not describe how constrained firms would respond in the absence of constraints.
Identifying this counterfactual requires accounting jointly for wage-setting constraints and variation in the underlying labor supply relationship.
Developing empirical strategies that do so is an important direction for future work.

%% file: m_appendix_theory.tex
\section{Details on Theoretical Model}
\label{asec:model}

\subsection{Micro-Foundation of Labor Supply Equation} 
\label{asec:micr_labor_supply}


We consider a unit measure of homogeneous workers choosing among a finite set of firms $\J$.
The indirect utility for worker $i$ from working at firm $j$ is 
$
    V_{ij} = \psi_j W_j \xi_{ij} ,
$
where $\psi_j W_j$ is the expected wage and $\xi_{ij}$ is an idiosyncratic preference shock.
We assume that $\xi_{ij}$ is drawn independently for each worker-firm pair from a Fréchet distribution with shape $\eta>0$ and firm-specific scale parameter (or amenity level) $A_j>0$, so the cdf is $F_j(\xi) = \exp(-A_j \xi^{-\eta})$.

Each worker selects the firm that offers the highest utility.
Let $u_j = \psi_j W_j$ denote the systematic component of utility. 
Standard results imply that the probability of choosing firm $j$ is  
$P_j = \left(A_j u_j^\eta\right)/\left(\sum_{k \in \mathcal{J}} A_k u_k^\eta\right)$. 
\citep[e.g.,][]{Train2009}.
Substituting $u_j$ yields
$$
    P_j = \frac{A_j (\psi_j W_j)^\eta}{\sum_{k \in \mathcal{J}} A_k (\psi_k W_k)^\eta} .
$$
Given that the total measure of workers is assumed to be one, the labor supply to firm $j$ equals this probability.
We define an aggregate wage index $\aggW$ that summarizes the attractiveness of all employment options in the market as
$
    \aggW = \left( \sum_{k \in \mathcal{J}} A_k (\psi_k W_k)^\eta \right)^{1/\eta} .
$
Substituting this definition into the expression for $P_j$ gives the desired result:
$$
    H(\psi_jW_j,\aggW;A_j) = \frac{A_j (\psi_j W_j)^\eta}{\aggW^\eta} = A_j \left(\frac{\psi_j W_j}{\aggW}\right)^\eta ,
$$
where the shape parameter $\eta$ corresponds to the labor supply elasticity.
The notation $H(\cdot;A_j)$ reflects that heterogeneity in supply for a given wage arises only due to $A_j$.

\subsection{Wage Floor Constraint} 
\label{asec:derivations_wage_floor_const}

This appendix discusses the solutions to the firm optimization problem under wage floor constraints, i.e., $W \geq \W$.
For simplicity, we abstract from the subscript $j$.

\paragraph{Unconstrained Regime.}

To start with \textit{unconstrained} firms, we assume $W^*_U>\W$.
First, note that the labor supply must bind as, for a given wage, profits are strictly increasing in $L$, so $\psi^*=1$.
We can then write profits as $\Pi(W) = \Phi f(H\left(W,\aggW;A\right)) - W H\left(W,\aggW;A\right)$.
The first-order condition (FOC) that determines the optimal wage $W^*_U$ equates the marginal revenue product of labor (MRPL) to the marginal cost of labor (MCL):
$$
    \underbrace{\Phi f'(H(W^*_U,\aggW;A)) \frac{dH(W^*_U,\aggW;A)}{dW}}_{MRPL}
    = 
    \underbrace{H(W^*_U,\aggW;A) + W^*_U\frac{dH(W^*_U,\aggW;A)}{dW}}_{MCL} ,
$$
where we assumed the firm takes $\aggW$ as exogenous, as in the main text. 
We can rearrange to obtain the usual wage equation in monopsonistic labor markets
\begin{equation}
    \label{eq:firm_sol_uncons}
    W^*_U = \left(\frac{\eta}{\eta+1}\right) \Phi f'\left(H(W^*_U,\aggW;A)\right)
        = \mu \Phi f'\left(L^*_U\right) ,
\end{equation}
where the second equality defines $\mu\equiv\eta/(\eta+1)$ and replaces $L^*_U = H(W^*_U,\aggW;A)$.

\paragraph{Supply- and Demand-Constrained Regimes.}

In the case of \textit{supply-constrained} firms, both constraints are binding.
The firm equilibrium is given by the constraints, resulting in $W^*_S=\W$ and $L^*_S=H\left(\W,\aggW;A\right)$.

Finally, for \textit{demand-constrained} firms the wage constraint is binding, so $W^*_D=\W$, but the labor supply constraint is not, thus $L_D^*<H\left(\psi^*_D\W,\aggW;A\right)$.
The problem reduces to $\max_L\Pi=\Phi f(L) - \W L$, with FOC given by $\Phi f'(L^*_D) = \W$.
We can recover the hiring probability with $\psi^*_D = L^*_D / H\left(\psi^*_D\W,\aggW;A\right)$, resulting in $\psi^*_D = \left(L^*_D/A\right)^{1/(\eta+1)}\left(\aggW/\W\right)^{\eta/(\eta+1)}$.

\paragraph{Observed Markdown Factor and Labor Share.}

We derive the observed markdown factor $\tilde{\mu} \equiv W / (\Phi f'(L))$ and the labor share $\tilde{s}_L \equiv WL / (\Phi f(L))$ for the three regimes.
The labor share is a function of the observed markdown and output elasticity, $\tilde{s}_L = \tilde{\mu} \alpha(L)$.

For \textit{unconstrained} firms, \eqref{eq:firm_sol_uncons} applies.
Thus, $\tilde{\mu}_U = \mu = \eta/(\eta+1)$ and $\tilde{s}_{L,U} = \mu \alpha(L^*_U)$.
Under a homogeneous production function (Assumption \assumpfullref{assu:production_function}{a}), the labor share is constant at $\mu\alpha$.
For \textit{supply-constrained} firms, employment and wages are $L^*_S = H(\W,\aggW;A)$ and $\W$.
The observed markdown is then $\tilde{\mu}_S = \W / (\Phi f'(L^*_S))$, which is decreasing in productivity $\Phi$.
Similarly, the labor share $\tilde{s}_{L,S}$ declines with productivity as revenue increases while the wage bill remains constant.
For \textit{demand-constrained} firms, as the first-order condition holds, $\tilde{\mu}_D = 1$.
Consequently, the labor share is $\tilde{s}_{L,D} = \alpha(L^*_D)$.
Under Assumption \assumpfullref{assu:production_function}{a}, $\tilde{s}_{L,D} = \alpha$.

\subsection{Local Bargaining Constraint} 
\label{asec:derivations_local_bargaining}

This appendix discusses the solution when the constraint takes the form $W\geq g(\Upsilon,\W) = \kappa \Upsilon + (1-\kappa)\W$.
As discussed in the main text, $\Upsilon=\Phi f(L)/L$ is the average revenue product of labor and $\kappa\in(0,1)$ is the bargaining power of the local union.
We assume $\Upsilon>\W$, which guarantees that the ``negotiated wage'' $\kappa \Upsilon + (1-\kappa)\W$ is above the wage floor $\W$.
This is a natural restriction since firms operating with an average product below $\W$ would have negative profits.
We omit subscripts $j$ for simplicity.

The solution $(W^*,L^*)$ is characterized by three regimes, as illustrated in Appendix Figure \ref{fig:model_solutions_local_bargaining}.
First, firms in the unconstrained regime set wages according to Equation \eqref{eq:firm_sol_uncons}, and employment is given by the labor supply curve.
The solution for supply-constrained firms is again determined by the intersection of the two constraints, i.e., $W^*_S = \kappa \Phi f(L^*_S)/L^*_S + (1-\kappa)\W$ and $L^*_S=A \left(W^*_S\right)^\eta \aggW^{-\eta}$.
Demand-constrained firms choose employment on the labor demand curve.
Substituting the wage constraint into the profit function we find $\Pi = (1-\kappa)\left(\Phi f(L) - \W L\right)$, with FOC $\W = \Phi f'(L^*_D)$ and final wage $W^*_D =  \kappa \Phi f(L^*_D)/L^*_D + (1-\kappa)\W$.
Given our assumptions, employment is not distorted by the local union bargaining power $\kappa$.%
    \footnote{This result follows from the functional form we imposed on $g(\cdot,\cdot)$. 
    \citet{Holden1988}, for example, discusses a ``work-to-rule'' setting where the average product of labor is multiplied by a constant lower than 1, which results in distortions of employment levels for demand-constrained firms.}

\input{figures/model_solutions_local_bargaining}

Appendix Figure \ref{fig:model_solutions_local_bargaining} displays the case of a homogeneous production function (Assumption \assumpfullref{assu:production_function}{a}), so it is useful to prove that the demand-constrained wage is constant in this case.
First, recall that Euler's theorem for homogeneous functions states $L f'(L) = \alpha f(L)$, or $\Phi f(L)/L = \Phi f'(L) / \alpha$, where $\alpha\in(0,1)$ is the degree of homogeneity.
Additionally, in equilibrium the FOC $\Phi f'(L^*_D) = \W$ holds, so  $\Phi f(L^*_D)/L^*_D = \W / \alpha$ and we obtain
\begin{equation}
    \label{eq:wage_loc_barg_dem_cons_firms}
        W^*_D = \W\left(1+\kappa\frac{1-\alpha}{\alpha}\right) .
\end{equation}

Let us briefly discuss an interpretation of the three regimes when $\kappa>0$.
The unconstrained case corresponds to a highly productive firm that faces a relatively weak union.
The firm finds it optimal to pay a higher wage than the negotiated one to expand.
The constrained case corresponds to stronger unions that set a binding negotiated wage.
If the negotiated wage is not too high, then the firm will still find it optimal to locate on the labor supply curve.
However, if the wage is high enough, it is better for the firm to choose an employment level lower than the one implied by the labor supply constraint.

\paragraph{Threshold productivity levels.}

We can determine what regime a firm operates in by comparing its productivity $\Phi$ with threshold productivity levels $\upperPhi$ and $\lowerPhi$.
The lower threshold is determined by the productivity level at which the demand-constrained employment coincides with the labor supply constraint at the demand-constrained wage $W_D^*$ from \eqref{eq:wage_loc_barg_dem_cons_firms}.
The FOC for demand-constrained firms is $\W = \Phi f'(L)$. Thus,
\begin{equation*}
    \label{eq:local_bargaining_lower_thresh}
        \lowerPhi f'\left(H\left(W^*_D,\aggW;A\right)\right) 
            = \W 
        \implies \lowerPhi 
            = \frac{\W}{f'\left(H\left(W^*_D,\aggW;A\right)\right)} .
\end{equation*}

The upper threshold is given by the level of $\Phi$ at which the unconstrained wage equation equals the wage constraint.
Denoting the unconstrained employment by $L^*_U(\Phi)$ we have
\begin{equation*}
    \label{eq:local_bargaining_upper_thresh}
        \mu \upperPhi f'\left(L^*_U(\upperPhi)\right) 
            = \kappa \upperPhi\frac{f\left(L^*_U(\upperPhi)\right)}{L^*_U(\upperPhi)} 
              + (1-\kappa)\W 
        \implies \upperPhi 
            = \frac{(1-\kappa)\W}{\mu f'\left(L^*_U(\upperPhi)\right) - \kappa\frac{f\left(L^*_U(\upperPhi)\right)}{L^*_U(\upperPhi)} } ,
\end{equation*}
which is only defined when the denominator is positive.
By concavity we know that $f(L)/L>f'(L)$, so we will only have unconstrained firms whenever $\mu$ is sufficiently big relative to $\kappa$.
In other words, with strong unions ($\kappa\to 1$) there will not be any unconstrained firms.

We can more easily compare the thresholds in the case of a homogeneous production function.
If so, $f(L)/L = f'(L) / \alpha$ and we can write the upper threshold as
$$
\upperPhi = \frac{(1-\kappa)\W}
                  {\mu f'\left(L^*_U(\upperPhi)\right) 
                      - \kappa\frac{f'\left(L^*_U(\upperPhi)\right)}{\alpha}} 
          = \left(\frac{1-\kappa}{\mu-\frac{\kappa}{\alpha}}\right)\frac{\W}{f'(L^*_U(\upperPhi))} .
$$
We see that $\upperPhi$ is defined whenever $\mu>\kappa/\alpha$.
To compare with $\lowerPhi$ note that $L^*_U(\upperPhi)>H(W_D^*)$, implying $f'(L^*_U(\upperPhi))<f'(H\left(W^*_D\right))$. Then
$$
\upperPhi > \left(\frac{1-\kappa}{\mu-\frac{\kappa}{\alpha}}\right)\lowerPhi .
$$
Since $\mu<1$ and $\alpha\in(0,1)$, it follows that $(1-\kappa)/(\mu-\kappa/\alpha) > 1$, implying $\upperPhi>\lowerPhi$ and the thresholds are ranked as expected.

\input{figures/model_solution_markdown_lshare}

\paragraph{Observed markdown factor and labor share.}

Finally, we characterize the observed markdown $\tilde{\mu}$ and labor share $\tilde{s}_L$ under local bargaining.
For \textit{unconstrained} firms, the results are identical to the wage floor case ($\tilde{\mu}_U=\mu$).
However, \textit{demand-constrained} firms exhibit a markdown strictly larger than 1.
Using the wage equation \eqref{eq:wage_loc_barg_dem_cons_firms} derived for homogeneous production functions, and noting that the MRPL equals $\W$ in this regime, we obtain
$$
\tilde{\mu}_D = \frac{W^*_D}{\text{MRPL}} = \frac{\W\left(1+\kappa\frac{1-\alpha}{\alpha}\right)}{\W} = 1 + \kappa\left(\frac{1-\alpha}{\alpha}\right).
$$
Since $\alpha \in (0,1)$ and $\kappa > 0$, it follows that $\tilde{\mu}_D > 1$.
Correspondingly, the labor share is $\tilde{s}_{L,D} = \tilde{\mu}_D \alpha = \alpha + \kappa(1-\alpha)>\alpha$.
This implies that firms facing binding bargaining constraints pay wages above the marginal revenue product, rationalizing ``markdowns'' greater than one.
In the \textit{supply-constrained} regime, the markdown transitions from the high level $\tilde{\mu}_D$ to the structural parameter $\mu$ as productivity increases, as illustrated in Panel (b) of Figure \ref{fig:model_solution_markdown_lshare}.

\subsection{General Wage-Setting Constraint} 
\label{asec:derivations_general_wconstraint}

The wage-setting constraint now takes the general form $W \geq g\left(\Upsilon,\W\right)$. 
We assume that $\Upsilon\geq g\left(\Upsilon,\W\right)\geq\W$, so the negotiated wage must be below the average product, guaranteeing positive profits, and above the outside option $\W$.
We also assume that $\partial g(\Upsilon,\W)/\partial\Upsilon > 0$ and $\partial g(\Upsilon,\W)/\partial\W > 0$, so the negotiated wage is increasing in the average revenue product of labor and the outside option.
We omit the subscript $j$ for simplicity.


The firm may operate in one of three regimes, depending on its productivity.
\textit{Unconstrained} firms voluntarily pay above the negotiated wage $g(\cdot,\cdot)$.
The labor supply binds, and we have $L^*_U = H\left(W^*_U, \aggW; A\right)$.
The solution for $W^*_U$ is given by Equation \eqref{eq:firm_sol_uncons}, where the wage is a markdown relative to the marginal revenue product of labor $\Phi f'(L_U^*)$.

The firm is \textit{supply constrained} when both constraints bind. 
In this case, its optimal choice is determined by the intersection of the constraints, so that $W^*_S = g\left(\Upsilon(\Phi, L^*_S), \W\right)$ and $L^*_S = H\left(W^*_S, \aggW; A\right)$.
It is useful to compare the equilibrium wage with $\Phi f'(L_S^*)$.
By assumption $\Upsilon(\Phi, L^*_S)=\Phi f(L^*_S)/L^*_S\geq W_S^*\geq\W$, and by concavity $\Phi f(L^*_S)/L^*_S>\Phi f'(L^*_S)$.
Both the wage and the marginal revenue product are strictly bounded above by the average revenue product.
However, the ranking between $W^*_S$ and $\Phi f'(L^*_S)$ is ambiguous.

The firm is \textit{demand constrained} if the wage constraint binds but the labor supply constraint does not.
In this case, the firm's optimization problem is to choose $L$ to maximize profits
$\Pi = \Phi f(L) - g\left(\Upsilon(\Phi,L),\W\right) L$.
Define the elasticities
$$
    \lambda(L) = \frac{\partial g(\Upsilon(\Phi,L),\W)}{\partial \Upsilon}\frac{\Upsilon(\Phi,L)}{g(\Upsilon(\Phi,L),\W)}
    \quad\text{ and }\quad
    \chi(L) = \frac{\partial\Upsilon(\Phi,L)}{\partial L}\frac{L}{\Upsilon(\Phi,L)} . 
$$
The first-order condition that implicitly determines $L^*_D$ is then
\begin{equation}
    \label{eq:emp_demand_cons_gen_constraint}
    \Phi f'(L^*_D) = \left[1 + \lambda(L^*_D)\chi(L^*_D)\right] g(\Upsilon(\Phi,L^*_D),\W)
                   = \omega(L^*_D) g(\Upsilon(\Phi, L^*_D),\W) 
\end{equation}
where $\omega(L^*_D) \equiv 1 + \lambda(L^*_D)\chi(L^*_D)$.%
    \footnote{In the local bargaining case (Section \ref{asec:derivations_local_bargaining}) we have $\lambda = \kappa\Upsilon / \left(\kappa\Upsilon + (1-\kappa)\W\right)$ and $\chi = \left(\Phi f'(L) -\Upsilon\right)/\Upsilon$.
    Replacing $g(\cdot,\cdot)$ and these expressions in Equation \eqref{eq:emp_demand_cons_gen_constraint} yields the same FOC, $\Phi f'(L^*) = \W$.}
By assumption we have $\lambda(L)>0$ and, under concave production functions, $\chi(L^*)<0$ implying $\omega(L^*)<1$.
Demand-constrained firms unambiguously set employment to equate the marginal revenue product of labor to a \textit{fraction} $\omega(L^*)$ of the wage $g(\Upsilon,\W)$.
To see this more clearly, we can write \eqref{eq:emp_demand_cons_gen_constraint} as
$$
\underbrace{g(\Upsilon(\Phi,L^*),\W)}_{\text{wage}} = \underbrace{\frac{1}{\omega(L^*)}}_{\text{wedge}>1} \underbrace{\Phi f'(L^*)}_{MRPL} .
$$
The firm hires beyond the point where the marginal revenue product equals the wage. 
Because each additional worker dilutes the average product of labor, thereby relaxing the wage constraint, the effective marginal cost of labor is strictly lower than the final wage.

\subsection{Equilibrium}
\label{asec:equilibrium_def}

In this appendix, we close the theoretical model.
First, we show that all firms have positive operating profits $\Pi_j$ in equilibrium.
Second, we show that a labor market equilibrium with wage-setting constraints exists and is unique.

\subsubsection{Partial Equilibrium Profits} 

First, for unconstrained firms, we have
\begin{equation*}
        \Pi^*_U = \Phi f(L^*_U) - \mu \Phi f'\left(L^*_U\right) L^*_U
                 = \Phi \left[f(L^*_U) - \mu f'\left(L^*_U\right) L^*_U\right] .
\end{equation*}
We know that $\mu\in(0,1)$.
Additionally, by the concavity of $f(\cdot)$, we have $f(L) > f'(L) L$ for any $L>0$.
These facts imply that the expression in brackets is positive, and thus $\Pi^*_U>0$.

Second, consider firms for which the wage-setting constraint binds, regardless
of its specific form.
For these firms, profits can be written as
\[
    \Pi^* = \Phi f(L^*) - W^* L^*
          = L^* \left[ \Phi \frac{f(L^*)}{L^*} - W^* \right] ,
\]
where $L^*$ denotes equilibrium employment and $W^*$ the constrained wage.
By assumption, the constraint satisfies
$g(\Upsilon(\Phi,L^*),\W) \le \Upsilon(\Phi,L^*) = \Phi f(L^*)/L^*$, so profits
are weakly positive.
To establish strict positivity, start with demand-constrained firms.
Rearranging the FOC in Equation \eqref{eq:emp_demand_cons_gen_constraint} implies $g(\Upsilon(\Phi,L^*_D),\W) = \Phi f'(L^*_D)/\omega(L^*_D)$.
Since $f(\cdot)$ is strictly concave, $\Phi\frac{f(L^*_D)}{L^*_D}> \Phi f'(L^*_D)$, so $\omega(L^*_D)<1$ implies $\Phi f(L^*_D)/L^*_D > g(\Upsilon(\Phi,L^*_D),\W)$ and profits are positive.
An analogous argument applies to supply-constrained firms, for which both constraints bind and $g(\Upsilon(\Phi,L^*_S),\W) < \Upsilon(\Phi,L^*_S)$ by construction.

\subsubsection{General Equilibrium} 
\label{asec:general_equilibrium}

We assume that the economy is populated by a unit measure of workers and a discrete set of firms $\J$.
We define the equilibrium below.

\begin{definition}[Equilibrium]
    \label{prop:model_equilibrium}
    Given productivities $\{\Phi_j\}_{j\in\J}$ and strictly concave production technologies $\{f_j(\cdot)\}_{j\in\J}$, amenities $\{A_j\}_{j\in\J}$, continuous wage-setting constraints $\{g_j(\cdot,\cdot)\}_{j\in\J}$, and a wage floor $\W$, an equilibrium is an aggregate wage index $\aggW^*$ and  firm-level allocations $\{(W_j^*,L_j^*,\psi_j^*)\}_{j\in\J}$ such that:
    \begin{enumerate}[noitemsep]
        \item (Firm optimality)
        For each firm $j$, taking $\aggW^*$ as given, $(W_j^*,L_j^*,\psi_j^*)$ solves the firm optimization problem.
        \item (Consistent rationing)
        If $L_j^* < H(\psi_j^* W_j^*,\aggW^*;A_j)$, then the equilibrium hiring probability satisfies $\psi_j^* = L_j^*/H(\psi_j^* W_j^*,\aggW^*;A_j)$.
        \item (Labor market clearing) $\sum_{j\in\J} H(\psi_j^* W_j^*,\aggW^*;A_j) = 1 $.
    \end{enumerate}
\end{definition}

The following proposition establishes the existence and uniqueness of the equilibrium.
As in the main text, we maintain the assumption that $g_j(\Upsilon_j,\W)\in[\W, \Upsilon_j]$ is continuous and weakly increasing in both of its arguments.

\begin{proposition}[Existence and Uniqueness of Equilibrium]
    \label{prop:existence_uniqueness_general}
    Given primitives $\{\Phi_j\}_{j\in\J}$, $\{f_j(\cdot)\}_{j\in\J}$, $\{A_j\}_{j\in\J}$, $\{g_j(\cdot,\cdot)\}_{j\in\J}$, and $\{\W_j\}_{j\in\J}$, an equilibrium
    \[
    \left(
    \aggW^*,
    \{W_j^*,L_j^*,\psi_j^*\}_{j\in\J}
    \right)
    \]
    exists and is unique.
\end{proposition}

\begin{proof}[Proof of Proposition \ref{prop:existence_uniqueness_general}]
    Fix any $\aggW>0$.
    For any firm $j$, taking $\aggW$ as given, the labor supply function $H(\psi_j W_j,\aggW;A_j)$ is continuous and strictly increasing in the expected wage.
    The wage-setting constraint $W_j \ge g_j(\Upsilon_j(L_j),\W)$ is continuous and bounded.
    As the production function $f_j(L_j)$ is strictly concave ($f_j''(L_j)<0$), the firm's MRPL intersects the relevant marginal cost of labor (determined by either the labor supply curve or the wage-setting constraint) at a single crossing point.
    Therefore, the firm optimization problem admits a unique solution $(W_j^*(\aggW),L_j^*(\aggW),\psi_j^*(\aggW))$ for each $\aggW>0$.

    Next, we show that $H(\psi_j^* W_j^*,\aggW;A_j)$ is strictly decreasing in $\aggW$ for all $j$.
    Unconstrained and supply-constrained firms have a hiring probability of $\psi_j^*=1$.
    An increase in $\aggW$ shifts the labor supply curve inward.
    For unconstrained firms, this raises the marginal cost of labor, reducing optimal employment.
    For supply-constrained firms, employment is determined by the intersection of the inward-shifting labor supply curve and the non-increasing wage-setting constraint $g_j(\cdot)$, which strictly lowers employment $L_j^*$.
    Thus, $H(\psi_j^* W_j^*,\aggW;A_j) = L_j^*(\aggW)$ is strictly decreasing in $\aggW$.
    For demand-constrained firms, the wage constraint binds such that the firm rations employment ($\psi_j^* < 1$). 
    In this regime, the optimal employment level is independent of $\aggW$. 
    However, the number of workers attached to the firm, $H_j$, adjusts to satisfy the consistent rationing condition. Substituting $\psi_j^* = L_j^*/H_j$ into the supply equation yields $H_j = A_j (L_j^* W_j^* / (H_j \aggW))^\eta$. 
    Solving for $H_j$ implies $H_j(\aggW) \propto \aggW^{-\frac{\eta}{1+\eta}}$. 
    Since $L_j^*$ and $W_j^*$ are constant with respect to $\aggW$, $H_j(\aggW)$ is strictly decreasing.
    
    Define the aggregate labor supply as
    \[
    S(\aggW) = \sum_{j\in\J} H(\psi_j^*(\aggW) W_j^*(\aggW),\aggW;A_j).
    \]
    Since $H(\psi_j^* W_j^*,\aggW;A_j)$ is continuous and strictly decreasing in $\aggW$, $S(\aggW)$ is also continuous and strictly decreasing.    
    As $\aggW \to 0$, supply to each firm diverges and $S(\aggW) \to \infty$.
    As $\aggW \to \infty$, supply to each firm vanishes and $S(\aggW) \to 0$.
    By the intermediate value theorem, a unique $\aggW^*>0$ such that $S(\aggW^*) = 1$ exists.
    Given $\aggW^*$, the allocations $\{(W_j^*,L_j^*,\psi_j^*)\}_{j\in\J}$ are uniquely determined, which establishes existence and uniqueness of the equilibrium.
\end{proof}

The wage index $\aggW^*$ depends on model parameters, in particular the wage floor $\W$, productivities $\{\Phi_j\}_{j\in\J}$, and amenities $\{A_j\}_{j\in\J}$.
The value of $\aggW^*$ relative to $\W$ determines the threshold productivities and the constrained status of firms.
Since we assumed away fixed costs and profits are positive, all $|\J|$ firms will be active in equilibrium.
If, in equilibrium, there are demand-constrained firms, there will be labor rationing, i.e., $\sum_{j\in\J} L_j^* < 1$.

\subsection{Comparative Statics} 
\label{asec:comparative_statics}

\subsubsection{Wage Floor Constraint}
\label{asec:comparative_statics_wage_floor}

We start with the pure wage floor constraint discussed in Section \ref{asec:derivations_wage_floor_const}.
The goal is to derive the elasticities of $L^*\left(\Phi, \W, \aggW, A\right)$ and $W^*\left(\Phi, \W, \aggW, A\right)$ with respect to $\Phi$, where we abstract from the subscript $j$.
As in the main text, we define $\alpha(L) \equiv \frac{Lf'(L)}{f(L)} > 0$ as the employment elasticity of revenue and $\gamma(L) \equiv - \frac{Lf''(L)}{f'(L)} > 0$ as the negative of the employment elasticity of the marginal revenue product of labor.

Totally differentiating the equations that determine the unconstrained equilibrium with respect to $\ln\Phi$ yields
$$
    \frac{d\ln W^*_U}{d\ln\Phi} = \frac{1}{1 + \eta\gamma(L^*_U)} > 0
    \quad\text{and}\quad
    \frac{d\ln L^*_U}{d\ln\Phi} = \frac{\eta}{1 + \eta\gamma(L^*_U)} > 0.
$$
Next, let us analyze wage-constrained firms. 
For \textit{supply-constrained} firms, wages and employment are independent of the firm's productivity, so
$$
    \frac{d\ln W^*_S}{d\ln\Phi} = 0
    \quad\text{and}\quad
    \frac{d\ln L^*_S}{d\ln\Phi} = 0.
$$
Finally, for \textit{demand-constrained} firms, we differentiate the first-order condition to obtain
$$
    \frac{d\ln W^*_D}{d\ln\Phi} = 0
    \quad\text{and}\quad
    \frac{d\ln L^*_D}{d\ln\Phi} = \frac{1}{\gamma(L^*)} > 0.
$$
In this case, an increase in productivity allows the firm to hire more workers at $\W$, with the magnitude of the increase depending on the curvature of the production function.

We compute the ratio between the employment and wage derivatives, following the usual approach to derive labor supply elasticities.
For unconstrained firms we have $\eta_U = \eta$.
For supply-constrained firms, both employment and wage responses are zero, and thus the ratio is undefined.
For demand-constrained firms we have $\eta_D\to\infty$.

We continue by deriving the response of value added to the shock, which in this model is simply equal to equilibrium revenue $\VA^*\left(\Phi,\W\right)=\Phi f\left(L^*\left(\Phi,\W\right)\right)$, to find
\begin{equation}
    \label{eq:d_VA_d_Phi_general}
    \frac{d\ln \VA^*}{d\ln\Phi} = 1 + \frac{d\ln f(L^*)}{d\ln\Phi} = 1 + \alpha(L^*) \frac{d\ln L^*}{d\ln\Phi} .
\end{equation}
For an \textit{unconstrained} firm, we substitute the elasticity of labor to obtain
$$
\frac{d\ln \VA^*_U}{d\ln\Phi} = 1 + \frac{\eta\alpha(L^*_U)}{1 + \eta\gamma(L^*_U)}
                              = \frac{1 + \eta\left(\gamma(L^*_U) + \alpha(L^*_U)\right)}
                                     {1 + \eta\gamma(L^*_U)}.
$$
Value added increases due to the direct productivity effect and because the firm increases its employment in response to the shock.
For a \textit{supply-constrained} firm, employment does not react to the firm's own productivity shock, resulting in
$$
\frac{d\ln \VA^*_S}{d\ln\Phi} = 1.
$$
Finally, for a \textit{demand-constrained} firm, the response of value added is
$$
\frac{d\ln \VA^*_D}{d\ln\Phi} = 1 + \frac{\alpha(L^*_D)}{\gamma(L^*_D)} .
$$

We conclude by computing the rent-sharing induced by a productivity shock as the ratio of the wage response to the value-added response.
Using previous expressions we have
$$
\theta_U = \frac{1}{1 + \eta\left(\gamma(L^*) + \alpha(L^*)\right)} 
\quad\text{ and }\quad
\theta_S = \theta_D = 0 .
$$

\paragraph{Homogeneous production function.}

Assume that $f(L)=L^\alpha$ for some constant $\alpha\in(0,1)$.
Then, we have $\alpha(L)=\alpha$ and $\gamma(L)=1 - \alpha$.
It is straightforward to solve for the derivatives above, all of which turn out to be constant.
Interestingly, the rent-sharing elasticity for unconstrained firms becomes $\theta_U=1/(1+\eta)$.

\subsubsection{Local Bargaining Constraint}
\label{asec:comparative_statics_local_barg}

We now discuss the local bargaining case introduced in Section \ref{asec:derivations_local_bargaining}.
We start by deriving the elasticities of $L^*\left(\Phi, \W, \aggW, A\right)$ and $W^*\left(\Phi, \W, \aggW, A\right)$ with respect to $\Phi$ under the local bargaining constraint $g(\Upsilon,\W) = \kappa \Upsilon + (1-\kappa)\W$.
As before, define $\alpha(L) \equiv \frac{Lf'(L)}{f(L)} > 0$ and $\gamma(L) \equiv - \frac{Lf''(L)}{f'(L)} > 0$.
Also note that the elasticity $\lambda(L) = \left(\partial g(\Upsilon,\W)/\partial \Upsilon\right) \left(\Upsilon/g(\Upsilon,\W)\right) = \kappa\Upsilon/\left(\kappa\Upsilon + (1-\kappa)\W\right)$ is simply the share of the average product in the wage.

For \textit{unconstrained} firms, the solution is identical to the pure wage floor case.

For \textit{supply-constrained} firms, both constraints bind. The equilibrium is determined by $W^*_S = \kappa \Phi f(L^*_S)/L^*_S + (1-\kappa)\W$ and $L^*_S = A \left(W^*_S\right)^\eta \aggW^{-\eta}$.
Substituting the first equation into the second yields
$
L^*_S = A \left(\kappa \Phi \frac{f(L^*_S)}{L^*_S} + (1-\kappa)\W\right)^\eta \aggW^{-\eta}.
$
Taking logs and totally differentiating with respect to $\ln L^*_S$ and $\ln\Phi$:
$$
    d\ln L^*_S = \eta \ \lambda(L^*_S) \left[d\ln\Phi + \left(\alpha(L^*_S) - 1\right) d\ln L^*_S\right].
$$
Thus, the employment response is
\begin{equation}
    \label{eq:d_L_d_Phi_supcons_local_barg}
    \frac{d\ln L^*_S}{d\ln\Phi} = \frac{\eta\lambda(L^*_S)}{1 + \eta\lambda(L^*_S)\left(1-\alpha(L^*_S)\right)} > 0.
\end{equation}
For the wage response, differentiate $W^*_S = \kappa \Phi f(L^*_S)/L^*_S + (1-\kappa)\W$ to obtain
\begin{equation}
    \label{eq:diff_wage_local_barg_constraint}
    d\ln W^*_S = \lambda(L_S^*)\left[ d\ln\Phi + \left(\alpha(L_S^*) - 1\right) d\ln L_S^* \right].
\end{equation}
Solving for $d\ln L^*_S$ in terms of $d\ln\Phi$ in \eqref{eq:d_L_d_Phi_supcons_local_barg} and replacing yields
\begin{equation}
    \label{eq:d_W_d_Phi_supcons_local_barg}
    \frac{d\ln W^*_S}{d\ln\Phi} = \frac{\lambda(L_S^*)}{1 + \eta\lambda(L_S^*)\left(1-\alpha(L^*_S)\right)} > 0 .
\end{equation}
Unlike the pure wage floor case, supply-constrained firms' wages and employment both respond positively to productivity shocks. 
This occurs because higher productivity increases the average revenue product, which directly enters the wage constraint.

For \textit{demand-constrained} firms, recall that the optimal employment satisfies $\Phi f'(L^*_D) = \W$ and the wage is $W^*_D = \kappa \Phi f(L^*_D)/L^*_D + (1-\kappa)\W$. Differentiating the FOC yields
$$
    \frac{d\ln L^*_D}{d\ln\Phi} = \frac{1}{\gamma(L^*_D)} > 0.
$$
We can differentiate $W_D^*=\kappa\Phi f(L^*_D)/L_D^* + (1-\kappa)\W$ to obtain an expression analogous to \eqref{eq:diff_wage_local_barg_constraint} for demand-constrained firms.
Replacing in $d\ln L^*_D = d\ln\Phi / \gamma(L^*_D)$ we get
$$
    \frac{d\ln W^*_D}{d\ln\Phi} 
        = \lambda(L_D^*)
          \left(\frac{\alpha(L_D^*)+\gamma(L_D^*)-1}{\gamma(L_D^*)}\right) .
$$
As $\gamma(L_D^*)>0$, the sign of $\frac{d\ln W^*_D}{d\ln\Phi}$ depends on the sign of $\alpha(L_D^*)+\gamma(L_D^*)-1$, which boils down to whether the output elasticity of labor $\alpha(L)$ is increasing or decreasing in $L$.
As discussed before, under Assumption \assumpfullref{assu:production_function}{a} $\alpha(L_D^*)+\gamma(L_D^*)=1$ and the wage response is zero, and under Assumption \assumpfullref{assu:production_function}{b} $\alpha(L_D^*)+\gamma(L_D^*)>1$ and the wage response is positive.

Next, we compute the ratio of employment to wage responses.
For unconstrained firms, we have $\eta_U = \eta$, as in the pure wage floor case.
For supply-constrained firms, we also obtain $\eta_S = \eta$.
Finally, for demand-constrained firms we have
$$
    \eta_D = \frac{d\ln L^*_D/d\ln\Phi}{d\ln W^*_D/d\ln\Phi} = \frac{1}{\lambda(L^*_D)\left(\alpha(L_D^*)+\gamma(L_D^*) - 1\right)} .
$$
In the case of a homogeneous production function (Assumption \assumpfullref{assu:production_function}{a}), the denominator is zero and so $\eta_D\to\infty$.
More generally, provided that the production function exhibits ``weak curvature'' (i.e., $\alpha(L)+\gamma(L) \approx 1$), the denominator remains small.
Consequently, we expect $\eta_D$ to be large and, in particular, $\eta_D > \eta$.

We now derive the response of value added to productivity shocks.
As before, value added is $\VA^*\left(\Phi,\W\right)=\Phi f\left(L^*\left(\Phi,\W\right)\right)$ and so a general expression for the derivative is \eqref{eq:d_VA_d_Phi_general}.
Then, for \textit{unconstrained} firms the result is identical to the wage floor case.
For \textit{supply-constrained} firms we substitute the employment elasticity to get
\begin{equation*}
\frac{d\ln \VA^*_S}{d\ln\Phi} 
    = \frac{1 + \eta\lambda(L^*_S)}{1 + \eta\lambda(L^*_S)\left(1-\alpha(L^*_S)\right)} .
\end{equation*}
For \textit{demand-constrained} firms we get
$$
\frac{d\ln \VA^*_D}{d\ln\Phi} = 1 + \frac{\alpha(L^*_D)}{\gamma(L^*_D)} ,
$$
which is identical to the wage floor case since the FOC is the same.

Finally, we compute the ratio of wage and value-added responses for firms in each regime.
For \textit{unconstrained} firms we have $\theta_U = 1/\left(1 + \eta\left[\gamma(L^*_U) + \alpha(L^*_U)\right]\right)$, which is identical to the wage floor case.

For \textit{supply-constrained} firms:
\begin{equation*}
    \theta_S = \frac{\lambda(L_S^*)}{1 + \eta\lambda(L^*_S)} > 0.
\end{equation*}
Unlike the wage floor case, supply-constrained firms show a positive rent-sharing elasticity under local bargaining.%
    \footnote{Panel (d) of Appendix Figure \ref{fig:model_solutions_local_bargaining} illustrates this result, showing a positive slope for the optimal wage with respect to $\Phi$ in the range $\Phi\in(\lowerPhi, \upperPhi)$.}
We can now discuss the ranking $\theta_S < \theta_U$ discussed in the main text.
The condition $\theta_U>\theta_S$ holds whenever
$$
\frac{1-\lambda(L_S^*)}{\lambda(L_S^*)} > \eta \left(\alpha(L_U^*) + \gamma(L_U^*) - 1\right) .
$$
In the Cobb-Douglas case (Assumption \assumpfullref{assu:production_function}{a}), the right-hand side is zero, and since $\lambda(L_S^*) \in (0,1)$ the inequality holds trivially.
In the general case, the inequality holds provided that the curvature of the production function is not sufficiently large to offset the dampening effect of the bargaining share.
Thus, rent-sharing for supply-constrained firms is typically dampened relative to the unconstrained benchmark.

And, for \textit{demand-constrained} firms,
$$
\theta_D = \lambda(L_D^*)\frac{\gamma(L_D^*)+\alpha(L_D^*)-1}{\gamma(L_D^*) + \alpha(L^*_D)} .
$$
As discussed before, in the Cobb-Douglas case we obtain $\theta_D = 0$.
In the general case, $\theta_D$ remains small relative to $\theta_U$ unless the production function exhibits strong curvature.

\subsubsection{General Wage-Setting Constraint}

We briefly discuss the predictions for this case.
Recall that $\lambda(L) = \frac{\partial g(\Upsilon,\W)}{\partial \Upsilon}\frac{\Upsilon}{g(\Upsilon,\W)}$ is the elasticity of the wage constraint with respect to the average product, and note that $\frac{\partial \ln \Upsilon}{\partial \ln L} = \alpha(L)-1$.
For \textit{unconstrained} firms, the predictions are identical to the previous cases.
For \textit{supply-constrained} firms, the expressions are identical to \eqref{eq:d_L_d_Phi_supcons_local_barg} and \eqref{eq:d_W_d_Phi_supcons_local_barg}, subject to the general definition of $\lambda(L)$.
Thus, provided $\lambda(L)>0$, we again find positive but dampened rent-sharing ($\theta_S < \theta_U$).
For \textit{demand-constrained} firms, the exact expressions can be obtained by differentiating the system that determines $(W_D^*,L_D^*)$ given by Equation \eqref{eq:emp_demand_cons_gen_constraint} and the wage $W_D^*=g(\Upsilon(\Phi,L_D^*),\W)$.
Crucially, the qualitative predictions remain unchanged: under homogeneous production functions ($\alpha(L)$ constant), the wage response is zero, and the apparent labor supply elasticity diverges ($\eta_D \to \infty$), just as in the local bargaining case.

\setcounter{equation}{0} 
\section{Proofs}
\label{asec:proofs}

\begin{proof}[Proof of Lemma \ref{lem:ident_target_params}]
    We first derive the result for $\bar{\eta}$. 
    In the counterfactual scenario without constraints, all firms are on the labor supply curve, thus Equation \eqref{eq:labor_supply} implies $\Delta \ln L^*_{j} = \eta \Delta \ln W^*_{j} - \eta\Delta\ln\aggW$. 
    Taking conditional expectations and differencing between $Z_j=1$ and $Z_j=0$, the term involving the aggregate wage index $\aggW$ cancels out and we obtain
    $$
    \E\left[\Delta\ln L^*_{j}|Z_j=1\right] - \E\left[\Delta\ln L^*_{j}|Z_j=0\right] = \eta \left( \E\left[\Delta\ln W^*_{j}|Z_j=1\right] - \E\left[\Delta\ln W^*_{j}|Z_j=0\right] \right) .
    $$
    Dividing the employment response by the wage response yields $\bar{\eta} = \eta$.
    
    Next, we derive the result for $\bar{\theta}$. 
    From Result \ref{imp:cstats_uncons_firm}, the structural rent-sharing elasticity for firm $j$ is $\theta_j \equiv (1+\eta[\alpha_j(L^*_j)+\gamma_j(L^*_j)])^{-1}$. 
    For this proof only, define the auxiliary coefficients $\beta_j \equiv (1+\eta\gamma_j(L^*_j))^{-1}$ and $\delta_j \equiv (1+\eta[\alpha_j(L^*_j)+\gamma_j(L^*_j)])/(1+\eta\gamma_j(L^*_j))$, so that $\theta_j = \beta_j/\delta_j$. Using the comparative statics derived in Appendix \ref{asec:comparative_statics}, the unconstrained responses can be written as $\Delta\ln W^*_j \approx \beta_j \Delta\ln\Phi_j$ and $\Delta\ln \VA^*_j \approx \delta_j \Delta\ln\Phi_j$. 
    Substituting into Definition \ref{def:target_param}:
    \begin{equation}
    \label{eq:theta_S_intermediate}
    \bar{\theta}= \frac{\E\left[\beta_j \Delta\ln\Phi_j | Z_j=1\right] - \E\left[\beta_j \Delta\ln\Phi_j | Z_j=0\right]}
    {\E\left[\delta_j \Delta\ln\Phi_j | Z_j=1\right] - \E\left[\delta_j \Delta\ln\Phi_j | Z_j=0\right]} .
    \end{equation}
    
    To express this as a weighted average, let $\tau_j = (\alpha_j,\gamma_j)$ denote firm $j$'s production technology, which determines $\beta_j$, $\delta_j$, and $\theta_j$.
    Then, for notational convenience write the coefficients as $\beta(\tau_j)$, $\delta(\tau_j)$, $\theta(\tau_j)$.
    Define the $\tau$-specific first stage:
    $$
    \pi(\tau) \equiv \E[\Delta\ln\Phi_j|Z_j=1, \tau_j = \tau] - \E[\Delta\ln\Phi_j|Z_j=0, \tau_j = \tau] .
    $$
    Applying the law of iterated expectations to the first term in the numerator of \eqref{eq:theta_S_intermediate}:
    \begin{equation*}
        \E\left[\beta_j \Delta\ln\Phi_j | Z_j=1\right] 
        = \E[\beta(\tau_j) \E[\Delta\ln\Phi_j | Z_j=1, \tau_j]  ] .
    \end{equation*}
    Applying the same reasoning to all four terms in \eqref{eq:theta_S_intermediate}, the numerator becomes:
    \begin{align*}
    &\E\left[\beta(\tau_j) \E[\Delta\ln\Phi_j | Z_j=1, \tau_j] \right] - \E\left[\beta(\tau_j) \E[\Delta\ln\Phi_j | Z_j=0, \tau_j] \right] \\
    &\qquad = \E\left[\beta(\tau_j) \left( \E[\Delta\ln\Phi_j | Z_j=1, \tau_j] - \E[\Delta\ln\Phi_j | Z_j=0, \tau_j] \right)\right] \\
    &\qquad = \E\left[\beta(\tau_j) \pi(\tau_j)\right] .
    \end{align*}
    By analogous reasoning, the denominator equals $\E[\delta(\tau_j) \pi(\tau_j)]$ and so
    $$
    \bar{\theta} = \frac{\E[\beta(\tau_j) \pi(\tau_j)]}{\E[\delta(\tau_j) \pi(\tau_j)]} = \frac{\E[\theta(\tau_j) \delta(\tau_j) \pi(\tau_j)]}{\E[\delta(\tau_j) \pi(\tau_j)]} ,
    $$
    where we replaced $\beta(\tau) = \theta(\tau) \delta(\tau)$.
    Define the weight for a firm with technology $\tau_j$ as:
    $
    \omega(\tau_j) \equiv \delta(\tau_j) \pi(\tau_j)/\E[\delta(\tau_j) \pi(\tau_j)] .
    $
    By construction, $\E[\omega(\tau_j)] = 1$. The target parameter can therefore be written as:
    $$
    \bar{\theta} = \E[\omega(\tau_j) \theta(\tau_j)] ,
    $$
    confirming that $\bar{\theta}$ identifies a weighted average of the firm-specific rent-sharing elasticities $\theta(\tau_j)$. 
    The weight $\omega(\tau_j)$ is proportional to $\delta(\tau_j) \pi(\tau_j)$: firms receive more weight when their value added responds more elastically to productivity shocks (higher $\delta$) and when the instrument induces larger productivity shifts for their technology type (higher $\pi$).
    
    Finally, we consider the case under Assumption \assumpfullref{assu:production_function}{a}. Here, the output elasticity of labor is constant, implying $\alpha_j(L) + \gamma_j(L) = 1$ for all firms. Consequently, $\theta(\tau_j) = (1+\eta)^{-1}$ is identical across all firms and $\bar{\theta} = (1+\eta)^{-1}$ follows directly.
\end{proof}

\begin{proof}[Proof of Proposition~\ref{prop:id_no_constraints}]
    \textit{Part~\ref{prop:id_no_constraints:eta}.}
    Under Assumption~\assumpfullref{assu:absence_constraints}{b}, firms are either unconstrained or supply-constrained.
    For these firms, the labor supply equation~\eqref{eq:labor_supply} holds:
    \begin{equation*}
        \Delta\ln L_j = \Delta\ln A_j + \eta \Delta\ln W_j - \eta \Delta\ln \aggW .
    \end{equation*}
    Taking expectations conditional on $Z_j = z$ for $z \in \{0,1\}$ and differencing:
    \begin{equation}\label{eq:change_lsupply_shock}
        \begin{split}
            \E\left[\Delta\ln L_j \mid Z_j=1\right] - \E\left[\Delta\ln L_j \mid Z_j=0\right]
                & = \E\left[\Delta\ln A_j \mid Z_j=1\right] - \E\left[\Delta\ln A_j \mid Z_j=0\right] \\
                & + \eta \left(\E\left[\Delta\ln W_j \mid Z_j=1\right] - \E\left[\Delta\ln W_j \mid Z_j=0\right] \right) .
        \end{split}
    \end{equation}
    By Assumption~\ref{assu:labor_exogeneity} (exogeneity), the term involving labor supply shifters $\Delta \ln A_j$ equals zero.
    Supply-constrained firms may show no increase in the expected wage, if the shock is small, or some increase in the wage, if the shock is large and some firms move to the unconstrained region.
    In any case, they are on the labor supply curve and \eqref{eq:change_lsupply_shock} holds.
    Solving for $\eta$:
    \begin{equation*}
        \eta = \frac{\E\left[\Delta \ln L_{j} \mid Z_j=1\right] - \E\left[\Delta \ln L_{j} \mid Z_j=0\right]}
                    {\E\left[\Delta \ln W_{j} \mid Z_j=1\right] - \E\left[\Delta \ln W_{j} \mid Z_j=0\right]}
             = \eta^{CE} 
    \end{equation*}
    as desired.
    Note that Assumption~\ref{assu:relevant_shock} (relevance) ensures the denominator is non-zero, as the unconstrained firms are on the labor supply curve, so productivity shocks affect wages.

    \textit{Part~\ref{prop:id_no_constraints:theta}.}
    Under Assumption~\assumpfullref{assu:absence_constraints}{a} all firms are unconstrained.
    For simplicity, we focus on the case of homogeneous production functions (Assumption~\assumpfullref{assu:production_function}{a}).
    If so, the unconstrained equilibrium is determined by $W_j = \mu \Phi_j \alpha L_j^{\alpha-1}$ and $L_j = A_j(W_j/\aggW)^\eta$.

    Let us start with the labor supply elasticity.
    As Assumption~\assumpfullref{assu:absence_constraints}{a} implies Assumption~\assumpfullref{assu:absence_constraints}{b}, Part~\ref{prop:id_no_constraints:eta} applies directly and $\eta^{CE}$ identifies $\eta$.

    Next, we focus on the rent-sharing elasticity.
    Solving for the unconstrained equilibrium $(W^*_j, L^*_j)$ and replacing $\VA^*_j = \Phi_j \left(L_j^*\right)^\alpha$ yields
    \begin{equation*}
        \begin{split}
            \ln W^*_j   & = \frac{1}{1 + \eta(1-\alpha)} \ln\Phi_j 
                          - \frac{1-\alpha}{1 + \eta(1-\alpha)} \ln A_j + C^W , \\
            \ln \VA^*_j & = \frac{1+\eta}{1 + \eta(1-\alpha)} \ln\Phi_j 
                          + \frac{\alpha}{1 + \eta(1-\alpha)} \ln A_j + C^{\VA} , \\
        \end{split}
    \end{equation*}
    where $C^W$ and $C^{\VA}$ are constants that are common to all firms.
    The expected changes in wages and value added due to the shocks are then:
    \begin{equation*}
        \begin{split}
            \E[\ln W^*_j | Z_j=1] - \E[\ln W^*_j | Z_j=0] 
                & = \frac{1}{1 + \eta(1-\alpha)} \left(\E[\ln \Phi_j | Z_j=1] - \E[\ln \Phi_j | Z_j=0] \right) \\
                 & - \frac{1-\alpha}{1 + \eta(1-\alpha)} \left(\E[\ln A_j | Z_j=1] - \E[\ln A_j | Z_j=0] \right) , \\
            \E[\ln \VA^*_j | Z_j=1] - \E[\ln \VA^*_j | Z_j=0] 
                & = \frac{1+\eta}{1 + \eta(1-\alpha)} \left(\E[\ln \Phi_j | Z_j=1] - \E[\ln \Phi_j | Z_j=0] \right) \\
                 & + \frac{\alpha}{1 + \eta(1-\alpha)} \left(\E[\ln A_j | Z_j=1] - \E[\ln A_j | Z_j=0] \right) .  \\
        \end{split}
    \end{equation*}
    By Assumption~\ref{assu:labor_exogeneity} (exogeneity), the terms involving $\Delta \ln A_j$ vanish, so $\theta^{\text{CE}}$ yields
    \begin{equation}
        \begin{split}
        \theta^{\text{CE}}
          & = \frac{\frac{1}{1 + \eta(1-\alpha)} \left(\E[\ln \Phi_j | Z_j=1] - \E[\ln \Phi_j | Z_j=0] \right)}
                   {\frac{1+\eta}{1 + \eta(1-\alpha)} \left(\E[\ln \Phi_j | Z_j=1] - \E[\ln \Phi_j | Z_j=0] \right)} \\
          & = \left(\frac{1}{1+\eta(1-\alpha)}\right)\left(\frac{1+\eta(1-\alpha)}{1+\eta}\right) 
           = \frac{1}{1+\eta} = \theta ,
        \end{split}
    \end{equation}
    as desired. 
    Note that the second equality uses Assumption~\ref{assu:relevant_shock} (relevance) to ensure that $\E[\Delta\ln \Phi_j \mid Z_j=1] - \E[\Delta\ln \Phi_j \mid Z_j=0] \neq 0$.
\end{proof}

\begin{proof}[Proof of Proposition \ref{prop:id_failure}]
    We provide counterexamples where identification fails, starting with the labor supply elasticity.
    Under Assumption \assumpfullref{assu:presence_constraints}{b} (Strong Constraints), there are demand-constrained firms in the economy.
    For simplicity, we consider the case where $\rho_{uu} + \rho_{ss} + \rho_{dd} = 1$, with strictly positive shares of unconstrained ($u$), supply-constrained ($s$), and demand-constrained ($d$) firms.
    For this proof only, we introduce the indicators $UU_j, SS_j, DD_j\in\{0,1\}$, which are equal to 1 if firm $j$ belongs to the respective group.
    
    The average employment response is an average of the group-specific responses:
    \begin{equation*}
        \begin{split}   
            \E\left[\Delta \ln L_j | Z_j=1\right] & - \E\left[\Delta \ln L_j | Z_j=0\right] \\
              & = \rho_{dd} \left(\E\left[\Delta \ln L_j | Z_j=1, DD_j=1\right] - \E\left[\Delta \ln L_j | Z_j=0, DD_j=1\right] \right) + \rho_{ss}\cdot 0 \\
              & + \rho_{uu} \left(\E\left[\Delta \ln L_j^* | Z_j=1, UU_j=1\right] - \E\left[\Delta \ln L_j^* | Z_j=0, UU_j=1\right]\right) .
        \end{split}
    \end{equation*}
    Similarly, the average wage response is:
    \begin{equation*}
        \begin{split}   
            \E\left[\Delta \ln W_j | Z_j=1\right] & - \E\left[\Delta \ln W_j | Z_j=0\right] = \rho_{dd}\cdot 0 + \rho_{ss} \cdot 0 \\
              & + \rho_{uu} \left( \E\left[\Delta \ln W_j^* | Z_j=1, UU_j=1\right] - \E\left[\Delta \ln W_j^* | Z_j=0, UU_j=1\right] \right) . \\
        \end{split}
    \end{equation*}
    The conventional estimator $\eta^{CE}$ is the ratio of these expressions.
    Using the fact that, for unconstrained firms, the ratio of responses identifies $\eta$, we obtain
    \begin{equation}\label{eq:corollary_decomp_lsupply}
        \begin{split}
            \eta^{CE}
              & = \eta 
              + \frac{\rho_{dd}}{\rho_{uu}}  \frac{\E\left[\Delta \ln L_j | Z_j=1, DD_j=1\right] - \E\left[\Delta \ln L_j | Z_j=0, DD_j=1\right]}
                  {\E\left[\Delta \ln W_j^* | Z_j=1, UU_j=1\right] - \E\left[\Delta \ln W_j^* | Z_j=0, UU_j=1\right]} .
        \end{split}
    \end{equation}
    Note that $\eta^{CE}$ does not identify the structural parameter $\eta$.
    The additional term, which arises from the employment response of demand-constrained firms, introduces bias.
    
    We now turn to the rent-sharing elasticity, relying on \assumpfullref{assu:presence_constraints}{a} (Weak Constraints).
    To show that demand-constrained firms are not needed for this result, we consider $\rho_{ss} + \rho_{uu} = 1$ with $\rho_{ss} > 0$.
    Since supply-constrained firms do not adjust wages, the average wage effect is:
    \begin{equation*}
        \begin{split}   
            \E\left[\Delta \ln W_j \right.& \left.| Z_j=1\right] - \E\left[\Delta \ln W_j | Z_j=0\right] \\
              & = \rho_{uu} \left(\E\left[\Delta \ln W_j^* | Z_j=1, UU_j=1\right] - \E\left[\Delta \ln W_j^* | Z_j=0, UU_j=1\right]\right) .
        \end{split}
    \end{equation*}
    However, value added responds for both groups. 
    For supply-constrained firms, value added moves with productivity even if employment is fixed. 
    The average value-added effect is:
    \begin{equation*}
    \begin{split}        
        \E\left[\Delta \ln \VA_j | Z_j=1\right] & - \E\left[\Delta \ln \VA_j | Z_j=0\right] \\ 
          & = \rho_{ss} \left(\E\left[\Delta \ln \VA_j | Z_j=1, SS_j=1\right] - \E\left[\Delta \ln \VA_j | Z_j=0, SS_j=1\right]\right) \\
          & + \rho_{uu}\left(\E\left[\Delta \ln \VA_j^* | Z_j=1, UU_j=1\right] - \E\left[\Delta \ln \VA_j^* | Z_j=0, UU_j=1\right]\right) .
    \end{split}
    \end{equation*}
    Constructing the estimator $\theta^{\text{CE}}$, we find:
    \begin{equation}
        \label{eq:corollary_decomp_rshare}
        \begin{split}            
            \theta^{\text{CE}}
                & = \iota \times 
                  \frac{\E\left[\Delta \ln W_j^* | Z_j=1, UU_j=1\right] 
                      - \E\left[\Delta \ln W_j^* | Z_j=0, UU_j=1\right]}
                   {\E\left[\Delta \ln \VA_j^* | Z_j=1, UU_j=1\right] 
                      - \E\left[\Delta \ln \VA_j^* | Z_j=0, UU_j=1\right]} \\
                & = \iota \ \theta
        \end{split}
    \end{equation}
    where the attenuation factor is
    $$
    \iota = \frac{\rho_{uu}\left(\E\left[\Delta \ln \VA_j^* | Z_j=1, UU_j=1\right] - \E\left[\Delta \ln \VA_j^* | Z_j=0, UU_j=1\right]\right)}
                     {\left\{\begin{matrix}
                         \rho_{uu}\left(\E\left[\Delta \ln \VA_j^* | Z_j=1, UU_j=1\right] - \E\left[\Delta \ln \VA_j^* | Z_j=0, UU_j=1\right]\right) \\
                     + \rho_{ss} \ \left(\E\left[\Delta \ln \VA_j | Z_j=1, SS_j=1\right] - \E\left[\Delta \ln \VA_j | Z_j=0, SS_j=1\right]\right)
                     \end{matrix}\right\}} \in (0,1) .
    $$
    The attenuation is driven by the responsiveness of value added among supply-constrained firms in the denominator.
    Thus, we conclude $\theta^{\text{CE}}<\theta$, as desired.
\end{proof}

\begin{proof}[Proof of Corollary \ref{coro:bias_formula}]
    We proceed by obtaining expressions for the different components in Equation \eqref{eq:corollary_decomp_lsupply}.
    Using the model's comparative statics with $f(L)=L^\alpha$, we have
    \begin{equation}\label{eq:proof_cor_lsupp_employment}
        \begin{split}
            \E\left[\Delta \right. & \ln L_j | \left. Z_j=1, DD_j=1\right] - \E\left[\Delta \ln L_j | Z_j=0, DD_j=1\right] \\
            & \approx \frac{1}{1-\alpha}
                \left(\E\left[\Delta \ln \Phi_j | Z_j=1, DD_j=1\right] - \E\left[\Delta \ln \Phi_j | Z_j=0, DD_j=1\right]\right) \\
            & = \frac{1}{1-\alpha} \zeta_{dd}
        \end{split}
    \end{equation}
    and
    \begin{equation}\label{eq:proof_cor_lsupp_wage}
        \begin{split}
            \E\left[\Delta \right. & \left.  \ln W_j^* |Z_j=1, UU_j=1\right] - \E\left[\Delta \ln W_j^* | Z_j=0, UU_j=1\right] \\
            & \approx \frac{1}{1+\eta(1-\alpha)}
                \left(\E\left[\Delta \ln \Phi_j | Z_j=1, UU_j=1\right] - \E\left[\Delta \ln \Phi_j | Z_j=0, UU_j=1\right]\right)  \\
            & = \frac{1}{1+\eta(1-\alpha)} \zeta_{uu},
        \end{split}
    \end{equation}
    where we introduced the notation 
    $$
        \zeta_{uu} = \E\left[\Delta \ln \Phi_j | Z_j=1, UU_j=1\right] 
                     - \E\left[\Delta \ln \Phi_j | Z_j=0, UU_j=1\right] , 
    $$
    and $\zeta_{dd}$ is defined analogously but conditioning on $DD_j=1$.
    Plugging \eqref{eq:proof_cor_lsupp_employment} and \eqref{eq:proof_cor_lsupp_wage} into \eqref{eq:corollary_decomp_lsupply} we get Equation \eqref{eq:bias_lsupply} in the corollary.
    Noting that $\alpha>0$, $\eta>0$, $\rho_{dd}\geq0$, and noting that $\operatorname{sgn}\left(\zeta_{dd}\right) = \operatorname{sgn}\left(\zeta_{uu}\right)$, we confirm that $\eta^{CE}-\eta\geq0$.
\end{proof}

\section{Additional Results for Local Bargaining Constraint}
\label{asec:additional_results_local_bargaining}

This appendix presents additional results for the setting with local bargaining constraints.

\begin{corollary}[Bias Decomposition Formula: Local Bargaining]
    \label{coro:bias_formula_bargaining}
    Consider the local bargaining constraint $g_j(\Upsilon_j,\W) = \kappa_j \Upsilon_j + (1-\kappa_j)\W$ under Assumption \assumpfullref{assu:production_function}{a} (homogeneous production functions).
    Let Assumptions \ref{assu:relevant_shock} and \ref{assu:labor_exogeneity} hold.
    Furthermore, assume $\rho_{uu}+\rho_{ss}+\rho_{dd}=1$ and  $\operatorname{sgn}(\zeta_{uu})=\operatorname{sgn}(\zeta_{ss})=\operatorname{sgn}(\zeta_{dd})$.
    Then, the bias of the conventional labor supply elasticity estimand is
    \begin{equation}\label{eq:bias_lsupply_bargaining}
        \eta^{CE} - \eta 
          = \frac{\rho_{dd} \frac{1}{1-\alpha}\frac{\zeta_{dd}}{\zeta_{uu}}}
              {\rho_{uu} \frac{1}{1+\eta(1-\alpha)} + \rho_{ss} \frac{\lambda}{1+\eta\lambda(1-\alpha)}\frac{\zeta_{ss}}{\zeta_{uu}}}
                         > 0,
    \end{equation}
    where we assumed $\lambda=\kappa_j\Upsilon_j/ \left(\kappa_j \Upsilon_j + (1-\kappa_j)\W\right)$ is a constant, for simplicity, and $\zeta_{dd}$, $\zeta_{ss}$, and $\zeta_{uu}$ are the expected changes in productivity due to the shock in each group.
\end{corollary}

\begin{proof}[Proof of Corollary \ref{coro:bias_formula_bargaining}]
    We decompose the estimator $\eta^{CE}$ into the contributions from the different groups of firms.
    Under the local bargaining constraint, firms in the unconstrained ($u$) and supply-constrained ($s$) regimes satisfy the labor supply condition so, for them, $\Delta \ln L_j = \eta \Delta \ln W_j$.
    In contrast, demand-constrained ($d$) firms are rationed. Under Assumption \assumpfullref{assu:production_function}{a} (homogeneous production functions), the wage for demand-constrained firms does not respond to productivity shocks; see Appendix \ref{asec:comparative_statics_local_barg}, where we discuss the comparative statics for the local bargaining constraint.
    Following the proof of Corollary \ref{coro:bias_formula}, we can then write
    \begin{equation}\label{eq:app_bias_local_barg}
        \eta^{CE}
          = \eta 
          + \rho_{dd} \frac{\E\left[\Delta \ln L_j | Z_j=1, DD_j=1\right] - \E\left[\Delta \ln L_j | Z_j=0, DD_j=1\right]}
              {\E\left[\Delta \ln W_j^* | Z_j=1, ON_j=1\right] - \E\left[\Delta \ln W_j^* | Z_j=0, ON_j=1\right]} ,
    \end{equation}
    where $ON_j$ is a vector that equals one if the firm is ``on'' the labor supply curve (either unconstrained $UU_j=1$ or supply-constrained $SS_j=1$).
    
    We construct the elements of $\eta^{CE}$.
    The average employment response among demand-constrained firms is given by  \eqref{eq:proof_cor_lsupp_employment}.
    The average wage response for $ON_j=1$ firms is
    \begin{equation*}\label{eq:proof_cor_lsupp_wage_local}
        \begin{split}
            \E\left[\Delta \right. & \left.  \ln W_j^* |Z_j=1, ON_j=1\right] - \E\left[\Delta \ln W_j^* | Z_j=0, ON_j=1\right] \\
            & \approx \rho_{uu}\frac{1}{1+\eta(1-\alpha)}
                \left(\E\left[\Delta \ln \Phi_j | Z_j=1, UU_j=1\right] - \E\left[\Delta \ln \Phi_j | Z_j=0, UU_j=1\right]\right)  \\
            & \quad + \rho_{ss}\frac{\lambda}{1+\eta\lambda(1-\alpha)}
                \left(\E\left[\Delta \ln \Phi_j | Z_j=1, SS_j=1\right] - \E\left[\Delta \ln \Phi_j | Z_j=0, SS_j=1\right]\right) \\
            & = \rho_{uu}\frac{1}{1+\eta(1-\alpha)}\zeta_{uu}
               + \rho_{ss}\frac{\lambda}{1+\eta\lambda(1-\alpha)}\zeta_{ss} ,
        \end{split}
    \end{equation*}
    where $\lambda=\left(\kappa\Upsilon\right)/\left(\kappa\Upsilon+(1-\kappa)\W\right)$ is the share of the negotiated wage corresponding to the average productivity, assumed constant for simplicity.
    Then, we can write \eqref{eq:app_bias_local_barg} as
    \begin{equation*}
        \eta^{CE} - \eta 
          = \frac{\rho_{dd} \frac{1}{1-\alpha}\frac{\zeta_{dd}}{\zeta_{uu}}}
              {\rho_{uu} \frac{1}{1+\eta(1-\alpha)} + \rho_{ss} \frac{\lambda}{1+\eta\lambda(1-\alpha)}\frac{\zeta_{ss}}{\zeta_{uu}}} ,
    \end{equation*}
    as desired.
    Given that $\alpha>0$ and $\eta>0$, and the assumptions on the signs of $\zeta_{dd}$, $\zeta_{ss}$, and $\zeta_{uu}$, it is easy to see that $\eta^{CE} - \eta>0$.
\end{proof}

Unlike in the wage floor case, supply-constrained firms mitigate the bias somewhat as they respond in both employment and wages.

We solve for $\rho_{dd}$ in equation \eqref{eq:bias_lsupply_bargaining}, assuming that the shock is equally strong across groups so $\zeta_{dd}/\zeta_{uu}=\zeta_{ss}/\zeta_{uu}=1$.
Recalling that $\rho_{dd}+\rho_{ss}+\rho_{uu}=1$, we get
\begin{equation}
    \label{eq:share_cons_local_barg}
        \rho_{dd} 
          = \frac{\Lambda}{1+\Lambda\frac{\lambda}{1+\eta\lambda(1-\alpha)}}\left[\rho_{uu}\left(\frac{1}{1+\eta(1-\alpha)}\right)+(1-\rho_{uu})\frac{\lambda}{1+\eta\lambda(1-\alpha)}\right] 
\end{equation}
where $\Lambda=(\eta^{CE}-\eta)(1-\alpha)$.
Analogous to the wage floor case discussed in Section \ref{sec:m_framework_ident_failure}, we use equation \eqref{eq:share_cons_local_barg} to quantify the share of demand-constrained firms, after plugging in values for $\eta^{CE}$, $\eta$, and $\alpha$, and given the share of unconstrained firms $\rho_{uu}$ in the economy.

%% file: figures/model_solutions_local_bargaining.tex
\begin{figure}[ht!]
    \centering
    \caption{Solution to Firm Optimization Problem with Local Bargaining.}
    \label{fig:model_solutions_local_bargaining}

    \pgfmathsetmacro{\A}{1}
    \pgfmathsetmacro{\eta}{2.6}
    \pgfmathsetmacro{\alpha}{0.65}
    \pgfmathsetmacro{\minW}{0.93}
    \pgfmathsetmacro{\kappa}{0.17} 
    \pgfmathsetmacro{\mu}{\eta / (\eta+1)}
    
    \pgfmathsetmacro{\phiZero}{3.85}  
    \pgfmathsetmacro{\phiOne}{2.3}    
    \pgfmathsetmacro{\phiTwo}{0.9}    

    \tikzset{
        solutionplot/.style={
            scale=2.4, 
            font=\small, 
            declare function={
                invSupply(\l) = \A * (\l)^(1/\eta);
                mcl(\l) = \A * ((\eta+1)/\eta) * (\l)^(1/\eta);
                mrpl(\l,\p) = \p * \alpha * (\l)^(\alpha - 1);
                arpl(\l,\p) = \p * (\l)^(\alpha - 1);
                rawBargain(\l,\p) = \kappa * arpl(\l,\p) + (1-\kappa) * \minW;
                wageConstraint(\l,\p) = max( rawBargain(\l,\p), \minW );
                wageCurve0(\l) = (arpl0(\l) > minW) ? (kappaUncons * arpl0(\l) + (1-kappaUncons) * minW) : minW;
                lOptUnc(\p) = ( (\p * \alpha * \mu) / \A )^(\eta / (1 + \eta*(1-\alpha)));
                wOptUnc(\p) = \A * (lOptUnc(\p))^(1/\eta);
                arplUnc(\p) = (\p)^(1 / (1 + \eta*(1-\alpha))) * ( \A / (\alpha * \mu))^(\eta*(1-\alpha) / (1 + \eta*(1-\alpha)));
                lOptDem(\p) = (\p * \alpha / \minW)^(1/(1-\alpha));
                wOptDem(\p) = wageConstraint(lOptDem(\p), \p);
                phiFromL(\l) = ( invSupply(\l) - (1-\kappa)*\minW ) / ( \kappa * (\l)^(\alpha-1) );
            }
        },
        wageplot/.style={
            scale=2.5,
            font=\small,
        }
    }

    \def\xmax{2.8}
    \def\ymax{2.15}
    \def\plotdomStart{0.2}
    \def\plotdomStartZeroMRPL{1.6}
    \def\plotdomStartZeroWageCons{0.5}
    \def\plotdomStartOneMRPL{0.4}
    \def\plotdomStartOneWageCons{0.25}
    \def\plotdomStartTwoMRPL{0.10}
    \def\plotdomStartTwoWageCons{0.07}
    \def\plotdomEndZero{\xmax-0.35}
    \def\plotdomEndOne{\xmax-0.35}
    \def\plotdomEndTwo{\xmax-0.9}

    \begin{subfigure}[t]{0.5\textwidth}
        \centering
        \caption{Unconstrained, $\textcolor{UnconsColor}{\Phi_0}$}
        \label{fig:loc_unconstrained}
        \begin{tikzpicture}[solutionplot]
            \draw[->, thick] (0,0) -- (\xmax, 0) node[below] {$L,H$};
            \draw[->, thick] (0,0) -- (0, \ymax) node[left] {$W$};

            \draw[dashed, color = CoolRed] (0, \minW) node[left] {$\W$} -- (\xmax-0.35, \minW);
            
            \draw[domain=\plotdomStart:\xmax-0.35, smooth, variable=\l, thick, name path=supply] 
                plot ({\l}, {invSupply(\l)}) node[font=\small, anchor=west] {$W(H)$};
            \draw[domain=\plotdomStart:\xmax-0.35, smooth, variable=\l, thick, color=orange] 
                plot ({\l}, {mcl(\l)}) node[font=\small, anchor=west] {$MCL$};
                
            \draw[domain=\plotdomStartZeroMRPL:\plotdomEndZero, smooth, variable=\l, thick, color=UnconsColor]
                plot ({\l}, {mrpl(\l, \phiZero)}); 
            \node[font=\small, left, color=UnconsColor] at ({\plotdomStartZeroMRPL}, {mrpl(\plotdomStartZeroMRPL, \phiZero) - .1}) {$MRPL(\Phi_0)$};

            \draw[domain=\plotdomStartZeroWageCons:\plotdomEndZero, smooth, variable=\l, thick, color=CoolRed]
                plot ({\l}, {wageConstraint(\l, \phiZero)});
            \node[font=\small, above, color=CoolRed] at (\plotdomStartZeroWageCons, {wageConstraint(\plotdomStartZeroWageCons, \phiZero)}) {$g(\Upsilon,\W)$};

            \pgfmathsetmacro{\lx}{lOptUnc(\phiZero)}
            \pgfmathsetmacro{\wx}{wOptUnc(\phiZero)}
            \pgfmathsetmacro{\mx}{mcl(\lx)}

            \fill[black] (\lx, \mx) circle (1.1pt); 
            \fill[black] (\lx, \wx) circle (1.1pt); 
            
            \draw[dashed] (\lx, 0) node[below] {$L^*_0$} -- (\lx, \mx);
            \draw[dashed] (0, \wx) node[left] {$W^*_0$} -- (\lx, \wx);

        \end{tikzpicture}
    \end{subfigure}%
    \begin{subfigure}[t]{0.5\textwidth}
        \centering
        \caption{Supply-Constrained, $\textcolor{SupConsColor}{\Phi_1}$}
        \label{fig:loc_supply_constrained}
        \begin{tikzpicture}[solutionplot]
            \draw[->, thick] (0,0) -- (\xmax, 0) node[below] {$L,H$};
            \draw[->, thick] (0,0) -- (0, \ymax) node[left] {$W$};

            \draw[dashed, color = CoolRed] (0, \minW) node[left, yshift=-3pt] {$\W$} -- (\xmax-0.35, \minW);
            
            \draw[domain=\plotdomStart:\xmax-0.35, smooth, variable=\l, thick, name path=supply] 
                plot ({\l}, {invSupply(\l)}) node[font=\small, anchor=west] {$W(H)$};
            \draw[domain=\plotdomStart:\xmax-0.35, smooth, variable=\l, thick, color=orange] 
                plot ({\l}, {mcl(\l)}) node[font=\small, anchor=west] {$MCL$};

            \draw[domain=\plotdomStartOneMRPL:\plotdomEndOne, smooth, variable=\l, thick, color=SupConsColor]
                plot ({\l}, {mrpl(\l, \phiOne)});
            \node[font=\small, right, color=SupConsColor] at ({\plotdomStartOneMRPL}, {mrpl(\plotdomStartOneMRPL, \phiOne)}) {$MRPL(\Phi_1)$};

            \draw[domain=\plotdomStartOneWageCons:\plotdomEndOne, smooth, variable=\l, thick, color=CoolRed, name path=constraint]
                plot ({\l}, {wageConstraint(\l, \phiOne)}); 
            \node[font=\small, above, color=CoolRed] at ({\plotdomStartOneWageCons+.1}, {wageConstraint(\plotdomStartOneWageCons, \phiOne)}) {$g(\Upsilon,\W)$};

            \pgfmathsetmacro{\lxLatent}{lOptUnc(\phiOne)}
            \pgfmathsetmacro{\wxLatent}{wOptUnc(\phiOne)}
            \pgfmathsetmacro{\mxLatent}{mcl(\lxLatent)}
            
            \node[fill=gray, shape=diamond, minimum size=6pt, inner sep=0pt] at (\lxLatent, \mxLatent) {};
            \node[fill=gray, shape=diamond, minimum size=6pt, inner sep=0pt] at (\lxLatent, \wxLatent) {};
            \draw[dashed, gray] (\lxLatent, 0) -- (\lxLatent, \mxLatent);

            \path[name intersections={of=supply and constraint, by=eqSup}];
            \fill[black] (eqSup) circle (1.1pt);
            
            \draw[dashed] (eqSup) -- (eqSup |- 0, 0) node[below] {$L^*_1$};
            \draw[dashed] (eqSup) -- (0, 0 |- eqSup) node[left] {$W^*_1$};

        \end{tikzpicture}
    \end{subfigure} \\
    \begin{subfigure}[t]{0.5\textwidth}
        \centering
        \caption{Demand-Constrained, $\textcolor{DemConsColor}{\Phi_2}$}
        \label{fig:loc_demand_constrained}
        \begin{tikzpicture}[solutionplot]
            \draw[->, thick] (0,0) -- (\xmax, 0) node[below] {$L,H$};
            \draw[->, thick] (0,0) -- (0, \ymax) node[left] {$W$};

            \draw[dashed, color = CoolRed, name path=minWageLine] (0, \minW) -- (\xmax-0.35, \minW); 
            
            \draw[domain=\plotdomStart:\xmax-0.35, smooth, variable=\l, thick, name path=supply] 
                plot ({\l}, {invSupply(\l)}) node[font=\small, anchor=west] {$W(H)$};
            \draw[domain=\plotdomStart:\xmax-0.35, smooth, variable=\l, thick, color=orange] 
                plot ({\l}, {mcl(\l)}) node[font=\small, anchor=west] {$MCL$};

            \draw[domain=\plotdomStartTwoMRPL:\plotdomEndTwo, smooth, variable=\l, thick, color=DemConsColor, name path=mrpl]
                plot ({\l}, {mrpl(\l, \phiTwo)});
            \node[font=\small, above, color=DemConsColor] at ({\plotdomStartTwoMRPL+.35}, {mrpl(\plotdomStartTwoMRPL, \phiTwo)}) {$MRPL(\Phi_2)$};

            \draw[domain=\plotdomStartTwoWageCons:\plotdomEndTwo, smooth, variable=\l, thick, color=CoolRed, name path=constraint]
                plot ({\l}, {wageConstraint(\l, \phiTwo)}) node[font=\small, anchor=south west] {$g(\Upsilon,\W)$};

            \pgfmathsetmacro{\lxLatent}{lOptUnc(\phiTwo)}
            \pgfmathsetmacro{\wxLatent}{wOptUnc(\phiTwo)}
            \pgfmathsetmacro{\mxLatent}{mcl(\lxLatent)}
            \node[fill=gray, shape=diamond, minimum size=5pt, inner sep=0pt] at (\lxLatent, \mxLatent) {};
            \node[fill=gray, shape=diamond, minimum size=5pt, inner sep=0pt] at (\lxLatent, \wxLatent) {};

            \pgfmathsetmacro{\lxDem}{lOptDem(\phiTwo)}
            \pgfmathsetmacro{\wxDem}{wageConstraint(\lxDem, \phiTwo)}
            
            \coordinate (L_determ) at (\lxDem, \minW);
            \fill[black] (L_determ) circle (1.1pt);
            
            \coordinate (Eq_Point) at (\lxDem, \wxDem);
            \fill[black] (Eq_Point) circle (1.1pt);

            \draw[dashed] (Eq_Point) -- (\lxDem, 0) node[below] {$L^*_2$};
            \draw[dashed] (Eq_Point) -- (0, \wxDem) node[left] {$W^*_2$};
            \draw[dotted, thick] (L_determ) -- (Eq_Point);

            \pgfmathsetmacro{\hxDem}{pow(\wxDem / \A, \eta)}
            \coordinate (H_Point) at (\hxDem, \wxDem);
            
            \fill[gray] (H_Point) circle (1.1pt);
            \draw[dashed] (Eq_Point) -- (H_Point); 
            \draw[dashed] (H_Point) -- (\hxDem, 0) node[below] {$H^*_2$};

        \end{tikzpicture}
    \end{subfigure}%
    \begin{subfigure}[t]{0.5\textwidth}
        \centering
        \caption{Optimal Wage and Labor Input}
        \label{fig:loc_outcomes}
        \begin{tikzpicture}[solutionplot]
            \draw[->, thick] (0,0) -- (\xmax,0) node[below] {$\Phi$};
            \draw[->, thick] (0,0) -- (0,\ymax) node[left] {$W,L$};

            \draw[dashed, color = CoolRed] (0, \minW) node[left] {$\W$} -- (\xmax-0.35, \minW);
            
            \pgfmathsetmacro{\wFlat}{\minW * (1+\kappa*(1-\alpha)/\alpha)}
            
            \def\pMin{0.5}
            \pgfmathsetmacro{\pThreshLow}{ \minW * pow(\wFlat, \eta*(1-\alpha)) / (\alpha * pow(\A, \eta*(1-\alpha))) }
            \pgfmathsetmacro{\pThreshHigh}{ pow((1-\kappa)/(\mu*\alpha-\kappa), 1+\eta*(1-\alpha)) * pow(\mu*\alpha / \A, \eta*(1-\alpha)) * pow(\minW, 1 + \eta*(1-\alpha)) }
            \def\pMax{4.65}

            \draw (\pMin/2, 0) -- ++(0,-2pt) node[below] {$\Phi_{\min}$};
            \draw (\pThreshLow/2, 0) -- ++(0,-2pt) node[below] {$\lowerPhi$};
            \draw (\pThreshHigh/2, 0) -- ++(0,-2pt) node[below] {$\upperPhi$};
            \draw[dashed, gray] (\pMin/2, 0) -- (\pMin/2, \ymax);
            \draw[dashed, gray] (\pThreshLow/2, 0) -- (\pThreshLow/2, \ymax);
            \draw[dashed, gray] (\pThreshHigh/2, 0) -- (\pThreshHigh/2, \ymax);

            \draw[domain=\pMin:\pThreshLow, thick, gray] 
                plot ({\x/2}, {\wFlat});

            \pgfmathsetmacro{\lStart}{ (\wFlat / \A)^\eta }
            \pgfmathsetmacro{\lEnd}{ lOptUnc(\pThreshHigh) }
            
            \draw[smooth, variable=\t, domain=\lStart:\lEnd, thick, gray]
                plot ({phiFromL(\t)/2}, {invSupply(\t)});

            \draw[domain=\pThreshHigh:\pMax, thick, gray] 
                plot ({\x/2}, {wOptUnc(\x)});
            \node[right, font=\small, gray, above] at ({\pMax/2}, {wOptUnc(\pMax)-.02}) {$W^*(\Phi,\cdot)$};

            \draw[domain=\pMin:\pThreshLow, thick, dashed] 
                plot ({\x/2}, {lOptDem(\x)*0.75});
            
            \draw[smooth, variable=\t, domain=\lStart:\lEnd, thick, dashed]
                plot ({phiFromL(\t)/2}, {\t*0.75});
            
            \draw[domain=\pThreshHigh:\pMax, thick, dashed] 
                 plot ({\x/2}, {lOptUnc(\x)*0.75});
            \node[right, font=\small] at ({\pMax/2-.75}, {lOptUnc(\pMax)*0.72}) {$L^*(\Phi,\cdot)$};

        \end{tikzpicture}
    \end{subfigure}
    
    \vspace{2mm}
    \begin{singlespace}
    \begin{minipage}{1\textwidth} \footnotesize\singlespacing
        Notes: 
        The figure illustrates the equilibrium of the model in Section \ref{sec:m_framework_theory} assuming a local bargaining constraint and a homogeneous production function with decreasing returns to scale in labor.
        Panel (a) shows an unconstrained firm for which the bargaining constraint is not binding.
        Panel (b) shows a supply-constrained firm for which both constraints are binding.
        Panel (c) shows a demand-constrained firm for which only the bargaining constraint is binding.
        In the three panels, $W(H)$ stands for the inverse labor supply curve, $MCL$ for the marginal cost 
        of labor, $MRPL$ for the marginal revenue product of labor, and $g(\Upsilon, \W)$ for the bargaining constraint.
        The black-filled circles indicate the equilibrium point, and the gray-filled diamonds in Panels (b) and (c) show the latent equilibrium point in the absence of the bargaining constraint.
        Panel (d) shows the optimal wage (solid line) and optimal labor choice (dashed line) as functions of productivity $\Phi$, where $\Phi_{\min}$ indicates the minimum level of productivity.
    \end{minipage}
    \end{singlespace}
\end{figure}

%% file: figures/model_solution_markdown_lshare.tex
\begin{figure}[hbt!]
    \centering
    \caption{Markdowns and Labor Shares with Wage-Setting Constraints.}
    \label{fig:model_solution_markdown_lshare}

    \def\xmax{2.75}
    \def\ymax{1.9}
    \def\markdownShift{1.25} 
    \def\lshareShift{1.0}   

    \begin{subfigure}[t]{0.5\textwidth}
        \centering
        \caption{Wage Floor Constraint}
        \label{fig:markdown_wagefloor}
        
        \begin{tikzpicture}[scale=2.5]
            \pgfmathsetmacro{\A}{1}
            \pgfmathsetmacro{\etaParam}{2}
            \pgfmathsetmacro{\alphaParam}{0.6}
            \pgfmathsetmacro{\minW}{0.97}
            \pgfmathsetmacro{\phiMin}{0.65}
            \pgfmathsetmacro{\phiMax}{4.7}
            \pgfmathsetmacro{\mkdwnfactor}{\etaParam / (\etaParam+1)}
            
            \pgfmathsetmacro{\phiLowerThresh}{(\minW/\A)^(1+\etaParam*(1-\alphaParam)) * \A / (\alphaParam)}
            \pgfmathsetmacro{\phiUpperThresh}{\phiLowerThresh / \mkdwnfactor}

            \tikzset{declare function={
                lbar = (\minW/\A)^(\etaParam);
                mrplPhi(\p) = \p * \alphaParam * (lbar)^(\alphaParam-1);
                mdSup(\p) = \minW / mrplPhi(\p);
            }}

            \draw[->, thick] (0,0) -- (\xmax,0) node[below] {$\Phi$};
            \draw[->, thick] (0,0) -- (0,\ymax-0.2) node[left] {$\tilde{\mu}, \tilde{s}_L$};

            \draw[dotted, gray] (0, 1*\markdownShift) -- (\xmax, 1*\markdownShift);
            \node[anchor=east, gray] at (0, 1*\markdownShift) {$1$};
            
            \draw[dotted, gray] (0, \mkdwnfactor*\markdownShift) -- (\xmax, \mkdwnfactor*\markdownShift);
            \node[anchor=east, gray] at (0, \mkdwnfactor*\markdownShift) {$\mu$};

            \draw[dotted, gray] (0, \alphaParam*\lshareShift) -- (\xmax, \alphaParam*\lshareShift);
            \node[anchor=east, color=blue] at (0, \alphaParam*\lshareShift) {$\alpha$};
            
            \draw[dotted, gray] (0, \mkdwnfactor*\alphaParam*\lshareShift) -- (\xmax, \mkdwnfactor*\alphaParam*\lshareShift);
            \node[anchor=east, color=blue] at (0, \mkdwnfactor*\alphaParam*\lshareShift) {$\mu\alpha$};

            \draw[dashed, gray] (\phiMin/2, 0) -- (\phiMin/2, \ymax-.3);
            \draw (\phiMin/2, 0) -- ++(0,-2pt) node[below] {$\Phi_{\text{min}}$};
            \draw[dashed, gray] (\phiLowerThresh/2, 0) -- (\phiLowerThresh/2, \ymax-.3);
            \draw (\phiLowerThresh/2, 0) -- ++(0,-2pt) node[below] {$\underline{\Phi}^{WF}$};
            \draw[dashed, gray] (\phiUpperThresh/2, 0) -- (\phiUpperThresh/2, \ymax-.3);
            \draw (\phiUpperThresh/2, 0) -- ++(0,-2pt) node[below] {$\overline{\Phi}^{WF}$};

            \draw[thick, black] (\phiMin/2, 1*\markdownShift) -- (\phiLowerThresh/2, 1*\markdownShift);
            
            \draw[domain=\phiLowerThresh:\phiUpperThresh, smooth, variable=\p, thick, black] 
                plot ({\p/2}, {mdSup(\p)*\markdownShift});
                
            \draw[thick, black] (\phiUpperThresh/2, \mkdwnfactor*\markdownShift) -- (\xmax-.2, \mkdwnfactor*\markdownShift);

            \node[above, font=\small] at (\phiMax/2 - .15, \mkdwnfactor*\markdownShift) {Markdown Factor};

            \draw[thick, blue] (\phiMin/2, \alphaParam*\lshareShift) -- (\phiLowerThresh/2, \alphaParam*\lshareShift);
            
            \draw[domain=\phiLowerThresh:\phiUpperThresh, smooth, variable=\p, thick, blue] 
                plot ({\p/2}, {\alphaParam * mdSup(\p)*\lshareShift});
            
            \draw[thick, blue] (\phiUpperThresh/2, \alphaParam*\mkdwnfactor*\lshareShift) -- (\xmax-.2, \alphaParam*\mkdwnfactor*\lshareShift);

            \node[below, color=blue] at (\phiMax/2-.03, \alphaParam*\mkdwnfactor*\lshareShift) {Labor Share};

        \end{tikzpicture}
    \end{subfigure}%
    \begin{subfigure}[t]{0.5\textwidth}
        \centering
        \caption{Local Bargaining Constraint}
        \label{fig:markdown_bargaining}
        
        \begin{tikzpicture}[scale=2.5]
            \pgfmathsetmacro{\A}{1}
            \pgfmathsetmacro{\etaParam}{2}
            \pgfmathsetmacro{\alphaParam}{0.6}
            \pgfmathsetmacro{\minW}{0.97}
            \pgfmathsetmacro{\phiMin}{0.65}
            \pgfmathsetmacro{\phiMax}{4.7}
            \pgfmathsetmacro{\kappaParam}{0.17}
            \pgfmathsetmacro{\mkdwnfactor}{\etaParam / (\etaParam+1)}

            \pgfmathsetmacro{\WageMarkup}{(1+\kappaParam*(1-\alphaParam)/\alphaParam)}
            \pgfmathsetmacro{\wFlat}{\minW * \WageMarkup}
            \pgfmathsetmacro{\mdDemVal}{\WageMarkup} 

            \pgfmathsetmacro{\pThreshLow}{ \minW * pow(\wFlat, \etaParam*(1-\alphaParam)) / (\alphaParam * pow(\A, \etaParam*(1-\alphaParam))) }
            \pgfmathsetmacro{\pThreshHigh}{ pow((1-\kappaParam)/(\mkdwnfactor*\alphaParam-\kappaParam), 1+\etaParam*(1-\alphaParam)) * pow(\mkdwnfactor*\alphaParam / \A, \etaParam*(1-\alphaParam)) * pow(\minW, 1 + \etaParam*(1-\alphaParam)) }

            \pgfmathsetmacro{\lStart}{ (\wFlat / \A)^\etaParam }
            \pgfmathsetmacro{\lEnd}{ ( (\pThreshHigh * \alphaParam * \mkdwnfactor) / \A )^(\etaParam / (1 + \etaParam*(1-\alphaParam))) }

            \pgfmathsetmacro{\wageOne}{\minW * (\alphaParam * (1-\kappaParam)) / (\alphaParam - \kappaParam)}
            \pgfmathsetmacro{\laborOne}{pow(\wageOne / \A, \etaParam)}
            \pgfmathsetmacro{\phiOne}{\wageOne / (\alphaParam * pow(\laborOne, \alphaParam - 1))}

            \tikzset{declare function={
                wSupply(\l) = \A * (\l)^(1/\etaParam);
                phiFromL(\l) = ( wSupply(\l) - (1-\kappaParam)*\minW ) / ( \kappaParam * (\l)^(\alphaParam-1) );
                mdBargainFromL(\l) = wSupply(\l) / ( (\alphaParam/\kappaParam) * (wSupply(\l) - (1-\kappaParam)*\minW) );
            }}

            \draw[->, thick] (0,0) -- (\xmax,0) node[below] {$\Phi$};
            \draw[->, thick] (0,0) -- (0,\ymax-0.2) node[left] {$\tilde{\mu}, \tilde{s}_L$};

            \draw[dotted, gray] (0, 1*\markdownShift) -- (\xmax, 1*\markdownShift);
            \node[anchor=east, gray] at (0, 1*\markdownShift) {$1$};
            
            \draw[dotted, gray] (0, \mkdwnfactor*\markdownShift) -- (\xmax, \mkdwnfactor*\markdownShift);
            \node[anchor=east, gray] at (0, \mkdwnfactor*\markdownShift) {$\mu$};

            \draw[dotted, gray] (0, \alphaParam*\lshareShift) -- (\xmax, \alphaParam*\lshareShift);
            \node[anchor=east, color=blue] at (0, \alphaParam*\lshareShift) {$\alpha$};

            \draw[dotted, gray] (0, \mkdwnfactor*\alphaParam*\lshareShift) -- (\xmax, \mkdwnfactor*\alphaParam*\lshareShift);
            \node[anchor=east, color=blue] at (0, \mkdwnfactor*\alphaParam*\lshareShift) {$\mu\alpha$};
            
            \draw[dashed, gray] (\phiMin/2, 0) -- (\phiMin/2, \ymax-.3);
            \draw (\phiMin/2, 0) -- ++(0,-2pt) node[below] {$\Phi_{\text{min}}$};
            \draw[dashed, gray] (\pThreshLow/2, 0) -- (\pThreshLow/2, \ymax-.3);
            \draw (\pThreshLow/2, 0) -- ++(0,-2pt);
            \node[below] at (\pThreshLow/2-.07, -.05) {$\underline{\Phi}^{LB}$};
            \draw[dashed, gray] (\pThreshHigh/2, 0) -- (\pThreshHigh/2, \ymax-.3);
            \draw (\pThreshHigh/2, 0) -- ++(0,-2pt);
            \node[below] at (\pThreshHigh/2, -.05) {$\overline{\Phi}^{LB}$};

            \draw[dashed, gray] (\phiOne/2, 0) -- (\phiOne/2, 1*\markdownShift);
            \draw (\phiOne/2, 0) -- ++(0,-2pt); 
            \node[below] at (\phiOne/2+.16, -.06) {$\Phi_{\tilde{\mu}=1}$};
            \fill[black] (\phiOne/2, 1*\markdownShift) circle (0.6pt);

            \draw[thick, black] (\phiMin/2, \mdDemVal*\markdownShift) -- (\pThreshLow/2, \mdDemVal*\markdownShift);

            \draw[dotted, gray] (0, \mdDemVal*\markdownShift) -- (\xmax, \mdDemVal*\markdownShift);
            \node[anchor=west, gray] at (\pThreshLow/2+.05, \mdDemVal*\markdownShift) {$1+\kappa\left(\frac{1-\alpha}{\alpha}\right)$};

            \draw[smooth, variable=\t, domain=\lStart:\lEnd, thick, black]
                plot ({phiFromL(\t)/2}, {mdBargainFromL(\t)*\markdownShift});

            \draw[thick, black] (\pThreshHigh/2, \mkdwnfactor*\markdownShift) -- (\xmax-.2, \mkdwnfactor*\markdownShift);

            \node[above, font=\small] at (\phiMax/2 - .05, \mkdwnfactor*\markdownShift + .05) {Markdown Factor};

            \draw[thick, blue] (\phiMin/2, \alphaParam*\mdDemVal*\lshareShift) -- (\pThreshLow/2, \alphaParam*\mdDemVal*\lshareShift);
            
            \draw[dotted, gray] (0, \alphaParam*\mdDemVal*\lshareShift) -- (\xmax, \alphaParam*\mdDemVal*\lshareShift);
            \node[anchor=west, color=blue] at (\pThreshLow/2+.1, \alphaParam*\mdDemVal*\lshareShift) {$\alpha + \kappa(1-\alpha)$};

            \draw[smooth, variable=\t, domain=\lStart:\lEnd, thick, blue]
                plot ({phiFromL(\t)/2}, {\alphaParam * mdBargainFromL(\t)*\lshareShift});

            \draw[thick, blue] (\pThreshHigh/2, \alphaParam*\mkdwnfactor*\lshareShift) -- (\xmax-.2, \alphaParam*\mkdwnfactor*\lshareShift);

            \node[below, color=blue] at (\phiMax/2-.03, \alphaParam*\mkdwnfactor*\lshareShift) {Labor Share};

        \end{tikzpicture}
    \end{subfigure}

    \vspace{2mm}
    \begin{singlespace}
    \begin{minipage}{1\textwidth} \footnotesize\singlespacing
        Notes: 
        This figure compares the theoretical predictions for the observed markdown factor $\tilde{\mu} = W / \text{MRPL}$ (black lines) and the labor share $\tilde{s}_L = W L / P Y$ (blue lines) as a function of firm productivity $\Phi$.
        Panel (a) presents the case of a pure wage floor constraint. 
        Here, demand-constrained firms ($\Phi \leq \underline{\Phi}^{WF}$) exhibit a markdown of 1, which declines with productivity in the supply-constrained regime, and equals $\mu$ in the unconstrained regime ($\Phi > \overline{\Phi}^{WF}$).
        Panel (b) presents the case of local bargaining.
        Demand-constrained firms ($\Phi \leq \underline{\Phi}^{LB}$) exhibit a markdown strictly greater than 1, equal to $1+\kappa\frac{1-\alpha}{\alpha}$, and a labor share strictly greater than the output elasticity $\alpha$, equal to $\alpha + \kappa(1-\alpha)$. 
        The superscripts $WF$ and $LB$ denote the wage floor and local bargaining models, respectively. 
    \end{minipage}
    \end{singlespace}
\end{figure}

%% file: m_appendix_empirical.tex
\setcounter{equation}{0} 
\setcounter{table}{0}
\setcounter{figure}{0}


\section{Evidence from Colombia}
\label{asec:colombia}

This appendix presents descriptive and reduced-form evidence for Colombian manufacturing firms, complementing the results for Portugal and Norway discussed in the main text. We include Colombia because its national MW is highly binding relative to other OECD economies, offering a distinct institutional setting in which to assess the relevance of wage-setting constraints, even though the underlying survey data do not allow for the same worker-level sample restrictions used in Portugal and Norway.

\input{figures/colombia_wages_empl_VA}

In Colombia, the MW is high and the role of collective bargaining is limited. 
In 2023, the MW was equivalent to approximately 90\% of the average wage, the highest ratio among OECD countries \citep{OECDEarnings2025}. Around half of the workforce is employed in the informal sector \citep{delgado2024worker}, where the high MW is not enforced.

\input{figures/lms_by_constraints_colombia}

We use the \textit{Encuesta Anual Manufacturera} (EAM), a longitudinal survey that provides annual information on firms in Colombia's manufacturing sector, covering firms with 10 or more employees, from 2000 to 2017. 
Unlike in Portugal and Norway, we cannot track workers' information over time, nor do we have access to firm customs data.
Thus, we measure the firm wage as the mean wage of permanent workers and employment as the count of permanent workers, and restrict the sample to firms with non-negative value added present in at least seven years of each cohort window. 
Appendix Table~\ref{tab:cohort_size} shows the number of firms in each year-cohort for our internal instrument design.
Panel (a) of Appendix Figure~\ref{fig:colombia_wages_empl_VA} shows the relationship between firm value added and average wages and employment.

Using all firms, we obtain a labor supply elasticity of $\CoLSelastAllBeta$ ($\text{SE}=\CoLSelastAllSE$) and a rent-sharing elasticity of $\CoRentShAllBeta$ ($\text{SE}=\CoRentShAllSE$) for Colombia, based on the reduced-form effects of a demand shock on value added, wages, and employment shown in Panel (b) of Appendix Figure~\ref{fig:colombia_wages_empl_VA}.%
    \footnote{Our labor supply elasticity estimate is higher than the value of 2.5 reported by \cite{amodio2024measuring} using similar data but identifying demand shocks through exchange rate-driven variation in export sales.
    Beyond the choice of shock, differences may also reflect their focus on exporting plants and on all types of workers (not only permanent workers) when measuring plant-level wages and employment.}

We further examine heterogeneity in shock responses by firms' constrained status. Appendix Figure~\ref{fig:lms_by_constraints_colombia} shows that wage and employment responses differ by constrained status (Panels a and b), and that the implied labor supply and rent-sharing elasticities vary accordingly as the threshold used to define constrained firms changes (Panels c and d).

\setcounter{equation}{0} 
\section{Details on Data Sources}
\label{asec:data}

\subsection{Portugal}

We focus on individuals aged 18 to 65, excluding apprentices, workers in agriculture, mining, or fishing, those in skilled agricultural occupations, and those with unidentified occupations.
We also exclude workers based in the islands (\textit{Azores} and \textit{Madeira}) or abroad, those without reported earnings, and those with invalid worker identifiers. 
When workers work for multiple firms in a given year, we use the wage from the highest-wage firm in the year. 

For the internal instrument, we define the mean firm wage for ``stayer'' workers who remain with the same firm for at least seven years within each eight-year cohort window, though we consider an alternative definition using ``incumbent'' workers. 
Additionally, we keep only firms with at least two stayers per year that appear continuously in QP over the eight years of each cohort. 
The resulting number of firms in each cohort used in the estimation with the internal instrument is reported in Appendix Table~\ref{tab:cohort_size}.

For the external instrument, we similarly define the firm wage using stayers.
Additionally, we apply several sample restrictions.
We restrict to firms that exported in 2005--2007
and whose cumulative real exports over the period accounted for at least 1\% of the cumulative real revenue. 
Following \cite{garin2024responsive}, we restrict to firms whose share of exports to Angola or Spain between 2005 and 2007 is less than 90\% of total exports.%
    \footnote{We exclude product–country shifts that are missing, either because they are the only suppliers to that market and are therefore mechanically dropped, or because the import demand shift is undefined in the post period.
    This results in the exclusion of a small number of firms from the sample.}

\paragraph{Estimating CBA wage floors.}

While our baseline definition of constrained firms relies on the MW, we consider an alternative approach based on the CBA wage floor.
To this end, we estimate the wage floors following \cite{cardoso2005contractual}.
We define the wage floor as the modal base wage within each CBA--job title--year cell.
We follow the standard sample restrictions in \cite{CardCardoso2022} for 2004--2017, and further exclude CBAs with fewer than 50 workers in a given year and drop CBA-1-digit job title-year cells with fewer than 10 workers. 
We retain modal floors for which the mode doesn't exceed 1.25 times the wage of the 10th percentile of the CBA--job title--year wage distribution, and the floor lies within a 25\% range of the median floor in each CBA--job title. 
We winsorize the wage-floor distribution at the 1st and 99th percentiles.
Overall, we identify 1,026 CBA wage floors. 

\paragraph{Minimum wage hikes.}

For the analysis in Section \ref{subsec:variation_constraints} in Portugal, we focus on the period 2012--2017. 
Starting from the restricted data we construct a firm-level dataset. 
We focus on full-time workers who have worked at least 150 hours in total per month.
To construct firm mean wages, we focus on stayers who remain with the same firm for at least five years during the 2012-2017 period. 
We retain only firms with at least two stayers. 
We then keep firms that have appeared continuously in QP for the six-year window. 
For firms with non-negative VA, we define the internal demand shock as the difference between 2016 and 2015, comparing firms in the top half of the VA growth distribution in each two-digit industry.

\subsection{Norway}

We focus on the worker's main job, defined as the job in which a worker obtains the highest cash earnings during a year. 
We drop spells lasting fewer than 31 days or involving fewer than 4 hours of work per week. 
We measure average hourly labor earnings as annual wage-cash earnings within the job spell relative to the total hours of work in the spell during the year. We restrict our sample to the private sector. 

For the internal instrument, we proceed analogously to Portugal.
We focus on the mean wages of stayers (workers staying in the same firm for at least seven years within each eight-year cohort window). 
We restrict the sample to firms with at least two stayers per year, observed continuously over the time window, and with non-negative value added. 
Appendix Table~\ref{tab:cohort_size} reports the number of firms per cohort used for the internal instrument.

When using the external instrument, we restrict the sample to firms that exported in at least three years prior to or including 2007 and whose cumulative real exports over 2005–2007 accounted for at least 1\% of their cumulative real revenue during the same period.%
    \footnote{Similarly to Portugal, we exclude country-products with missing shifts, resulting in dropping a small number of firms.}

\setcounter{equation}{0} 
\setcounter{table}{0}
\setcounter{figure}{0}


\section{Impacts of Demand Shocks and Minimum Wage Hikes}
\label{asec:triple_did}

We provide additional evidence on MW hikes and develop the triple-differences design used for Portugal in Section~\ref{subsec:variation_constraints}.
We first present event-study estimates, then separate MW and demand effects before combining them to compare shock responsiveness across groups.

\subsection{Event-study evidence on MW hikes}

\input{figures/mwshock_pretends_portugal}

We estimate the effects of MW hikes across the firm groups defined in Section~\ref{subsec:data_var_constr}.
We restrict the sample to firms without a 2016 demand shock ($Z_j=0$), where $Z_j$ indicates the internal instrument constructed using 2015--2016 value-added growth.
We study effects over 2012--2017, with 2013 as the reference year, before the October 2014 hike:
\begin{equation} \label{eq:mw_shock}
    \ln Y_{jt}
    = \sum_{G\neq\text{UUU}} \sum_{\tau\neq2013}
    \lambda^{G}_{Y,\tau}
    \mathbf{1}\{G_j=G\} \mathbf{1}\{t=\tau\}
    + \delta^Y_{k(j)t} + \omega^Y_j + \upsilon^Y_{jt},
\end{equation}
where $\omega^Y_j$ are firm effects, $\delta^Y_{k(j)t}$ are industry-by-year effects, and $\upsilon^Y_{jt}$ is the residual.
The coefficients $\lambda^{G}_{Y,\tau}$ measure changes in $Y\in\{W,L\}$ relative to UUU firms and the 2013 baseline.%
    \footnote{Wages are mean hourly total wages of stayers, and employment counts full-time workers.}

The MW hikes raise wages most among firms initially closest to the MW.
Panel (a) of Appendix Figure~\ref{fig:mw_c_uc_did} shows wage increases for SC firms of approximately \VcMWWageSCSixteenPct\% in 2016 and \VcMWWageSCSeventeenPct\% in 2017, as well as an increase following the October 2014 hike.
For UCC firms, wage effects are smaller but become significant in 2017.
For UUC firms, wages appear to increase in 2017 but the effect remains insignificant.
Panel (b) shows small and statistically insignificant employment responses in 2016 and 2017 for all three exposed groups.

\subsection{Constructing the triple-differences specification}

We first isolate the MW component using firms without a demand shock.
Appendix Table~\ref{tab:tripleDiff_portugal} compares 2017 with 2014, so we restrict the sample to these two years.
Columns 1, 4, and 7 compare SC and UCC firms with UUU firms, which remain unconstrained at the 2017 MW:
\begin{equation*}
    \ln Y_{jt} = \lambda^{SC}_{Y} \mathbf{1}\{t=2017\} \mathbf{1}\{G_j=\text{SC}\}
               + \lambda^{UCC}_{Y} \mathbf{1}\{t=2017\} \mathbf{1}\{G_j=\text{UCC}\}
               + \delta^Y_{k(j)t} + \omega^Y_j + \upsilon^Y_{jt}.
\end{equation*}
The coefficients measure differential changes in $Y\in\{W,L,\VA\}$ relative to UUU firms.

The MW-only estimates show significant wage increases for both SC and UCC firms, with a larger increase for SC (Column 1 of Appendix Table~\ref{tab:tripleDiff_portugal}).
Column 4 shows a positive employment effect for SC, significant at the 10\% level, but no significant effect for UCC.

We next estimate the demand response among ``just'' unconstrained firms (UUU).
Columns 2, 5, and 8 of Appendix Table~\ref{tab:tripleDiff_portugal} show estimates of:
\begin{equation*}
    \ln Y_{jt} = \beta^{UUU}_{Y} \mathbf{1}\{t=2017\} Z_j
               + \delta^Y_{k(j)t} + \omega^Y_j + \upsilon^Y_{jt}.
\end{equation*}
The estimates imply increases of approximately \VcRFDemandWageUUUPct\% in wages, \VcRFDemandEmpUUUPct\% in employment, and \VcRFDemandVAUUUPct\% in value added by 2017, with implied elasticities similar to our baseline.

We then combine MW and demand shocks to compare responsiveness across groups.
The joint specification in Columns 3, 6, and 9 of Appendix Table~\ref{tab:tripleDiff_portugal} includes SC, UCC, and UUU firms and shows estimates of the following specification:
\begin{equation*}
\begin{split}
    \ln Y_{jt} & = \gamma_{Y}^{SC} \ \mathbf{1}\{t=2017\} \ \mathbf{1}\{G_j=\text{SC}\}
               + \gamma_{Y}^{UCC} \ \mathbf{1}\{t=2017\} \ \mathbf{1}\{G_j=\text{UCC}\} \\
              & \quad +\beta^{UUU}_{Y} \ Z_j \ \mathbf{1}\{t=2017\} \\
               & \quad + \beta_{Y}^{SC,UUU} \ Z_j \mathbf{1}\{t=2017\} \ \mathbf{1}\{G_j=\text{SC}\} \\
              & \quad + \beta_{Y}^{UCC,UUU} \ Z_j \ \mathbf{1}\{t=2017\} \ \mathbf{1}\{G_j=\text{UCC}\} \\
              & \quad + \delta_{k(j)t}^{Y} + \omega_j^{Y} + \upsilon_{jt}^{Y}.
\end{split}
\end{equation*}
This equation simply extends Equation~(\ref{eq:wage_effect_triple_diff}) to include SC firms.%
    \footnote{Firm fixed effects absorb the time-invariant group and shock indicators and their interactions.
    Industry-by-year fixed effects absorb the common post-period indicator.}

The joint specification separates MW-related changes from demand responses.
The omitted category is UUU firms without a demand shock.
For unshocked firms, $\gamma^{SC}_{Y}$ and $\gamma^{UCC}_{Y}$ capture changes relative to UUU, reported in the SC and UCC rows.
The coefficient $\beta^{UUU}_{Y}$ gives the demand response for UUU, while the responses for SC and UCC are $\beta^{UUU}_{Y}+\beta^{SC,UUU}_{Y}$ and $\beta^{UUU}_{Y}+\beta^{UCC,UUU}_{Y}$.
Thus, these coefficients report the triple differences: the change between shocked and unshocked firms within each group, relative to that of the UUU group.

\input{tables/tripleDiff_portugal}

The joint estimates show smaller wage responses among constrained firms.
Column 3 of Appendix Table~\ref{tab:tripleDiff_portugal} implies responses of approximately \VcRFJointWageSCTotalPct\% for SC and \VcRFJointWageUCCTotalPct\% for UCC, compared with \VcRFJointWageUUUPct\% for UUU.
Employment responsiveness in column (6) appears lower for SC, like in our baseline analysis (see Appendix Figure~\ref{fig:tripleDiff_portugal_emp_VA}).
Value-added responses do not differ significantly across groups (Column 9).

%% file: figures/colombia_wages_empl_VA.tex
\begin{figure}[H]
    \centering
    \caption{Employment, Wages, and Value Added: Cross-Sectional Relationship and Responses to Shocks.}
    \label{fig:colombia_wages_empl_VA}
    
    \begin{subfigure}[t]{0.53\textwidth}
        \centering
        \caption{Cross-Sectional Relationship}
        \label{fig:hockey_stick_colombia}
        \includegraphics[width=\textwidth]{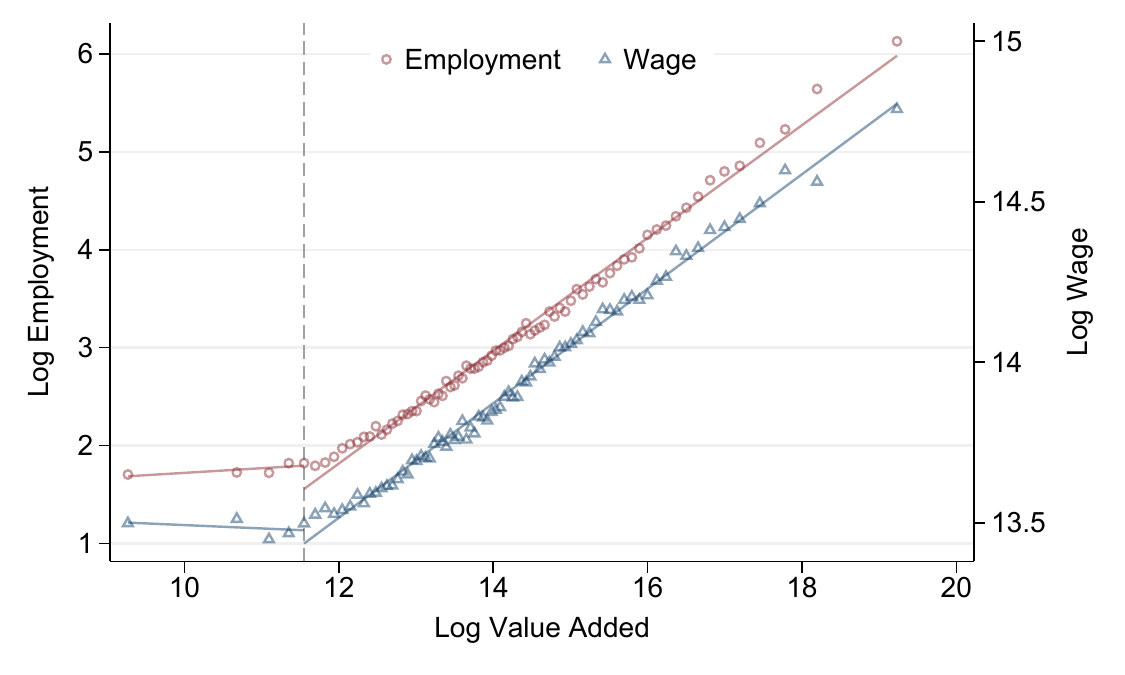}
    \end{subfigure}%
    \begin{subfigure}[t]{0.47\textwidth}
        \centering
        \caption{Firm Responses to Demand Shocks}
        \label{fig:baseline_estimates_colombia}
        \includegraphics[width=\textwidth]{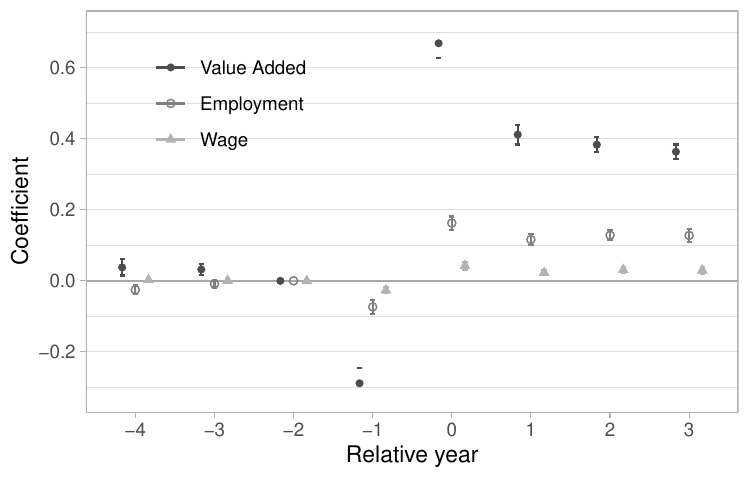}
    \end{subfigure}
    \vspace{2mm}
    \begin{singlespace}
    \begin{minipage}{1\textwidth} \footnotesize
        Notes: 
        The figure shows the relationships of log wages and log employment with log value added (VA), cross-sectionally and following VA shocks, for Colombian manufacturing firms.
        Panel (a) shows the cross-sectional relationships, residualizing each variable with respect to two-digit industry and year effects, for the 2012 cohort.
        We construct 80 equal-sized bins of log VA and compute means of log employment and log wages within each bin.
        Solid lines correspond to linear fits on the binned data.
        Panel (b) shows estimates of the effects of a demand shock, using the internal shock design, on log VA, log wage, and log employment.
        The wage variable corresponds to the mean wage of permanent workers.
    \end{minipage}
    \end{singlespace}
\end{figure}

%% file: figures/lms_by_constraints_colombia.tex
\begin{figure}[hbt!]
    \centering
    \caption{Demand Shocks by Constrained Status and Implied Elasticities.}
    \label{fig:lms_by_constraints_colombia}
    \label{fig:elast_vary_threshold_colombia}

    \begin{subfigure}{.44\textwidth}
        \centering
        \caption{Mean Wage}
        \includegraphics[width=.99\textwidth]{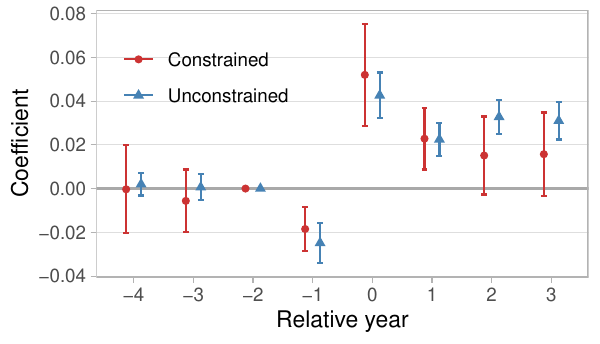}
    \end{subfigure}%
    \begin{subfigure}{.44\textwidth}
        \centering
        \caption{Employment}
        \includegraphics[width=.99\textwidth]{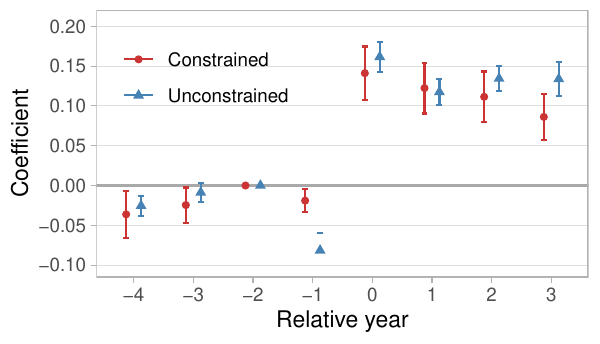}
    \end{subfigure} \\

    \begin{subfigure}{.44\textwidth}
        \centering
        \caption{Labor Supply Elasticity}
        \includegraphics[width=.96\textwidth]{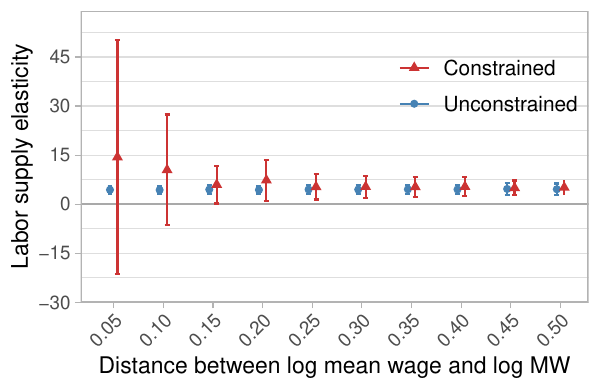}
    \end{subfigure}%
    \begin{subfigure}{.44\textwidth}
        \centering
        \caption{Rent-Sharing Elasticity}
        \includegraphics[width=.96\textwidth]{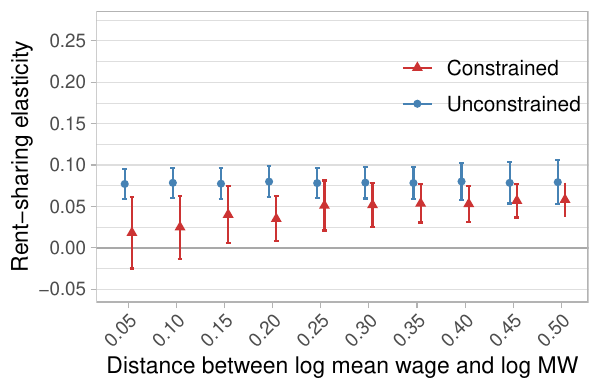}
    \end{subfigure}

    \vspace{2mm}
    \begin{singlespace}
    \begin{minipage}{1\textwidth} \footnotesize
        Notes:
        Panels (a) and (b) show estimates of the effects of a demand shock on log firm mean wage and log employment by constrained status of firms, using the internal instrument design.
        Constrained firms are those with an average wage within 15\% of the national MW. 
        Panels (c) and (d) show estimates of implied labor supply and rent-sharing elasticities varying the threshold used to define constrained firms based on proximity to the national MW.
    \end{minipage}
    \end{singlespace}
\end{figure}

%% file: figures/mwshock_pretends_portugal.tex
\begin{figure}[hbt!]
    \caption{Wage and Employment Responses to MW Hikes in Portugal.}
    \centering
    \label{fig:mw_c_uc_did}
        \begin{subfigure}[t]{0.48\textwidth}
        \centering
        \caption{Mean Wage}
        \includegraphics[width=\linewidth]{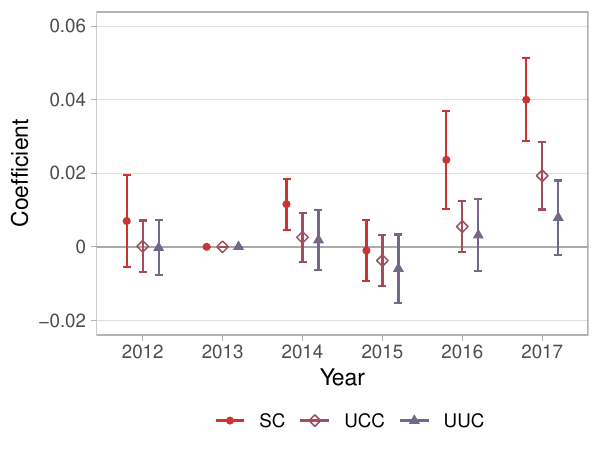}
        \label{fig:mw_c_did}
        \end{subfigure}
        \hfill
        \begin{subfigure}[t]{0.48\textwidth}
        \centering
        \caption{Employment}
        \includegraphics[width=\linewidth]{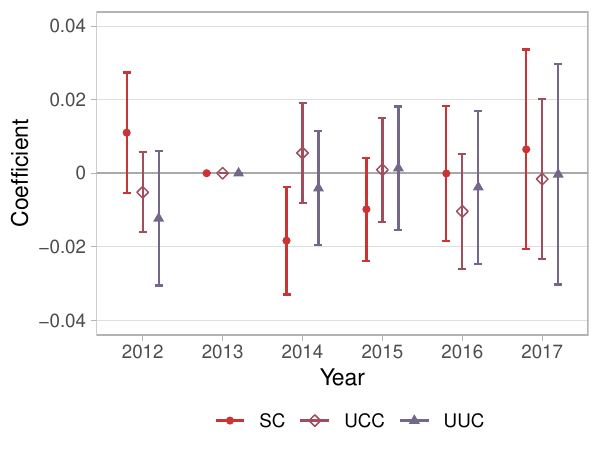}
        \label{fig:mw_uc_did}
    \end{subfigure}
    \vspace{2mm}
    \begin{minipage}{1\textwidth} \footnotesize 
        Notes:
        The figure shows dynamic wage and employment responses to MW hikes in Portugal among firms without a 2016 demand shock ($Z_j=0$).
        Panels (a) and (b) report effects on log mean hourly total wages of stayers and log full-time employment, respectively.
        Estimates compare SC, UCC, and UUC firms with UUU firms, using 2013 as the reference year.
        Constraint groups are defined using mean base wage-to-MW ratios averaged over 2014--2015, as illustrated in Figure~\ref{fig:firm_wages_portugal}.
        Regressions include firm and industry-by-year fixed effects.
        Bars indicate 95\% confidence intervals based on standard errors clustered by two-digit industry.
    \end{minipage}
\end{figure}

%% file: tables/tripleDiff_portugal.tex
\begin{landscape}
\begin{singlespace}
\begin{table}[hbt!]
    \centering
    \caption{Firm Wage, Employment, and Value Added Responses to Minimum Wage and Demand Shocks: Variation in Constraints in Portugal.}
    \label{tab:tripleDiff_portugal}
    \begin{tabular}{l*{9}{c}}
        \toprule
        & \multicolumn{3}{c}{Mean Wage} & \multicolumn{3}{c}{Employment} & \multicolumn{3}{c}{Value Added} \\
        \cmidrule(lr){2-4} \cmidrule(lr){5-7} \cmidrule(lr){8-10}
        & (1) & (2) & (3) & (4) & (5) & (6) & (7) & (8) & (9) \\
        & MW & Demand & Joint & MW & Demand & Joint & MW & Demand & Joint \\
        & Shock & Shock &  & Shock & Shock &  & Shock & Shock &  \\
        \midrule
        SC & 0.030*** &  & 0.032*** & 0.023* &  & 0.022* & 0.030 &  & 0.034 \\
         & (0.007) &  & (0.006) & (0.012) &  & (0.012) & (0.028) &  & (0.024) \\[1mm]
        UCC & 0.019*** &  & 0.022*** & -0.009 &  & -0.010 & 0.014 &  & 0.023 \\
         & (0.006) &  & (0.006) & (0.009) &  & (0.009) & (0.018) &  & (0.017) \\[1mm]
        UUU, Demand shock &  & 0.028*** & 0.028*** &  & 0.151*** & 0.152*** &  & 0.280*** & 0.280*** \\
         &  & (0.006) & (0.006) &  & (0.017) & (0.017) &  & (0.018) & (0.018) \\[1mm]
        SC, Demand shock &  &  & -0.021*** &  &  & -0.038* &  &  & 0.008 \\
         &  &  & (0.008) &  &  & (0.020) &  &  & (0.040) \\[1mm]
        UCC, Demand shock &  &  & -0.018* &  &  & -0.004 &  &  & -0.032 \\
         &  &  & (0.009) &  &  & (0.013) &  &  & (0.026) \\
        \midrule
        $N$ & 16,486 & 17,206 & 34,626 & 16,486 & 17,206 & 34,626 & 16,486 & 17,206 & 34,626 \\
        Firm FE & Yes & Yes & Yes & Yes & Yes & Yes & Yes & Yes & Yes \\
        Industry $\times$ Year FE & Yes & Yes & Yes & Yes & Yes & Yes & Yes & Yes & Yes \\
        \bottomrule
    \end{tabular}

    \vspace{2mm}
    \begin{minipage}{.95\linewidth} \footnotesize
        Notes:
        The table shows the effects of MW increases and a demand shock on log mean hourly wages (Columns 1--3), log employment (Columns 4--6), and log value added (Columns 7--9) of Portuguese firms.
        All columns compare 2017 with 2014.
        Groups are defined using firms' mean base wage-to-MW ratios averaged over 2014--2015, as described in Section~\ref{subsec:data_var_constr}.
        The demand-shock indicator $Z_j$ equals one for firms with above-median value-added growth between 2015 and 2016 within their two-digit industry.
        MW shock columns compare SC and UCC firms with UUU firms among firms without a demand shock ($Z_j=0$).
        Demand shock columns compare firms with and without a demand shock within the UUU group.
        Joint columns include SC, UCC, and UUU firms, with UUU firms without a demand shock ($Z_j=0$) as the reference group.
        Mean wages are measured for stayers in all three wage specifications.
        All specifications include firm and industry-by-year fixed effects.
        Standard errors in parentheses are clustered at the two-digit industry level.
        {*} \(p<0.1\), {**} \(p<0.05\), {***} \(p<0.01\).
    \end{minipage}

\end{table}
\end{singlespace}
\end{landscape}

%% file: m_appendix_policy.tex
\setcounter{equation}{0} 
\setcounter{table}{0}
\setcounter{figure}{0}

\section{General Equilibrium Model}
\label{asec:ge_model}

This appendix describes the general equilibrium model from Section \ref{sec:m_policy}.
We first specify how we extend the theoretical framework in Section \ref{sec:m_framework_theory}.
We then describe the data and calibration approach.
Finally, we explain how we compute the optimal MW under correct and biased beliefs about the labor supply elasticity.
We calibrate ``true economies'' to different values of the labor market elasticity $\eta$, and study how biased beliefs matter for different levels of $\eta$.

\subsection{Theoretical Model}

We partition the labor market into local labor markets $k$, each endowed with a mass of firms that are heterogeneous in their productivity.
We normalize the potential worker population to 1.
Workers decide whether to enter the labor market and, if entering, they choose a local labor market $k$ and a firm $j$.

\paragraph{Firms.}

Firms produce according to $Y_j=\Phi_j L_j^\alpha$, with $0<\alpha<1$, and face a common minimum wage $\W$.
Firms are located in a sector-region cell $c$ and, within the cell, a local labor market $k$.
Specifically, we separate each cell into a high- and low-productivity market, and assign firms permanently to each, based on their productivity.
This allows weaker substitution across low- and high-productivity firms than within each local market.

Each firm solves the same optimization problem as in the main text, with a labor supply curve that depends on the local labor market $k$ (specified below).
We maintain the assumption that firms are atomistic.
The firm's solution has three regimes as in Appendix \ref{asec:derivations_wage_floor_const}.
Unconstrained firms pay $W_j=\left(\eta/(1+\eta)\right)\alpha\Phi_jL_j^{\alpha-1}$.
Supply-constrained firms pay $\W$.
Demand-constrained firms pay $\W$, choose $L_j=(\alpha\Phi_j/\W)^{1/(1-\alpha)}$, and ration jobs ($\psi_j<1$).

Each cell is endowed with a continuum of firms $M_c$ with productivity drawn from
\begin{equation}\label{eq:ge_productivity}
    F_c(\Phi)=1-\left(\frac{\Phi}{\Phi_c^{\min}}\right)^{-q_s} ,
\end{equation}
if $\Phi\geq\Phi_c^{\min}$ and 0 otherwise.
The shape $q_s$ varies across sectors $s$ and minimum productivity $\Phi_c^{\min}$ varies across sector-regions $c$.
Firm masses, productivity, and local market membership remain fixed when the minimum wage changes, and we abstract from entry and exit.

\paragraph{Labor supply.}

Labor supply to firm $j$ in market $k$ is
\begin{equation}\label{eq:ge_firm_supply}
    H_j=B_k\left[A_c(\Phi_j)\psi_jW_j\right]^\eta,
    \qquad \psi_j=\frac{L_j}{H_j},
\end{equation}
where $A_c(\Phi_j)$ is a firm amenity and $B_k$ is an endogenous market labor supply term.
This follows the structure in, e.g., \citet{Hermo2025} or \citet{AzkarateAskasuaZerecero2025}.
We additionally allow amenities to vary with productivity to match the empirical cross-sectional relationship between employment and wages.
This is important because the model imposes a linear cross-sectional relationship between $\ln L_j$ and $\ln W_j$ with slope $\eta$, which is rejected by the data.
Specifically, we assume a continuous amenity schedule that is constant at low productivity, declines at rate $d$ in the middle, and declines at rate one in the upper tail:
\begin{equation}\label{eq:ge_amenities}
    A_c(\Phi)=
    \begin{cases}
        1, & \Phi\leq\Phi_{1c},\\
        (\Phi/\Phi_{1c})^{-d}, & \Phi_{1c}<\Phi\leq\Phi_{2c},\\
        (\Phi_{2c}/\Phi_{1c})^{-d}(\Phi/\Phi_{2c})^{-1}, & \Phi>\Phi_{2c},
    \end{cases}
\end{equation}
where $\Phi_{1c}$ and $\Phi_{2c}$ are cell-specific productivity thresholds.%
    \footnote{With constant amenities, $\log L_j$ among unconstrained firms is linear in $\log W_j$ with slope $\eta$.
    Under \eqref{eq:ge_amenities}, the unconstrained log employment-wage slope is $\eta$ below $\Phi_{1c}$, $\eta(1-d)/[1+\eta(1-\alpha)d]$ between $\Phi_{1c}$ and $\Phi_{2c}$, and zero above $\Phi_{2c}$.}

Let $F_k$ denote the market's productivity distribution.%
    \footnote{$F_k$ is the distribution within $c$ conditional on the high- or low-productivity segment of the cell.}
The market wage index is
\begin{equation}\label{eq:ge_wage_index_market}
    \mathcal{W}_k 
       = \left\{ M_k \int\left[A_c(\Phi)\psi(\Phi)W(\Phi)\right]^\eta dF_k(\Phi)
                \right\}^{1/\eta} ,
\end{equation}
and the aggregate wage index is $\mathcal{W}=\left[\sum_k(\mathcal{A}_k\mathcal{W}_k)^\sigma\right]^{1/\sigma}$, where the cross-market elasticity $\sigma\in(1,\eta)$ governs substitution across markets and scalars $\mathcal{A}_k$ give workers' preferences for local labor market $k$.%
    \footnote{A micro-foundation for \eqref{eq:ge_firm_supply} in which workers face idiosyncratic preference shocks when choosing firm $j$ in market $k$ yields
    $V=\Gamma(1-1/\eta)\Gamma(1-1/\sigma)\mathcal{W}$, where $\Gamma(\cdot)$ is the gamma function \citep[see, e.g.,][]{Hermo2025}.}
Conditional on entry to the labor market, the share of workers choosing market $k$ is
$s_k=(\mathcal{A}_k\mathcal{W}_k)^\sigma/\mathcal{W}^\sigma$.
Labor market entry depends on the expected value of working, $V$, relative to an outside option $\bar v$:
\begin{equation}\label{eq:ge_participation}
    \iota(V)=\frac{1}{1+(\bar v/V)^\zeta} .
\end{equation}

\paragraph{General equilibrium.}

An equilibrium combines optimal firm choices and worker participation and allocation with labor market clearing, given the MW and the model parameters.
We distinguish baseline parameters, $\alpha$, $\sigma$, and $M_c$, from parameters estimated conditional on $\eta$: productivity shape parameters $q_s$ and minima $\Phi_c^{\min}$, amenity parameters $d$, $\Phi_{1c}$, and $\Phi_{2c}$, market preferences $\mathcal A_k$, and participation parameters $\zeta$ and $\bar v$.

Given $\eta$ and these parameters, an equilibrium consists of firm wages $W_j$, employment $L_j$, labor supply $H_j$, hiring probabilities $\psi_j$, and market labor supply terms $B_k$ that satisfy the firm choices above, Equations \eqref{eq:ge_firm_supply}--\eqref{eq:ge_participation}, and labor market clearing.
Labor market clearing imposes that the measure of workers supplying labor to firms in market $k$, $M_k \int B_k \left[A_c(\Phi)\psi(\Phi)W(\Phi)\right]^\eta d F_k(\Phi) = B_k \mathcal W_k^\eta$, equals the share of workers supplying labor to the market, $\iota(V) s_k$.
The equality implies that
\begin{equation}
    \label{eq:labor_market_shifter}
    B_k = \frac{\iota(V)s_k}{\mathcal W_k^\eta} .
\end{equation}

For each candidate MW, we solve jointly for the labor supply terms $B_k$ that clear all markets.
Aggregate employment is $L(\W)=\sum_k M_k\int L_j(\Phi;\W)dF_k(\Phi)$, where the expression for $L_j$ depends on the constraint status of each firm.

\subsection{Data and Baseline Parameters}

We calibrate the model to QP data from 2009--2011 and moments from our internal instrument sample in Portugal.
We use firms from 2009--2011 observed for at least four consecutive years and firm-years with positive employment and a nonmissing hourly base wage.
We group firms into 14 broad sectors and four regions, yielding 56 sector-region cells, and split those into wages above and below 1.5 times the MW, yielding 112 local markets.%
    \footnote{We use a 1.5 multiplier following the findings in \citet{Oliveira2023}, who finds spillover effects of the MW up to 45\% above it in Portugal during 2006--2019.
    The separation isolates firms with wages above 1.5 times the MW from direct spillovers, though they could still be affected by cross-market spillovers.}
We pool the three years and use firm-year frequencies to determine $M_c$, normalized to sum to one.

We calibrate the reference MW ($\W_0$) once to match a 29\% constrained share (Table \ref{tab:lms_descriptives}) and hold it fixed across true economies, that is, across values of the ``true'' elasticity $\eta$ used in the calibration.
We multiply annual hourly minimum wages by a common factor so that 29\% of firm-years satisfy $\log(W_{jt}/\W_t)\leq0.15$.
We use the annual real MW (transformed to hours using 173 hours per month) adjusted by this factor to construct constrained shares and the ratio of mean wage to the MW.
The reference $\W_0$ equals the mean annual real hourly MW over 2009--2011, adjusted by the same factor.

The baseline parameters $\alpha$, $\sigma$, and $M_c$ are also fixed across true economies and do not vary with $\eta$.
We set the cross-market elasticity to $\sigma=1.5$ and estimate $\alpha$ once using $\eta=5$, close to Portugal's estimate.
For unconstrained firms, the wage markdown implies a labor share of $\mu\alpha$.
We compute the aggregate labor share among unconstrained firms (firm-years with $\log(W_{jt}/\W_t)>0.15$) to obtain $\WelfLaborShare$, and divide by $\mu=5/6$ to get $\alpha=\WelfAlpha$.
The firm masses $M_c$ are the observed firm-year shares described above.
After calibrating the MW and $\alpha$, we trim the top 0.1\% of firm wages within each sector before computing these shares and estimating the remaining parameters.

\subsection{Estimation and Calibration Conditional on \texorpdfstring{$\eta$}{eta}}

For each true $\eta\in\{3,4,5,6,7\}$, we estimate firm heterogeneity and calibrate the economy to the same empirical moments, keeping the baseline parameters fixed.

\paragraph{Amenities.}

To recover the parameters $d$, $\Phi_{1c}$, and $\Phi_{2c}$ in Equation \eqref{eq:ge_amenities}, we estimate the piecewise-linear cross-sectional relationship between log employment and log wages implied by the model for unconstrained firms.
Specifically, let $x_{jt}=\log(W_{jt}/\W_0)$ and estimate
\begin{equation}\label{eq:ge_employment_profile}
    \log \E[L_{jt}\mid x_{jt},c,t]
    =\lambda_{ct}+\eta\min\{x_{jt},\underline{x}\}
    +\beta\min\{\max\{x_{jt}-\underline{x},0\},\overline{x}-\underline{x}\},
\end{equation}
where $\lambda_{ct}$ are cell-by-year effects and $0<\underline{x}<\overline{x}$ are common log wage cutoffs.
Consistent with the amenity schedule, the slope is restricted to $\eta$ below $\underline{x}$ and to zero above $\overline{x}$.
We estimate the middle slope $\beta$ and both cutoffs jointly.%
    \footnote{We use firm-level Poisson pseudo-maximum likelihood, so the specification targets mean employment conditional on wages rather than mean log employment.
    We search cutoffs over a two-dimensional grid and refine them continuously from several starting points, selecting the converged solution with the highest Poisson likelihood.
    For each cutoff pair we fit the model to obtain $\beta$.}
We then invert the theoretical middle-segment slope $\beta = \eta(1-d)/[1+\eta(1-\alpha)d]$ using $\eta$ and $\alpha$ to recover $d$, and map the estimated wage cutoffs $\underline{x}$ and $\overline{x}$ into the productivity thresholds $\Phi_{1c}$ and $\Phi_{2c}$.
Thus $d$, $\Phi_{1c}$, and $\Phi_{2c}$ vary across true economies.
Appendix Figure \ref{fig:struc_model_empl_wage} shows the implied cross-sectional relationship for wages above the MW.

\input{figures/struc_model_empl_wage}

\vspace{-3mm}

\paragraph{Productivity.}

Conditional on $\eta$ and amenities, we estimate one Pareto shape $q_s$ per sector by matching its mean firm wage relative to the MW, weighting firms equally.
Following the Pareto assumption, each cell's constrained share is $p_c = 1 - \left(\Phi_c^{\min}/\upperPhi_c\right)^{q_s}$, where $\upperPhi_c$ is the upper threshold from Section~\ref{sec:m_framework_theory}, thus $\left(\upperPhi_c/\Phi_c^{\min}\right) = (1-p_c)^{-1/q_s}$.
Given this ratio and the amenity schedule, we integrate the implied wage schedule to obtain an equation for the ratio of the sector mean wage to the MW as a function of $q_s$ only.
We choose $q_s$ to equate this prediction to its empirical counterpart.
The productivity levels $\Phi_c^{\min}$ and $\upperPhi_c$ are determined jointly with baseline labor supply in the next subsection.

\paragraph{Baseline allocation.}

We calibrate the remaining parameters, $\Phi_c^{\min}$, $\mathcal A_k$, $\zeta$, and $\bar v$, to the same set of data moments for each $\eta$, allowing their values to differ across true economies.

First, we choose $\Phi_c^{\min}$ and baseline $B_k$ to match each cell's constrained firm share and employment share, and an aggregate employment-to-population ratio of $L_0 = \WelfEmploymentRate$.%
    \footnote{The employment-to-population ratio is the 2009--2011 average employment rate among people aged 15--64, from the Eurostat table ``Employment and labour force by sex and age - annual data'' \texttt{lfsi\_emp\_a} \href{https://ec.europa.eu/eurostat/api/dissemination/statistics/1.0/data/lfsi_emp_a?geo=PT&age=Y15-64&sex=T&unit=PC_POP&indic_em=EMP_LFS&sinceTimePeriod=2009&untilTimePeriod=2011}{(API query)}.}
We assume a common baseline $B_k$ for the two markets within each cell and separate them at the productivity where the baseline wage reaches $1.5 \W_0$.
The resulting allocation determines baseline participation, $\iota_0 = \sum_k M_k\int H_j(\Phi) dF_k(\Phi)$.

We then compute the market wage indices using Equation \eqref{eq:ge_wage_index_market} and choose $\mathcal A_k$ to support this allocation through market clearing in Equation \eqref{eq:labor_market_shifter}.
We normalize their common scale so that $V_0=1$, since scaling all $\mathcal A_k$ and $\bar v$ proportionally leaves participation and market shares unchanged.
Finally, we set $\zeta=\WelfParticipationElasticity/(1-\iota_0)$ to match a participation elasticity of \WelfParticipationElasticity, and choose $\bar v$ to reproduce the implied $\iota_0$ at $V_0=1$.%
    \footnote{The participation elasticity, which we use for calibration, is $d\log\iota/d\log V=\zeta(1-\iota)$.
    \citet{Chetty2011} recommend a value of 0.25 for the extensive-margin elasticity.
    We vary the participation-elasticity target and cross-market labor supply elasticity $\sigma$ in sensitivity checks.}

\input{tables/struc_model_calibration}

Appendix Table \ref{tab:struc_model_calibration} summarizes targeted and untargeted moments for $\eta=5$.
The model matches cell constrained shares and employment shares, and sector mean wage ratios, by construction.
It also closely reproduces the share of employment at constrained firms: 10.05\% in the model compared with 10.15\% in the data.
The model predicts a demand-constrained firm share of 8.52\%, compared with \WelfDemandConstrainedData\% from the empirical exercise in Section \ref{subsec:resu_byconst}.
The fit is less precise for other distributional moments. 

\subsection{Optimal Minimum Wages and Biased Beliefs}

For each true elasticity $\eta\in\{3,4,5,6,7\}$, we reestimate the amenity profile and sector productivity shapes, and recalibrate the baseline allocation to the same empirical targets.
The production elasticity $\alpha$, reference MW, cross-market elasticity, employment-rate target, and participation-elasticity target remain fixed across these economies.

Within a given true economy, let $\mathcal{P}_\eta$ collect all primitives other than $\eta$, and write aggregate employment as $L(\W;\eta,\mathcal{P}_\eta)$.
The true employment-maximizing MW solves
\begin{equation*}
    \W_\eta^*\in\operatorname*{arg\,max}_{\W\geq0}
    L(\W;\eta,\mathcal{P}_\eta).
\end{equation*}
The planner instead perceives $\widetilde\eta=\eta(1+b)$ and chooses
\begin{equation*}\label{eq:ge_true_optimum}
    \widetilde\W_\eta(b)\in\operatorname*{arg\,max}_{\W\geq0}L(\W;\widetilde\eta,\mathcal{P}_\eta).
\end{equation*}
Only the perceived elasticity changes; the remaining parameters $\mathcal P_\eta$ remain at their true-economy values.
We then evaluate the chosen minimum wage in the true economy to obtain realized employment, $L(\widetilde\W_\eta(b);\eta,\mathcal{P}_\eta)$.

Let $L_\eta^{\mathrm{nb}}$ denote employment at a nonbinding minimum wage.
The share of attainable employment gains forgone is
\begin{equation*}\label{eq:ge_unrealized_gain}
    \frac{L(\W_\eta^*;\eta,\mathcal{P}_\eta)
          -L(\widetilde\W_\eta(b);\eta,\mathcal{P}_\eta)}
         {L(\W_\eta^*;\eta,\mathcal{P}_\eta)-L_\eta^{\mathrm{nb}}}.
\end{equation*}
Thus the loss is measured relative to the employment gain achievable in the true economy, starting from a nonbinding policy.
We evaluate biases from $-50\%$ to $50\%$ in $0.5\%$ intervals.%
    \footnote{To find each optimum, we search a broad minimum-wage grid scaled by the lowest wage in the no-minimum-wage equilibrium, refine every grid-local employment maximum, and check that employment is lower in the distant tail.
    Firm outcomes are integrated analytically over each productivity segment, including the unbounded upper tail, and market clearing is imposed at every policy evaluation.}

%% file: figures/struc_model_empl_wage.tex
\begin{figure}[hbt!]
    \centering
    \caption{Firm Wages and Employment: Data versus Calibrated Model.}
    \label{fig:struc_model_empl_wage}
    \includegraphics[width=.55\linewidth]{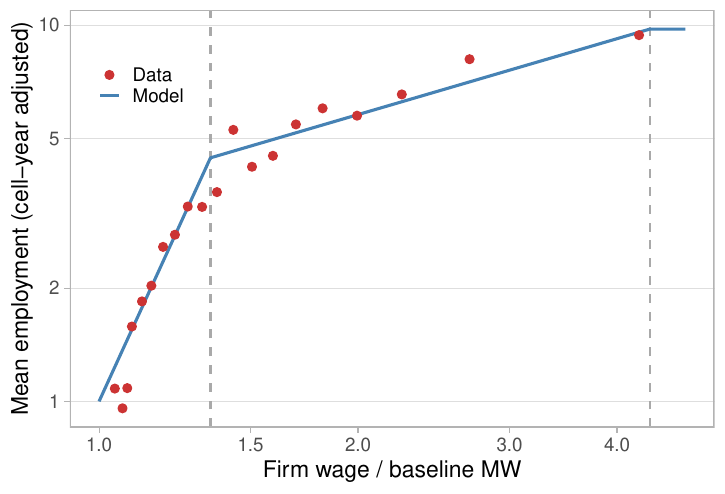}
    \vspace{2mm}
    \begin{minipage}{1\textwidth} \footnotesize
        Notes: 
        The figure compares the cross-sectional relationship between employment and wages between the data and the calibrated model.
        The sample contains firms paying above the annual MW.
        Points show mean employment divided by the fitted cell-by-year effect in levels, in 20 wage bins. 
        The horizontal coordinate is the geometric mean wage relative to the baseline MW. 
        Dashed lines mark the estimated wage cutoffs.
    \end{minipage}
\end{figure}

%% file: tables/struc_model_calibration.tex
\begin{table}[hbt!]
    \centering
    \caption{General Equilibrium Model: Baseline Parameters and Estimates.}
    \label{tab:struc_model_calibration}
    \small

    \begin{tabular}{@{}lcc@{}}
        \toprule
        & Data / target & Model \\
        \midrule
        \multicolumn{3}{l}{\textbf{Panel A: Targeted moments}} \\[1mm]
        \hspace{3mm} Constrained firms (\%) & 29.0 & 29.0 \\
        \hspace{3mm} Mean firm wage / MW & 1.618 & 1.618 \\
        \hspace{3mm} Sector-region cell employment shares: range (\%) & 0.1--14.0 & 0.1--14.0 \\
        \hspace{3mm} Sector-region cell firm shares: range (\%) & 0.0--11.2 & 0.0--11.2 \\
        \hspace{3mm} Employment / working-age population (\%) & 62.0 & 62.0 \\
        \midrule
        \multicolumn{3}{l}{\textbf{Panel B: Non-targeted moments}} \\[1mm]
        \hspace{3mm} Employment at constrained firms (\%) & 10.1 & 10.0 \\
        \hspace{3mm} Demand-constrained firms (\%) & 3.9 & 8.5 \\
        \hspace{3mm} Worker-weighted median wage / MW & 1.625 & 1.431 \\
        \hspace{3mm} Worker-weighted P90 wage / MW & 3.245 & 4.283 \\
        \hspace{3mm} Employment at firms paying above twice the MW (\%) & 32.0 & 32.0 \\
        \hspace{3mm} Firm P90 wage / MW & 2.420 & 2.124 \\
        \hspace{3mm} Firm employment P90 / P10 & 19.00 & 10.95 \\
        \bottomrule
    \end{tabular}

    \vspace{2mm}
    \begin{minipage}{1\textwidth} \footnotesize
        Notes: The table illustrates model calibration and fit.
        Panel A shows targeted moments, used to calibrate the model, and Panel B shows untargeted moments. 
    \end{minipage}
\end{table}

%% file: tables/cohort_size.tex
\begin{table}[H]
    \centering
    \caption{Number of Firms Used in Estimations By Country and Cohort.}
    \label{tab:cohort_size}
    
    \begin{tabular}{@{}lccc@{}}
        \toprule
         & Portugal & \ Norway \ & Colombia \\ \midrule
        2001 &        & 26,739 &        \\
        2002 &        & 27,866 &        \\
        2003 &        & 28,809 &        \\
        2004 &        & 29,730 & 5,012 \\
        2005 &        & 30,192 & 5,022 \\
        2006 &        & 30,518 & 5,247 \\
        2007 &        & 30,598 & 5,253 \\
        2008 & 51,966 & 31,076 & 5,382 \\
        2009 & 51,989 & 31,966 & 5,377 \\
        2010 & 50,494 & 32,802 & 5,358 \\
        2011 & 51,531 & 33,949 & 5,799 \\
        2012 & 52,506 & 34,647 & 6,764 \\
        2013 & 55,487 & 34,879 & 7,420 \\
        2014 & 56,984 & 36,720 & 7,026 \\
        2015 &        & 38,120 &        \\
        2016 &        & 39,302 &        \\
       \bottomrule
    \end{tabular}

    \vspace{2mm}
    \begin{minipage}{1\textwidth} \footnotesize
        Notes:
        The table shows the number of firms per cohort that we use for the internal shock design.
    \end{minipage}
\end{table}

%% file: tables/exporter_descriptives.tex
\begin{table}[ht!]
    \centering
    \caption{Descriptive Statistics for Exporting and Non-Exporting Firms.}
    \label{tab:exporter_descriptives}
    \begin{tabular}{@{}lccccccc@{}}
        \toprule
        & \multicolumn{3}{c}{Portugal} & \multicolumn{3}{c}{Norway} \\
        \cmidrule(lr){2-4} \cmidrule(lr){5-7}
        &  &  & Non- &  & & Non- \\
         & All & Exporter & Exporter & All & Exporter & Exporter \\
        \cmidrule(lr){2-7}
        & (1) & (2) & (3) & (4) & (5) & (6) \\
        \midrule
        \textit{Number of firms} & 52,814 & 4,095 & 48,719 & 31,539 & 2,967 & 28,572 \\
        \hspace{3mm} (Share exporting) &   & (0.08) & (0.92) &   & (0.09) & (0.91) \\
        \addlinespace[1mm]
        \textit{Log value added} & & & & & & \\
        \hspace{3mm} Mean & 12.22 & 13.57 & 12.11 & 15.64 & 16.76 & 15.52 \\
        \hspace{3mm} SD & 1.32 & 1.39 & 1.25 & 1.26 & 1.44 & 1.18 \\
        \addlinespace[1mm]
        \textit{Log Employment} & & & & & & \\
        \hspace{3mm} Mean & 2.22 & 3.46 & 2.11 & 2.37 & 3.22 & 2.28 \\
        \hspace{3mm} SD & 1.08 & 1.20 & 1.00 & 1.13 & 1.28 & 1.08 \\
        \addlinespace[1mm]
        \textit{Log Wages} & & & & & & \\
        \hspace{3mm} Mean & 1.64 & 1.77 & 1.63 & 5.47 & 5.60 & 5.46 \\
        \hspace{3mm} SD & 0.44 & 0.45 & 0.44 & 0.34 & 0.30 & 0.35 \\
        \addlinespace[1mm]
        \bottomrule
    \end{tabular}

    \vspace{2mm}
    \begin{minipage}{1\textwidth} \footnotesize
    Notes: 
    The table shows descriptive statistics for key firm-level variables in 2007.
    The sample is restricted to firms with non-negative value added, at least two stayers per year,
    and observed continuously over the cohort window, where stayers are workers who remain with
    the same firm for at least seven years within each eight-year cohort window.
    ``All'' refers to all firms satisfying these restrictions.
    ``Exporter'' refers to firms that exported in at least three years prior to or including 2007
    and whose cumulative real exports over 2005--2007 accounted for at least 1\% of their cumulative
    real revenue over the same period. For Portugal, we additionally exclude firms whose share of
    exports to Angola or Spain over 2005--2007 exceeded 90\% of total exports, following
    \cite{garin2024responsive}.
    ``Non-Exporter'' refers to all remaining firms in the sample.
    The wage variable corresponds to the mean hourly wage of stayers,
    measured in Euros (Portugal) and Norwegian Kroner (Norway).
\end{minipage}
\end{table}

%% file: figures/model_variation_constraints_design.tex
\begin{figure}[ht!]
    \centering
    \caption{Triple-Difference Design with Wage Floor Hike and Demand Shock.}
    \label{fig:model_variation_constraints_design}

    \def\xmax{2.7}
    \def\ymax{1.5}
    
    \pgfmathsetmacro{\tmin}{0}
    \pgfmathsetmacro{\tmax}{2.4}
    \pgfmathsetmacro{\tstar}{1.3}
    
    \pgfmathsetmacro{\yzero}{0.2}
    \pgfmathsetmacro{\factorDiffEmpPlot}{0.9}
    \pgfmathsetmacro{\factorDiffWagePlot}{3.5}
    
    \pgfmathsetmacro{\A}{1}
    \pgfmathsetmacro{\eta}{3.5}
    \pgfmathsetmacro{\alpha}{0.42}
    \pgfmathsetmacro{\mu}{\eta / (\eta+1)}
    
    \pgfmathsetmacro{\phiMax}{4.2}
    \pgfmathsetmacro{\phiMin}{0.5}
    \pgfmathsetmacro{\shockFactor}{1.75}
    \pgfmathsetmacro{\empRescale}{.75}
    
    \pgfmathsetmacro{\minWpre}{0.4} 
    \pgfmathsetmacro{\minWpost}{0.75} 

    \pgfmathsetmacro{\phiLowerThreshPre}{(\minWpre/\A)^(1+\eta*(1-\alpha)) * \A / (\alpha)}
    \pgfmathsetmacro{\phiUpperThreshPre}{\phiLowerThreshPre / \mu}
    
    \pgfmathsetmacro{\phiLowerThreshPost}{(\minWpost/\A)^(1+\eta*(1-\alpha)) * \A / (\alpha)}
    \pgfmathsetmacro{\phiUpperThreshPost}{\phiLowerThreshPost / \mu}

    \pgfkeys{/pgf/declare function={
        lstarPhi(\p) = ((\p*\alpha/\A) * \mu)^(\eta/(1+\eta*(1-\alpha)));
        wstarPhi(\p) = \A*(lstarPhi(\p))^(1/\eta);
        lstarConstrainedPhiPre(\p) = (\minWpre/\A)^(\eta);
        lstarDemandPhiPre(\p) = (\p*\alpha/\minWpre)^(1/(1-\alpha));
        lstarConstrainedPhiPost(\p) = (\minWpost/\A)^(\eta);
        lstarDemandPhiPost(\p) = (\p*\alpha/\minWpost)^(1/(1-\alpha));
    }}

    \pgfmathsetmacro{\phiCC}{(\phiLowerThreshPre + \phiMin)/2} 
    \pgfmathsetmacro{\phiUC}{(\phiUpperThreshPre * (1/3) + \phiUpperThreshPost * (2/3))}
    \pgfmathsetmacro{\phiUU}{2 * \phiUpperThreshPost}

    \pgfmathsetmacro{\wageUCpre}{wstarPhi(\phiUC)}
    \pgfmathsetmacro{\wageUU}{wstarPhi(\phiUU)}
    \pgfmathsetmacro{\wageUCpostRaw}{wstarPhi(\shockFactor*\phiUC)}
    \pgfmathsetmacro{\wageUCpost}{max(\minWpost, \wageUCpostRaw)}
    \pgfmathsetmacro{\wageUUpost}{wstarPhi(\shockFactor*\phiUU)}
    \pgfmathsetmacro{\wageDiffUCpost}{\yzero + \factorDiffWagePlot*(ln(\wageUCpost) - ln(\minWpost))}
    \pgfmathsetmacro{\wageDiffUUpost}{\yzero + \factorDiffWagePlot*(ln(\wageUUpost) - ln(\wageUU))}

    \pgfmathsetmacro{\empCCpre}{\empRescale*lstarDemandPhiPre(\phiCC)}
    \pgfmathsetmacro{\empUCpre}{\empRescale*lstarPhi(\phiUC)}
    \pgfmathsetmacro{\empUUpre}{\empRescale*lstarPhi(\phiUU)}
    \pgfmathsetmacro{\empCCpostNoShock}{\empRescale*lstarDemandPhiPost(\phiCC)}
    \pgfmathsetmacro{\empUCpostNoShock}{\empRescale*lstarConstrainedPhiPost(\phiUC)} 
    \pgfmathsetmacro{\empUUpostNoShock}{\empRescale*lstarPhi(\phiUU)} 
    \pgfmathsetmacro{\empCCpostWithShock}{lstarDemandPhiPost(\shockFactor*\phiCC)}
    \pgfmathsetmacro{\wageCheck}{wstarPhi(\shockFactor*\phiUC)}
    \pgfmathsetmacro{\empUCpostWithShock}{(\wageCheck > \minWpost) ? \empRescale*lstarPhi(\shockFactor*\phiUC) : \empRescale*lstarConstrainedPhiPost(\shockFactor*\phiUC)}
    \pgfmathsetmacro{\empUUpostWithShock}{\empRescale*lstarPhi(\shockFactor*\phiUU)}
    \pgfmathsetmacro{\empDiffCCpost}{\yzero + \factorDiffEmpPlot*(ln(\empCCpostWithShock) - ln(\empCCpostNoShock))}
    \pgfmathsetmacro{\empDiffUCpost}{\yzero + \factorDiffEmpPlot*(ln(\empUCpostWithShock) - ln(\empUCpostNoShock))}
    \pgfmathsetmacro{\empDiffUUpost}{\yzero + \factorDiffEmpPlot*(ln(\empUUpostWithShock) - ln(\empUUpostNoShock))}

    \pgfmathsetmacro{\empShift}{.25}
    \pgfmathsetmacro{\empCCpre}{\empShift + \empCCpre}
    \pgfmathsetmacro{\empUCpre}{\empShift + \empUCpre}
    \pgfmathsetmacro{\empUUpre}{\empShift + \empUUpre}
    \pgfmathsetmacro{\empCCpostNoShock}{\empShift + \empCCpostNoShock}
    \pgfmathsetmacro{\empUCpostNoShock}{\empShift + \empUCpostNoShock}
    \pgfmathsetmacro{\empUUpostNoShock}{\empShift + \empUUpostNoShock}
    \pgfmathsetmacro{\empCCpostWithShock}{\empShift + \empCCpostWithShock}
    \pgfmathsetmacro{\empUCpostWithShock}{\empShift + \empUCpostWithShock}
    \pgfmathsetmacro{\empUUpostWithShock}{\empShift + \empUUpostWithShock}

    \tikzset{
        unifiedplot/.style={
            scale=2.5,
            font=\small,
        }
    }
    
    \begin{minipage}{.5\textwidth} \centering
        Wage (Left Column)
    \end{minipage}%
    \begin{minipage}{.5\textwidth} \centering
        Employment (Right Column)
    \end{minipage} \\
    \vspace{2mm}
    \begin{minipage}{.95\textwidth} \centering
        \textbf{(a)} Evolution of Outcomes: Firms Without a Demand Shock
    \end{minipage} \\
    \begin{subfigure}[t]{0.5\textwidth}
        \centering
        \begin{tikzpicture}[unifiedplot]
            \draw[->, thick] (0,0) -- (\xmax,0) node[below left] {Time};
            \draw[->, thick] (0,0) -- (0, \ymax) node[left] {$W$};
            
            \draw[dashed, lightgray, thin] (\tmin, \minWpre) -- (\tmax, \minWpre);
            \node[font=\small, anchor=east] at (0, \minWpre) {$\underline{W}^{\text{pre}}$};
            \draw[dashed, lightgray, thin] (\tmin, \minWpost) -- (\tmax, \minWpost);
            \node[font=\small, anchor=east] at (0, \minWpost) {$\underline{W}^{\text{post}}$};

            \draw[dashed, lightgray, thin] (\tstar, 0) node[below, gray] {$t^*$} -- (\tstar, \ymax);

            \draw[CoolRed, very thick] (\tmin, \minWpre) -- (\tstar, \minWpre);
            \draw[CoolRed, very thick] (\tstar, .993*\minWpost) -- (\tmax, .993*\minWpost);
            \node[font=\small, anchor=north west, CoolRed] at (\tmax, \minWpost) {CC};

            \draw[CoolMix, very thick, densely dashed] (\tmin, \wageUCpre) -- (\tstar, \wageUCpre);
            \draw[CoolMix, very thick, densely dashed] (\tstar, 1.007*\minWpost) -- (\tmax, 1.007*\minWpost);
            \node[font=\small, anchor=south west, CoolMix] at (\tmax, \minWpost) {UC};
            
            \draw[CoolBlue, very thick] (\tmin, \wageUU) -- (\tmax, \wageUU);
            \node[font=\small, anchor=west, CoolBlue] at (\tmax, 1.075*\wageUU) {UU};
        \end{tikzpicture}
    \end{subfigure}%
    \begin{subfigure}[t]{0.5\textwidth}
        \centering
        \begin{tikzpicture}[unifiedplot]
            \draw[->, thick] (0,0) -- (\xmax,0) node[below left] {Time};
            \draw[->, thick] (0,0) -- (0, \ymax) node[left] {$L$};
            
            \draw[dashed, lightgray, thin] (\tstar, 0) node[below, gray] {$t^*$} -- (\tstar, \ymax);

            \draw[CoolRed, very thick] (\tmin, \empCCpre) -- (\tstar, \empCCpre);
            \draw[CoolRed, very thick] (\tstar, \empCCpostNoShock) -- (\tmax, \empCCpostNoShock);
            \node[font=\small, anchor=west, CoolRed] at (\tmax, \empCCpostNoShock) {CC};

            \draw[CoolMix, very thick, densely dashed] (\tmin, \empUCpre) -- (\tstar, \empUCpre);
            \draw[CoolMix, very thick, densely dashed] (\tstar, \empUCpostNoShock) -- (\tmax, \empUCpostNoShock);
            \node[font=\small, anchor=west, CoolMix] at (\tmax, \empUCpostNoShock) {UC};
            
            \draw[CoolBlue, very thick] (\tmin, \empUUpre) -- (\tmax, \empUUpre);
            \node[font=\small, anchor=west, CoolBlue] at (\tmax, \empUUpre) {UU};
        \end{tikzpicture}
    \end{subfigure} \\
    \vspace{2mm}
    \begin{minipage}{.95\textwidth} \centering
        \textbf{(b)} Evolution of Outcomes: Firms With a Demand Shock
    \end{minipage} \\
    \begin{subfigure}[t]{0.5\textwidth}
        \centering
        \begin{tikzpicture}[unifiedplot]
            \draw[->, thick] (0,0) -- (\xmax,0) node[below left] {Time};
            \draw[->, thick] (0,0) -- (0, \ymax) node[left] {$W$};
            
            \draw[dashed, lightgray, thin] (\tmin, \minWpre) -- (\tmax, \minWpre);
            \node[font=\small, anchor=east] at (0, \minWpre) {$\underline{W}^{\text{pre}}$};
            \draw[dashed, lightgray, thin] (\tmin, \minWpost) -- (\tmax, \minWpost);
            \node[font=\small, anchor=east] at (0, \minWpost) {$\underline{W}^{\text{post}}$};

            \draw[dashed, lightgray, thin] (\tstar, 0) node[below, gray] {$t^*$} -- (\tstar, \ymax);

            \draw[CoolRed, very thick] (\tmin, \minWpre) -- (\tstar, \minWpre);
            \draw[CoolRed, very thick] (\tstar, \minWpost) -- (\tmax, \minWpost);
            \node[font=\small, anchor=west, CoolRed] at (\tmax, \minWpost) {CC};

            \draw[CoolMix, very thick, densely dashed] (\tmin, \wageUCpre) -- (\tstar, \wageUCpre);
            \draw[CoolMix, very thick, densely dashed] (\tstar, \wageUCpost) -- (\tmax, \wageUCpost);
            \node[font=\small, anchor=west, CoolMix] at (\tmax, \wageUCpost) [yshift=5pt] {UC};
            
            \draw[CoolBlue, very thick] (\tmin, \wageUU) -- (\tstar, \wageUU);
            \draw[CoolBlue, very thick] (\tstar, \wageUUpost) -- (\tmax, \wageUUpost);
            \node[font=\small, anchor=west, CoolBlue] at (\tmax, \wageUUpost) {UU};
        \end{tikzpicture}
    \end{subfigure}%
    \begin{subfigure}[t]{0.5\textwidth}
        \centering
        \begin{tikzpicture}[unifiedplot]
            \draw[->, thick] (0,0) -- (\xmax,0) node[below left] {Time};
            \draw[->, thick] (0,0) -- (0, \ymax) node[left] {$L$};
            
            \draw[dashed, lightgray, thin] (\tstar, 0) node[below, gray] {$t^*$} -- (\tstar, \ymax);

            \draw[CoolRed, very thick] (\tmin, \empCCpre) -- (\tstar, \empCCpre);
            \draw[CoolRed, very thick] (\tstar, \empCCpostWithShock) -- (\tmax, \empCCpostWithShock);
            \node[font=\small, anchor=west, CoolRed] at (\tmax, \empCCpostWithShock) {CC};

            \draw[CoolMix, very thick, densely dashed] (\tmin, \empUCpre) -- (\tstar, \empUCpre);
            \draw[CoolMix, very thick, densely dashed] (\tstar, \empUCpostWithShock) -- (\tmax, \empUCpostWithShock);
            \node[font=\small, anchor=west, CoolMix] at (\tmax, \empUCpostWithShock) [yshift=5pt] {UC};
            
            \draw[CoolBlue, very thick] (\tmin, \empUUpre) -- (\tstar, \empUUpre);
            \draw[CoolBlue, very thick] (\tstar, \empUUpostWithShock) -- (\tmax, \empUUpostWithShock);
            \node[font=\small, anchor=west, CoolBlue] at (\tmax, \empUUpostWithShock) {UU};
        \end{tikzpicture}
    \end{subfigure} \\
    \vspace{2mm}
    \begin{minipage}{.95\textwidth} \centering
        \textbf{(c)} Differential Evolution of Outcomes (Demand Shock vs.\ No Demand Shock)
    \end{minipage} \\
    \begin{subfigure}[t]{0.5\textwidth}
        \centering
        \begin{tikzpicture}[unifiedplot]
            \draw[->, thick] (0,0) -- (\xmax,0) node[below left] {Time};
            \draw[->, thick] (0,0) -- (0, \ymax);
            \node[left, align=center] at (0, .925*\ymax) {$\Delta \ln W$};
            
            \draw[dashed, lightgray, thin] (\tmin, \yzero) node[left, black] {$0$} -- (\tmax, \yzero);
            \draw[dashed, lightgray, thin] (\tstar, 0) node[below, gray] {$t^*$} -- (\tstar, \ymax);

            \draw[CoolRed, very thick] (\tmin, .925*\yzero) -- (\tstar, .925*\yzero);
            \draw[CoolRed, very thick] (\tstar, .925*\yzero) -- (\tmax, .925*\yzero);
            \node[font=\small, anchor=west, CoolRed] at (\tmax, .925*\yzero) {CC};

            \draw[CoolMix, very thick, densely dashed] (\tmin, \yzero) -- (\tstar, \yzero);
            \draw[CoolMix, very thick, densely dashed] (\tstar, \wageDiffUCpost) -- (\tmax, \wageDiffUCpost);
            \node[font=\small, anchor=west, CoolMix] at (\tmax, \wageDiffUCpost) {UC};
            
            \draw[CoolBlue, very thick] (\tmin, 1.075*\yzero) -- (\tstar, 1.075*\yzero);
            \draw[CoolBlue, very thick] (\tstar, \wageDiffUUpost) -- (\tmax, \wageDiffUUpost);
            \node[font=\small, anchor=west, CoolBlue] at (\tmax, \wageDiffUUpost) {UU};
        \end{tikzpicture}
    \end{subfigure}%
    \begin{subfigure}[t]{0.5\textwidth}
        \centering
        \begin{tikzpicture}[unifiedplot]
            \draw[->, thick] (0,0) -- (\xmax,0) node[below left] {Time};
            \draw[->, thick] (0,0) -- (0, \ymax);
            \node[left, align=center] at (0, .925*\ymax) {$\Delta \ln L$};
            
            \draw[dashed, lightgray, thin] (\tmin, \yzero) node[left, black] {$0$} -- (\tmax, \yzero);
            \draw[dashed, lightgray, thin] (\tstar, 0) node[below, gray] {$t^*$} -- (\tstar, \ymax);

            \draw[CoolRed, very thick] (\tmin, .925*\yzero) -- (\tstar, .925*\yzero);
            \draw[CoolRed, very thick] (\tstar, \empDiffCCpost) -- (\tmax, \empDiffCCpost);
            \node[font=\small, anchor=west, CoolRed] at (\tmax, \empDiffCCpost) {CC};

            \draw[CoolMix, very thick, densely dashed] (\tmin, \yzero) -- (\tstar, \yzero);
            \draw[CoolMix, very thick, densely dashed] (\tstar, \empDiffUCpost) -- (\tmax, \empDiffUCpost);
            \node[font=\small, anchor=west, CoolMix] at (\tmax, \empDiffUCpost) {UC};
            
            \draw[CoolBlue, very thick] (\tmin, 1.075*\yzero) -- (\tstar, 1.075*\yzero);
            \draw[CoolBlue, very thick] (\tstar, \empDiffUUpost) -- (\tmax, \empDiffUUpost);
            \node[font=\small, anchor=west, CoolBlue] at (\tmax, \empDiffUUpost) {UU};
        \end{tikzpicture}
    \end{subfigure}

    \vspace{2mm}
    \begin{singlespace}
    \begin{minipage}{1\textwidth} \footnotesize\singlespacing
        Notes: 
        The figure illustrates the predictions of the model with a wage floor constraint introduced in          Section \ref{sec:m_framework_theory}
        for wage and employment outcomes when a wage floor hike 
        from $\W^{\pre}$ to $\W^{\post}$ at time $t^*$ coincides with a positive demand shock.
        Panels (a) and (b) show the evolution of outcomes for firms that do not receive
        a demand shock and firms that do receive it, respectively.
        Panel (c) plots the difference in outcomes between firms with and without the shock.
        As in Figure \ref{fig:model_outcomes_mw_change}, we distinguish between three types 
        of firms based on their pre-shock wage: 
        those constrained at both wage floors (CC), those unconstrained before but 
        constrained after the increase (UC), and those unconstrained at either 
        wage floor (UU).
        For each outcome, the y-scale of Panels (a) and (b) is constant, whereas the 
        y-scale of Panel (c) is not.
    \end{minipage}
    \end{singlespace}
\end{figure}

%% file: figures/CBA_wagefloors_portugal.tex
\begin{figure}[H]
    \caption{Wage Floors in Collective Bargaining Agreements in Portugal.}
    \label{fig:cba_rel_floor_cushion_hist}
    \centering
    
    \begin{subfigure}[t]{0.6\textwidth}
        \centering
        \caption{Relative CBA Wage Floor}
        \includegraphics[width=\linewidth]{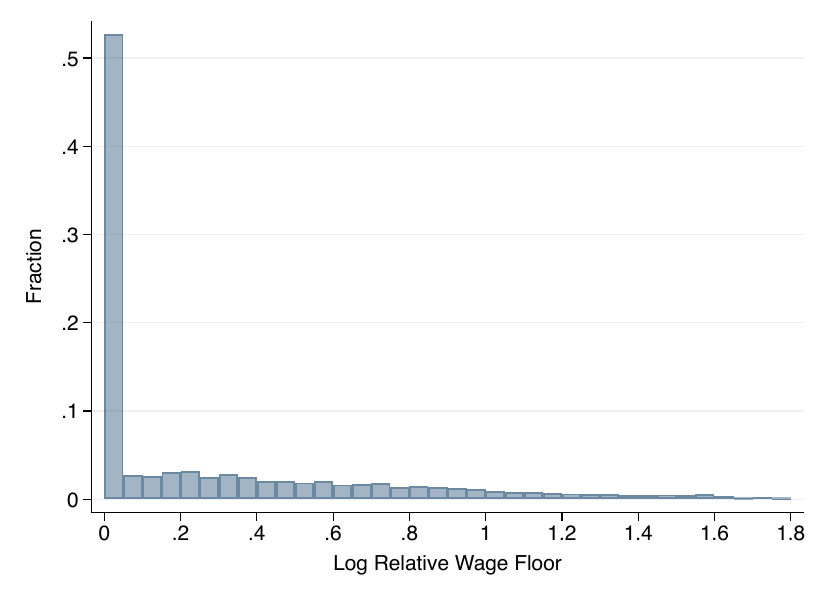}
        \label{fig:cba_rel_floor_hist}
    \end{subfigure}
    \hfill \\
    \begin{subfigure}[t]{0.6\textwidth}
        \centering
        \caption{Wage Cushions}
        \includegraphics[width=\linewidth]{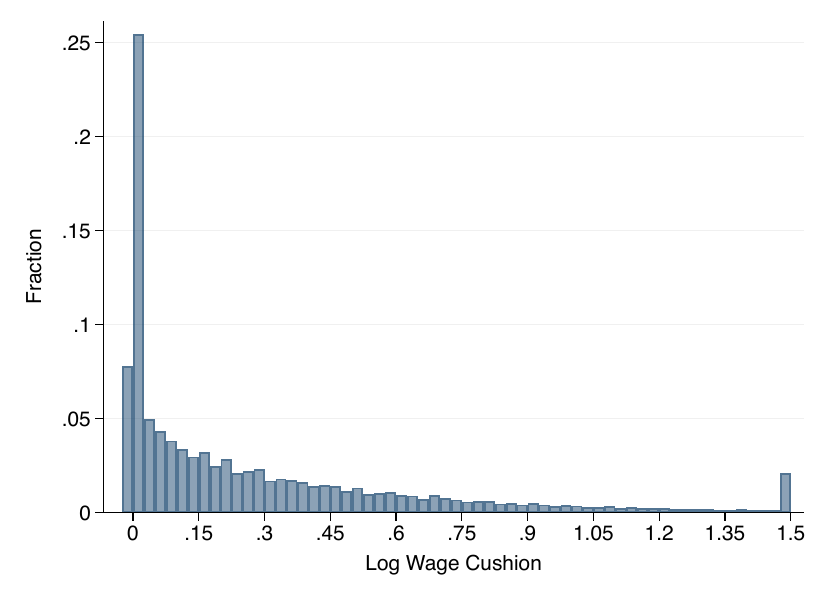}
        \label{fig:wage_cushion_hist}
    \end{subfigure}

    \vspace{2mm}
    \begin{minipage}{1\textwidth} \footnotesize
        Notes: 
        This figure shows the distribution of relative CBA wage floors (panel a) and wage cushions (panel b) in Portugal. After applying our sample restrictions, we further drop CBAs with fewer than 50 workers per year, and CBA--job title--year cells with fewer than 10 workers. The relative wage floor is defined as $\log(\text{CBA wage floor}/\text{MW})$. The wage cushion is defined as $\log(\text{base wage}_i/\text{CBA wage floor})$, and it is censored at 1.5. CBA wage floors are identified from the modal base wage within each CBA--job title--year cell, retaining the mode only when it lies near the bottom of the wage distribution.
    \end{minipage}
\end{figure}

%% file: figures/mw_evolution_portugal.tex
\begin{figure}[H]
    \centering
    \caption{Evolution of the National Minimum Wage and the Share of Minimum-Wage Workers in Portugal.}
    \label{fig:mw_workers_time}

    \begin{subfigure}{1\textwidth}\centering
        \caption{Nominal and Real MW}
        \includegraphics[width=0.6\textwidth]{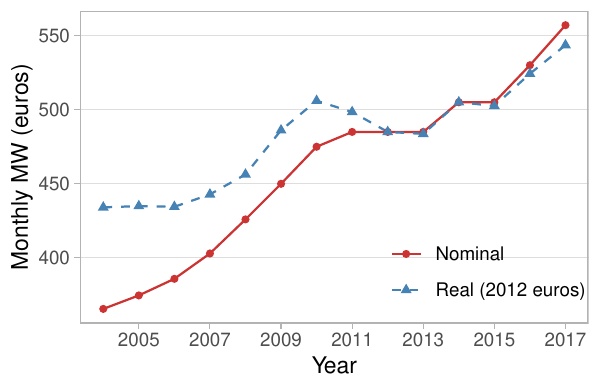}
    \end{subfigure} \\
    \begin{subfigure}{1\textwidth}\centering
        \caption{Share of Workers at MW}
        \includegraphics[width=0.6\textwidth]{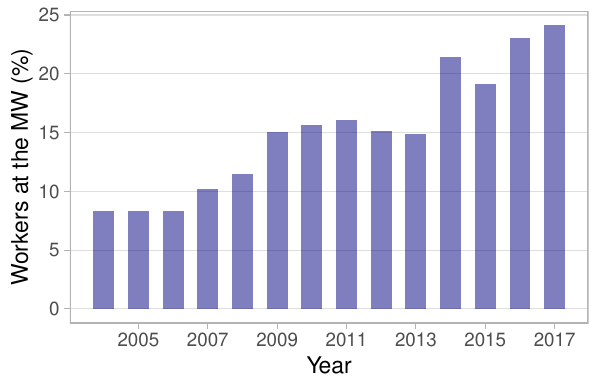}
    \end{subfigure} 
    
    \vspace{2mm}
    \begin{singlespace}
    \begin{minipage}{1\textwidth} \footnotesize 
        Notes:
        This figure shows the evolution of the national minimum wage and the share of minimum-wage workers in Portugal between 2004 and 2017.
        Panel~(a) shows the nominal monthly MW and the real MW in 2012 euros.
        Panel~(b) shows the share of workers whose monthly base wage, excluding supplementary and bonus payments, is at or below the MW.
        We include full-time workers (at least 150 monthly hours) aged 18 to 65 in all firms in each year, without restricting to stayers or to firms observed in multiple years.
        The 2014 MW is the one set in October 2014 (505 euros), when the QP is collected.
    \end{minipage}
    \end{singlespace}
\end{figure}

%% file: figures/institutions_world.tex
\begin{figure}[ht!] 
    \centering 
    \caption{Wage-Setting Regimes around the World.} \label{fig:institutions_world} 
    
    \begin{subfigure}{.99\textwidth} 
        \centering 
        \caption{World} 
        \includegraphics[width=.99\textwidth]{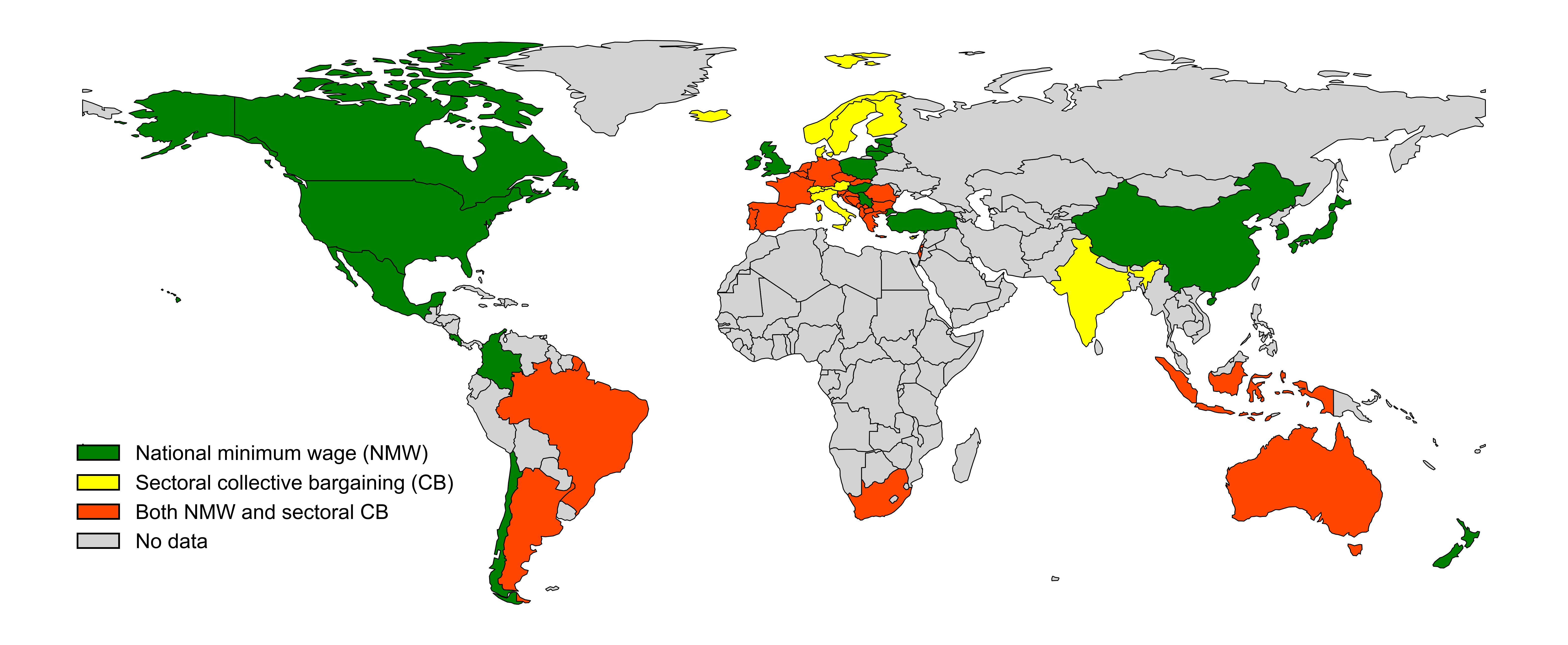} 
    \end{subfigure} \\ 
    \begin{subfigure}{.99\textwidth} 
        \centering 
        \caption{Europe} 
        \includegraphics[width=.99\textwidth]{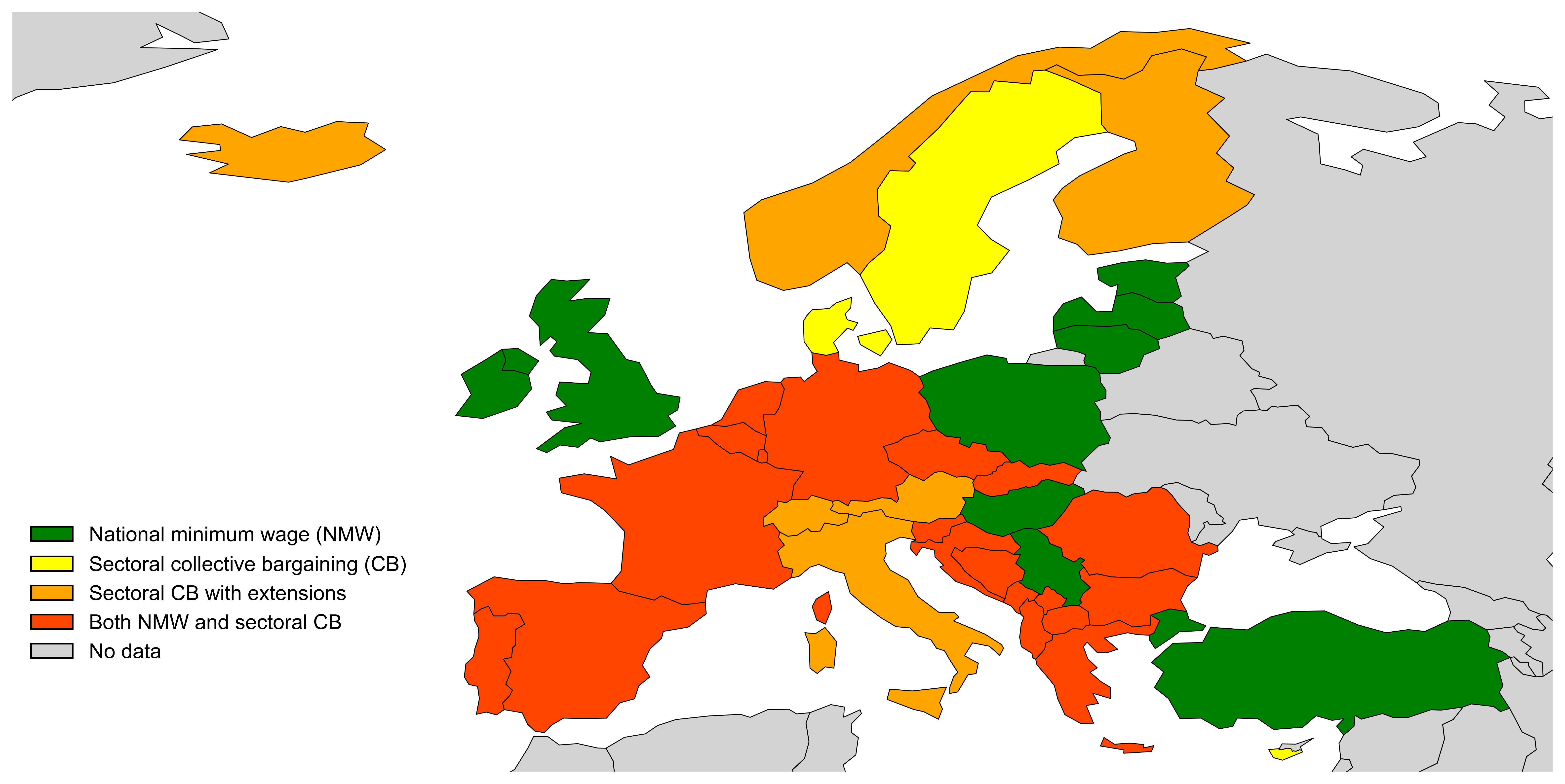}    
    \end{subfigure} 
    
    \vspace{2mm} 
    \begin{singlespace}
    \begin{minipage}{1\textwidth} \footnotesize 
    Notes: Data are from ICTWSS \citep{OECDVisser2023}. 
    The map displays the predominant wage-setting institutions in different countries in 2019.
    Panel (a) shows all countries, and 
    Panel (b) shows a zoomed-in view of Europe.
    In Panel (a) we split the countries into three groups:
    countries with a national minimum wage (NMW) in green, 
    countries with any sectoral collective bargaining (CB) in orange, and 
    countries with both institutions in red. 
    In Panel (b) we also indicate whether countries with sectoral CB also 
    experienced extensions of the negotiated wage floors to non-covered workers.
    \end{minipage}
    \end{singlespace}
\end{figure}

%% file: figures/firm_wages_portugal.tex
\begin{figure}[H]
    \centering
    \caption{Distribution of Firm Mean Wages in Portugal.}
    \label{fig:firm_wages_portugal}

    \begin{subfigure}{0.5\textwidth}
        \centering
        \caption{Ratio of Mean Wage to MW, 2014--2015}
        \includegraphics[width=\textwidth]{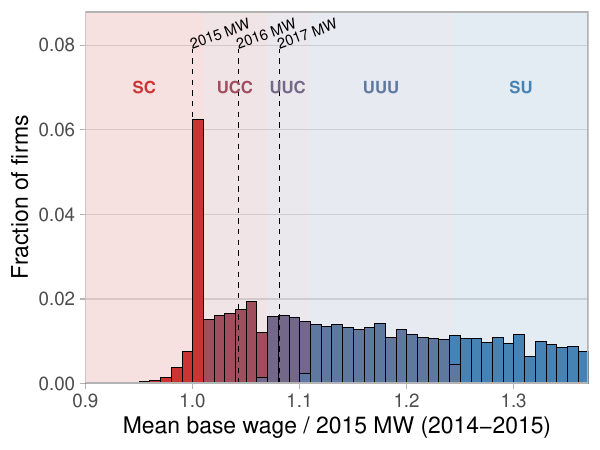}
    \end{subfigure}%
    \begin{subfigure}{0.5\textwidth}
        \centering
        \caption{Mean Firm Wage, 2015}
        \includegraphics[width=\textwidth]{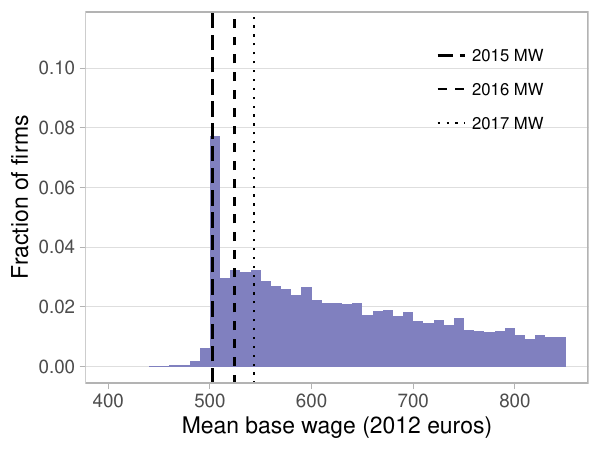}
    \end{subfigure} \\
    \begin{subfigure}{0.5\textwidth}
        \centering
        \caption{Mean Firm Wage, 2016}
        \includegraphics[width=\textwidth]{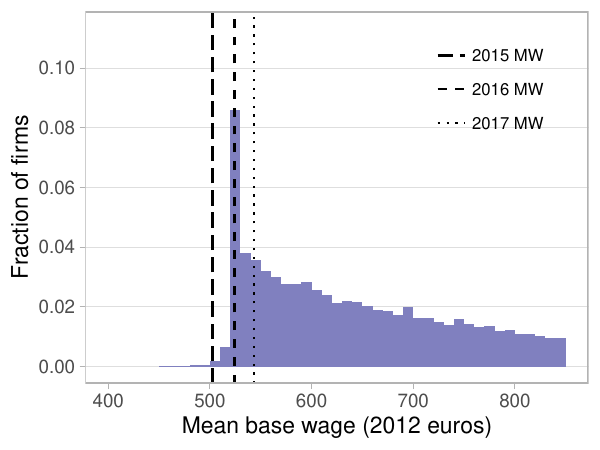}
    \end{subfigure}%
    \begin{subfigure}{0.5\textwidth}
        \centering
        \caption{Mean Firm Wage, 2017}
        \includegraphics[width=\textwidth]{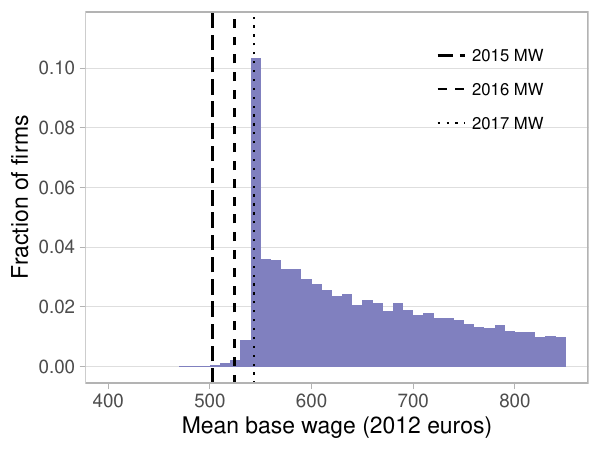}
    \end{subfigure}

    \vspace{2mm}
    \begin{singlespace}
    \begin{minipage}{1\textwidth} \footnotesize
        Notes: 
        The figure shows the distribution of firm mean wages across time relative to the MW levels.
        Panel (a) shows the average ratio of the mean firm wage to each year's MW in 2014--2015, and illustrates our definition of groups used in our analysis.
        Panels (b), (c), and (d) show the distribution of the mean firm wage in 2015, 2016, and 2017, respectively.
        The mean firm wage is computed using the base wage for full-time workers 
        (with over 150 monthly working hours) aged between 18 and 65. 
        The sample includes firms with non-negative value added. 
        Vertical lines highlight the MW levels of different years in the plots.
    \end{minipage}
    \end{singlespace}
\end{figure}

%% file: figures/export_countries.tex
\begin{figure}[H]
    \centering
    \caption{Trends in Total Exports for Different Countries.}
    \label{fig:exports_growth_countries}

    \begin{subfigure}{0.6\textwidth}
        \centering
                \caption{EU and the US}
        \includegraphics[width=\textwidth]{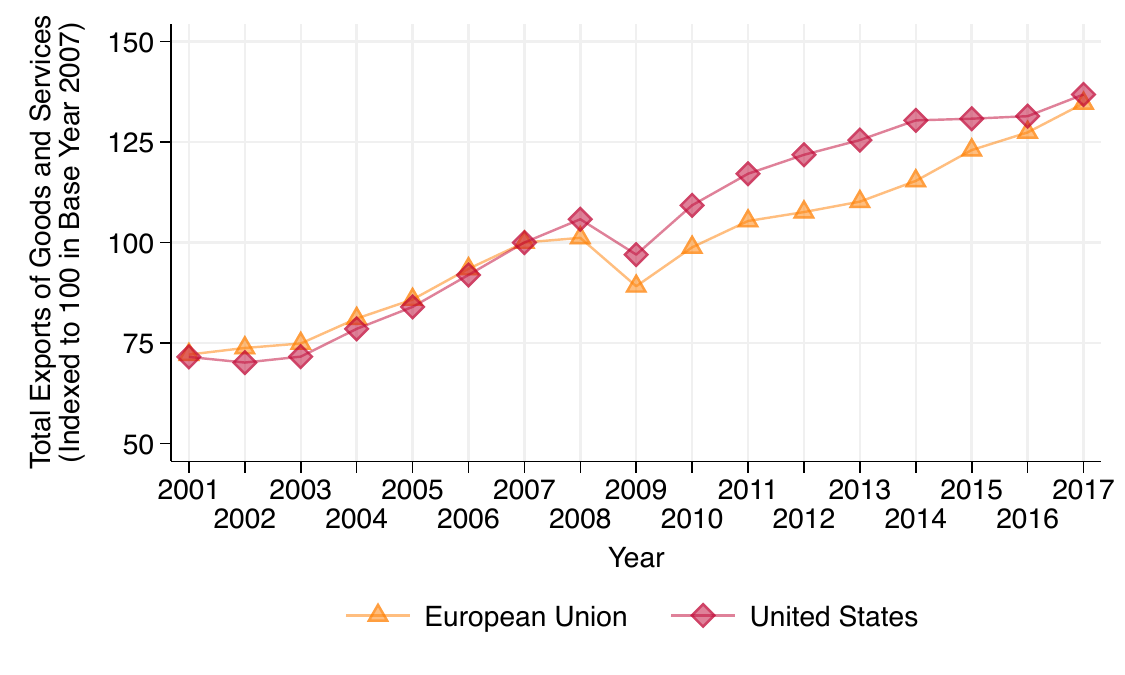}
    \end{subfigure} \\
    \begin{subfigure}{0.6\textwidth}
        \centering
                \caption{Portugal}
        \includegraphics[width=\textwidth]{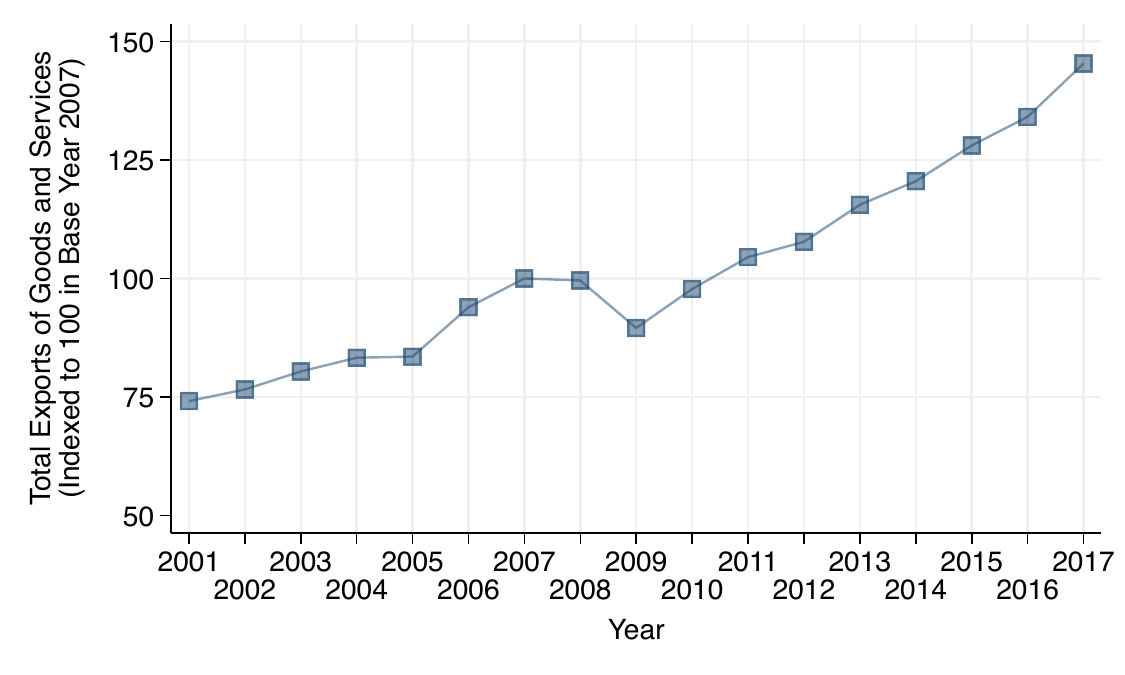}
    \end{subfigure} \\
    \begin{subfigure}{0.6\textwidth}
        \centering
       \caption{Norway}
        \includegraphics[width=\textwidth]{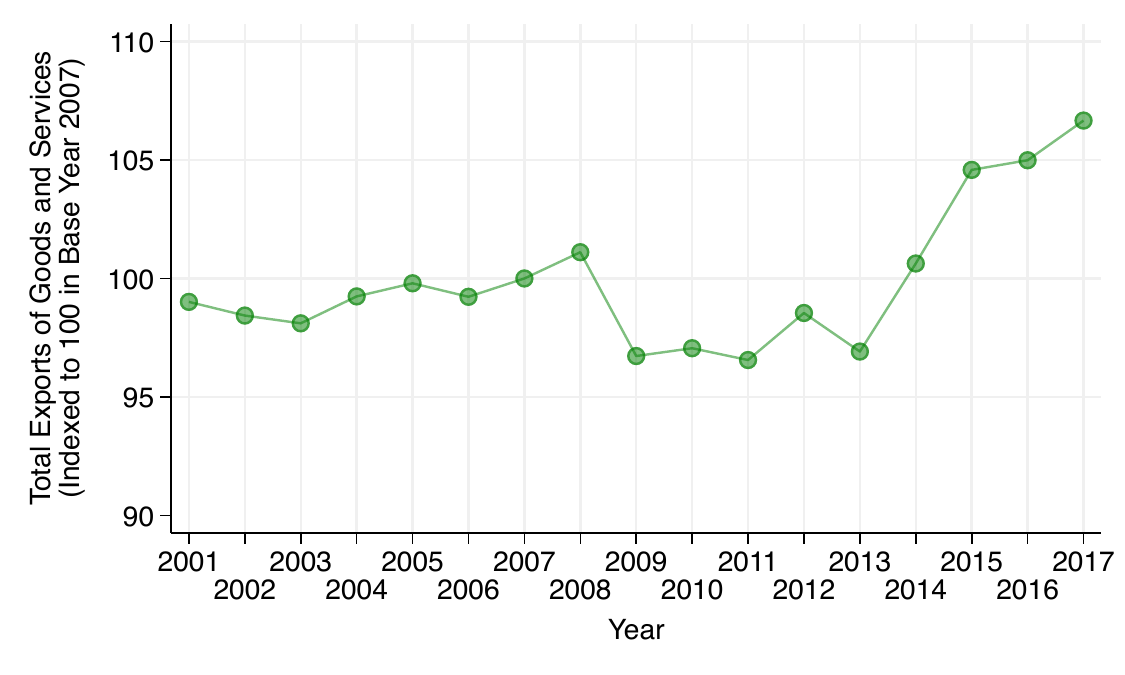}
    \end{subfigure}

    \vspace{2mm}
    \begin{singlespace}
    \begin{minipage}{1\textwidth} \footnotesize
        Notes: Total values of exports of goods and services are reported in constant 2015 USD and obtained from the World Bank. We index all values to 100 in 2007. 
        Panel (a) compares the European Union (EU) and the United States (US), while panels (b) and (c) consider Portugal and Norway, respectively.
    \end{minipage}
    \end{singlespace}
\end{figure}

%% file: figures/lms_stayers_incumbents.tex
\begin{figure}[ht!]
    \centering
    \caption{Wage Responses to Demand Shocks: Stayers versus Incumbents.}
    \label{fig:lms_stayers_incumbents}
    
    \begin{subfigure}{.55\textwidth}
        \centering
        \caption{Portugal}
        \includegraphics[width=.99\textwidth]{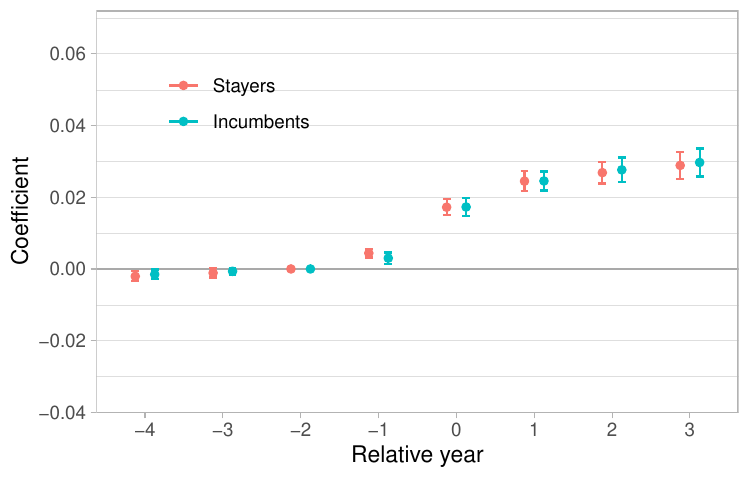}
    \end{subfigure} \\
    \begin{subfigure}{.55\textwidth}
        \centering
        \caption{Norway}
        \includegraphics[width=.99\textwidth]{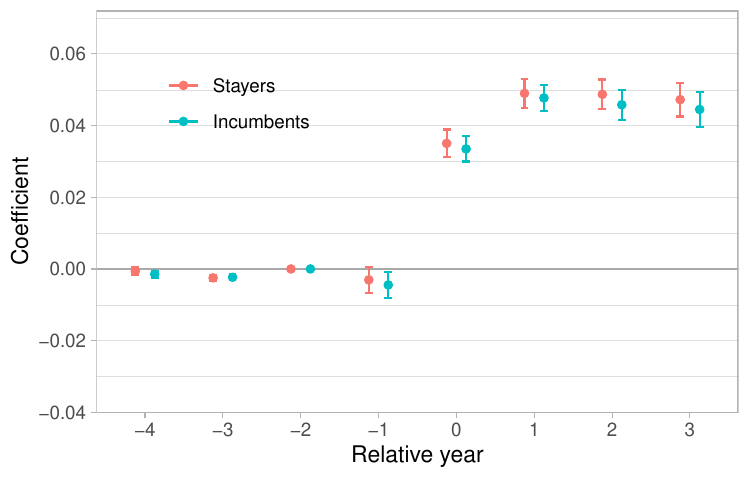}
    \end{subfigure}

    \vspace{2mm}
    \begin{singlespace}
    \begin{minipage}{1\textwidth} \footnotesize
        Notes:
        The figure shows estimates of the effects of exposure to a demand shock using the internal instrument (\(Z^{\text{Int}}_j\)) on the wage of stayers and incumbents.
        Stayers are workers who remain 
        in the firm for 7 years within the sample window, and incumbents are the subset of workers who remain in the firm in the first three periods.
        The sample includes firms that have positive employment for 7 years in the window and at least 2 stayers.
    \end{minipage}
    \end{singlespace}
\end{figure}

%% file: figures/global_lselast_cba.tex
\begin{figure}[H]
    \centering
    \caption{Estimates of Firm-Level Labor Supply Elasticities and Collective Bargaining Coverage in Developing Countries.}
    \label{fig:global_ls_elast_cba_cov}
    
    \includegraphics[width=0.7\textwidth]{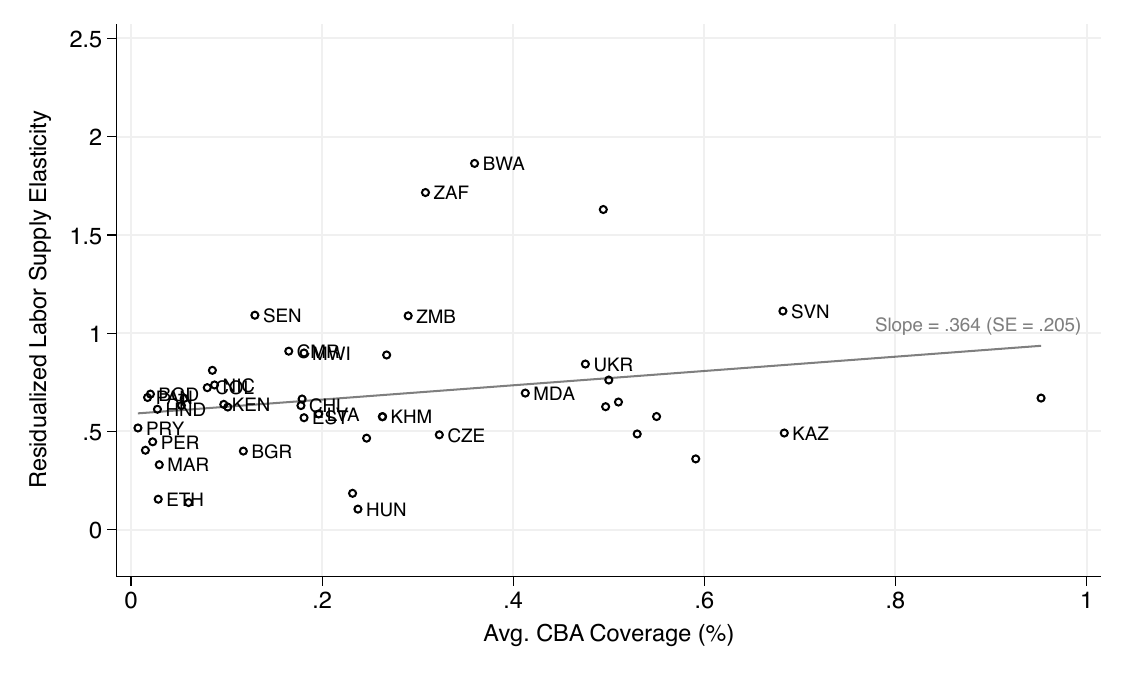}

    \vspace{2mm}
    \begin{singlespace}
    \begin{minipage}{1\textwidth} \footnotesize
        Notes: 
        Estimates of the median firm markdown are taken from \citet{AmodioEtAl2024} and transformed into labor supply elasticities.
        These elasticities are residualized with respect to two polynomials in GDP per capita and self-employment rates, as well as a dummy for unemployment protection, and winsorized at the 5th and 95th percentiles. 
        Data on the average CBA coverage are from the International Labour Organization’s IRData database, which combines survey- and administrative-based sources. 
        The average coverage rate is computed using available information for 2010–2019 and is defined as the share of employees covered by collective bargaining relative to those with the right to be covered. 
        The resulting sample covers 46 of the 81 countries with markdown measures.
    \end{minipage}
    \end{singlespace}
\end{figure}

%% file: figures/lms_by_constraints_va.tex
\begin{figure}[ht!]
    \centering
    \caption{Firm Value Added Responses to Demand Shocks: Exposure to Wage-Setting Constraints.}
    \label{fig:va_by_country}
    
    \begin{subfigure}{.5\textwidth}
        \centering
        \caption{Portugal}
        \includegraphics[width=.99\textwidth]{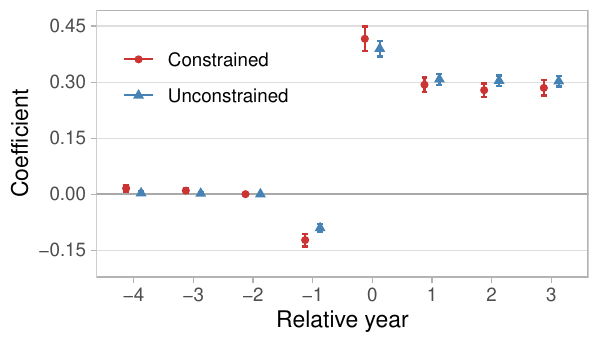}
    \end{subfigure}%
    \begin{subfigure}{.5\textwidth}
        \centering
        \caption{Norway}
        \includegraphics[width=.99\textwidth]{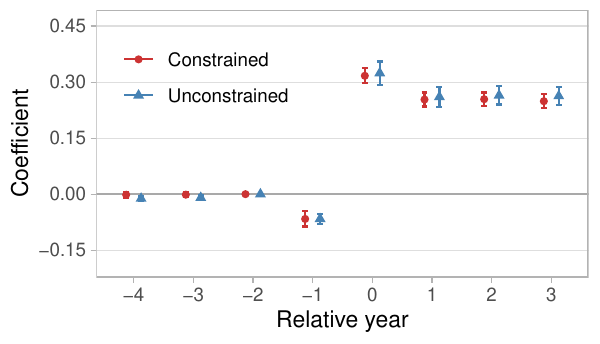}
    \end{subfigure}

    \vspace{2mm}
    \begin{singlespace}
    \begin{minipage}{1\textwidth} \footnotesize
        Notes:
        The figure shows estimates of the effects of exposure to a demand shock using the internal shock (\(Z^{\text{Int}}_j\)) on the firm value added by constrained status.
        Constrained refers to firms with an average wage within 15\% of the national MW for Portugal, or covered by a CBA for Norway, whereas Unconstrained refers to the rest of the firms.
        The internal shock (\(Z^{\text{Int}}_j\)) is defined as having value added growth between -1 and 0 above the median within the firm's 2-digit sector.
        For Portugal and Norway, the average firm wage is computed for stayers (workers who remain in the firm for 7 years within the sample window).
        The sample includes firms that have positive employment for 7 years in the window and, for Portugal and Norway, at least 2 stayers.
    \end{minipage}
    \end{singlespace}
\end{figure}

%% file: figures/elast_vary_threshold.tex
\begin{figure}[hbt!]
    \centering
    \caption{Implied Labor Supply and Rent-Sharing Elasticities in Portugal: Varying the Definition of Constrained Firms.}
    \label{fig:elast_vary_threshold}

    \begin{subfigure}{.5\textwidth}
        \centering
        \caption{Labor Supply}
        \includegraphics[width=.96\textwidth]{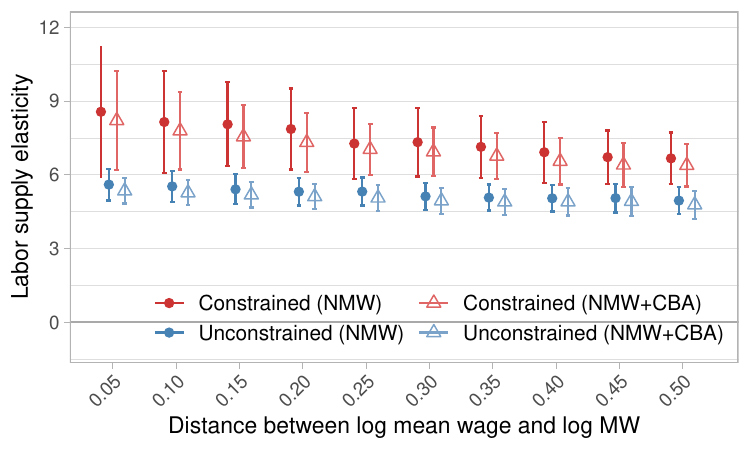}
    \end{subfigure}%
    \begin{subfigure}{.5\textwidth}
        \centering
        \caption{Rent-Sharing}
        \includegraphics[width=.96\textwidth]{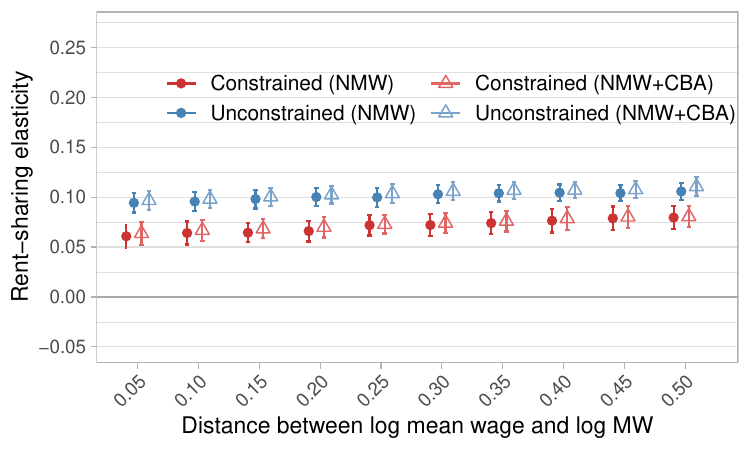}
    \end{subfigure}

    \vspace{2mm}
    \begin{singlespace}
    \begin{minipage}{1\textwidth} \footnotesize
        Notes:
        The figure shows estimates of implied labor supply and rent-sharing elasticities for Portugal using the internal shock design, varying the threshold used to define constrained firms.
        We report estimates using two definitions: (i) proximity to the national MW (NMW), with filled circles, and (ii) proximity to CBA wage floors (NMW+CBA), with hollow triangles.
    \end{minipage}
    \end{singlespace}
\end{figure}

%% file: figures/lms_by_constraints_portugal_combined.tex
\begin{figure}[ht!]
    \centering
    \caption{Firm Response by the Type of Wage-Setting Constraints in Portugal.}
    \label{fig:lms_by_constraints_portugal_combined}
    
    \begin{subfigure}{.5\textwidth}
        \centering
        \caption{Value Added}
        \includegraphics[width=.99\textwidth]{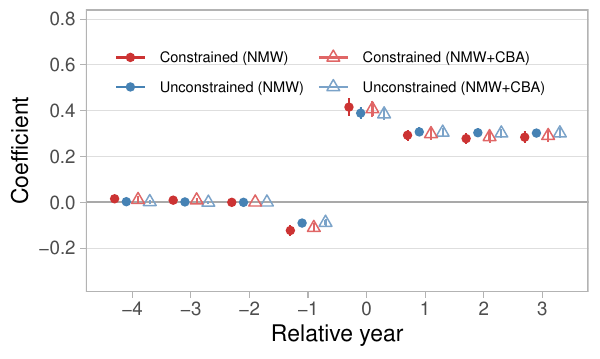}
    \end{subfigure}%
    \begin{subfigure}{.5\textwidth}
        \centering
        \caption{Mean Wage}
        \includegraphics[width=.99\textwidth]{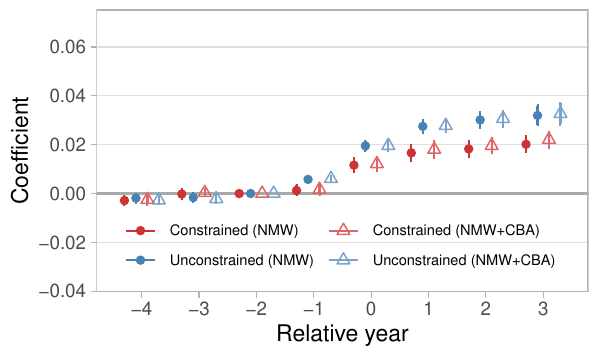}
    \end{subfigure}\\

    \begin{subfigure}{.5\textwidth}
        \centering
        \caption{Employment}
        \includegraphics[width=.99\textwidth]{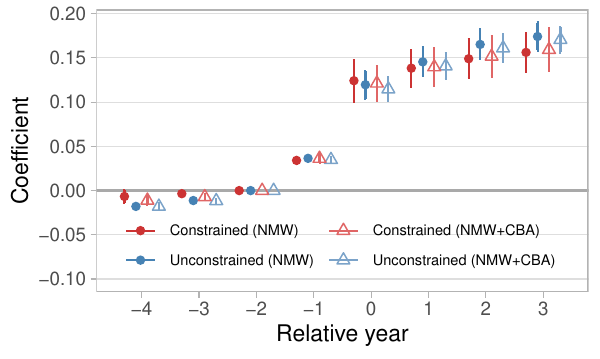}
    \end{subfigure}

    \vspace{2mm}
    \begin{singlespace}
    \begin{minipage}{1\textwidth} \footnotesize
        Notes:
        The figure shows estimates of the effects of a demand shock, using the internal shock design, on firm-level outcomes by constrained status, where constraints combine MW and CBA wage floors.
        Constrained refers to firms with an average wage within 15\% of the national MW (NMW) or of the CBA wage floor (NMW+CBA), whereas Unconstrained refers to the rest of the firms.
        We estimate the effect of a treatment indicator for having value-added growth between periods $-1$ and 0 above the median within the firm’s 2-digit sector.
    \end{minipage}
    \end{singlespace}
\end{figure}

%% file: figures/wage_floor_effects.tex
\begin{figure}[ht!]
    \centering
    \caption{Mean Wage Floor Responses to Internal Shocks.}
    \label{fig:wage_floor_effects}

    \begin{subfigure}{.55\textwidth}
        \centering
        \caption{Portugal}
        \includegraphics[width=.99\textwidth]{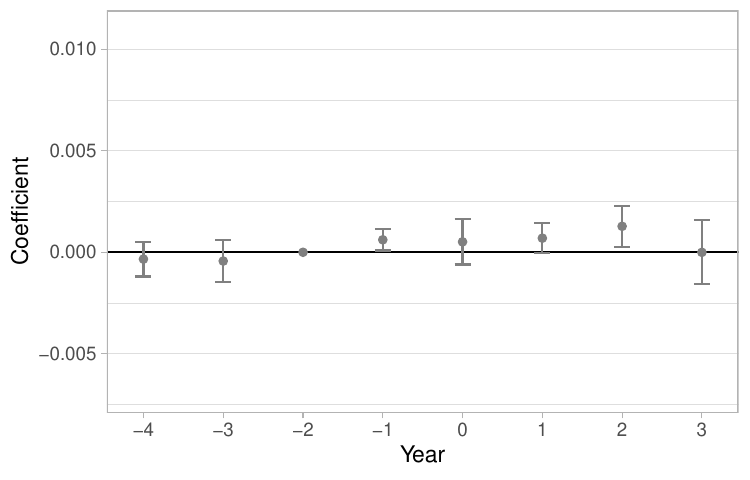}
    \end{subfigure} \\
    
    \begin{subfigure}{.55\textwidth}
        \centering
        \caption{Norway}
        \includegraphics[width=.99\textwidth]{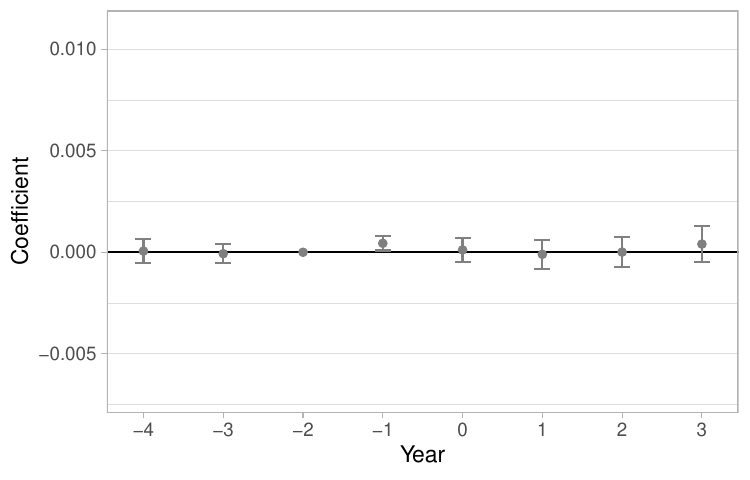}
    \end{subfigure}

    \vspace{2mm}
    \begin{singlespace}
    \begin{minipage}{1\textwidth} \footnotesize
        Notes:
        The figure shows estimates of the effects of exposure to a demand shock, using an internal shock design, on the mean wage floor among stayers in the firm.
        For Portugal, we focus on firms with non-missing CBA wage floors, as described in Appendix \ref{asec:data}, and for Norway, we focus on firms covered by CBAs.
        Stayers are workers who remain in the firm for at least 7 years within the sample window.
        The sample includes firms with at least 2 stayers within the sample window.
    \end{minipage}
    \end{singlespace}
\end{figure}

%% file: figures/export_shock_by_constraints.tex
\begin{figure}[ht!]
    \centering
    \caption{Firm Responses to Demand Shocks: External Shock and Exposure to Wage-Setting Constraints.}
    \label{fig:export_shock_by_constraints}
    
    \begin{subfigure}{.5\textwidth}
        \centering
        \caption{Portugal: Value Added}
        \includegraphics[width=.8\textwidth]{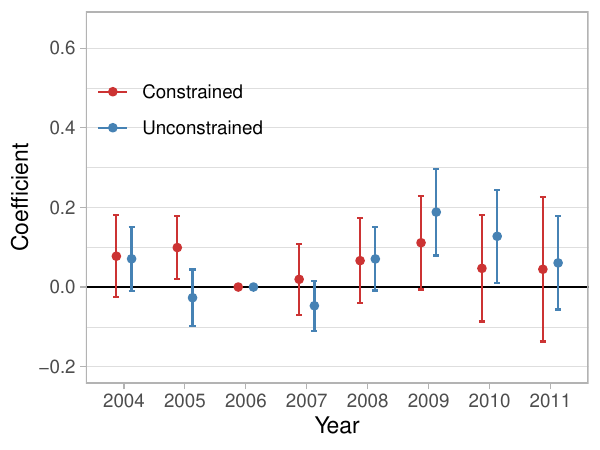}
    \end{subfigure}%
    \begin{subfigure}{.5\textwidth}
        \centering
        \caption{Norway: Value Added}
        \includegraphics[width=.8\textwidth]{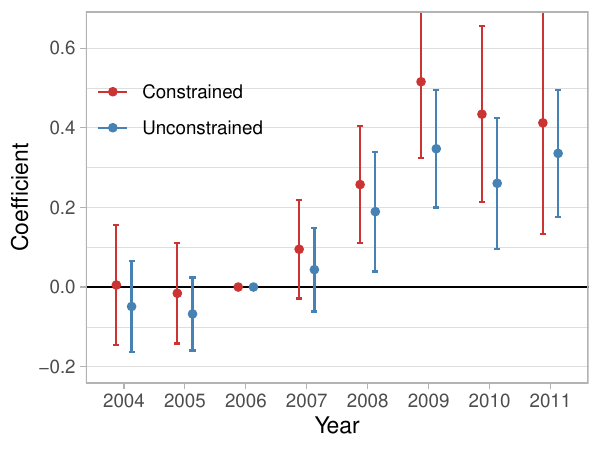}
    \end{subfigure} \\
    \begin{subfigure}{.5\textwidth}
        \centering
        \caption{Portugal: Mean Wage}
        \includegraphics[width=.8\textwidth]{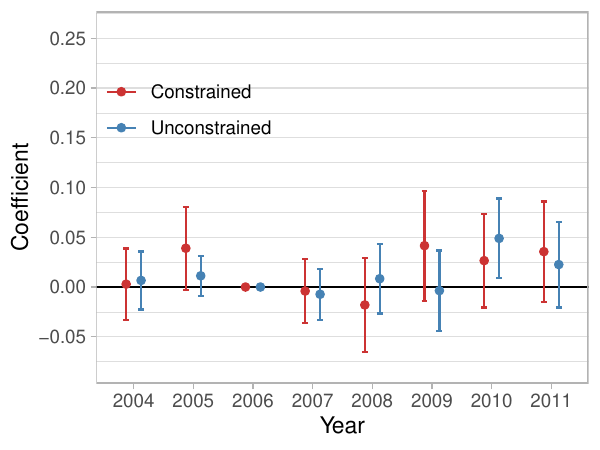}
    \end{subfigure}%
    \begin{subfigure}{.5\textwidth}
        \centering
        \caption{Norway: Mean Wage}
        \includegraphics[width=.8\textwidth]{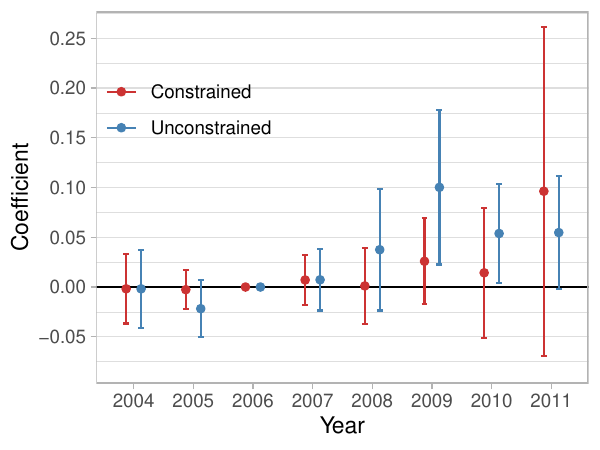}
    \end{subfigure} \\
    \begin{subfigure}{.5\textwidth}
        \centering
        \caption{Portugal: Employment}
        \includegraphics[width=.8\textwidth]{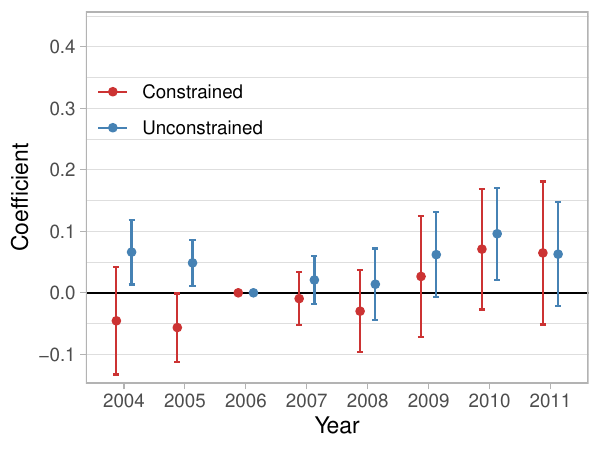}
    \end{subfigure}%
    \begin{subfigure}{.5\textwidth}
        \centering
        \caption{Norway: Employment}
        \includegraphics[width=.8\textwidth]{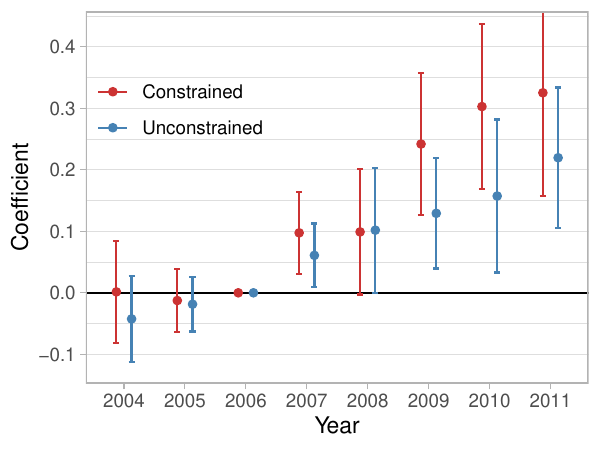}
    \end{subfigure} \\
    
    \vspace{2mm}
    \begin{singlespace}
    \begin{minipage}{1\textwidth} \footnotesize
        Notes: 
        The figure shows estimates of the effects of a demand shock, using an external shock design, on firm-level outcomes, by constrained status of firms.
        The left column shows the effects for Portugal, whereas the right column shows the effects for Norway.
        We restrict to firms exporting for at least three years up to 2007, with exports exceeding 1\% of their revenue in 2005–2007, and, for Portugal, we further exclude firms that primarily export to Angola or Spain.
        Panels (a)-(b) show effects on value added, panels (c)-(d) show wage effects, whereas panels (e)-(f) show employment effects.
        In each case, we estimate the effect of an increase in the demand for a firm's exports, defined as the mean change in world import demand for the firm's 2005--2007 exports.
        The sample includes exporting firms that have positive employment for 7 years in the window and at least 2 stayers.
        In Portugal, firms are classified as constrained using a looser threshold of approximately 50 percent above the MW to ensure a sufficient sample.
    \end{minipage}
    \end{singlespace}
\end{figure}

%% file: figures/elast_const_external.tex
\begin{figure}[ht!]
    \centering
    \caption{Estimated Firm-Level Elasticities: External Shock and Exposure to Wage-Setting Constraints
             in Norway.}
    \label{fig:elast_const_external}


    \begin{subfigure}{.5\textwidth}
        \centering
        \caption{Labor Supply Elasticity} 
        \includegraphics[width=.99\textwidth]{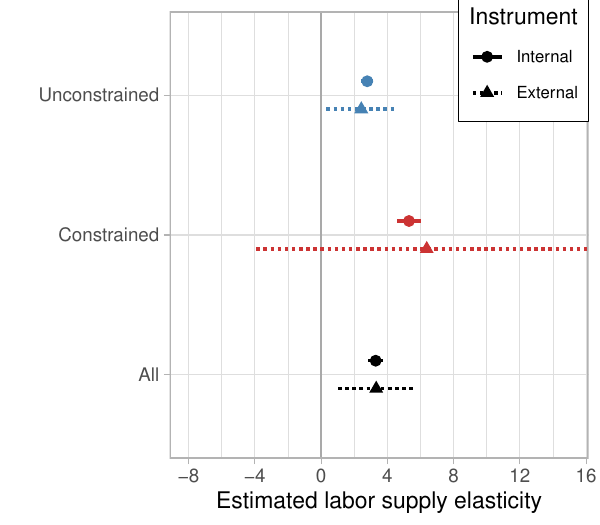}
    \end{subfigure}%
    \begin{subfigure}{.5\textwidth}
        \centering
        \caption{Rent-Sharing Elasticity} 
        \includegraphics[width=.99\textwidth]{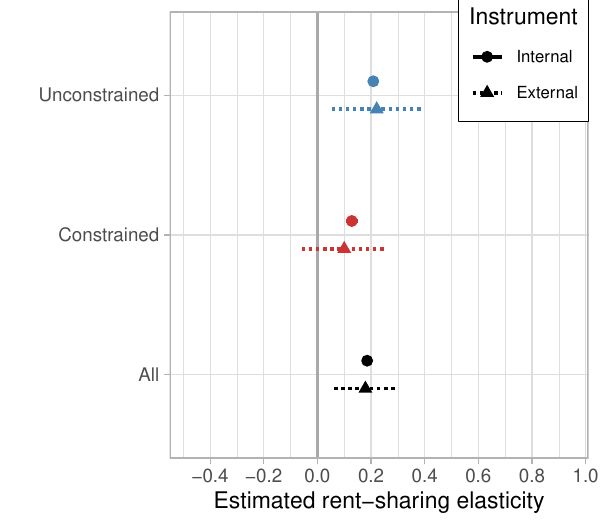}
    \end{subfigure}

    \vspace{2mm}
    \begin{singlespace}
    \begin{minipage}{1\textwidth} \footnotesize
        Notes: 
        The figure shows estimated elasticities implied by reduced-form effects of a demand shock on value added, wages, and employment, using the external shock design, by constrained status.
        Constrained firms are those covered by CBAs. 
        We restrict to firms exporting for at least three years up to 2007, with exports exceeding 1\% of their revenue in 2005–2007. 
        The left column shows labor supply elasticities, and the right column shows rent-sharing elasticities.
        To construct the mean reduced-form responses, we use the dynamic estimates in Appendix Figure~\ref{fig:export_shock_by_constraints} and average the post-periods 1 through 3.
    \end{minipage}
    \end{singlespace}
\end{figure}

%% file: figures/tripleDiff_portugal_emp_VA.tex
\begin{figure}[H]
    \centering
    \caption{Employment and Value Added Responses to Demand Shocks: Variation in Constraints in Portugal.}
    \label{fig:tripleDiff_portugal_emp_VA}

    \begin{subfigure}{.5\textwidth}
        \centering
        \caption{Value Added}
        \includegraphics[width=.99\textwidth]{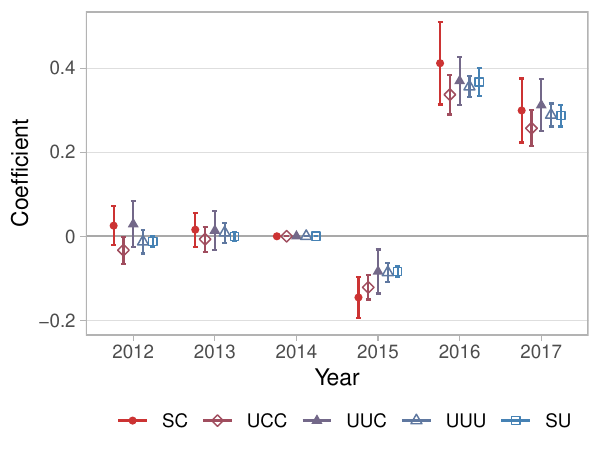}
    \end{subfigure}%
    \begin{subfigure}{.5\textwidth}
        \centering
        \caption{Employment}
        \includegraphics[width=.99\textwidth]{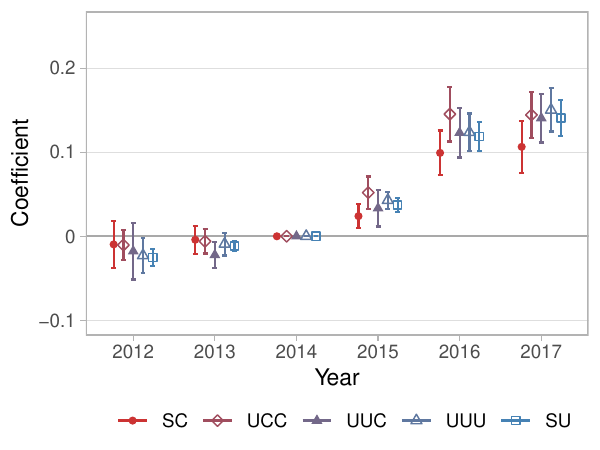}
    \end{subfigure} \\
    
    \begin{subfigure}{.5\textwidth}
        \centering
        \caption{Value Added Relative to UUU}
        \includegraphics[width=.99\textwidth]{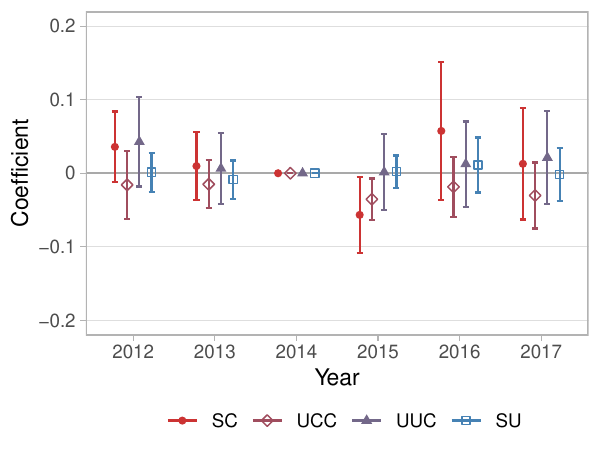}
    \end{subfigure}%
    \begin{subfigure}{.5\textwidth}
        \centering
        \caption{Employment Relative to UUU}
        \includegraphics[width=.99\textwidth]{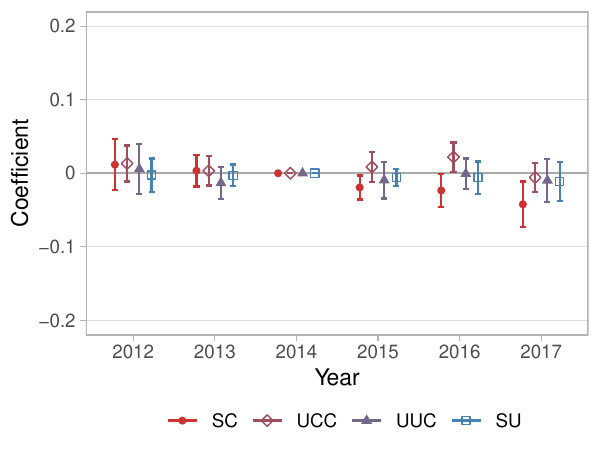}
    \end{subfigure}

    \vspace{2mm}
    \begin{singlespace}
    \begin{minipage}{1\textwidth} \footnotesize
        Notes: The figure shows estimates of the effects of a demand shock, using the internal shock design, by constrained group.
        The estimation approach follows Figure~\ref{fig:tripleDiff_portugal_wage}, but using employment and value added as outcomes.
        Panels (a) and (b) show event-study coefficients estimated separately for each group, and Panels (c) and (d) show differences with respect to the ``just'' unconstrained group (UUU).
    \end{minipage}
    \end{singlespace}
\end{figure}

%% file: figures/tripleDiff_portugal_var_thresh.tex
\begin{figure}[H]
    \centering
    \caption{Wage Responses to Demand Shocks at Shifted Minimum-Wage Thresholds.}
    \label{fig:tripleDiff_portugal_var_thresh}

    \includegraphics[width=.6\textwidth]{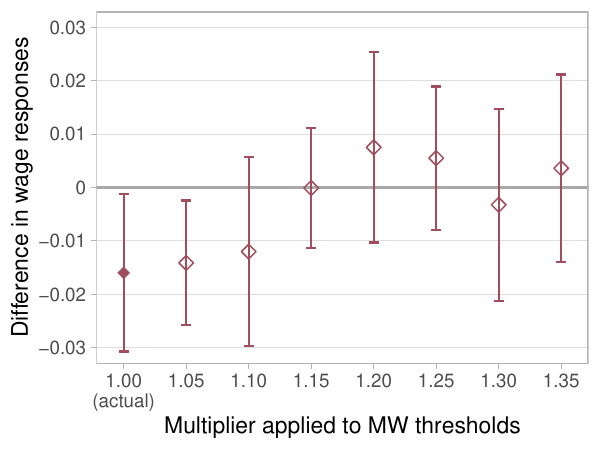}

    \vspace{2mm}
    \begin{singlespace}
    \begin{minipage}{1\textwidth} \footnotesize
        Notes:
        The figure shows the difference in wage responses to the 2015--2016 demand shock between UCC and UUU firms, measured in 2017 relative to 2014.
        We re-estimate the event-study version of Equation~(\ref{eq:wage_effect_triple_diff}) after multiplying all endpoints of the wage intervals defining these groups by the factor shown on the horizontal axis.
        A multiplier of one reproduces the baseline classification based on mean base wage-to-MW ratios averaged over 2014--2015, illustrated in Figure~\ref{fig:firm_wages_portugal}.
        Higher multipliers compare firms farther above the actual MW thresholds.
        Bars indicate 95\% confidence intervals based on standard errors clustered by two-digit industry.
    \end{minipage}
    \end{singlespace}
\end{figure}

%% file: figures/struc_model_res_sensitivity.tex
\begin{figure}[hbt!]
    \centering
    \caption{Minimum Wage Policy under a Mismeasured Labor Supply Elasticity: Sensitivity.}
    \label{fig:struc_model_res_sensitivity}
    \begin{subfigure}{.99\textwidth}        
        \caption{Cross-Market Labor Supply Elasticity $\sigma$}
        \includegraphics[width=0.99\linewidth]{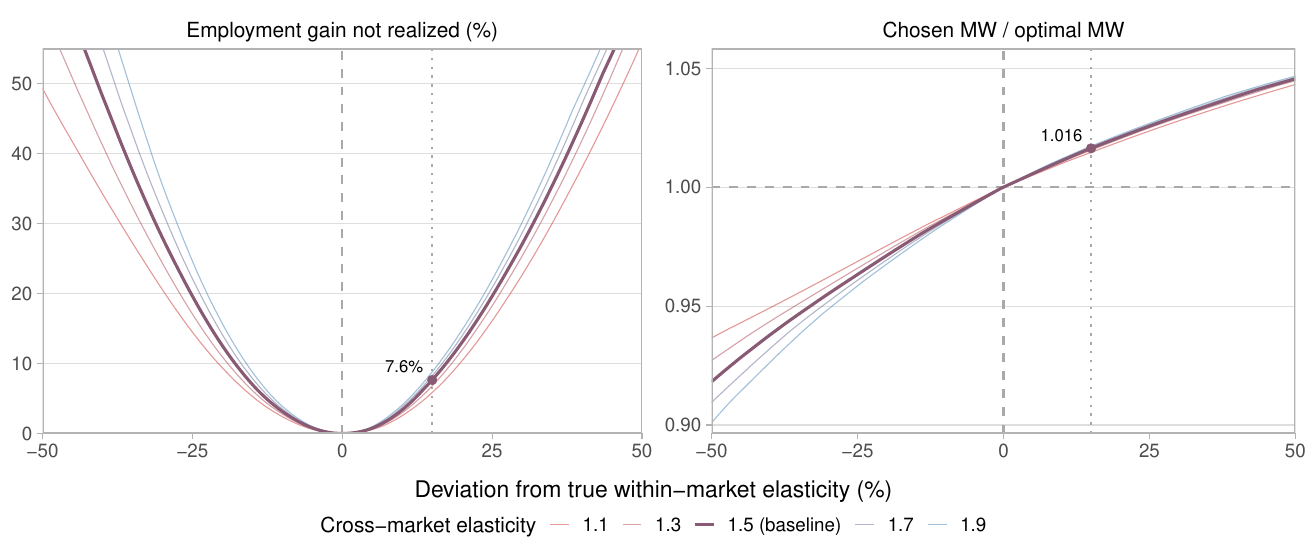}
    \end{subfigure} \\ 
    \vspace{1mm}
    \begin{subfigure}{.99\textwidth}
        \caption{Participation Elasticity (Extensive-Margin $\zeta$)}    
        \includegraphics[width=0.99\linewidth]{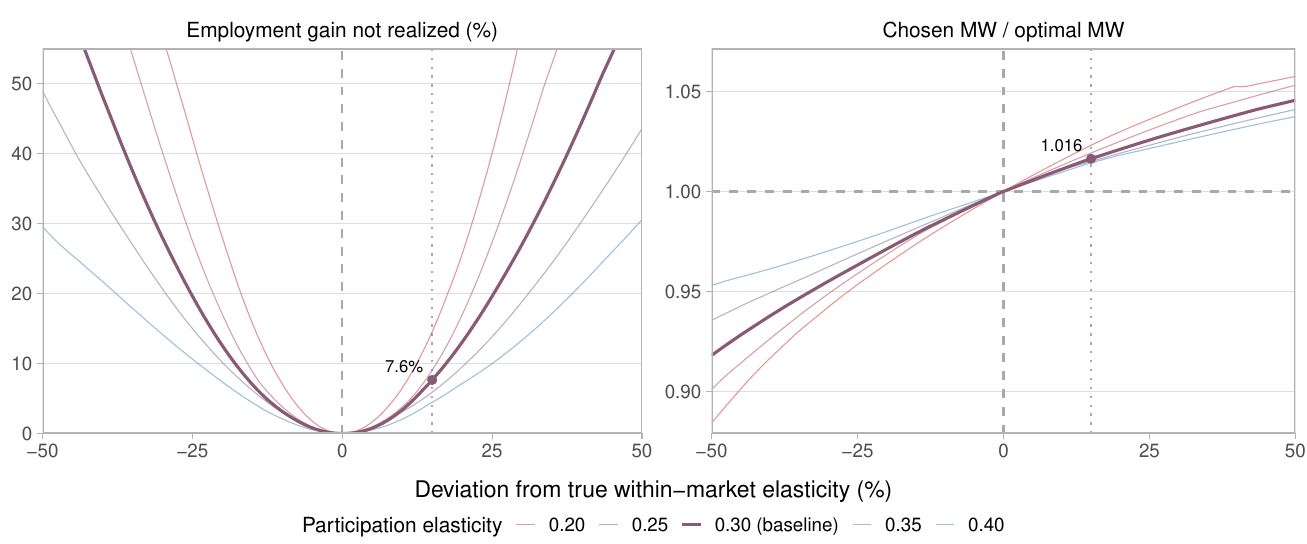}
    \end{subfigure}

    \vspace{2mm}
    \begin{minipage}{1\textwidth} \footnotesize
        Notes:
        The figure shows simulation results from the general equilibrium model calibrated to Portuguese data, for a labor supply elasticity $\eta=5$.
        We re-estimate the baseline economy in Figure \ref{fig:struc_model_results}, either changing the value of the cross-market elasticity $\sigma$, in Panel (a), or the value of the participation elasticity $\zeta$, in Panel (b).
        The horizontal axis shows the deviation from the true elasticity $b$, in percent.
        The dotted line marks a \WelfBiasPct\ percent deviation, and the labeled point shows the baseline economy.
    \end{minipage}
\end{figure}

%% file: manuscript.bbl
\begin{thebibliography}{14}
\newcommand{\enquote}[1]{``#1''}
\expandafter\ifx\csname natexlab\endcsname\relax\def\natexlab#1{#1}\fi

\bibitem[\protect\citeauthoryear{Amodio, Brancati, Brummund, de~Roux, and
  Di~Maio}{Amodio et~al.}{2024}]{AmodioEtAl2024}
\textsc{Amodio, F., E.~Brancati, P.~Brummund, N.~de~Roux, and M.~Di~Maio}
  (2024): \enquote{{Labor Market Power and Self-Employment Around the World},}
  CEPR Discussion Paper 18828, {Centre for Economic Policy Research (CEPR)},
  Paris and London.

\bibitem[\protect\citeauthoryear{Amodio and De~Roux}{Amodio and
  De~Roux}{2024}]{amodio2024measuring}
\textsc{Amodio, F. and N.~De~Roux} (2024): \enquote{Measuring Labor Market
  Power in Developing Countries: Evidence from {Colombian} Plants,}
  \emph{Journal of Labor Economics}, 42, 949--977.

\bibitem[\protect\citeauthoryear{Azkarate-Askasua and
  Zerecero}{Azkarate-Askasua and Zerecero}{2025}]{AzkarateAskasuaZerecero2025}
\textsc{Azkarate-Askasua, M. and M.~Zerecero} (2025): \enquote{Union and Firm
  Labor Market Power,} Working Paper.

\bibitem[\protect\citeauthoryear{Card and Cardoso}{Card and
  Cardoso}{2022}]{CardCardoso2022}
\textsc{Card, D. and A.~R. Cardoso} (2022): \enquote{Wage flexibility under
  sectoral bargaining,} \emph{Journal of the European Economic Association},
  20, 2013--2061.

\bibitem[\protect\citeauthoryear{Cardoso and Portugal}{Cardoso and
  Portugal}{2005}]{cardoso2005contractual}
\textsc{Cardoso, A.~R. and P.~Portugal} (2005): \enquote{Contractual wages and
  the wage cushion under different bargaining settings,} \emph{Journal of Labor
  Economics}, 23, 875--902.

\bibitem[\protect\citeauthoryear{Chetty, Guren, Manoli, and Weber}{Chetty
  et~al.}{2011}]{Chetty2011}
\textsc{Chetty, R., A.~Guren, D.~Manoli, and A.~Weber} (2011): \enquote{Are
  Micro and Macro Labor Supply Elasticities Consistent? A Review of Evidence on
  the Intensive and Extensive Margins,} \emph{American Economic Review}, 101,
  471–75.

\bibitem[\protect\citeauthoryear{Delgado-Prieto}{Delgado-Prieto}{2024}]{delgado2024worker}
\textsc{Delgado-Prieto, L.} (2024): \enquote{Worker Responses to Immigration
  Across Firms: Evidence from Colombia,} Tech. rep., Technical Report.

\bibitem[\protect\citeauthoryear{Garin and Silv{\'e}rio}{Garin and
  Silv{\'e}rio}{2024}]{garin2024responsive}
\textsc{Garin, A. and F.~Silv{\'e}rio} (2024): \enquote{How responsive are
  wages to firm-specific changes in labour demand? Evidence from idiosyncratic
  export demand shocks,} \emph{Review of Economic Studies}, 91, 1671--1710.

\bibitem[\protect\citeauthoryear{Hermo}{Hermo}{2025}]{Hermo2025}
\textsc{Hermo, S.} (2025): \enquote{Collective Bargaining Networks and the
  Propagation of Shocks,} \emph{Working paper}.

\bibitem[\protect\citeauthoryear{Holden}{Holden}{1988}]{Holden1988}
\textsc{Holden, S.} (1988): \enquote{Local and Central Wage Bargaining,}
  \emph{Scandinavian Journal of Economics}, 90, 93--99.

\bibitem[\protect\citeauthoryear{{OECD}}{{OECD}}{2025}]{OECDEarnings2025}
\textsc{{OECD}} (2025): \enquote{Earnings and Wages Database: Minimum relative
  to average wages of full-time workers,}
  \url{https://data-explorer.oecd.org/}, accessed: August 2, 2025.

\bibitem[\protect\citeauthoryear{{OECD and AIAS}}{{OECD and
  AIAS}}{2023}]{OECDVisser2023}
\textsc{{OECD and AIAS}} (2023): \enquote{Institutional Characteristics of
  Trade Unions, Wage Setting, State Intervention and Social Pacts, version
  1.1,} \emph{OECD Publishing, Paris}.

\bibitem[\protect\citeauthoryear{Oliveira}{Oliveira}{2023}]{Oliveira2023}
\textsc{Oliveira, C.} (2023): \enquote{The minimum wage and the wage
  distribution in Portugal,} \emph{Labour Economics}, 85, 102459.

\bibitem[\protect\citeauthoryear{Train}{Train}{2009}]{Train2009}
\textsc{Train, K.~E.} (2009): \emph{Discrete Choice Methods with Simulation},
  Cambridge: Cambridge University Press, 2nd ed.

\end{thebibliography}


\begin{thebibliography}{72}
\newcommand{\enquote}[1]{``#1''}
\expandafter\ifx\csname natexlab\endcsname\relax\def\natexlab#1{#1}\fi

\bibitem[\protect\citeauthoryear{Abowd and Lemieux}{Abowd and
  Lemieux}{1993}]{AbowdLemieux1993}
\textsc{Abowd, J.~A. and T.~Lemieux} (1993): \enquote{The Effects of Product
  Market Competition on Collective Bargaining Agreements: The Case of Foreign
  Competition in Canada,} \emph{Quarterly Journal of Economics}, 108,
  983--1014.

\bibitem[\protect\citeauthoryear{Adamopoulou, Manaresi, Rachedi, and
  Yurdagul}{Adamopoulou et~al.}{forthcoming}]{AdamopoulouEtAlForthcoming}
\textsc{Adamopoulou, E., F.~Manaresi, O.~Rachedi, and E.~Yurdagul}
  (forthcoming): \enquote{Minimum Wages and Insurance within the Firm,}
  \emph{Journal of Labor Economics}.

\bibitem[\protect\citeauthoryear{Addison, Portugal, and
  de~Almeida~Vilares}{Addison et~al.}{2023}]{addison2023union}
\textsc{Addison, J.~T., P.~Portugal, and H.~de~Almeida~Vilares} (2023):
  \enquote{Union membership density and wages: The role of worker, firm, and
  job-title heterogeneity,} \emph{Journal of Econometrics}, 233, 612--632.

\bibitem[\protect\citeauthoryear{Agostinelli, Ferraro, Sorrenti, and
  Treuren}{Agostinelli et~al.}{2025}]{AgostinelliEtAl2025}
\textsc{Agostinelli, F., D.~Ferraro, G.~Sorrenti, and L.~Treuren} (2025):
  \enquote{Employment Relationships, Wage Setting, and Labor Market Power,}
  Working Paper 34439, National Bureau of Economic Research.

\bibitem[\protect\citeauthoryear{Ahlfeldt, Roth, and Seidel}{Ahlfeldt
  et~al.}{2023}]{AhlfeldtEtAl2023}
\textsc{Ahlfeldt, G.~M., D.~Roth, and T.~Seidel} (2023): \enquote{Optimal
  minimum wages in spatial economies,} Tech. rep., CEPR Discussion Paper No.
  DP16913.

\bibitem[\protect\citeauthoryear{Amodio, Brancati, Brummund, de~Roux, and
  Di~Maio}{Amodio et~al.}{2024}]{AmodioEtAl2024}
\textsc{Amodio, F., E.~Brancati, P.~Brummund, N.~de~Roux, and M.~Di~Maio}
  (2024): \enquote{{Labor Market Power and Self-Employment Around the World},}
  CEPR Discussion Paper 18828, {Centre for Economic Policy Research (CEPR)},
  Paris and London.

\bibitem[\protect\citeauthoryear{Amodio, Brancati, de~Roux, and Di~Maio}{Amodio
  et~al.}{2025}]{AmodioEtAl2025}
\textsc{Amodio, F., E.~Brancati, N.~de~Roux, and M.~Di~Maio} (2025):
  \enquote{{Labor Market Institutions and Wage-Setting Power: Evidence from
  Latin America and the Caribbean},} CEPR Discussion Paper 20539.

\bibitem[\protect\citeauthoryear{Andrews, Barnhard, and Carlson}{Andrews
  et~al.}{2026}]{AndrewsBarnhardCarlson2026}
\textsc{Andrews, I., H.~Barnhard, and J.~Carlson} (2026): \enquote{True and
  Pseudo-True Parameters,} \emph{Econometric Theory}, 1–33.

\bibitem[\protect\citeauthoryear{Arnold}{Arnold}{2019}]{Arnold2019}
\textsc{Arnold, D.} (2019): \enquote{Mergers and acquisitions, local labor
  market concentration, and worker outcomes,} Working Paper.

\bibitem[\protect\citeauthoryear{Azar and Marinescu}{Azar and
  Marinescu}{2024}]{AzarMarinescu2024}
\textsc{Azar, J. and I.~Marinescu} (2024): \enquote{Monopsony power in the
  labor market,} in \emph{Handbook of Labor Economics}, Elsevier, vol.~5,
  761--827.

\bibitem[\protect\citeauthoryear{Azkarate-Askasua and
  Zerecero}{Azkarate-Askasua and Zerecero}{2025}]{AzkarateAskasuaZerecero2025}
\textsc{Azkarate-Askasua, M. and M.~Zerecero} (2025): \enquote{Union and Firm
  Labor Market Power,} Working Paper.

\bibitem[\protect\citeauthoryear{Balke and Lamadon}{Balke and
  Lamadon}{2022}]{BalkeLamadon2022}
\textsc{Balke, N. and T.~Lamadon} (2022): \enquote{Productivity Shocks,
  Long-Term Contracts, and Earnings Dynamics,} \emph{American Economic Review},
  112, 2139–77.

\bibitem[\protect\citeauthoryear{Balsvik, Jensen, and Salvanes}{Balsvik
  et~al.}{2015}]{balsvik2015}
\textsc{Balsvik, R., S.~Jensen, and K.~G. Salvanes} (2015): \enquote{{Made in
  China, sold in Norway: Local labor market effects of an import shock},}
  \emph{Journal of Public Economics}, 127, 137--144.

\bibitem[\protect\citeauthoryear{Bassier and Budlender}{Bassier and
  Budlender}{2025}]{BassierBudlender2025}
\textsc{Bassier, I. and J.~Budlender} (2025): \enquote{When do employers share?
  Rent sharing, monopsony and minimum wages,} CEP Discussion Paper 2134.

\bibitem[\protect\citeauthoryear{Berger, Herkenhoff, and Mongey}{Berger
  et~al.}{2022}]{BHM2022}
\textsc{Berger, D., K.~Herkenhoff, and S.~Mongey} (2022): \enquote{Labor market
  power,} \emph{American Economic Review}, 112, 1147--1193.

\bibitem[\protect\citeauthoryear{Berger, Herkenhoff, and Mongey}{Berger
  et~al.}{2025}]{BergerEtAl2025}
---\hspace{-.1pt}---\hspace{-.1pt}--- (2025): \enquote{Minimum Wages,
  Efficiency, and Welfare,} \emph{Econometrica}, 93, 265--301.

\bibitem[\protect\citeauthoryear{Bhuller, Moene, Mogstad, and Vestad}{Bhuller
  et~al.}{2022}]{BhullerEtAl2022}
\textsc{Bhuller, M., K.~O. Moene, M.~Mogstad, and O.~L. Vestad} (2022):
  \enquote{Facts and fantasies about wage setting and collective bargaining,}
  \emph{Journal of Economic Perspectives}, 36, 29--52.

\bibitem[\protect\citeauthoryear{Borusyak, Hull, and Jaravel}{Borusyak
  et~al.}{2022}]{BorusyakEtAl2022}
\textsc{Borusyak, K., P.~Hull, and X.~Jaravel} (2022):
  \enquote{Quasi-experimental shift-share research designs,} \emph{Review of
  Economic Studies}, 89, 181--213.

\bibitem[\protect\citeauthoryear{Bronfenbrenner}{Bronfenbrenner}{1956}]{Bronfenbrenner1956}
\textsc{Bronfenbrenner, M.} (1956): \enquote{Potential Monopsony in Labor
  Markets,} \emph{ILR Review}, 9, 577--588.

\bibitem[\protect\citeauthoryear{Caldwell, Dube, and Naidu}{Caldwell
  et~al.}{forthcoming}]{CaldwellDubeNaidu2026}
\textsc{Caldwell, S., A.~Dube, and S.~Naidu} (forthcoming): \enquote{Monopsony
  Makes It Big,} \emph{Journal of Economic Literature}.

\bibitem[\protect\citeauthoryear{Card and Cardoso}{Card and
  Cardoso}{2022}]{CardCardoso2022}
\textsc{Card, D. and A.~R. Cardoso} (2022): \enquote{Wage flexibility under
  sectoral bargaining,} \emph{Journal of the European Economic Association},
  20, 2013--2061.

\bibitem[\protect\citeauthoryear{Card, Cardoso, Heining, and Kline}{Card
  et~al.}{2018}]{CCHK2018}
\textsc{Card, D., A.~R. Cardoso, J.~Heining, and P.~Kline} (2018):
  \enquote{Firms and labor market inequality: Evidence and some theory,}
  \emph{Journal of Labor Economics}, 36, S13--S70.

\bibitem[\protect\citeauthoryear{Card, Devicienti, and Maida}{Card
  et~al.}{2014}]{CardDevicientiMaida2013}
\textsc{Card, D., F.~Devicienti, and A.~Maida} (2014): \enquote{Rent-sharing,
  Holdup, and Wages: Evidence from Matched Panel Data,} \emph{Review of
  Economic Studies}, 81, 84--111.

\bibitem[\protect\citeauthoryear{Cardoso and Portugal}{Cardoso and
  Portugal}{2005}]{cardoso2005contractual}
\textsc{Cardoso, A.~R. and P.~Portugal} (2005): \enquote{Contractual wages and
  the wage cushion under different bargaining settings,} \emph{Journal of Labor
  Economics}, 23, 875--902.

\bibitem[\protect\citeauthoryear{Chan, Mattana, Salgado, and Xu}{Chan
  et~al.}{2025}]{ChanMattanaSalgadoXu2023}
\textsc{Chan, M., E.~Mattana, S.~Salgado, and M.~Xu} (2025): \enquote{{Dynamic
  Wage Setting: The Role of Monopsony Power and Employment Costs},} Working
  paper.

\bibitem[\protect\citeauthoryear{Chetty}{Chetty}{2009}]{Chetty2009}
\textsc{Chetty, R.} (2009): \enquote{Sufficient Statistics for Welfare
  Analysis: A Bridge Between Structural and Reduced-Form Methods,} \emph{Annual
  Review of Economics}, 1, 451--488.

\bibitem[\protect\citeauthoryear{Deb, Eeckhout, Patel, and Warren}{Deb
  et~al.}{2024}]{DebEtAl2024}
\textsc{Deb, S., J.~Eeckhout, A.~Patel, and L.~Warren} (2024):
  \enquote{Walras–Bowley Lecture: Market Power and Wage Inequality,}
  \emph{Econometrica}, 92, 603--636.

\bibitem[\protect\citeauthoryear{Dhyne, Kikkawa, Komatsu, Mogstad, and
  Tintelnot}{Dhyne et~al.}{2025}]{DhyneEtAl2024}
\textsc{Dhyne, E., A.~K. Kikkawa, T.~Komatsu, M.~Mogstad, and F.~Tintelnot}
  (2025): \enquote{Firm Responses and Wage Effects of Foreign Demand Shocks
  with Fixed Labor Costs and Monopsony,} \emph{American Economic Review}, 115,
  4328–68.

\bibitem[\protect\citeauthoryear{Diegmann, Muller, and Schoefer}{Diegmann
  et~al.}{2026}]{DiegmannAl2026}
\textsc{Diegmann, A., S.~Muller, and B.~Schoefer} (2026): \enquote{Off the
  Labor Supply Curve: The Zero Employer Size Wage Effect Within Large Firms,}
  Tech. rep., NBER Working Paper No. 35265.

\bibitem[\protect\citeauthoryear{Dodini, Stansbury, and Willén}{Dodini
  et~al.}{2026}]{dodini2026}
\textsc{Dodini, S., A.~Stansbury, and A.~Willén} (2026): \enquote{{Who Pays
  for Unions?}} \emph{Quarterly Journal of Economics}, 141, 3095--3159.

\bibitem[\protect\citeauthoryear{Dube, Giuliano, and Leonard}{Dube
  et~al.}{2019}]{DubeEtAl2019}
\textsc{Dube, A., L.~Giuliano, and J.~Leonard} (2019): \enquote{Fairness and
  Frictions: The Impact of Unequal Raises on Quit Behavior,} \emph{American
  Economic Review}, 109, 620–63.

\bibitem[\protect\citeauthoryear{Dube and Lindner}{Dube and
  Lindner}{2024}]{DubeLindner2024}
\textsc{Dube, A. and A.~Lindner} (2024): \enquote{Minimum wages in the 21st
  century,} \emph{Handbook of Labor Economics}, 5, 261--383.

\bibitem[\protect\citeauthoryear{Faia, Lochner, and Schoefer}{Faia
  et~al.}{2026}]{FaiaAl2026}
\textsc{Faia, E., B.~Lochner, and B.~Schoefer} (2026): \enquote{Monopsony,
  Markdowns, and Minimum Wages,} Tech. rep., NBER Working Paper No. 34699.

\bibitem[\protect\citeauthoryear{Garin and Silv{\'e}rio}{Garin and
  Silv{\'e}rio}{2024}]{garin2024responsive}
\textsc{Garin, A. and F.~Silv{\'e}rio} (2024): \enquote{How responsive are
  wages to firm-specific changes in labour demand? Evidence from idiosyncratic
  export demand shocks,} \emph{Review of Economic Studies}, 91, 1671--1710.

\bibitem[\protect\citeauthoryear{Gaulier and Zignago}{Gaulier and
  Zignago}{2010}]{gaulier2010baci}
\textsc{Gaulier, G. and S.~Zignago} (2010): \enquote{BACI: International Trade
  Database at the Product-Level. The 1994- 2007 Version,} Working Papers
  2010-23, CEPII.

\bibitem[\protect\citeauthoryear{Grigsby, Hurst, and Yildirmaz}{Grigsby
  et~al.}{2021}]{GrigsbyEtAl2021}
\textsc{Grigsby, J., E.~Hurst, and A.~Yildirmaz} (2021): \enquote{Aggregate
  Nominal Wage Adjustments: New Evidence from Administrative Payroll Data,}
  \emph{American Economic Review}, 111, 428–71.

\bibitem[\protect\citeauthoryear{Guiso, Pistaferri, and Schivardi}{Guiso
  et~al.}{2005}]{Guiso2005}
\textsc{Guiso, L., L.~Pistaferri, and F.~Schivardi} (2005): \enquote{{Insurance
  within the Firm},} \emph{Journal of Political Economy}, 113, 1054--1087.

\bibitem[\protect\citeauthoryear{G{\"u}rtzgen}{G{\"u}rtzgen}{2009}]{Gurtzgen2009}
\textsc{G{\"u}rtzgen, N.} (2009): \enquote{Rent-Sharing and Collective
  Bargaining Coverage: Evidence from Linked Employer--Employee Data,}
  \emph{Scandinavian Journal of Economics}, 111, 323--349.

\bibitem[\protect\citeauthoryear{Hazell, Patterson, Sarsons, and Taska}{Hazell
  et~al.}{forthcoming}]{HazellEtAl2025}
\textsc{Hazell, J., C.~Patterson, H.~Sarsons, and B.~Taska} (forthcoming):
  \enquote{National Wage Setting,} \emph{American Economic Review}.

\bibitem[\protect\citeauthoryear{Hermo}{Hermo}{2025}]{Hermo2025}
\textsc{Hermo, S.} (2025): \enquote{Collective Bargaining Networks and the
  Propagation of Shocks,} \emph{Working paper}.

\bibitem[\protect\citeauthoryear{Holden}{Holden}{1988}]{Holden1988}
\textsc{Holden, S.} (1988): \enquote{Local and Central Wage Bargaining,}
  \emph{Scandinavian Journal of Economics}, 90, 93--99.

\bibitem[\protect\citeauthoryear{Holmlund and Zetterberg}{Holmlund and
  Zetterberg}{1991}]{HolmlundZetterberg1991}
\textsc{Holmlund, B. and J.~Zetterberg} (1991): \enquote{{Insider Effects in
  Wage Determination: Evidence from Five Countries},} \emph{European Economic
  Review}, 35, 1009--1034.

\bibitem[\protect\citeauthoryear{Hummels, J{\o}rgensen, Munch, and
  Xiang}{Hummels et~al.}{2014}]{hjmx2014}
\textsc{Hummels, D., R.~J{\o}rgensen, J.~Munch, and C.~Xiang} (2014):
  \enquote{The wage effects of offshoring: Evidence from Danish matched
  worker-firm data,} \emph{American Economic Review}, 104, 1597--1629.

\bibitem[\protect\citeauthoryear{J{\"a}ger, Naidu, and Schoefer}{J{\"a}ger
  et~al.}{2025}]{JagerEtAl2024}
\textsc{J{\"a}ger, S., S.~Naidu, and B.~Schoefer} (2025): \enquote{{Collective
  Bargaining, Unions, and the Wage Structure: An International Perspective},}
  \emph{Handbook of Labor Economics}, 6, 229--372.

\bibitem[\protect\citeauthoryear{J{\"a}ger, Schoefer, Young, and
  Zweimüller}{J{\"a}ger et~al.}{2020}]{JagerEtAl2020}
\textsc{J{\"a}ger, S., B.~Schoefer, S.~Young, and J.~Zweimüller} (2020):
  \enquote{Wages and the Value of Nonemployment,} \emph{Quarterly Journal of
  Economics}, 135, 1905--1963.

\bibitem[\protect\citeauthoryear{J{\'a}{\~n}ez and
  Delgado-Prieto}{J{\'a}{\~n}ez and
  Delgado-Prieto}{2024}]{JaniezDelgadoPrieto2024}
\textsc{J{\'a}{\~n}ez, {\'A}. and L.~Delgado-Prieto} (2024): \enquote{Monopsony
  Power and Firm Organization,} Working Paper.

\bibitem[\protect\citeauthoryear{Kline}{Kline}{2024}]{Kline2024}
\textsc{Kline, P.} (2024): \enquote{{Firm Wage Effects},} \emph{Handbook of
  Labor Economics}, 5, 115--181.

\bibitem[\protect\citeauthoryear{Kline}{Kline}{2025}]{Kline2025}
---\hspace{-.1pt}---\hspace{-.1pt}--- (2025): \enquote{{Labor Market Monopsony:
  Fundamentals and Frontiers},} \emph{Handbook of Labor Economics}, 6,
  655--728.

\bibitem[\protect\citeauthoryear{Kline, Petkova, Williams, and Zidar}{Kline
  et~al.}{2019}]{KlineEtAl2019}
\textsc{Kline, P., N.~Petkova, H.~Williams, and O.~Zidar} (2019): \enquote{{Who
  Profits from Patents? Rent-Sharing at Innovative Firms},} \emph{Quarterly
  Journal of Economics}, 134, 1343--1404.

\bibitem[\protect\citeauthoryear{Kroft, Luo, Mogstad, and Setzler}{Kroft
  et~al.}{2025}]{KroftEtAl2025}
\textsc{Kroft, K., Y.~Luo, M.~Mogstad, and B.~Setzler} (2025):
  \enquote{Imperfect Competition and Rents in Labor and Product Markets: The
  Case of the Construction Industry,} \emph{American Economic Review}, 115,
  2926–69.

\bibitem[\protect\citeauthoryear{Lamadon, Mogstad, and Setzler}{Lamadon
  et~al.}{2022}]{LMS2022}
\textsc{Lamadon, T., M.~Mogstad, and B.~Setzler} (2022): \enquote{Imperfect
  competition, compensating differentials, and rent sharing in the US labor
  market,} \emph{American Economic Review}, 112, 169--212.

\bibitem[\protect\citeauthoryear{Manning}{Manning}{1987}]{Manning1987}
\textsc{Manning, A.} (1987): \enquote{An Integration of Trade Union Models in a
  Sequential Bargaining Framework,} \emph{The Economic Journal}, 97, 121--139.

\bibitem[\protect\citeauthoryear{Manning}{Manning}{2003}]{Manning2003Ch12}
---\hspace{-.1pt}---\hspace{-.1pt}--- (2003): \enquote{The Minimum Wage and
  Trade Unions,} in \emph{Monopsony in Motion: Imperfect Competition in Labor
  Markets}, Princeton University Press, 325--359.

\bibitem[\protect\citeauthoryear{Manning}{Manning}{2021}]{Manning2021}
---\hspace{-.1pt}---\hspace{-.1pt}--- (2021): \enquote{Monopsony in Labor
  Markets: A Review,} \emph{ILR Review}, 74, 3--26.

\bibitem[\protect\citeauthoryear{Manski}{Manski}{2021}]{Manski2021}
\textsc{Manski, C.~F.} (2021): \enquote{Econometrics for Decision Making:
  Building Foundations Sketched by {Haavelmo} and {Wald},} \emph{Econometrica},
  89, 2827--2853.

\bibitem[\protect\citeauthoryear{McDonald and Solow}{McDonald and
  Solow}{1981}]{McDonaldSolow1981}
\textsc{McDonald, I.~M. and R.~M. Solow} (1981): \enquote{Wage Bargaining and
  Employment,} \emph{The American Economic Review}, 71, 896--908.

\bibitem[\protect\citeauthoryear{Moene}{Moene}{1988}]{Moene1988}
\textsc{Moene, K.~O.} (1988): \enquote{Union's Threats and Wage Determination,}
  \emph{The Economic Journal}, 98, 471--483.

\bibitem[\protect\citeauthoryear{Moene and Wallerstein}{Moene and
  Wallerstein}{1997}]{MoeneWallerstein1997}
\textsc{Moene, K.~O. and M.~Wallerstein} (1997): \enquote{Pay Inequality,}
  \emph{Journal of Labor Economics}, 15, 403--430.

\bibitem[\protect\citeauthoryear{OECD}{OECD}{2019}]{OECD2019collective}
\textsc{OECD} (2019): \enquote{Collective bargaining systems and worker voice
  arrangements in OECD countries,} \emph{Negotiating Our Way Up}, 22.

\bibitem[\protect\citeauthoryear{{OECD and AIAS}}{{OECD and
  AIAS}}{2023}]{OECDVisser2023}
\textsc{{OECD and AIAS}} (2023): \enquote{Institutional Characteristics of
  Trade Unions, Wage Setting, State Intervention and Social Pacts, version
  1.1,} \emph{OECD Publishing, Paris}.

\bibitem[\protect\citeauthoryear{Olden and Møen}{Olden and
  Møen}{2022}]{OldenMoen2022}
\textsc{Olden, A. and J.~Møen} (2022): \enquote{The triple difference
  estimator,} \emph{The Econometrics Journal}, 25, 531--553.

\bibitem[\protect\citeauthoryear{Olsson}{Olsson}{2024}]{Olsson2024}
\textsc{Olsson, M.} (2024): \enquote{Labor cost adjustments during the Great
  Recession: Wages, separations, and labor market frictions,} \emph{Available
  at SSRN}.

\bibitem[\protect\citeauthoryear{Prager and Schmitt}{Prager and
  Schmitt}{2021}]{PragerSchmitt2021}
\textsc{Prager, E. and M.~Schmitt} (2021): \enquote{Employer consolidation and
  wages: Evidence from hospitals,} \emph{American Economic Review}, 111,
  397--427.

\bibitem[\protect\citeauthoryear{Robinson}{Robinson}{1933}]{robinson1933economics}
\textsc{Robinson, J.} (1933): \emph{The Economics of Imperfect Competition},
  London: Macmillan.

\bibitem[\protect\citeauthoryear{Rusinek and Rycx}{Rusinek and
  Rycx}{2013}]{RusinekRycx2013}
\textsc{Rusinek, M. and F.~Rycx} (2013): \enquote{Rent-Sharing under Different
  Bargaining Regimes: Evidence from Linked Employer–Employee Data,}
  \emph{British Journal of Industrial Relations}, 51, 28--58.

\bibitem[\protect\citeauthoryear{Sokolova and Sorensen}{Sokolova and
  Sorensen}{2021}]{SokolovaSorensen2021}
\textsc{Sokolova, A. and T.~Sorensen} (2021): \enquote{Monopsony in Labor
  Markets: A Meta-Analysis,} \emph{ILR Review}, 74, 27--55.

\bibitem[\protect\citeauthoryear{Tortarolo and Zarate}{Tortarolo and
  Zarate}{2018}]{TortaroloZarate2018}
\textsc{Tortarolo, D. and R.~D. Zarate} (2018): \enquote{Measuring Imperfect
  Competition in Product and Labor Markets. An Empirical Analysis using
  Firm-level Production Data,} Working paper 2018/03, CAF.

\bibitem[\protect\citeauthoryear{Van~Reenen}{Van~Reenen}{1996}]{VanReenen1996}
\textsc{Van~Reenen, J.} (1996): \enquote{{The Creation and Capture of Rents:
  Wages and Innovation in a Panel of U. K. Companies},} \emph{Quarterly Journal
  of Economics}, 111, 195--226.

\bibitem[\protect\citeauthoryear{Vera}{Vera}{2022}]{Vera2022}
\textsc{Vera, M.~D.} (2022): \enquote{Firm-level productivity and demand shocks
  in imperfectly competitive labor markets: implications for wage dynamics,}
  Working Paper.

\bibitem[\protect\citeauthoryear{Volpe}{Volpe}{2026}]{Volpe2026}
\textsc{Volpe, O.} (2026): \enquote{Job Preferences, Labor Market Power, and
  Inequality,} Working Paper.

\bibitem[\protect\citeauthoryear{Wong}{Wong}{2026}]{Wong2025}
\textsc{Wong, H.~C.} (2026): \enquote{Understanding High-Wage Firms: Monopoly,
  Monopsony, and Bargaining Power,} \emph{American Economic Review}, 116,
  2504–41.

\bibitem[\protect\citeauthoryear{Yeh, Macaluso, and Hershbein}{Yeh
  et~al.}{2022}]{YehMacalusoHershbein2022}
\textsc{Yeh, C., C.~Macaluso, and B.~Hershbein} (2022): \enquote{Monopsony in
  the US Labor Market,} \emph{American Economic Review}, 112, 2099--2138.

\end{thebibliography}
